\documentclass[]{tADP2e}
\usepackage{longtable}

\newcommand{\dif}{\mathrm{d}}%
\renewcommand{\Eins}{\mathbf{1}}%
\newcommand{\fdif}{\operatorname{\delta}}
\newcommand{\Fdif}[2]{\frac{\fdif\!#1}{\fdif\!#2}}
\newcommand{\ii}{i}%
\newcommand{\rw}{\mathbf{\tilde{r}}}
\newcommand{\wn}{k}

\newcommand{\Nabla}{\nabla_{\vec{r}}}%
\newcommand{\NablaS}{\nabla_{\vec{r}'}}%
\newcommand{\NablaW}{\nabla_{\rw}}%
\newcommand{\NablaI}[1]{\nabla_{\vec{r}_{#1}}}%
\newcommand{\pdif}[2]{\frac{\partial#1}{\partial#2}}%
\newcommand{\trace}{\operatorname{tr}}%
\newcommand{\etal}{et\,al.}%
\newcommand{\ie}{i.\,e.}%
\newcommand{\eg}{e.\,g.}%
\newcommand{\etc}{etc.}%
\newcommand{\vs}{vs.\ }%
\newcommand{\abs}[1]{\lvert#1\rvert}%
\newcommand{\norm}[1]{\lvert#1\rvert}%
\newcommand{\kint}{\int\!\!\!\:\!\dif\mathbf{\wn}\,}%
\newcommand{\vint}{\int\!\!\!\:\!\dif\mathbf{r}\,}%
\newcommand{\vints}{\int\!\!\!\:\!\dif\mathbf{r}'\:\!}%
\newcommand{\vintI}[1]{\int\!\!\!\:\!\dif\mathbf{r}_{#1}\,}%
\newcommand{\vintn}[1]{\int\!\!\!\:\!\dif\mathbf{r}_{#1}\,}%
\newcommand{\vintw}{\int\!\!\!\:\!\dif\rw}
\newcommand{\uu}{\mathbf{\hat{u}}}%
\newcommand{\pp}{\mathbf{\hat{p}}}%
\newcommand{\nn}{\mathbf{\hat{n}}}%
\newcommand{\uint}{\int\!\!\!\:\!\dif\uu\,}%
\newcommand{\uintn}[1]{\int\!\!\!\:\!\dif\uu_{#1}\,}%
\newcommand{\ZT}[1]{\textquotedblleft#1\textquotedblright}%
\newcommand{\kB}{k_{\mathrm{B}}}%
\newcommand{\kBT}{k_{\mathrm{B}}T}%

\newcommand{\Drho}{D_{\mathrm{T}}}
\newcommand{\rhoM}{\rho_{0}}
\newcommand{\rhoR}{\rho_{\mathrm{ref}}}
\newcommand{\rhoq}{\rhoR}
\newcommand{\rhoL}{\rhoM}
\newcommand{\rhoN}{\rhoM}
\newcommand{\rhoqA}{\bar{\rho}_{\mathrm{A}}}
\newcommand{\rhoqB}{\bar{\rho}_{\mathrm{B}}}
\newcommand{\nnq}{\rhoM}
\newcommand{\psiSH}{\tilde{\psi}}
\newcommand{\psiPF}{\psi}
\newcommand{\PsiW}{\tilde{\Psi}}%
\newcommand{\FSH}{\tilde{\mathcal{F}}}
\newcommand{\Fq}{\bar{\mathcal{F}}}
\newcommand{\gw}{\tilde{g}}
\newcommand{\hw}{\tilde{h}}
\newcommand{\gcw}{\tilde{g}_{\mathrm{c}}}
\newcommand{\Fho}{\mathcal{F}_{\mathrm{ho}}}
\newcommand{\Us}{U_{\mathrm{s}}}
\newcommand{\cII}{c^{(2)}}
\newcommand{\cIIw}{\tilde{c}^{(2)}}
\newcommand{\ipd}{R_{\mathrm{p}}}
\newcommand{\KDN}{\hat{\psi}}%
\newcommand{\CRP}[9]{(Reproduced with permission from #1, \textit{#2}, #3 #4 (#5), #6, no.\ #7, DOI: #8 $\copyright$ #5 by #9.)}%
\newcommand{\CRPE}[9]{(Reproduced from #1, \textit{#2}, #3 #4 (#5), #6, no.\ #7, DOI: #8 $\copyright$ #5 by #9.)}%

\renewcommand{\vec}{\mathbf}

\begin{document}

\doi{10.1080/0001873YYxxxxxxxx}
\issn{1460-6976}
\issnp{0001-8732}  \jvol{00} \jnum{00} \jyear{2012} \jmonth{XXX}

\markboth{Taylor \& Francis and I. T. Consultant}{Advances in Physics}

\title{Phase-field-crystal models for condensed matter dynamics on atomic length and diffusive time scales: an overview}

\author{Heike Emmerich$^{\mathrm{a}\ast}$\thanks{$^\ast$Corresponding authors. 
Emails: heike.emmerich@uni-bayreuth.de, hlowen@thphy.uni-duesseldorf.de, and grana@szfki.hu
\vspace{6pt}}, Hartmut L\"owen$^{\mathrm{b}\ast}$, Raphael Wittkowski$^{\rm b}$, Thomas Gruhn$^{\rm a}$,
\\Gyula I. T\'oth$^{\rm c}$, Gy\"orgy Tegze$^{\rm c}$, and L\'aszl\'o Gr\'an\'asy$^{\mathrm{c,d}\ast}$\\
$^{\rm a}$\textit{Lehrstuhl f\"ur Material- und Prozesssimulation, Universit\"at Bayreuth, D-95440 Bayreuth, Germany},\\
$^{\rm b}$\textit{Institut f\"ur Theoretische Physik II, Weiche Materie, Heinrich-Heine-Universit\"at D\"usseldorf, D-40225 D\"usseldorf, Germany},\\
$^{\rm c}$\textit{Institute for Solid State Physics and Optics, Wigner Research Centre for Physics, \\
P.O. Box 49, H-1525 Budapest, Hungary},\\
$^{\rm d}$\textit{BCAST, Brunel University, Uxbridge, Middlesex, UB8 3PH, U.K.}}

\maketitle

\begin{abstract}
Here, we review the basic concepts and applications of the \textit{phase-field-crystal} (PFC) method, which is one of the latest 
simulation methodologies in materials science for problems, where atomic- and microscales are tightly coupled. 
The PFC method operates on atomic length and diffusive time scales, and thus constitutes a computationally efficient alternative 
to molecular simulation methods. Its intense development in materials science started fairly recently following the work by 
Elder \etal\ [Phys.\ Rev.\ Lett.\ 88 (2002), p.\ 245701]. Since these initial studies, dynamical density functional theory and 
thermodynamic concepts have been linked to the PFC approach to serve as further theoretical fundaments for the latter.  
In this review, we summarize these methodological development steps as well as the most important applications of the 
PFC method with a special focus on the interaction of development steps taken in hard and soft matter physics, 
respectively. Doing so, we hope to present today's state of the art in PFC modelling as well as the potential, which might still arise 
from this method in physics and materials science in the nearby future.
\bigskip

\begin{keywords} 
phase-field-crystal (PFC) models, static and dynamical density functional theory (DFT and DDFT), 
condensed matter dynamics of liquid crystals and polymers, nucleation and pattern formation, simulations in materials science, 
colloidal crystal growth and growth anisotropy 
\end{keywords}
\bigskip

\newpage
\centerline{\bfseries Table of contents}\medskip

\hbox to \textwidth{\hsize\textwidth\vbox{\hsize18pc
\hspace*{-12pt} {1.}    Introduction\\
{2.}    From density functional theory\\
\hspace*{6pt}    to phase-field-crystal models\\
\hspace*{10pt}{2.1.} Density functional theory\\
\hspace*{10pt}{2.2.} Dynamical density functional theory\\
\hspace*{24pt} {2.2.1.}   Basic equations\\
\hspace*{24pt} {2.2.2.}   Brownian dynamics: Langevin\\
\hspace*{47pt} and Smoluchowski picture\\
\hspace*{24pt} {2.2.3.}   Derivation of DDFT\\
\hspace*{24pt} {2.2.4.}   Application of DDFT to\\
\hspace*{47pt} colloidal crystal growth\\
\hspace*{10pt}{2.3.} Derivation of the PFC model\\
\hspace*{24pt}  for isotropic particles from DFT\\
\hspace*{24pt} {2.3.1.}   Free-energy functional\\
\hspace*{24pt} {2.3.2.}   Dynamical equations\\
\hspace*{24pt} {2.3.3.}   Colloidal crystal growth:\\
\hspace*{47pt} DDFT versus PFC modelling\\
{3.}    Phase-field-crystal modelling\\
\hspace*{6pt}    in condensed matter physics\\
\hspace*{10pt}{3.1.}   The original PFC model and its\\
\hspace*{24pt}    generalisations\\
\hspace*{24pt} {3.1.1.}   Single-component PFC models\\
\hspace*{24pt} {3.1.2.}   Binary PFC models\\
\hspace*{24pt} {3.1.3.}   PFC models for liquid crystals\\
\hspace*{24pt} {3.1.4.}   Numerical methods\\
\hspace*{24pt} {3.1.5.}   Coarse-graining the PFC\\ 
\hspace*{47pt} models\\
\hspace*{10pt}{3.2.}   Phase diagrams the PFC models\\
 \hspace*{24pt} realize\\
\hspace*{24pt} {3.2.1.}    Phase diagram of single-compo-\\
\hspace*{47pt}   nent and binary systems\\ 
\hspace*{24pt} {3.2.2.}    Phase diagram of two-\\
\hspace*{47pt}   dimensional liquid crystals\\
\hspace*{10pt}{3.3.}   Anisotropies in the PFC models\\
\hspace*{24pt} {3.3.1.}    Free energy of the liquid-solid\\
\hspace*{47pt}    interface\\
\hspace*{24pt} {3.3.2.}    Growth anisotropy\\
\hspace*{10pt}{3.4.}   Glass formation\\
\hspace*{10pt}{3.5.}   Phase-field-crystal modelling of foams\\
\hspace*{10pt}{3.6.}   Coupling to hydrodynamics\\
}
\hspace{-24pt}\vbox{\noindent\hsize18pc
{4.}    Phase-field-crystal models applied\\
\hspace*{6pt}    to nucleation and pattern formation\\
\hspace*{6pt}    in metals\\
\hspace*{10pt}{4.1.}    Properties of nuclei from\\
\hspace*{24pt}    extremum principles\\
\hspace*{24pt} {4.1.1.}    Homogeneous nucleation\\
\hspace*{24pt} {4.1.2.}    Heterogeneous nucleation\\
\hspace*{10pt}{4.2.}    Pattern formation\\
\hspace*{24pt} {4.2.1.}    PFC modelling of surface\\ 
\hspace*{47pt}   patterns\\
\hspace*{24pt} {4.2.2.}    Pattern formation in binary\\ 
\hspace*{47pt}   solidification\\
\hspace*{10pt}{4.3.}    Phenomena in the solid state\\
\hspace*{24pt} {4.3.1.}    Dislocation dynamics and grain\\ 
\hspace*{47pt}   boundary melting\\
\hspace*{24pt} {4.3.2.}    Crack formation and\\ 
\hspace*{47pt}   propagation\\ 
\hspace*{24pt} {4.3.3.}    Strain-induced morphologies\\ 
\hspace*{24pt} {4.3.4.}    Kirkendall effect\\ 
\hspace*{24pt} {4.3.5.}    Density/solute trapping\\ 
\hspace*{24pt} {4.3.6.}    Vacancy/atom transport in the\\ 
\hspace*{47pt}   VPFC model\\
{5.}    Phase-field-crystal modelling\\
\hspace*{6pt}    in soft matter physics\\
\hspace*{10pt}{5.1.}   Application to colloids\\
\hspace*{24pt} {5.1.1.}    Nucleation in colloidal\\ 
\hspace*{47pt}    crystal aggregation\\
\hspace*{24pt} {5.1.2.}    Pattern formation in colloidal\\ 
\hspace*{47pt}    crystal aggregation\\
\hspace*{24pt} {5.1.3.}    Colloid patterning\\
\hspace*{10pt}{5.2.}   Application to polymers\\
\hspace*{24pt} {5.2.1.}    Self-consistent field theory\\ 
\hspace*{47pt}    for polymer systems\\
\hspace*{24pt} {5.2.2.}    Phase-field methods\\ 
\hspace*{47pt}    for polymer systems\\
\hspace*{24pt} {5.2.3.}    Comparison of the methods\\ 
\hspace*{10pt}{5.3.}   Application to liquid crystals\\
{6.}    Summary and outlook\\
Appendix \\
Acknowledgements\\
References\\
List of abbreviations\\
}}
\end{abstract}

\newpage
\section{Introduction}
\begin{figure}
\begin{center}%
\includegraphics[height=0.3\linewidth]{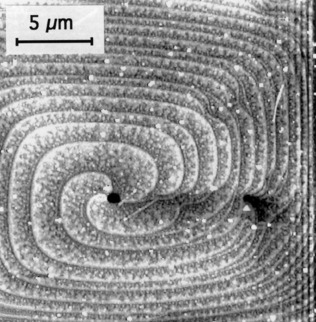}
\includegraphics[height=0.3\linewidth]{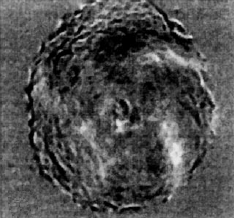}
\includegraphics[height=0.3\linewidth]{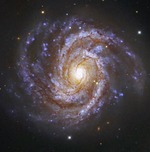}
\end{center}%
\caption{\label{fig:Galaxy}Pattern formation on microscopic to cosmic length scales. From left to right: multiple 
spiralling nanoscale terraces starting from a central heterogeneity. 
\nocite{Klemenz1998}\CRP{C.\ Klemenz}{Hollow cores and step-bunching effects on (001) YBCO surfaces grown by liquid-phase epitaxy}{J.\ Cryst.\ Growth}{187}{1998}{221-227}{2}{10.1016/S0022-0248(97)00866-X}{Elsevier}  
Cellular slime mould self-organized into a five-arm spiral structure.
\nocite{VasievSW1997}\CRP{B.\ Vasiev, F.\ Siegert, and C.\ Weijer}{Multiarmed spirals in excitable media}{Phys.\ Rev.\ Lett.\ }{78}{1997}{2489-2492}{12}{10.1103/PhysRevLett.78.2489}{the American Physical Society}
Messier $100$, a multi-arm spiral galaxy in the Virgo Supercluster, $60$ million light-years from earth. 
(Credit: ESO/IDA/Danish 1.5m/R.\ Gendler, J.-E.\ Ovaldsen, C.\ C.\ Th{\"o}ne, and C.\ F{\'e}ron.)}
\end{figure}

Pattern formation has been observed in complex systems from microscopic to cosmic scales (for examples, see figure \ref{fig:Galaxy}), 
a phenomenon that has been exciting the fantasy of humanity for a long time. Nonequilibrium systems in physics, 
chemistry, biology, mathematics, cosmology, and other fields produce an amazingly rich and visually fascinating 
variety of spatiotemporal behaviour. Experiments and simulations show that many of such  systems -- reacting 
chemicals, bacteria colonies, granular matter, plasmas -- often display analogous dynamical behaviour. The wish 
to find the origin of the common behaviour has been driving the efforts for finding unifying schemes that allow 
the assigning of many of these processes into a few universality classes. Pattern formation and the associated 
nonlinear dynamics have received a continuous attention of the statistical physics community over the past decades. 
Reviews of the advances made in different directions are available in the literature and range from early works 
on critical dynamics \cite{HohenbergH1977} via phase-separation \cite{gunton1983} and pattern formation in 
nonequilibrium systems \cite{CrossH1993,Seul1995} to recent detailed treatments of the field in books 
(\eg, references \cite{ChaikinL1995,pismen2006,cross2010}). 

\enlargethispage{8pt}
In particular, Seul and Andelman \cite{Seul1995} described pattern formation 
on the mesoscale as manifestation of modulated structures. Within this approach, 
the modulated phases are stabilized by competing attractive and repulsive interactions,  
which favour inhomogeneities characterized by a certain modulation length scale. The 
modulations are described by a single scalar order parameter. As outlined in reference \cite{Seul1995}, 
the idea of Seul and Andelman can be applied to a large variety of systems ranging 
from Langmuir films over semiconductor surfaces and magnet garnets to polyelectrolyte
solutions. Furthermore, the pioneering theories of spontaneous domain formation in 
magnetic materials and in the intermediate state of type I superconductors has been  
reinterpreted within this framework.

In the past decade, special attention has been paid to a similar model, whose mathematical formulation has 
been laid down decades earlier to address hydrodynamic instabilities \cite{PhysRevA.15.319} and to describe 
the transition to the antiferromagnetic state in liquid $^3$He or to a non-uniform state in cholesteric liquid crystals 
\cite{Brazowskii1975}, whereas recently it has been employed for the modelling of crystallization in undercooled liquids 
on the atomic scale \cite{ElderKHG2002}. This approach is known to the materials science community 
as the phase-field-crystal (PFC) model \cite{ElderKHG2002}, and has proven to be an amazingly 
efficient tool for addressing crystalline self-organization/pattern formation on the \textit{atomistic} scale. 

The PFC approach attracts attention owing to a unique situation: the crystallization of liquids is 
traditionally addressed on this scale by the density functional theory (DFT) \cite{Oxtoby1991,Singh1991,Loewen1994a}, 
whose best developed non-perturbative version, known as the fundamental-measure theory (FMT) \cite{Rosenfeld1989}, leads 
to unprecedented accuracy for such properties as the liquid-solid interfacial free energy 
\cite{PhysRevE.74.021603,HaertelOREHL2012} or the nucleation barrier \cite{PhysRevE.74.021603}. 
However, handling of large systems is hampered by the complexity of such models. In turn, the PFC 
model, being a simplistic DFT itself, incorporates most of the essential 
physics required to handle freezing: it is atomistic, anisotropies and elasticity are automatically 
there, the system may choose from a variety of periodic states [such as body-centred cubic (bcc), face-centred cubic (fcc), and hexagonal close packed (hcp)] besides 
the homogeneous fluid, \etc\  
The free-energy functional is fairly simple having the well-known Swift-Hohenberg (SH) form 
\begin{equation}
\FSH = \vintw \,\bigg( \frac{\psiSH}{2} \Big( \!-\beta + \big(\wn_0^2+\NablaW^2\big)^2 \Big) \psiSH 
+ \frac{\psiSH^4}{4} \bigg) \,,
\label{eq0}
\end{equation}
where $\psiSH$ is the reduced particle density and $\beta$ a reduced temperature, while $\wn_0$ is the 
absolute value of the wave number vector the system prefers. (In equation \eqref{eq0} all quantities are dimensionless.) 
This together with the assumption of overdamped conservative (diffusive) dynamics (a major deviation from the non-conservative dynamics 
of the SH model) leads to a relatively simple equation of motion (EOM) 
that, in turn, allows the handling of a few times $10^7$ atoms on the diffusive time scale. 
Such abilities can be further amplified by the amplitude equation versions \cite{PhysRevE.72.020601} 
obtained by renormalisation group theory, which combined with advanced numerics \cite{PhysRevE.76.056706} allows 
for the handling of relatively big chunks of material, while retaining all the atomic scale physics. 
Such a coarse-grained PFC model, relying on equations of motion for the amplitudes and phases, 
can be viewed as a physically motivated continuum model akin to the highly successful and 
popular phase-field (PF) models \cite{HoytAK2003,GranasyPW2004,Karma2005,Emmerich2008,ProvatasE2010}, which however 
usually contain \textit{ad hoc} assumptions. 
Accordingly, the combination of the PFC model with coarse graining establishes a link between DFT and 
conventional PF models, offering a way for deriving the latter on physical grounds.  

\begin{table}[b]
\centering
\caption{A non-exclusive collection of phenomena addressed using PFC techniques.}
\label{tab:no1}
\begin{tabular}{l l}
\hline
\hline
\textbf{Phenomena} & \textbf{References} \\
\hline
Liquid-solid transition:\\ 
-- dendrites & \cite{ElderPBSG2007,JPhysCondMat.20.404205,JPhysCondMat.22.364101,tegze.phd.thesis,Tang2011146}\\
-- eutectics & \cite{ElderPBSG2007,JPhysCondMat.22.364101,tegze.phd.thesis,PhysRevE.81.011602}\\
-- homogeneous nucleation & \cite{Tang2011146,JPhysCondMat.22.364101,PhysRevE.81.011602,JPhysCondMat.22.364104,PhilosMag.91.123}\\
-- heterogeneous nucleation & \cite{Tang2011146,PhysRevE.81.011602,JPhysCondMat.21.464108,PhysRevLett.107.175702}\\
-- grain-boundary melting & \cite{PhysRevB.77.224114,PhysRevB.78.184110}\\
-- fractal growth & \cite{C0SM00944J,PhysRevLett.106.195502}\\
-- crystal anisotropy & \cite{PhysRevB.76.184107,PhysRevE.79.011607,PhysRevLett.103.035702,JPhysCondMat.21.464109,PhilosMag.91.123,tms2011sandiego,C0SM00944J}\\
-- density/solute trapping & \cite{C0SM00944J,PhysRevLett.106.195502,tms2012orlando}\\
-- glass formation & \cite{PhysRevE.77.061506,PhysRevLett.106.175702,PhysRevLett.107.175702}\\
-- surface alloying & \cite{PhysRevLett.105.126101,prl.unpub.elder}\\
Colloid patterning & \cite{PhilosMag.91.123}\\
Grain-boundary dynamics & \cite{ElderG2004}\\
Crack propagation & \cite{ElderG2004}\\
Elasticity, plasticity, dislocation dynamics & \cite{ElderKHG2002,ElderG2004,StefanovicHP2006,PhysRevLett.101.158701,ChanTDDG2010,PhysRevB.81.214201}\\
Kirkendall effect & \cite{PhilosMag.91.151}\\
Vacancy transport & \cite{PhysRevE.79.035701}\\
Liquid phase separation with colloid\\ 
accumulation at phase boundaries & \cite{AlandLV2011}\\
Liquid crystals & \cite{Loewen2010,WittkowskiLB2010,AchimWL2011}\\
Formation of foams & \cite{PhysRevE.81.065301}\\
\hline
\hline
\end{tabular}
\end{table}

In its simplest formulation, defined above, the PFC model consists of only a single model parameter $\beta$ (provided 
that length is measured in $\wn_0^{-1}$ units). Still it has a fairly complex phase diagram in three spatial dimensions (3D), which 
has stability domains for the bcc, fcc, and hcp structures, as opposed to the single triangular crystal structure 
appearing in 2D. Introducing additional model parameters, recent extensions of the PFC model either aim at further 
controlling of the predicted crystal structure or attempt to refine the description of real materials. Other extensions 
address binary systems, yet others modify the dynamics via considering further modes of density relaxation 
besides the diffusive one, while adopting a free energy that ensures particle conservation and allows
assigning inertia to the particles. In a few cases, PFC models tailored to specific applications have reached 
the level of being quantitative. Via the PFC models, a broad range of exciting phenomena became 
accessible for atomistic simulations (see Tab.\ \ref{tab:no1}), a situation that motivates our review of the present 
status of PFC modelling.

While Tab.\ \ref{tab:no1} contains a fairly impressive list, it is expected to be only the beginning 
of the model's employment in materials science and engineering. For example, 
true knowledge-based tailoring of materials via predictive PFC calculations 
is yet an open vision, for which a number of difficulties need to be overcome. 
We are going to review a few of the most fundamental ones of these open issues. 
For example, the PFC models still have to establish themselves as widely accepted 
simulation tools in materials engineering/design, which requires methodological advances 
in various directions such as (a) ensuring the quantitativeness of PFC predictions for 
practically relevant (multi-phase multi-component) materials and (b) a consistent 
extension of PFC modelling to some essential circumstances such as non-isothermal 
problems, coupling to hydrodynamics, or handling of non-spherical molecules.

So far, only limited reviews of PFC modelling are available \cite{ProvatasE2010,EmmerichGL2011}. 
Therefore, we give a comprehensive overview of PFC modelling in the present review. 
Especially, we present the main achievements of PFC modelling and demonstrate the potential 
these models offer for addressing problems in physics and materials science. 
We pay special attention to the similarities and differences of development steps 
taken in hard and soft matter physics, respectively. The rest of our review article is structured as 
follows: in section \ref{chap:II}, we present a detailed theoretical derivation of the PFC model 
on the basis of dynamical density functional theory (DDFT). Section \ref{chap:III} is devoted to
some essential features of the PFC model and its generalisations including the realization 
of different crystal lattices, the predicted phase diagrams, anisotropy, and some specific
issues such as glass formation, application to foams, and the possibility for coupling 
to hydrodynamics. Section \ref{chap:IV} addresses nucleation and pattern formation in metallic 
alloys, whereas section \ref{chap:V} deals with the application of the PFC models to prominent 
soft matter systems. Finally, in section \ref{chap:VI}, we offer a few concluding remarks and 
an outlook to probable developments in the near future.

\section{\label{chap:II}From density functional theory to phase-field-crystal models}
\begin{figure}
\centering\includegraphics[width=0.8\linewidth]{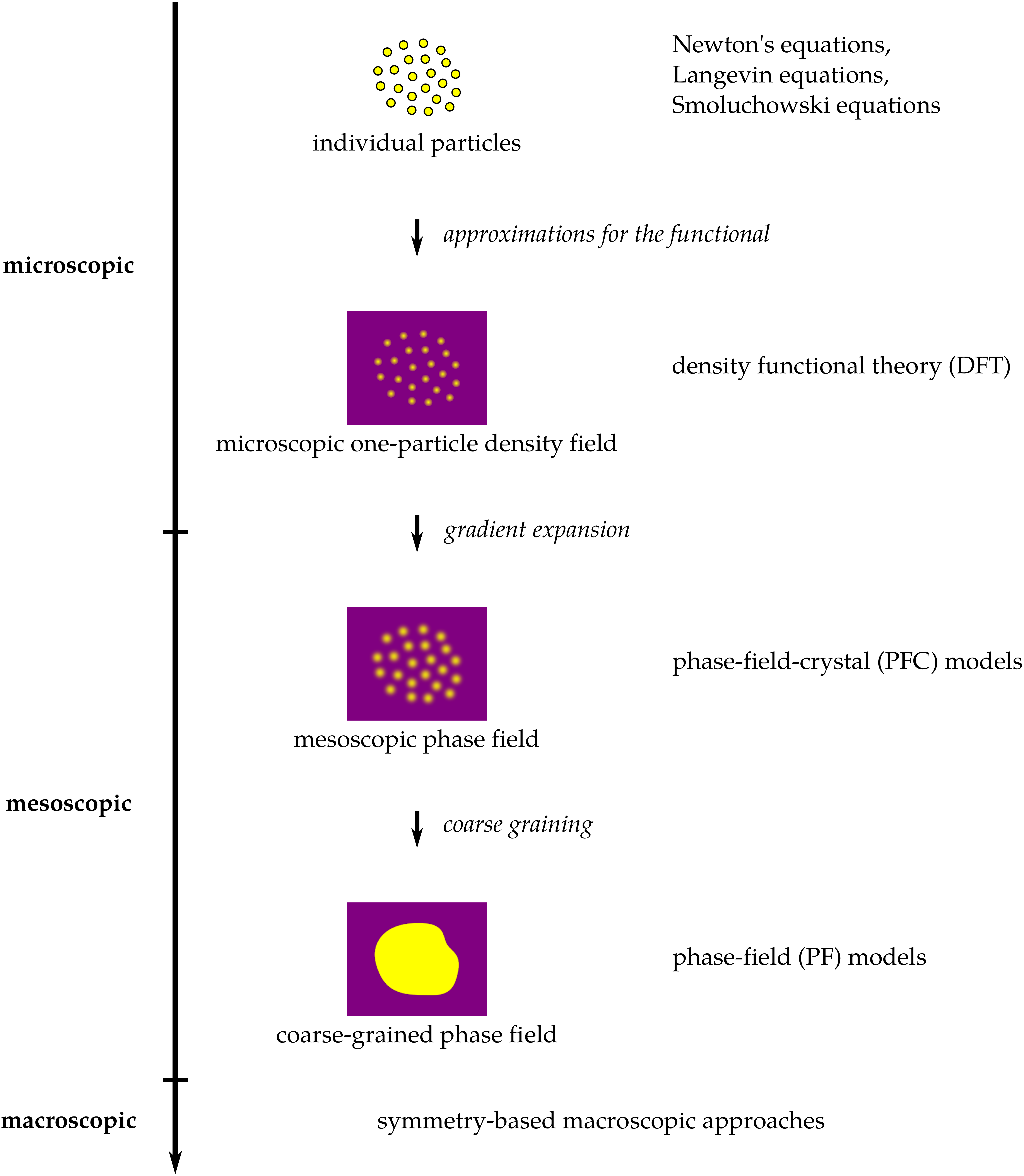}%
\caption{\label{fig:Skalen}Levels of description with the corresponding methods and theories (schematic).} 
\end{figure}

Freezing and crystallization phenomena are described best on the most fundamental level of individual particles, 
which involves the microscopic size and interaction length scale of the particles (see figure \ref{fig:Skalen}).
The individual dynamics of the particles happens correspondingly on a microscopic time scale. 
In the following, two different classes of materials, namely \textit{molecular} and \textit{colloidal} materials, need clear distinction. 
The former comprise metals as well as molecular insulators and semiconductors. We consider these molecular systems as classical particles, 
where the quantum-mechanical nature of the electrons merely enters via effective molecular force fields. 
The corresponding molecular dynamics is governed by Newton's second law. Hence the length scale is atomic (about a few Angstroms) and the 
typical time scale is roughly a picosecond.

The latter material class of colloidal systems involves typically mesoscopic particles immersed in a molecular viscous fluid as a solvent  
that are interacting via effective forces \cite{HansenL2000}. These colloidal suspensions have a dimension typically in the range between 
a nanometre and a micrometer and are therefore classical particles. Thus, the corresponding \ZT{microscopic} length scale describing their 
extension and interaction range is much bigger than for the molecular materials. The individual particle dynamics is 
\textit{Brownian motion} \cite{Einstein1905c,FreyK2005}, \ie, it is completely overdamped\footnote{It is interesting to note that there are 
also mesoscopic particle systems with Newtonian dynamics, which are virtually undamped. 
These are realized in so-called complex plasmas \cite{MorfillI2009,MorfillIBL2010}, where dust particles are dispersed and levitated in a plasma.} 
superimposed with stochastic kicks of the solvent. The corresponding coarse-grained Brownian time scale upon which individual particle motion 
occurs is much longer (about a microsecond) \cite{Pusey1991}.

In terms of static equilibrium properties (such as structural correlations and phase transitions), both metals and colloids can just be regarded as 
classical interacting many-body systems. For this purpose, density functional theory (DFT) was developed \cite{Evans1979,Loewen1994a,Singh1991}: 
DFT is a microscopic theory, \ie, it starts with the (effective) interparticle interactions and predicts the free energy and the static 
many-body correlations. In principle, DFT is exact, but for practical applications one has to rely on approximations.

In the past years, it has become clear that DFT is an ideal theoretical framework to justify and to derive the free-energy functional of 
coarse-grained models as the PFC approach \cite{ElderPBSG2007,vanTeeffelenBVL2009}. PFC models keep the microscopic length scale, but describe 
the microscopically structured density field in a very rough way, for example, by keeping only its first Fourier modes for a crystal. 
Although some microscopic details are lost, the basic picture of the crystal is kept and much larger system sizes can be explored numerically. 
The PFC models are superior to simple PF models, which work with a single order parameter on a more coarse-grained regime. 
Finally, there are also phenomenological hydrodynamic approaches that are operating on the macroscopic length and time scale.

This pretty transparent hierarchy of length scales for static equilibrium properties gets more complex for the dynamics. 
In order to discuss this in more detail, it is advantageous to start with the colloidal systems first. 
Here, the individual dynamics is already dissipative and overdamped: the \ZT{microscopic} equations governing the colloidal Brownian dynamics 
are either the \textit{Langevin equation} for the individual particle trajectories
or the \textit{Smoluchowski equation} for the time evolution of the many-body probability density \cite{Dhont1996,DoiE2007}. 
Both approaches are stochastically equivalent \cite{Risken1996}. In the end of the past century, it has been shown that there is a 
dynamic generalisation of DFT, the DDFT, which describes the time evolution of the many-body system 
within the time-dependent one-body density as a generalized deterministic diffusion equation. This provides a significant simplification of the 
many-body problem. Unfortunately, DDFT is not on the same level as the Smoluchowski or Langevin picture since an additional 
adiabaticity approximation is needed to derive it. This approximation implies, that the one-body density is a slowly relaxing variable and all 
higher density correlations relax much faster to thermodynamic equilibrium \cite{EspanolL2009}. 
Fortunately, the adiabaticity approximation is reasonable for many practical applications except for situations, where fluctuations play a 
significant role. Now, DDFT can be used as an (approximate) starting point to derive the dynamics of a PFC model systematically 
\cite{vanTeeffelenBVL2009}. This also points to alternative dynamical equations, which can be implemented within a numerically similar effort as 
compared to ordinary PFC equations, but are a bit closer to DDFT.

For Newtonian dynamics, on the other hand, intense research is going on to derive a similar kind of DDFT 
\cite{Archer2006,Archer2009,MarconiM2010,MarconiTCM2008}. Still the diffusive (or model B) dynamics for a conserved order-parameter field can be 
used as an effective dynamics on mesoscopic time scales with an effective friction. 
Then, the long-time self-diffusion coefficient sets the time scale of this process. One should, however, point out that the PFC dynamics for 
molecular systems is dynamically more coarse-grained than for their colloidal counterparts.

\subsection{\label{subsubsec:DFT}Density functional theory}
Density functional theory (DFT) is a microscopic theory for inhomogeneous complex fluids in equilibrium \cite{Evans1979,Loewen1994a,Roth2010,Singh1991} that needs only the 
particle interactions and the underlying thermodynamic conditions as an input. The central idea is to express the free energy of the many-body system 
as a functional of the inhomogeneous one-body density. As it stands originally, DFT is a theory for static quantities. 
Most of the actual applications of DFT are for spherically symmetric pairwise interactions between classical particles (mostly hard spheres) 
\cite{Evans1979,RothELK2002,Singh1991}, but they can also be generalized to anisotropic interactions (as relevant for non-spherical hard bodies or 
molecules) \cite{GrafL1999,HansenGoosM2009,Loewen2010b,PoniewierskiH1988}. One of the key applications of DFT concerns the equilibrium freezing and 
melting \cite{Loewen1994a,RamakrishnanY1979,RosenfeldSLT1997,Singh1991} including the fluid-solid interface 
\cite{CurtinA1986,OhnesorgeLW1991,OhnesorgeLW1993,OhnesorgeLW1994}. Further information about DFT and a detailed historic overview can be found in 
several articles and books like references \cite{CurtinA1986,Evans1992,HansenMD2006,Kalikmanov2010,Loewen2002}.

More recently, static DFT was generalized towards time-dependent processes in non\-e\-qui\-li\-bri\-um. 
The extended approach is called dynamical density functional theory (DDFT). DDFT was first derived in 1999 for isotropic Brownian particles by 
Marconi and Tarazona \cite{MarconiT1999,MarconiT2000} starting from the Langevin picture of individual particle trajectories. 
An alternate derivation based on the Smoluchowski picture was presented in 2004 by Archer and Evans \cite{ArcherE2004}. 
In both schemes an additional \textit{adiabaticity approximation} is needed: correlations of high order in nonequilibrium are approximated by those 
in equilibrium for the same one-body density. These derivations were complemented by a further approach on the basis of a 
projection operator technique \cite{EspanolL2009}. The latter approach sheds some light on the adiabaticity approximation: 
it can be viewed by the assumption that the one-body density relaxes much slower than any other density correlations of higher order. 
DDFT can be flexibly generalized towards more complex situations including mixtures \cite{Archer2005}, active particles \cite{WensinkL2008}, 
hydrodynamic interactions \cite{RexL2008,RexL2009}, shear flow \cite{Rauscher2010} and non-spherical particles \cite{RexWL2007,WittkowskiL2011}. 
However, as already stated above, it is much more difficult to derive a DDFT for Newtonian dynamics, where inertia and flow effects invoke a 
treatment of the momentum density field of the particles \cite{Archer2006,Archer2009,MarconiM2010,MarconiTCM2008}.

In detail, DFT gives access to the free energy for a system of $N$ classical particles, whose centre-of-mass positions are defined through the 
vectors $\vec{r}_{i}$ with $i\in\{1,\dotsc,N\}$, by the \textit{one-particle density} $\rho(\vec{r})$, which provides the probability to find a 
particle at position $\vec{r}$. Its microscopic definition is 
\begin{equation}
\rho(\vec{r}) = \bigg\langle\sum^{N}_{i=1} \delta(\vec{r}-\vec{r}_{i})\bigg\rangle 
\end{equation}
with the normalised classical canonical (or grand canonical) ensemble-average $\langle\,\cdot\,\rangle$. At given temperature $T$ and 
chemical potential $\mu$, the particles are interacting via a pairwise (two-body) potential $U_{2}(\vec{r}_{1}-\vec{r}_{2})$. 
Furthermore, the system is exposed to an external (one-body) potential $U_{1}(\vec{r})$ (describing, for example, gravity or system boundaries), 
which gives rise in general to an inhomogeneous one-particle density $\rho(\vec{r})$.
DFT is based on the following variational theorem:
\begin{quote}
\textit{There exists a unique grand canonical free-energy functional $\Omega(T,\mu,[\rho(\vec{r})])$  
of the one-particle density $\rho(\vec{r})$, which becomes minimal for the equilibrium one-particle density $\rho(\vec{r})$:
\begin{equation}
\Fdif{\Omega(T,\mu,[\rho(\vec{r})])}{\rho(\vec{r})}=0 \;.
\label{eq:DFT_Theorem}
\end{equation}
If the grand canonical functional $\Omega(T,\mu,[\rho(\vec{r})])$ is evaluated at the equilibrium one-par\-ti\-cle density $\rho(\vec{r})$, 
it is the real equilibrium grand canonical free energy of the inhomogeneous system.}
\end{quote}
Hence, DFT establishes a basis for the determination of the equilibrium one-particle density field $\rho(\vec{r})$ of an arbitrary classical 
many-body system. However, in practice, the exact form of the grand canonical free-energy density functional $\Omega(T,\mu,[\rho(\vec{r})])$ 
is not known and one has to rely on approximations. Via a Legendre transform, the grand canonical functional can be expressed by an equivalent 
Helmholtz free-energy functional $\mathcal{F}(T,[\rho(\vec{r})])$,
\begin{equation}
\Omega(T,\mu,[\rho(\vec{r})])=\mathcal{F}(T,[\rho(\vec{r})]) - \mu\!\vint \rho(\vec{r}) \;,
\label{eq:OmegaLT}
\end{equation}
with $V$ denoting the system volume. The latter is conveniently split into three contributions:
\begin{equation}
\mathcal{F}(T,[\rho(\vec{r})])=\mathcal{F}_{\mathrm{id}}(T,[\rho(\vec{r})]) +\mathcal{F}_{\mathrm{exc}}(T,[\rho(\vec{r})])+\mathcal{F}_{\mathrm{ext}}(T,[\rho(\vec{r})])\;.
\label{eq:F_Zerlegung}%
\end{equation}
Here, $\mathcal{F}_{\mathrm{id}}(T,[\rho(\vec{r})])$ is the (exact) \textit{ideal gas free-energy functional} \cite{Evans1979} 
\begin{equation}
\mathcal{F}_{\mathrm{id}}(T,[\rho(\vec{r})])=\kBT\!\vint\rho(\vec{r})\big(\ln\!\big(\Lambda^{3}\rho(\vec{r})\big)-1\big) \,,
\label{eq:F_id}
\end{equation}
where $\kB$ is the Boltzmann constant and $\Lambda$ the thermal de Broglie wavelength. 
The second term on the right-hand-side of equation \eqref{eq:F_Zerlegung} is the \textit{excess free-energy functional} 
$\mathcal{F}_{\mathrm{exc}}(T,[\rho(\vec{r})])$ describing the excess free energy over the exactly known ideal-gas functional. 
It incorporates all correlations due to the pair interactions between the particles. In general, it is not known explicitly and therefore needs 
to be approximated appropriately \cite{Evans1979,Singh1991}. The last contribution is the \textit{external free-energy functional} \cite{Evans1979}
\begin{equation}
\mathcal{F}_{\mathrm{ext}}(T,[\rho(\vec{r})])=\vint\rho(\vec{r})U_{1}(\vec{r})\;.
\label{eq:F_ext}%
\end{equation}
A formally exact expression for $\mathcal{F}_{\mathrm{exc}}(T,[\rho(\vec{r})])$ is gained by a functional Taylor expansion in the density variations 
$\Delta\rho(\vec{r})=\rho(\vec{r})-\rhoR$ around a homogeneous reference density 
$\rhoR$ \cite{Evans1979,RamakrishnanY1979}:
\begin{equation}
\mathcal{F}_{\mathrm{exc}}(T,[\rho(\vec{r})]) = 
\mathcal{F}^{(0)}_{\mathrm{exc}}(\rhoR)+\kBT \sum^{\infty}_{n=1}\frac{1}{n!}\mathcal{F}^{(n)}_{\mathrm{exc}}(T,[\rho(\vec{r})]) 
\label{eq:FTEg}%
\end{equation}
with 
\begin{equation}
\mathcal{F}^{(n)}_{\mathrm{exc}}(T,[\rho(\vec{r})]) 
= -\!\vintn{1}\:\!\dotsi\!\!\vintn{n}\,c^{(n)}(\vec{r}_{1},\dotsc,\vec{r}_{n}) \prod^{n}_{i=1}\Delta\rho(\vec{r}_{i}) \;.
\label{eq:Fexcng}%
\end{equation}
Here, $c^{(n)}(\vec{r}_{1},\dotsc,\vec{r}_{n})$ denotes the $n$th-order direct correlation function \cite{HansenMD2006} in the homogeneous 
reference state given by
\begin{equation}
c^{(n)}(\vec{r}_{1},\dotsc,\vec{r}_{n})
=-\frac{1}{\kBT}\:\!\frac{\delta^{n}\mathcal{F}_{\mathrm{exc}}(T,[\rho(\vec{r})])}{\delta\rho(\vec{r}_{1})\dotsb
\delta\rho(\vec{r}_{n})}\bigg\rvert_{\rhoR}
\label{eq:cn}%
\end{equation}
depending parametrically on $T$ and $\rhoR$. 

In the functional Taylor expansion \eqref{eq:FTEg}, the constant zeroth-order contribution is irrelevant and the first-order contribution 
corresponding to $n=1$ is zero.\footnote{This follows from the representation \eqref{eq:Fexcng} under consideration of the translational 
and rotational symmetries of the isotropic bulk fluid that also apply to the direct correlation function $c^{(1)}(\vec{r}_{1})=const.$} 
The higher-order terms are nonlocal and do not vanish in general.

In the simplest nontrivial approach, the functional Taylor expansion is truncated at second order. The resulting approximation is known as 
the \emph{Ramakrishnan-Yussouff theory} \cite{RamakrishnanY1979} 
\begin{equation}
\mathcal{F}_{\mathrm{exc}}(T,[\rho(\vec{r})]) 
= -\frac{1}{2}\:\!\kBT\!\vintn{1}\!\!\vintn{2}\,c^{(2)}(\vec{r}_{1}-\vec{r}_{2}) \Delta\rho(\vec{r}_{1})\Delta\rho(\vec{r}_{2})  
\label{eq:RYg}%
\end{equation}
and predicts the freezing transition of hard spheres both in three \cite{RamakrishnanY1979} and 
two spatial dimensions (2D) \cite{vanTeeffelenLHL2006}.\footnote{More refined approaches include also the third-order term \cite{Barrat1987} with an
approximate triplet direct correlation function \cite{BarratHP1987,BarratHP1988}.} 
The Ramakrishnan-Yussouff approximation needs the fluid direct pair-correlation function $c^{(2)}(\vec{r}_{1}-\vec{r}_{2})$ as an input.  
For example, $c^{(2)}(\vec{r}_{1}-\vec{r}_{2})$ can be gained from liquid integral equation theory, which links 
$c^{(2)}(\vec{r}_{1}-\vec{r}_{2})$ to the pair-interaction potential $U_{2}(\vec{r}_{1}-\vec{r}_{2})$. 
Well-known analytic approximations for the direct pair-correlation function include the second-order \textit{virial expression} \cite{vanRoijBMF1995}
\begin{equation}
c^{(2)}(\vec{r}_{1}-\vec{r}_{2})=\exp\!\bigg(\!-\frac{U_{2}(\vec{r}_{1}-\vec{r}_{2})}{\kBT}\bigg)-1 \;.
\end{equation}
The resulting \textit{Onsager functional} for the excess free energy becomes asymptotically exact in the low density limit \cite{HansenMD2006}. 
An alternative is the \textit{random-phase} or \textit{mean-field approximation}
\begin{equation}
c^{(2)}(\vec{r}_{1}-\vec{r}_{2})=-\frac{U_{2}(\vec{r}_{1}-\vec{r}_{2})}{\kBT} \;.
\label{eq:random_phase_approximation}%
\end{equation}
For bounded potentials, this mean-field approximation becomes asymptotically exact at high densities 
\cite{LikosHLL2002,LikosLWL2001,Louis2001,RexWL2007}. Non-perturbative expressions for the excess free-energy functional for colloidal particles 
are given by \textit{weighted-density approximations} \cite{CurtinA1985,CurtinA1986,DentonA1989,GrafL1999,PoniewierskiH1988} or follow 
from FMT \cite{HansenGoosM2009,HansenGoosM2010}. 
FMT was originally introduced in 1989 by Rosenfeld for isotropic particles \cite{Evans1992,Rosenfeld1989,RosenfeldSLT1996,RosenfeldSLT1997} and 
then refined later \cite{RothELK2002,Tarazona2000} -- for a review, see reference \cite{Roth2010}. 
For hard spheres, FMT provides an excellent approximation for the excess free-energy functional with an unprecedented accuracy. 
It was also generalized to arbitrarily shaped particles \cite{HansenGoosM2009,HansenGoosM2010,Rosenfeld1994}.

\subsection{\label{subsec:DDFT}Dynamical density functional theory}
 
\subsubsection{Basic equations}
Dynamical density functional theory (DDFT) is the time-dependent analogue of static DFT and can be classified as \textit{linear-response theory}. 
In its basic form, it describes the slow dissipative nonequilibrium relaxation dynamics of a system of $N$ Brownian particles close to 
thermodynamic equilibrium or the behaviour in a time-dependent external potential $U_{1}(\vec{r},t)$. 
Now a time-dependent one-particle density field is defined via
\begin{equation}
\rho(\vec{r},t) = \bigg\langle\sum^{N}_{i=1} \delta\big(\vec{r}-\vec{r}_{i}(t)\big)\bigg\rangle \,,
\end{equation}
where $\langle\,\cdot\,\rangle$ denotes the normalised classical canonical noise-average over the particle trajectories and $t$ is the time variable.

This one-particle density is conserved and its dynamics is assumed to be dissipative via the generalized (deterministic) diffusion equation 
\begin{equation}
\pdif{\rho(\vec{r},t)}{t}=\frac{D_{\mathrm{T}}}{\kBT}\,\Nabla\!\cdot\!\bigg(\rho(\vec{r},t)\Nabla
\Fdif{\mathcal{F}(T,[\rho(\vec{r},t)])}{\rho(\vec{r},t)}\bigg) \,.
\label{eq:DDFTs}
\end{equation}
Here, $D_{\mathrm{T}}$ denotes a (short-time) translational diffusion coefficient for the Brownian system. 
Referring to equations \eqref{eq:DFT_Theorem} and \eqref{eq:OmegaLT}, the functional derivative in the DDFT equation can be interpreted as an 
inhomogeneous chemical potential 
\begin{equation}
\mu(\vec{r},t)=\Fdif{\mathcal{F}(T,[\rho(\vec{r},t)])}{\rho(\vec{r},t)}
\end{equation}
such that the DDFT equation \eqref{eq:DDFTs} corresponds to a generalized Fick's law of particle diffusion. 
As already mentioned, DDFT was originally invented \cite{ArcherE2004,MarconiT1999} for colloidal particles, which exhibit Brownian motion, 
but is less justified for metals and atomic systems whose dynamics are ballistic \cite{Archer2006,Archer2009}.

\subsubsection{Brownian dynamics: Langevin and Smoluchowski picture} 
The DDFT equation \eqref{eq:DDFTs} can be derived \cite{MarconiT1999} from Langevin equations  
that describe the stochastic motion of the $N$ isotropic colloidal particles in an incompressible liquid of 
viscosity $\eta$ at low Reynolds number (Stokes limit). In the absence of hydrodynamic interactions between the Brownian particles, 
these coupled Langevin equations for the positions $\vec{r}_{i}(t)$ of the colloidal spheres with radius $R_{\mathrm{s}}$ describe completely 
overdamped motion plus stochastic noise \cite{Dhont1996,Risken1996}:
\begin{equation}
\dot{\vec{r}}_{i} = \xi^{-1} (\vec{F}_{i} + \vec{f}_{i})\;,\qquad i=1,\dotsc,N \;.
\label{eq:LEs}%
\end{equation}
Here, $\xi$ is the Stokesian friction coefficient ($\xi=6\pi\eta R_{\mathrm{s}}$ for spheres of radius $R_{\mathrm{s}}$ with 
stick boundary conditions) and 
\begin{equation}
\vec{F}_{i}(t)=-\NablaI{i} U(\vec{r}_{1},\dotsc,\vec{r}_{N},t)
\end{equation}
are the deterministic forces caused by the total potential 
\begin{equation}
U(\vec{r}_{1},\dotsc,\vec{r}_{N},t)=U_{\mathrm{ext}}(\vec{r}_{1},\dotsc,\vec{r}_{N},t)+U_{\mathrm{int}}(\vec{r}_{1},\dotsc,\vec{r}_{N}) 
\end{equation}
with 
\begin{equation}
U_{\mathrm{ext}}(\vec{r}_{1},\dotsc,\vec{r}_{N},t)=\sum^{N}_{i=1}U_{1}(\vec{r}_{i},t) 
\end{equation}
and 
\begin{equation}
U_{\mathrm{int}}(\vec{r}_{1},\dotsc,\vec{r}_{N})=\sum^{N}_{\begin{subarray}{c}i,j=1\\i<j\end{subarray}}U_{2}(\vec{r}_{i}-\vec{r}_{j})\;.
\end{equation}
On top of these deterministic forces, also stochastic forces $\vec{f}_{i}(t)$ due to thermal fluctuations act on the Brownian particles. 
These random forces are modelled by Gaussian white noises with vanishing mean values
\begin{equation}
\langle\vec{f}_{i}(t)\rangle=\vec{0}
\end{equation}
and with Markovian second moments 
\begin{equation}
\langle\vec{f}_{i}(t_{1})\otimes\vec{f}_{j}(t_{2})\rangle= 2\:\!\xi\:\! \kBT \:\!\Eins\:\! \delta_{ij}\delta(t_{1}-t_{2}) \;,
\end{equation}
where $\otimes$ is the ordinary (dyadic) tensor product (to make the notation compact) 
and $\Eins$ denotes the $3\!\times\!3$-dimensional unit matrix. This modelling of the stochastic forces is dictated by the 
fluctuation-dissipation theorem, which for spheres yields the Stokes-Einstein relation $D_{\mathrm{T}}=\kBT/\xi$ \cite{Einstein1905c}, 
that couples the short-time diffusion coefficient $D_{\mathrm{T}}$ of the colloidal particles to the Stokes friction coefficient $\xi$.

An alternate description of Brownian dynamics is provided by the Smoluchowski picture, which is stochastically equivalent to the 
Langevin picture \cite{DoiE2007, Risken1996}. The central quantity in the Smoluchowski picture is the $N$-particle 
probability density $P(\vec{r}_{1},\dotsc,\vec{r}_{N},t)$ whose time evolution is described by the Smoluchowski equation 
\cite{Risken1996,Smoluchowski1916}
\begin{equation}
\pdif{}{t}P(\vec{r}_{1},\dotsc,\vec{r}_{N},t)=\hat{\mathcal{L}}\,P(\vec{r}_{1},\dotsc,\vec{r}_{N},t)
\label{eq:SEs}%
\end{equation}
with the Smoluchowski operator 
\begin{equation}
\hat{\mathcal{L}}=D_{\mathrm{T}}\sum^{N}_{i=1} 
\NablaI{i}\!\cdot\!\bigg(\NablaI{i}\frac{U(\vec{r}_{1},\dotsc,\vec{r}_{N},t)}{\kBT}+\NablaI{i}\bigg) \,.
\end{equation}
While the $N$-particle probability density $P(\vec{r}_{1},\dotsc,\vec{r}_{N},t)$ in this Smoluchowski equation is a highly nontrivial function 
for interacting particles, it is often sufficient to consider one-body or two-body densities. 
The one-particle probability density $P(\vec{r},t)$ is proportional to the one-particle number density $\rho(\vec{r},t)$. 
In general, all $n$-particle densities with $n\leqslant N$ can be obtained from the $N$-particle probability density 
$P(\vec{r}_{1},\dotsc,\vec{r}_{N},t)$ by integration over the remaining degrees of freedom:
\begin{equation}
\rho^{(n)}(\vec{r}_{1},\dotsc,\vec{r}_{n},t)=\frac{N!}{(N-n)!}\vintn{n+1}\:\!\dotsi\!\!\vintn{N}P(\vec{r}_{1},\dotsc,\vec{r}_{N},t) \;.
\end{equation}

\subsubsection{Derivation of DDFT}
We now sketch how to derive the DDFT equation \eqref{eq:DDFTs} from the Smoluchowski picture following the idea of 
Archer and Evans \cite{ArcherE2004}. Integrating the Smoluchowski equation \eqref{eq:SEs} over the positions of $N-1$ particles 
yields the exact equation 
\begin{equation}
\dot{\rho}(\vec{r},t)=D_{\mathrm{T}}\Nabla\!\cdot\!\bigg(\Nabla\rho(\vec{r},t)-
\frac{\bar{F}(\vec{r},t)}{\kBT}+\frac{\rho(\vec{r},t)}{\kBT}\Nabla U_{1}(\vec{r},t)\bigg)
\label{eq:rhogs}%
\end{equation}
for the one-particle density $\rho(\vec{r},t)$, where 
\begin{equation}
\bar{F}(\vec{r},t)=-\!\int\!\!\!\:\!\dif\mathbf{r}'\:\!\rho^{(2)}(\vec{r},\vec{r}'\!,t)\Nabla U_{2}(\vec{r}-\vec{r}') 
\label{eq:Kqts}%
\end{equation}
is an average force, that in turn depends on the nonequilibrium two-particle density $\rho^{(2)}(\vec{r}_{1},\vec{r}_{2},t)$. 
This quantity is approximated by an equilibrium expression. To derive this expression, we consider first the equilibrium state 
of equation \eqref{eq:rhogs}. This leads to  
\begin{equation}
\bar{F}(\vec{r})=\kBT\:\!\Nabla \rho(\vec{r})+\rho(\vec{r})\Nabla\overline{U}_{1}(\vec{r})\;,
\label{eq:abtI}%
\end{equation}
which is the first equation of the Yvon-Born-Green hierarchy, with a \ZT{substitute} external potential  
$\overline{U}_{1}(\vec{r})$. In equilibrium, DFT implies
\begin{equation}
\begin{split}
0&=\Fdif{\Omega(T,\mu,[\rho(\vec{r})])}{\rho(\vec{r})}=\Fdif{\mathcal{F}(T,[\rho(\vec{r})])}{\rho(\vec{r})}-\mu \\
&=\kBT\ln\!\:\!\big(\Lambda^{3}\rho(\vec{r})\big)+\Fdif{\mathcal{F}_{\mathrm{exc}}(T,[\rho(\vec{r})])}{\rho(\vec{r})}+\overline{U}_{1}(\vec{r})
\end{split}
\end{equation}
and after application of the gradient operator
\begin{equation}
0=\kBT\frac{\Nabla\rho(\vec{r})}{\rho(\vec{r})}
+\Nabla\Fdif{\mathcal{F}_{\mathrm{exc}}(T,[\rho(\vec{r})])}{\rho(\vec{r})}+\Nabla\overline{U}_{1}(\vec{r}) \;.
\label{eq:abtII}%
\end{equation}
A comparison of equations \eqref{eq:abtI} and \eqref{eq:abtII} yields 
\begin{equation}
\bar{F}(\vec{r})=-\rho(\vec{r})\Nabla\Fdif{\mathcal{F}_{\mathrm{exc}}(T,[\rho(\vec{r})])}{\rho(\vec{r})} \;.
\end{equation}
It is postulated, that this relation also holds in nonequilibrium. The nonequilibrium correlations are thus approximated by equilibrium ones 
at the same $\rho(\vec{r})$ via a suitable \ZT{substitute} equilibrium potential $\overline{U}_{1}(\vec{r})$. 
With this \textit{adiabatic approximation}, equation \eqref{eq:rhogs} becomes 
\begin{equation}
\dot{\rho}(\vec{r},t)=D_{\mathrm{T}}\Nabla\!\cdot\!\bigg(\!\Nabla\rho(\vec{r},t)
+\frac{\rho(\vec{r})}{\kBT}\Nabla\Fdif{\mathcal{F}_{\mathrm{exc}}(T,[\rho(\vec{r})])}{\rho(\vec{r})}
+\frac{\rho(\vec{r},t)}{\kBT}\Nabla U_{1}(\vec{r},t)\!\bigg) \:\!,
\end{equation}
which is the DDFT equation \eqref{eq:DDFTs}. 

It is important to note that the DDFT equation \eqref{eq:DDFTs} is a deterministic equation, \ie, there are no additional noise terms. 
If noise is added, there would be double counted fluctuations in the equilibrium limit of equation \eqref{eq:DDFTs} since $\mathcal{F}(T,[\rho])$ 
is the \textit{exact} equilibrium functional, which in principle includes all fluctuations. The drawback of the adiabatic approximation, 
on the other hand, is that a system is trapped for ever in a metastable state. This unphysical behaviour can be changed by adding noise on a 
phenomenological level though violating the caveat noted above. A pragmatic recipe is to add noise only when fluctuations are needed to push 
the system out of a metastable state or to regard a fluctuating density field as an initial density profile for subsequent deterministic 
time evolution via DDFT. In conclusion, the drawback of the adiabatic approximation is that DDFT is some kind of mean-field theory.
For example, DDFT as such is unable to predict nucleation rates. It is rather a realistic theory, if a systematic drive pushes the system, 
as occurs, for example, for crystal growth.

\subsubsection{\label{par:CCG}Application of DDFT to colloidal crystal growth} 
\begin{figure}
\centering\includegraphics[width=0.4\linewidth]{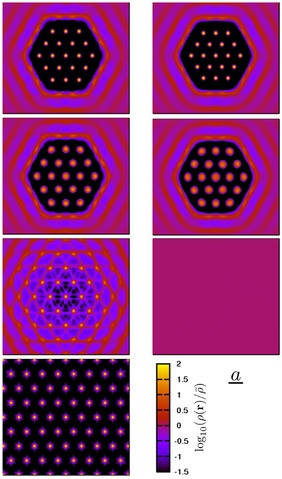}%
\caption{\label{fig:TLLa}Crystallization starting at a colloidal cluster.
The plots show DDFT results for the time-dependent density field. 
$A\rhoR=0.7$ (left column) and $A\rhoR=0.6$ (right column) at times 
$t/\tau_{\mathrm{B}}=0,0.001,0.1,1$ (from top to bottom) with the area $A$ of a unit cell 
of the imposed crystalline seed, the Brownian time $\tau_{\mathrm{B}}$, and the lattice constant $a=(2/(\sqrt{3}\rhoR))^{1/2}$.
For $A\rhoR=0.7$, the cluster is compressed in comparison to the stable bulk crystal, but there is still crystal growth possible, 
while the compression is too high for $A\rhoR=0.6$. The initial nucleus melts in this case. 
\nocite{vanTeeffelenLL2008}\CRPE{S.\ van Teeffelen, C.\ N.\ Likos, and H. L\"owen}{Colloidal crystal growth at externally imposed nucleation clusters}{Phys.\ Rev.\ Lett.\ }{100}{2008}{108302}{10}{10.1103/PhysRevLett.100.108302}{the American Physical Society}} 
\end{figure}

\begin{figure}
\centering\includegraphics[width=0.6\linewidth]{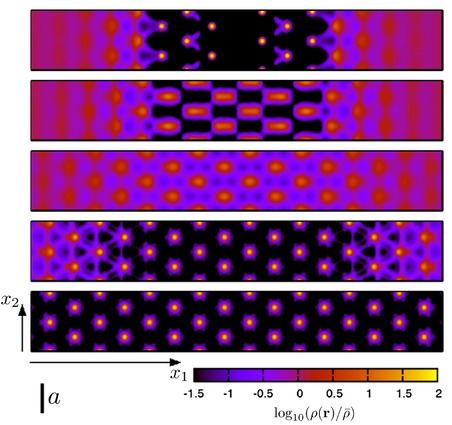}%
\caption{\label{fig:TLLc}Crystallization starting at two triple-rows of hexagonally crystalline particles that are separated 
by a gap. The contour plots show the density field of a growing crystal at times $t/\tau_{\mathrm{B}}=0,0.01,0.1,0.63,1$ (from top to bottom).  
\nocite{vanTeeffelenLL2008}\CRPE{S.\ van Teeffelen, C.\ N.\ Likos, and H. L\"owen}{Colloidal crystal growth at externally imposed nucleation clusters}{Phys.\ Rev.\ Lett.\ }{100}{2008}{108302}{10}{10.1103/PhysRevLett.100.108302}{the American Physical Society}} 
\end{figure}

An important application of DDFT is the description of colloidal crystal growth. In reference \cite{vanTeeffelenLL2008}, 
DDFT was applied to two-dimensional dipoles, whose dipole moments are perpendicular to a confining plane. 
These dipoles interact with a repulsive inverse power-law potential $U_{2}(\vec{r})=u_{0}r^{-3}$, where $r=\norm{\mathbf{r}}$ is the 
interparticle distance. This model can be realized, for example, 
by superparamagnetic colloids at a water-surface in an external magnetic field \cite{Loewen2001}. 
Figures \ref{fig:TLLa} and \ref{fig:TLLc} show DDFT results from reference \cite{vanTeeffelenLL2008}.

In figure \ref{fig:TLLa}, the time evolution of the one-particle density of an initial colloidal cluster of $19$ particles arranged in a 
hexagonal lattice is shown. This prescribed cluster is surrounded by an undercooled fluid and can act as a nucleation seed, 
if its lattice constant is chosen appropriately. The initial cluster either initiates crystal growth (left column in figure \ref{fig:TLLa}) 
or the system relaxes back to the undercooled fluid (right column).

A similar investigation is also possible for other initial configurations like rows of seed particles. Figure \ref{fig:TLLc} shows the 
crystallization process starting with six infinitely long particle rows of a hexagonal crystal, where a gap separates the first three rows 
from the remaining three rows. If this gap is not too big, the density peaks rearrange and a growing crystal front emerges.

\subsection{\label{sec:PFC_Isotropic}Derivation of the PFC model for isotropic particles from DFT}
Though approximate in practice, DFT and DDFT can be regarded to be a high level of microscopic description, which provides a framework to calibrate 
the more coarse-grained PFC approach. In this section, we at first describe the derivation for spherical interactions in detail and then focus more 
on anisotropic particles. There are two different aspects of the PFC modelling, which can be justified from DFT respectively DDFT, namely statics and 
dynamics. The static free energy used in the PFC model was first derived from DFT by Elder and co-workers in 2007 \cite{ElderPBSG2007}, 
while the corresponding dynamics was derived from DDFT by van Teeffelen and co-workers in 2009 \cite{vanTeeffelenBVL2009}. 
We follow the basic ideas of these works in the sequel.

\subsubsection{\label{sec:fef}Free-energy functional} 
For the static part, we first of all define  a scalar dimensionless order-parameter field $\psi(\vec{r})$\footnote{Notice that the 
order-parameter field $\psi(\vec{r})$ introduced here is not identical with the field $\psiSH(\rw)$ in equation \eqref{eq0}, 
although both fields are dimensionless.} 
by the relative density deviation 
\begin{equation}
\rho(\vec{r})=\rhoR\:\!(1+\psi(\vec{r})) 
\label{eq:prho}
\end{equation}
around the prescribed fluid reference density $\rhoR$. 
This relative density deviation $\psi(\vec{r})$ is considered to be small, $\abs{\psi(\vec{r})}\ll 1$, and slowly varying in space 
(on the microscale). The basic steps to derive the PFC free energy are threefold: 
i) insert the parametrisation \eqref{eq:prho} into the (microscopic) free-energy functional \eqref{eq:F_Zerlegung}, 
ii) Taylor-expand systematically in terms of powers of $\psi(\vec{r})$, 
iii) perform a gradient expansion \cite{Evans1979,LoewenBW1989,LoewenBW1990,Lutsko2006,OhnesorgeLW1991} of $\psi(\vec{r})$. 
Consistent with the assumption that density deviations are small, the Ramakrishnan-Yussouff approximation \eqref{eq:RYg} is used as a convenient 
approximation for the free-energy functional.

For the local ideal gas free-energy functional \eqref{eq:F_id} this yields\footnote{This Taylor approximation of the logarithm has the 
serious consequence that the non-negative-density constraint $\rho(\vec{r})\geqslant 0$ gets lost in the PFC model.} 
\begin{equation}
\mathcal{F}_{\mathrm{id}}[\psi(\vec{r})] = F_{0} +\rhoR\,\kB T\!\vint\bigg(\psi+\frac{\psi^{2}}{2}
-\frac{\psi^{3}}{6}+\frac{\psi^{4}}{12}\bigg)
\label{eq:approx_Fid}
\end{equation}
with the irrelevant constant $F_{0} = \rhoR V\kB T(\ln(\Lambda^{3}\rhoR)-1)$. The Taylor expansion is performed 
up to fourth order, since this is the lowest order which enables the formation of stable crystalline phases. 
The nonlocal Ramakrishnan-Yussouff approximation \eqref{eq:RYg} for the approximation of the excess free-energy functional 
$\mathcal{F}_{\mathrm{exc}}[\psi(\vec{r})]$ is gradient-expanded to make it local. 
For this purpose, it is important to note that -- in the fluid bulk reference state -- the direct pair-correlation function 
$c^{(2)}(\vec{r}_{1}-\vec{r}_{2})$ entering into the Ramakrishnan-Yussouff theory has the same symmetry as the 
interparticle interaction potential $U_{2}(\vec{r}_{1}-\vec{r}_{2})$. For radially symmetric interactions (\ie, spherical particles), 
there is both \textit{translational} and \textit{rotational} invariance implying
\begin{equation}
c^{(2)}(\vec{r}_{1},\vec{r}_{2})\equiv c^{(2)}(\vec{r}_{1}-\vec{r}_{2})\equiv c^{(2)}(r)
\label{eq:cII}%
\end{equation}
with the relative distance $r=\abs{\vec{r}_{1}-\vec{r}_{2}}$. Then, as a consequence of equation \eqref{eq:cII}, 
the Ramakrishnan-Yussouff approximation is a convolution integral. 
Consequently, a Taylor expansion of the Fourier transform $\tilde{c}^{(2)}(\vec{k})$ of the direct correlation function 
in Fourier space (around the wave vector $\vec{k}=\vec{0}$)
\begin{equation}
\tilde{c}^{(2)}(\vec{k})=\tilde{c}^{(2)}_{0}+\tilde{c}^{(2)}_{2}\vec{k}^{2}+\tilde{c}^{(2)}_{4}\vec{k}^{4}+\dotsb
\label{eq:cIIkGE}%
\end{equation}
with expansion coefficients $\tilde{c}^{(2)}_{i}$ becomes a gradient expansion in real space
\begin{equation}
c^{(2)}(\vec{r})=c^{(2)}_{0}-c^{(2)}_{2}\Nabla^{2}+c^{(2)}_{4}\Nabla^{4} \mp\dotsb
\label{eq:cIIrGE}%
\end{equation}
with the gradient expansion coefficients $c^{(2)}_{i}$. Clearly, gradients of odd order vanish due to parity inversion symmetry 
$c^{(2)}(-\vec{r})=c^{(2)}(\vec{r})$ of the direct pair-correlation function.

The gradient expansion up to fourth order is the lowest one that makes stable periodic density fields possible. 
We finally obtain 
\begin{equation}
\mathcal{F}_{\mathrm{exc}}[\psi(\vec{r})] = F_{\mathrm{exc}} 
-\frac{\rhoR}{2}\:\!\kB T\!\vint\big(A_{1}\psi^{2}+A_{2}\psi\Nabla^{2}\psi+A_{3}\psi\Nabla^{4}\psi \big) 
\label{eq:approx_Fexc}%
\end{equation}
with the irrelevant constant $F_{\mathrm{exc}}=\mathcal{F}^{(0)}_{\mathrm{exc}}(\rhoR)$ and the coefficients
\begin{equation}
A_{1}=4\pi\rhoR\int^{\infty}_{0}\!\!\!\!\!\!\!\dif r\, r^{2}c^{(2)}(r)\;,\quad
A_{2}=\frac{2}{3}\pi\rhoR\int^{\infty}_{0}\!\!\!\!\!\!\!\dif r\, r^{4}c^{(2)}(r)\;,\quad
A_{3}=\frac{\pi\rhoR}{30}\int^{\infty}_{0}\!\!\!\!\!\!\!\dif r\, r^{6}c^{(2)}(r) 
\end{equation}
that are moments of the fluid direct correlation function $c^{(2)}(r)$. 

Finally, the external free-energy functional \eqref{eq:F_ext} can be written as 
\begin{equation}
\mathcal{F}_{\mathrm{ext}}[\psi(\vec{r})] = F_{\mathrm{ext}} 
+\rhoR\!\vint\psi(\vec{r}) U_{1}(\vec{r})
\label{eq:approx_Fext}%
\end{equation}
with the irrelevant constant $F_{\mathrm{ext}}=\rhoR \int\!\!\dif\vec{r}\, U_{1}(\vec{r})$. We add as a comment here that this external part is 
typically neglected in most of the PFC calculations. Altogether, we obtain
\begin{equation}
\begin{split}
\mathcal{F}[\psi(\vec{r})] =\rhoR\,\kB T\!\vint 
&\bigg(A'_{1}\psi^{2}+A'_{2}\psi\Nabla^{2}\psi +A'_{3}\psi\Nabla^{4}\psi -\frac{\psi^{3}}{6}+\frac{\psi^{4}}{12}\bigg)
\end{split}
\label{eq:HFEF}
\end{equation}
for the total Helmholtz free-energy functional and the scaled coefficients 
\begin{equation}
A'_{1}=\frac{1}{2}\:\!(1-A_{1})\;,\qquad A'_{2}=-\frac{1}{2}A_{2}\;,\qquad A'_{3}=-\frac{1}{2}A_{3}
\label{eq:EM_As}%
\end{equation}
are used for abbreviation, where the coefficient $A'_{2}$ should be positive in order to favour non-uniform phases and the last coefficient 
$A'_{3}$ is assumed to be positive for stability reasons. By comparison of equation \eqref{eq:HFEF} with the original PFC model \eqref{eq0}, 
that was initially proposed on the basis of general symmetry considerations in reference \cite{ElderKHG2002}, 
analytic expressions can be assigned to the unknown coefficients in the original PFC model: 
when we write the order-parameter field in equation \eqref{eq0} as $\psiSH(\vec{r})=\alpha(1-2\psi(\vec{r}))$ 
with a constant $\alpha$ and neglect constant contributions as well as terms linear in $\psi(\vec{r})$ in the free-energy density, 
we obtain the relations  
\begin{equation}
\alpha=\frac{1}{\sqrt{24 A'_{3}}}\;,\quad
\FSH=\frac{1}{12\:\!\rhoR\:\!\kBT A'^{2}_{3}}\mathcal{F}\;,\quad 
\beta=\frac{1}{8 A'_{3}}-\frac{A'_{1}}{A'_{3}}+\frac{A'^{2}_{2}}{4 A'^{2}_{3}}\;,\quad 
\wn_{0}=\sqrt{\frac{A'_{2}}{2 A'_{3}}}  
\end{equation}
between the coefficients in equations \eqref{eq0} and \eqref{eq:HFEF}.

\subsubsection{\label{subsubsec:DE_PFC}Dynamical equations} 
We turn to the dynamics of the PFC model and derive it here from DDFT. Inserting the representation \eqref{eq:prho} for the 
one-particle density field into the DDFT equation \eqref{eq:DDFTs}, we obtain for the dynamical evolution of the order-parameter field 
$\psi(\vec{r},t)$
\begin{equation}
\pdif{\psi(\vec{r},t)}{t}=D_{\mathrm{T}}\Nabla\!\cdot\!\bigg(\!(1+\psi)\Nabla\bigg(2A'_{1}\psi 
+ 2A'_{2}\Nabla^{2}\psi + 2A'_{3}\Nabla^{4}\psi - \frac{\psi^{2}}{2} + \frac{\psi^{3}}{3}\bigg)\!\!\bigg)\,.
\label{eq:EM_Dynamik}%
\end{equation}
This dynamical equation (called PFC1 model in reference \cite{vanTeeffelenBVL2009}) still differs from the original dynamical equation of the PFC model. 
The latter can be gained by a further constant-mobility approximation (CMA), where the space- and time-dependent mobility 
$D_{\mathrm{T}}\rho(\vec{r},t)$ in the DDFT equation is replaced by the constant mobility $D_{\mathrm{T}}\rhoR$. 
The resulting dynamical equation (called PFC2 model in reference \cite{vanTeeffelenBVL2009}) coincides with the original PFC dynamics given by
\begin{equation}
\pdif{\psi(\vec{r},t)}{t}=D_{\mathrm{T}}\Nabla^{2}\bigg(2A'_{1}\psi + 2A'_{2}\Nabla^{2}\psi 
+ 2A'_{3}\Nabla^{4}\psi - \frac{\psi^{2}}{2} + \frac{\psi^{3}}{3}\bigg)
\label{eq:EM_Dynamik_CMA}%
\end{equation}
for the time-dependent translational density $\psi(\vec{r},t)$. We remark that this dynamical equation can also be derived from an 
equivalent dissipation functional $\mathfrak{R}$ known from linear irreversible thermodynamics \cite{deGrootM1984,MartinPP1972,Reichl1998}.
A further transformation of this equation to the standard form of the dynamic PFC model will be established in section \ref{sec:scpfcm}.

\subsubsection{Colloidal crystal growth: DDFT versus PFC modelling} 
\begin{figure}
\centering\includegraphics[width=0.575\linewidth]{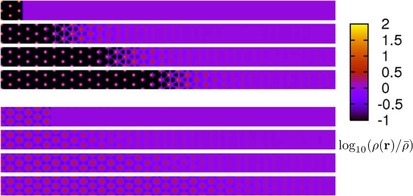}%
\caption{\label{fig:TBVLa}Colloidal crystal growth within DDFT (upper panel) and the PFC1 model (lower panel). 
The crystallization starts with an initial nucleus of $5$ and $11$ rows of hexagonally crystalline particles, respectively. 
The density field of the growing crystal is shown at times $t/\tau_{\mathrm{B}}=0,0.5,1,1.5$. 
\nocite{vanTeeffelenBVL2009}\CRPE{S.\ van Teeffelen, R.\ Backofen, A.\ Voigt, and H. L\"owen}{Derivation of the phase-field-crystal model for colloidal solidification}{Phys.\ Rev.\ E}{79}{2009}{051404}{5}{10.1103/PhysRevE.79.051404}{the American Physical Society}} 
\end{figure}

\begin{figure}
\centering\includegraphics[width=0.5\linewidth]{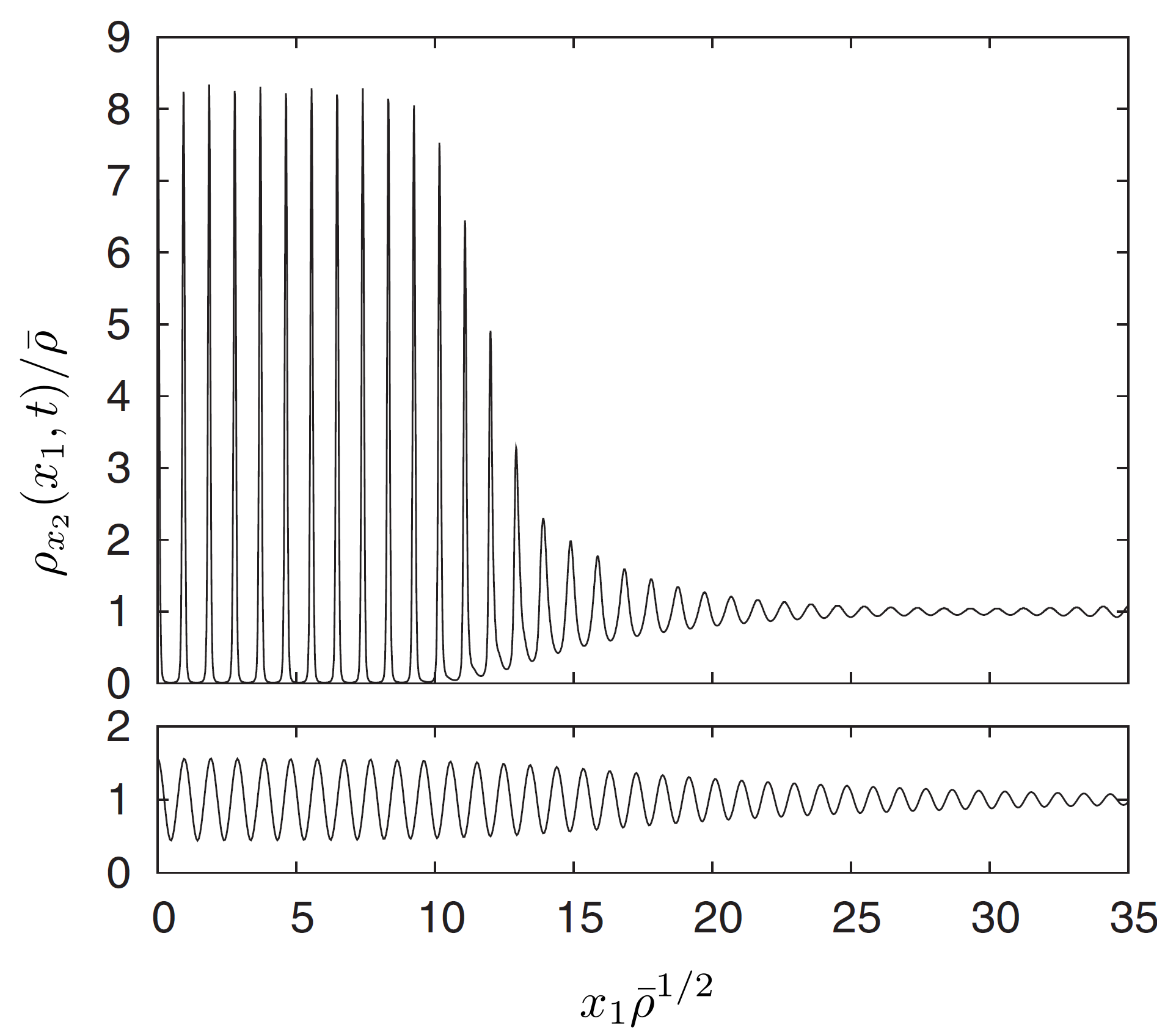}%
\caption{\label{fig:TBVLb}Comparison of DDFT (upper panel) and PFC1 (lower panel) results. 
For an analogous situation as in figure \ref{fig:TBVLa}, this plot shows the laterally averaged density 
$\rho_{x_{2}}(x_{1},t)=\langle\rho(\vec{r},t)\rangle_{x_{2}}$ at $t=\tau_{\mathrm{B}}$. 
\nocite{vanTeeffelenBVL2009}\CRPE{S.\ van Teeffelen, R.\ Backofen, A.\ Voigt, and H. L\"owen}{Derivation of the phase-field-crystal model for colloidal solidification}{Phys.\ Rev.\ E}{79}{2009}{051404}{5}{10.1103/PhysRevE.79.051404}{the American Physical Society}} 
\end{figure}

\begin{figure}
\centering\includegraphics[width=0.5\linewidth]{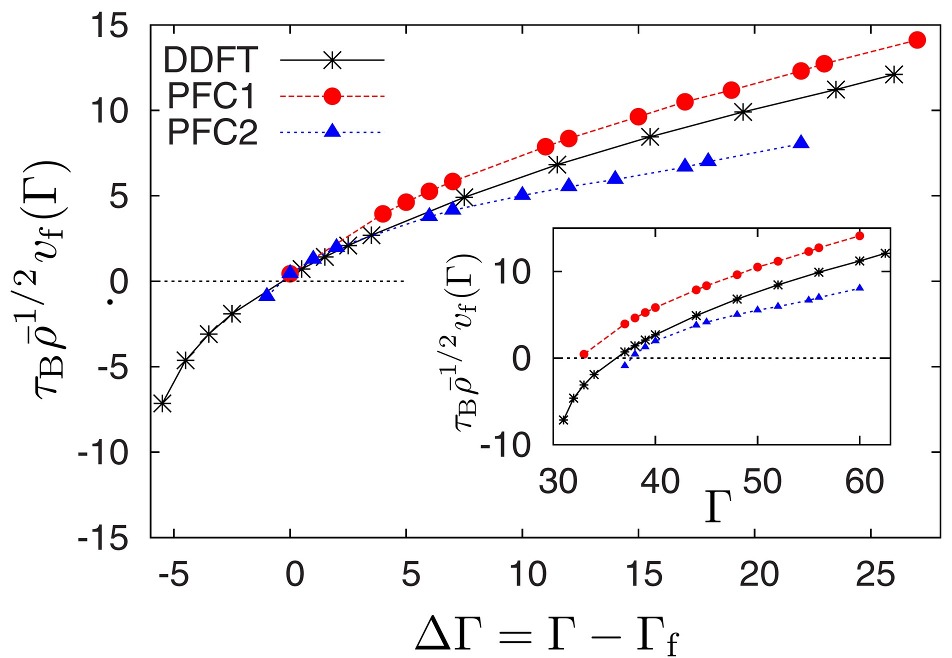}%
\caption{\label{fig:TBVLc}Comparison of DDFT, the PFC1 model, and the PFC2 model \cite{vanTeeffelenBVL2009}.
The plot shows the velocity $v_{\mathrm{f}}$ of a crystallization front in the $(11)$-direction in dependence of the relative 
coupling constant $\Delta\Gamma=\Gamma-\Gamma_{\mathrm{f}}$ with the total coupling constant $\Gamma$ and the 
coupling constant of freezing $\Gamma_{\mathrm{f}}$. 
In the inset, the velocity $v_{\mathrm{f}}$ is shown in dependence of $\Gamma$. 
\nocite{vanTeeffelenBVL2009}\CRPE{S.\ van Teeffelen, R.\ Backofen, A.\ Voigt, and H. L\"owen}{Derivation of the phase-field-crystal model for colloidal solidification}{Phys.\ Rev.\ E}{79}{2009}{051404}{5}{10.1103/PhysRevE.79.051404}{the American Physical Society}} 
\end{figure}

Results of the PFC1 model, the PFC2 model, and DDFT are compared for colloidal crystal growth in reference \cite{vanTeeffelenBVL2009}. 
Figures \ref{fig:TBVLa}-\ref{fig:TBVLc} show the differences for the example of a growing crystal front starting at the edge of a prescribed 
hexagonal crystal. The underlying colloidal systems are the same as in section \ref{par:CCG}. In figure \ref{fig:TBVLa}, the time evolution of the 
one-particle density is shown for DDFT and for the PFC1 model. The PFC2 model leads to results very similar to those for the PFC1 model and is 
therefore not included in this figure. 

Two main differences in the results of DDFT and of the PFC1 model are obvious: first, the density peaks are much higher and narrower in the 
DDFT results than for the PFC1 model. While these peaks can be approximated by Gaussians in the case of DDFT, they are much broader 
sinusoidal modulations for the PFC1 model. Secondly, also the width of the crystal front obtained within DDFT is considerably smaller than in 
the PFC approach.  

These qualitative differences can also be observed in figure \ref{fig:TBVLb}. There, the laterally averaged density   
$\rho_{x_{2}}(x_{1},t)=\langle\rho(\vec{r},t)\rangle_{x_{2}}$ associated with the plots in figure \ref{fig:TBVLa} is shown, 
where $\langle\,\cdot\,\rangle_{x_{2}}$ denotes an average with respect to $x_{2}$. A further comparison of DDFT and the PFC approaches is 
possible with respect to the velocity $v_{\mathrm{f}}$ of the crystallization front. The corresponding results are shown in figure \ref{fig:TBVLc} 
in dependence of the total coupling constant $\Gamma=u_{0}v^{3/2}/(\kBT)$ and the relative coupling constant 
$\Delta\Gamma=\Gamma-\Gamma_{\mathrm{f}}$, where $\Gamma_{\mathrm{f}}$ denotes the coupling constant of freezing. 
Due to the power-law potential of the considered colloidal particles, their behaviour is completely characterized by the dimensionless coupling 
parameter $\Gamma$. When plotted versus $\Delta\Gamma$, the growth velocity of the PFC1 model is in slightly better agreement than that of the 
PFC2 model. 

\section{\label{chap:III}Phase-field-crystal modelling in condensed matter physics}
The original PFC model has the advantage over most other microscopic techniques, such as molecular-dynamics (MD) simulations, 
that the time evolution of the system can be studied on the diffusive time scale making the long-time behaviour and the large-scale structures 
accessible \cite{ElderKHG2002,ElderG2004}. 
As already outlined in section \ref{subsec:DDFT}, we note that the diffusion-controlled relaxation dynamics the original PFC model 
assumes is relevant for micron-scale colloidal particles in carrier fluid \cite{vanTeeffelenBVL2009,vanTeeffelenLL2008}, where the self-diffusion 
of the particles is expected to be the dominant way of density relaxation. For normal liquids, the hydrodynamic mode of density relaxation is 
expected to dominate. The modified PFC (MPFC) model introduces linearised hydrodynamics, realized via incorporating a term proportional to the 
second time derivative of the particle density into the EOM \cite{StefanovicHP2006,PhysRevE.79.051110}, yielding a 
two-time-scale density relaxation: a fast acoustic process in addition to the long time diffusive relaxation of the original PFC model. 
A three-time-scale extension incorporates phonons into the PFC model \cite{PhysRevB.75.054301,EurPhysJB.66.329}. Another interesting group of 
models have been obtained by coarse-graining the PFC approaches \cite{PhysRevE.72.020601,PhysRevE.74.011601,PhysRevE.76.056706}, leading to 
equations of motion that describe the spatiotemporal evolution of the Fourier amplitudes and the respective phase information characterizing the 
particle density field. Combined with adaptive grid schemes, the amplitude equation models are expected to become a numerically especially efficient 
class of the PFC models of crystallization \cite{PhysRevE.76.056706}.

Finally, we address here recent advances in the modelling of molecules or liquid crystalline systems, which are composed of anisotropic particles. 
There is a large number of molecular and colloidal realizations of these non-spherical particles. The simplest non-spherical shape is 
rotationally symmetric about a certain axis (like rods, platelets, and dumbbells) and is solely described by an additional orientation vector. 
Liquid crystalline systems show an intricate freezing behaviour in equilibrium, where mesophases occur, that can possess both orientational 
and translational ordering. Here, we show that the microscopic DFT approach for liquid crystals provides an excellent starting point to derive 
PFC-type models for liquid crystals. This gives access to the phase diagram of liquid crystalline phases and to their dynamics promising a 
flourishing future to predict many fundamentally important processes on the microscopic level.

\subsection{\label{gaeopfcm}The original PFC model and its generalisations}
The original PFC model has several equivalent formulations and extensions that we review in this section. 
We first address the single-component PFC models. Then, an overview of their binary generalisations will be given. 
In both cases, complementing section \ref{chap:II}, we start with presenting different forms of the free-energy functional, 
followed by a summary of specific forms of the EOM and of the Euler-Lagrange equation (ELE). 
Finally, we review the numerical methods applied for solving the EOM and ELE as well as various approaches for the amplitude equations.

\subsubsection{\label{sec:scpfcm}Single-component PFC models}

\paragraph{\label{sec:tfe}The free energy} 

\subparagraph{The single-mode PFC model} 
The earliest formulation of the single-mode PFC (1M-PFC) model \cite{ElderKHG2002,ElderG2004} has been derived as a 
SH model with conserved dynamics to incorporate mass conservation. Accordingly, the dimensionless free energy of the 
heterogeneous system is given by the usual SH expression \eqref{eq0}. 
We note that in equations \eqref{eq0} and \eqref{eq8} the analogous quantities differ by only appropriate numerical factors 
originating from the difference in the length scales.

As already outlined in section \ref{sec:PFC_Isotropic}, the free energy of the earliest and simplest PFC model \cite{ElderKHG2002} 
has been re-derived \cite{ElderPBSG2007} from that of the 
perturbative DFT of Ramakrishnan and Yussouff \cite{RamakrishnanY1979}, in which the free-energy difference 
$\Delta\mathcal{F}=\mathcal{F}-\Fq$\footnote{To keep the notation simple, we ignore $\Fq$ and write $\mathcal{F}$ instead of $\Delta\mathcal{F}$ 
throughout this article.} of the crystal relative to a reference liquid of particle density $\rhoq$ and free energy $\Fq$ is 
expanded with respect to the local density difference $\Delta\rho(\vec{r})=\rho(\vec{r})-\rhoq$, 
while retaining the terms up to the two-particle term (see section \ref{sec:fef}):
\begin{equation}
\begin{split}
\frac{\mathcal{F}}{\kBT} = & \vint \bigg( \rho \ln\!\bigg(\frac{\rho}{\rhoq }\bigg) - \Delta\rho \bigg) 
- \frac{1}{2} \vintI{1}\!\!\!\vintI{2} \Delta\rho(\mathbf{r}_1) \cII(\mathbf{r}_1,\mathbf{r}_2)\Delta\rho(\mathbf{r}_2)  
+ \dotsb
\end{split}
\label{eq2}
\end{equation}
Fourier expanding the particle density, one finds that for the 
solid $\rho_{\mathrm{s}} = \rhoq\, ( 1 + \eta_{\mathrm{s}} + \sum_\mathbf{K} A_\mathbf{K} \exp(\ii\mathbf{K}\cdot\vec{r}) )$, 
where $\eta_{s}$ is the fractional density change upon freezing, while $\mathbf{K}$ are reciprocal lattice vectors (RLVs)    
and $A_\mathbf{K}$ the respective Fourier amplitudes. 
Introducing the reduced number density  
$\psi = (\rho - \rhoq )/\rhoq = \eta_{\mathrm{s}} + \sum_\mathbf{K} A_\mathbf{K} \exp(\ii\mathbf{K}\cdot\vec{r})$ 
one obtains
\begin{equation}
\begin{split}
\frac{\mathcal{F}}{\rhoq\,\kBT} = & \vint \big( (1+\psi)\ln(1+\psi)-\psi \big) \\
& - \frac{\rhoq}{2} \vintI{1}\!\!\!\vintI{2} \psi(\mathbf{r}_1)\cII(\norm{\mathbf{r}_1-\mathbf{r}_2})\psi(\mathbf{r}_2)  
+ \dotsb
\end{split}
\label{eq3}
\end{equation}
Expanding next $\cII(\norm{\mathbf{r}_1 - \mathbf{r}_2})$ in Fourier space, $\cIIw(k) \approx \cIIw_0 + \cIIw_2 k^2 + \cIIw_4 k^4 + \dotsb$, 
where $\cIIw(k)$ has its first peak at $k = 2\pi/\ipd$, the signs of the coefficients alternate. 
(Here, $\ipd$ is the inter-particle distance.) Introducing the dimensionless two-particle direct correlation function 
$c(k)=\rhoq \,\cIIw(k) \approx \sum_{j=0}^m c_{2j}k^{2j} = \sum_{j=0}^m b_{2j}(k \ipd)^{2j}$, which is related to the structure factor 
as $S(k)=1/(1-c(k))$, and integrating the second term on the right-hand-side of equation \eqref{eq3} with respect to $\mathbf{r}_2$ and finally replacing 
$\mathbf{r}_1$ by $\mathbf{r}$, the free-energy difference can be rewritten as
\begin{equation}
\frac{\mathcal{F}}{\rhoq\,\kBT}  \approx \vint \bigg( \!(1+\psi)\ln(1+\psi)-\psi
-\frac{\psi}{2}\bigg(\sum_{j=0}^m (-1)^j c_{2j} \Nabla^{2j} \bigg)\psi \bigg) \,.
\label{eq4}
\end{equation}
The reference liquid is not necessarily the initial liquid. Thus, we have here two parameters to control the driving force for solidification: 
the initial liquid number density $\nnq$ (corresponding to a reduced initial density of $\psi_0$) and the temperature $T$, 
if the direct correlation function depends on temperature. 
Taylor-expanding $\ln(1 + \psi)$ for small $\psi$ one obtains
\begin{equation}
\frac{\mathcal{F}}{\rhoq\,\kBT} \approx \vint \Bigg(\frac{\psi^2}{2}-\frac{\psi^3}{6}+\frac{\psi^4}{12}
-\frac{\psi}{2}\Bigg( \sum_{j=0}^m (-1)^j c_{2j} \Nabla^{2j} \Bigg)\psi \Bigg) \,.
\label{eq5}
\end{equation}
For $m = 2$, corresponding to the earliest version of the PFC model \cite{ElderKHG2002}, and taking the alternating sign of the 
expansion coefficients of $\cIIw_i$ into account, equation \eqref{eq4} transforms to the following form:
\begin{equation}
\frac{\mathcal{F}}{\rhoq\,\kBT} \approx \vint \bigg( \frac{\psi^2}{2}(1+\abs{b_0}) 
+ \frac{\psi}{2} \big( \abs{b_2}\ipd^2\Nabla^2 + \abs{b_4}\ipd^4\Nabla^4 \big) \psi 
-\frac{\psi^3}{6} + \frac{\psi^4}{12} \bigg) \,.
\label{eq6}
\end{equation}
Introducing the new variables
\begin{center}
\begin{tabular}{ll}
$B_{\mathrm{l}} = 1+\abs{b_0} = 1-c_0$ & [$=(1/\kappa)/(\rhoq\,\kB T)$, \small{where $\kappa$ is the compressibility}],\\
$B_{\mathrm{s}} = \abs{b_2}^2/(4\abs{b_4})$ & [$=K/(\rhoq\,\kB T)$, \small{where $K$ is the bulk modulus}],\\
$R = \ipd (2\abs{b_4}/\abs{b_2})^{1/2}$ & [$=$ \small{the new length scale $(x=R\tilde{x})$, which is now related to} \\ 
& \small{the position of the maximum of the Taylor expanded $\cIIw(k)$}],
\end{tabular}
\end{center}
and a multiplier $v$ for the $\psi^3$-term (that accounts for the zeroth-order contribution from three-particle correlations), 
one obtains the form used by Berry \etal\ \cite{PhysRevB.77.224114,PhysRevE.77.061506}:
\begin{equation}
\begin{split}
\mathcal{F} &= \vint f[\psi] \\
& = \rhoq\,\kBT \!\vint \bigg( \frac{\psi}{2} \big( B_{\mathrm{l}} + B_{\mathrm{s}}\big( 2R^2\Nabla^2 + R^4\Nabla^4\big) \big) \psi 
-v \frac{\psi^3}{6} + \frac{\psi^4}{12} \bigg) \,.
\end{split}
\label{eq7}
\end{equation}
Here, $f[\psi]$ denotes the full (dimensional) free-energy density.

\textit{The Swift-Hohenberg-type dimensionless form.} 
Introducing a new set of variables, 
$x=R\tilde{x}$, $\psi = (3B_{\mathrm{s}})^{1/2}\psiSH$, $\mathcal{F} = (3\rhoq\,\kBT R^d B_{\mathrm{s}}^2) \tilde{\mathcal{F}}$, 
where $d$ is the number of spatial dimensions, the free energy can be transcribed into the following dimensionless form:
\begin{equation}
\tilde{\mathcal{F}} = \vintw\, \bigg( \frac{\psiSH}{2} \big(\!-\epsilon + \big(1+\NablaW^2\big)^2\big) \psiSH 
+ p \frac{\psiSH^3}{3} + \frac{\psiSH^4}{4} \bigg) \,.
\label{eq8}
\end{equation}
Here, $p = -(v/2)(3/B_{\mathrm{s}})^{1/2} = -v(3\abs{b_4}/\abs{b_2}^2)^{1/2}$ and 
$\epsilon = -\Delta B/B_{\mathrm{s}} = -((1 + \abs{b_0})/(\abs{b_2}^2/(4\abs{b_4}))-1)$, 
while $\psiSH = \psi/(3B_{\mathrm{s}})^{1/2}$. The quantities involved in equation \eqref{eq8} are all dimensionless. 
Using the appropriate length unit, this expression becomes equivalent to equation \eqref{eq0} for $p=0$. 

Equation \eqref{eq8} suggests that the $m = 2$ PFC model contains only two dimensionless similarity parameters, 
$\epsilon$ and $p$, composed of the original model parameters. 
We note finally that even the third-order term can be eliminated. In the respective $p' = 0$ SH model \eqref{eq0}, 
the state $(\epsilon' = \epsilon + p^2/3, \psi' = \psiSH + p/3)$ corresponds to the state $(\epsilon,\psiSH)$ of the original $p \neq 0$ model. 
This transformation leaves the grand canonical potential difference, the ELE (see section \ref{sec:TELE}), 
and the EOM (see section \ref{sec:TEOM}) invariant. 
Accordingly, it is sufficient to address the case $p = 0$.

We stress here that in these models the approximation for the two-particle direct correlation function leads to a well defined wavelength 
of the density waves the system tends to realize (hence the name \ZT{single-mode PFC} (1M-PFC) model). Accordingly, any periodic density distribution 
that honours this wavelength represents a local minimum of the free energy. Indeed, the 1M-PFC model has stability domains for the 
bcc, fcc, and hcp structures (see section \ref{sec:Atctcs}). Furthermore, elasticity and crystal anisotropies are automatically present in the model.

Model parameters of the SH formulation have been deduced to fit to the properties of bcc Fe by Wu and Karma \cite{PhysRevB.76.184107}.

\subparagraph{The two-mode PFC model} An attempt has been made to formulate a free energy that prefers the fcc structure at small $\epsilon$ 
values \cite{PhysRevE.81.061601}, where a linear elastic behaviour persists. To achieve this, two well defined wavelengths were used 
(first and second neighbour RLVs), hence the name \textit{\ZT{two-mode PFC}} (2M-PFC) model. The respective free-energy functional contains two 
new parameters:
\begin{equation}
\tilde{\mathcal{F}} = \vintw\, \bigg( \frac{\psiSH}{2} \Big( \!-\epsilon 
+ \big(1+\NablaW^2\big)^2 \big( R_1 + \big(\wn^2_{\mathrm{rel}} + \NablaW^2\big)^2 \big)\!\Big)\psiSH + \frac{\psiSH^4}{4} \bigg) \,.
\label{eq9}
\end{equation}
Here, $R_1$ controls the relative stability of the fcc and bcc structures, while $\wn_{\mathrm{rel}}$ is the ratio of the two wave numbers 
($\wn_{\mathrm{rel}} = 2/\sqrt{3}$ for fcc using the $(111)$ and $(200)$ RLVs). Remarkably, the 1M-PFC model can be recovered for $R_1 \to \infty$. 
Model parameters have been deduced to fit to the properties of fcc Fe by Wu and Karma \cite{PhysRevE.81.061601}.

We note finally that the 1M-PFC and 2M-PFC models can be cast into a form that interpolates between them by varying a single parameter 
$\lambda = R_1/(1 + R_1) \in [0, 1]$ as follows \cite{PhysRevLett.107.175702}: 
\begin{equation}
\tilde{\mathcal{F}} = \vintw\, \bigg( \frac{\psiSH}{2} \Big(\!-\epsilon 
+ \big(1+\NablaW^2\big)^2 \big( \lambda +(1-\lambda) \big(\wn^2_{\mathrm{rel}} + \NablaW^2\big)^2 \big)\!\Big)\psiSH + \frac{\psiSH^4}{4} \bigg) \,. 
\label{eq10}
\end{equation}
This expression recovers the 1M-PFC model limit for $\lambda = 1$ ($R_1 \to \infty$).

\subparagraph{The eighth-order fitting PFC model} 
To approximate real bcc materials better, an eighth-order expansion of the Fourier transform of the direct correlation function around its 
maximum ($k = k_{\mathrm{m}}$) has been performed recently, leading to what is termed the eighth-order fitting version of the 
phase-field-crystal (EOF-PFC) model \cite{JaatinenAEAN2009}:
\begin{equation}
\cIIw(k) \approx \cIIw(k_{\mathrm{m}}) - \Gamma \bigg( \frac{k^2}{k_{\mathrm{m}}^2}-1 \bigg)^2 
- E_{\mathrm{B}} \bigg( \frac{k^2}{k_{\mathrm{m}}^2}-1 \bigg)^4 \,. 
\label{eq11}
\end{equation}
The expansion parameters were then fixed so that the position, height, and the second derivative of $\cIIw(k)$ are accurately recovered. 
This is ensured by
\begin{equation}
\Gamma = - \frac{k_{\mathrm{m}}^2 (\cIIw)''(k_{\mathrm{m}})}{8} \quad \text{and} \quad E_{\mathrm{B}} 
= \cIIw(k_{\mathrm{m}})-\cIIw(0)-\Gamma \;.
\label{eq12}
\end{equation}
With this choice of the model parameters and using relevant data for Fe from reference \cite{PhysRevB.76.184107}, they reported a fair agreement 
with MD simulation results for the volume change upon melting, the bulk moduli of the liquid and solid phases, 
and for the magnitude and anisotropy of the bcc-liquid interfacial free energy \cite{JaatinenAEAN2009}.

\subparagraph{\label{sec:Atctcs}Attempts to control the crystal structure in PFC models} 
Greenwood and co-workers \cite{GreenwoodPR2010,PhysRevE.83.031601} (GRP-PFC model) have manipulated the two-particle direct correlation function so that its peaks 
prefer the desired structural correlations -- an approach that enables them to study transitions between the bcc, fcc, hcp, and sc structures.
Wu, Plapp, and Voorhees \cite{JPhysCondMat.22.364102} have investigated the possibility to control crystal symmetries within the PFC method 
via tuning nonlinear resonances. They have proposed a general recipe for developing free-energy functionals that realize coexistence between the 
liquid and  periodic phases of desired crystal symmetries, and have illustrated this via presenting a free-energy functional that leads to 
square-lattice-liquid coexistence in 2D. A possible extension of the method to the 3D case for simple cubic (sc) structures has also been discussed.

\subparagraph{The vacancy PFC model} 
The vacancy PFC (VPFC) model is an important extension of the PFC model that adds a term to the free energy that penalizes the negative values 
of the particle density, allowing thus for an explicit treatment of vacancies \cite{PhysRevE.79.035701}:
\begin{equation}
\tilde{\mathcal{F}} = \vintw\, \bigg( \frac{\psiSH}{2} \Big(\! -\beta + \big(\wn^{2}_{0}+\NablaW^2\big)^2\Big)\psiSH + \frac{\psiSH^4}{4} 
+ h\big(\abs{\psiSH^3}-\psiSH^3\big) \bigg) \,. 
\label{eq13}
\end{equation}
Here, $h$ in the last term on the right-hand-side is a constant. The new term is a piecewise function that is zero for $\psiSH > 0$ and 
positive for $\psiSH < 0$. It is then possible to obtain a mixture of density peaks (particles) and vacant areas (where $\psiSH \approx 0$), 
resembling thus to snapshots of liquid configurations or crystalline structures with defects. This allows structural modelling of the fluid phase 
and is an important step towards combining the PFC model with fluid flow. The same approach has been used to address the dynamics of 
glasses \cite{PhysRevLett.106.175702}.

\subparagraph{The anisotropic PFC model}
Recently, Prieler \etal\ \cite{JPhysCondMat.21.464110} have extended the PFC approach  
by replacing the Laplacian in equation \eqref{eq0} by more general differential operators allowing spatial 
anisotropy. Doing so and setting $\tau=-(\wn_0^2-\beta)$ one arrives at the dimensionless free-energy functional 
of the so-called anisotropic PFC (APFC) model: 
\begin{eqnarray}
\tilde{\mathcal{F}}& = &\vintw\, \bigg( \frac{\psiSH}{2} 
\bigg( \!-\tau + a_{ij} \frac{\partial^2}{\partial \tilde{x}_i \partial \tilde{x}_j} 
+b_{ijkl}\frac{\partial^4}{\partial \tilde{x}_i \partial \tilde{x}_j\partial \tilde{x}_k \partial \tilde{x}_l}\bigg)\psiSH 
+ \frac{\psiSH^4}{4} \bigg) \,. 
\label{eq:APFC}
\end{eqnarray}
Here, $a_{ij}$ is a symmetric matrix and $b_{ijkl}$ is a tensor of rank $4$ with the 
symmetry of an elastic tensor: $i\leftrightarrow j, k\leftrightarrow l, (i,j)\leftrightarrow (k,l)$ \cite{JPhysCondMat.21.464110}.  
Choudhary \etal\ \cite{ChoudharyLEL2011,ChoudharyKE2012} proved that based on a functional of the form \eqref{eq:APFC} further 
crystal lattices can be assessed as hexagonal, bcc, and corresponding sheared structures, for which 
they have presented the elastic parameters and identified the stationary states.

\paragraph{\label{sec:TEOM}The equation of motion} 
In the PFC models, different versions of the EOM have been employed. In all cases, conservative dynamics is assumed on the ground that 
mass conservation needs to be satisfied. (The original SH model differs from the 1M-PFC model only in the EOM, 
for which the SH model assumes non-conserved dynamics.) Most of the PFC models rely on an overdamped conservative EOM 
(see, \eg, references \cite{ElderKHG2002,ElderPBSG2007,JPhysCondMat.20.404205,Tang2011146,tegze.phd.thesis,JPhysCondMat.22.364101,
PhysRevE.81.011602,JPhysCondMat.22.364104,PhilosMag.91.123,JPhysCondMat.21.464108,JPhysCondMat.21.464110,
PhysRevLett.108.025502,PhysRevB.77.224114,PhysRevB.78.184110,C0SM00944J,PhysRevLett.106.195502,JPhysCondMat.21.464109,
PhysRevE.77.061506,PhysRevE.81.061601,ElderG2004,vanTeeffelenBVL2009,PhysRevLett.107.175702}). 
In fact, this means that the particle density relaxes diffusively, a feature more characteristic to colloidal systems than to molten metals: 
in colloidal systems of particles floating in a carrier fluid, Brownian motion is the dominant mechanism of particle motion, 
whose properties are captured reasonably well by overdamped dynamics \cite{vanTeeffelenBVL2009,vanTeeffelenLL2008,C0SM00944J} 
(see also section \ref{subsec:DDFT}). 
On the contrary, in molten metals a density deficit can be reduced by a hydrodynamic flow of particles. 
Apparently, a proper dynamics of the solid (presence of phonons) requires three time scales \cite{PhysRevB.75.054301,EurPhysJB.66.329}. 
While the overdamped model has only the long diffusive time scale, the MPFC model realizes linearised hydrodynamics via adding 
a term proportional to the second time derivative of the density field. This new term leads to the appearance of an acoustic relaxation of 
the density (not true acoustic phonons) on a fast time scale, in addition to the slow diffusive relaxation at later stages 
\cite{StefanovicHP2006,PhysRevE.79.051110}. In a recent work, phonon dynamics, that acts on a third time scale, 
has also been introduced into the PFC model \cite{PhysRevB.75.054301}. 
It has been shown that there exists a scale window, in which the longitudinal part of the full three-scale model reduces to the MPFC model, 
whereas the linearised hydrodynamics of the latter converges to the diffusive dynamics of the original PFC model for sufficiently long  
times \cite{PhysRevB.75.054301,EurPhysJB.66.329}.

\subparagraph{The overdamped equation of motion} 
In the majority of the PFC simulations, an overdamped conserved dynamics is assumed [that is analogous to the DDFT EOM \eqref{eq:DDFTs} for 
colloidal systems, however, assuming here a constant mobility coefficient, $M_\rho = \rhoM\Drho/(\kBT)$]. Accordingly, the (dimensional) 
EOM has the form
\begin{equation}
\frac{\partial\rho}{\partial t} 
= \Nabla\!\cdot\!\bigg( M_\rho\:\! \Nabla \frac{\delta \mathcal{F}}{\delta\rho} \bigg) + \zeta_\rho \;, 
\label{eq14}
\end{equation}
where $\zeta_\rho$ stands for the fluctuations of the density flux, whose correlator reads as 
$\langle\zeta_\rho(\mathbf{r},t)\zeta_\rho(\mathbf{r}',t') \rangle=-2M_\rho \kBT \Nabla^2 \delta(\mathbf{r}-\mathbf{r}') \delta(t-t')$. 
(For a discretised form of the conserved noise see reference \cite{PhysRevE.60.3614}.)

Changing from variable $\rho$ to $\psi$, introducing $M_\psi = ((1+\psi_0) \Drho /(\rhoq\,\kBT))$, scaling time and 
distance as $t=\tau\tilde{t}$ and $x=\ipd\tilde{x}$, where $\tau = \ipd /(\Drho (1+ \psi_0))$, 
and inserting the free energy from equation \eqref{eq6}, one obtains the following dimensionless EOM:
\begin{equation}
\frac{\partial \psi}{\partial \tilde{t}} =  \NablaW^2 \bigg( \psi(1+\abs{b_0}) + \sum_{j=1}^m \abs{b_{2j}} \NablaW^{2j} \psi  
- \frac{\psi^2}{2} + \frac{\psi^3}{3} \bigg) + \zeta_\psi 
\label{eq15}
\end{equation}
with $\langle\zeta_\psi(\tilde{\mathbf{r}},\tilde{t})\zeta_\psi(\tilde{\mathbf{r}}',\tilde{t}')\rangle
=-(2/(\rhoq \ipd^d)) \NablaW^2 \delta(\tilde{\mathbf{r}}-\tilde{\mathbf{r}}') \delta(\tilde{t}-\tilde{t}')$. 
Analogously, the EOM corresponding to equation \eqref{eq7} has the form
\begin{equation}
\frac{\partial \psi}{\partial t} = \Nabla\!\cdot\!\bigg( M_\psi\:\! \rhoq\,\kBT\:\! \Nabla\bigg(\!  
\big( B_{\mathrm{l}} + B_{\mathrm{s}}\big(R^2\Nabla^2+R^4\Nabla^4\big)\big)\psi 
-v\frac{\psi^2}{2}+\frac{\psi^3}{3} \bigg)\! \bigg) + \zeta'_\psi \;,
\label{eq16}
\end{equation}
where $\langle\zeta'_\psi(\mathbf{r},t)\zeta'_\psi(\mathbf{r}',t')\rangle=-2M_\psi \kBT \Nabla^2 \delta(\mathbf{r}-\mathbf{r}') \delta(t-t')$.

\textit{Dimensionless form in Swift-Hohenberg fashion.} 
Introducing the variables $t = \tau \tilde{t}$, $x = R \tilde{x}$, 
and $\psi = (3B_{\mathrm{s}})^{1/2} \psiSH =(3B_{\mathrm{s}})^{1/2}(\psi' - p/3)$ into equation \eqref{eq16}, 
where $\tau = R^2/(B_{\mathrm{s}} M_\psi \rhoq\,\kBT)$, 
the EOM can be written in the form
\begin{equation}
\frac{\partial \psi'}{\partial \tilde{t}} = \NablaW^2 \Big(\!\Big(\!-\epsilon' + \big(1+\NablaW^2\big)^2 \Big)\psi' 
+ \psi'^3 \Big) + \zeta \;, 
\label{eq17}
\end{equation}
where $\epsilon'=\epsilon+p^2/3 = -(\Delta B - (v/2)^2)/B_{\mathrm{s}}=-((1+\abs{b_0})/(\abs{b_2}^2/(4\abs{b_4}))-(1+v^2(\abs{b_4}/\abs{b_2}^2)))$ 
and the dimensionless noise strength is 
$\alpha=2/(3B_{\mathrm{s}}^2\rhoq R^d)=2^{5-d/2}\abs{b_4}^{2-d/2}/(3\ipd^d\rhoq\, \abs{b_2}^{4-d/2})$, 
while the correlator for the dimensionless noise reads as 
$\langle \zeta(\tilde{\mathbf{r}},\tilde{t})\zeta(\tilde{\mathbf{r}}',\tilde{t}') \rangle=-\alpha\NablaW^2\delta(\tilde{\mathbf{r}}
-\tilde{\mathbf{r}}')\delta(\tilde{t}-\tilde{t}')$.

Summarizing, the dynamical $m=2$ 1M-PFC model has two dimensionless similarity parameters $\epsilon'$ and $\alpha$ composed of the 
original (physical) model parameters. This is the generic form of the $m = 2$ 1M-PFC model; some other formulations 
\cite{PhysRevB.77.224114,PhysRevE.77.061506} can be transformed into this form.

\subparagraph{The modified PFC model} 
Acoustic relaxation has been partly incorporated by applying an underdamped EOM. 
Stefanovic, Haataja, and Provatas \cite{StefanovicHP2006} have incorporated a second order time derivative into the EOM of their 
modified PFC (MPFC) model, which extends the previous PFC formalism by generating dynamics on two time scales:
\begin{equation}
\frac{\partial^2\rho}{\partial t^2}+\kappa\frac{\partial\rho}{\partial t}=\lambda^2\frac{\delta\mathcal{F}}{\delta\rho} \;,
\label{eq18}
\end{equation}
where $\kappa$ and $\lambda$ are constants. 
At early times, molecular positions relax fast, consistently with elasticity theory, whereas at late times 
diffusive dynamics dominates the kinetics of phase transformations, the diffusion of vacancies, 
the motion of grain boundaries, and dislocation climb. 
In other words, elastic interactions mediated by wave modes have been incorporated, 
which travel on time scales that are orders of magnitude slower than the molecular vibrations 
yet considerably faster than the diffusive time scale.
A similar EOM has been proposed for the VPFC model, whose free-energy functional forces the order parameter to be non-negative. 
The resulting approach dictates the number of atoms and describes the motion of each of them. Solution of the respective EOM might be viewed 
as essentially performing MD simulations on diffusive time scales \cite{PhysRevE.79.035701}. A similar approach has been 
adapted to study the dynamics of monatomic and binary glasses, however, using MPFT-type free-energy functionals and temperature-dependent 
noise terms \cite{PhysRevLett.106.175702}.

\paragraph{\label{sec:TELE}The Euler-Lagrange equation} 
The ELE can be used to study equilibrium features including the mapping of the phase diagram \cite{JPhysCondMat.22.364101} as well as  
the evaluation of the free energy of the liquid-solid interface \cite{JPhysCondMat.22.364101} and of the nucleation barrier 
\cite{PhysRevLett.107.175702}. We note that noise may influence the phase diagram and other physical properties. 
Therefore, the results from the ELE and EOM are expected to converge for $\zeta \to 0$. We also call attention to the fact that so far as the 
equilibrium results (obtained by ELE) are concerned, the SH and 1M-PFC models are equivalent.

Once the free-energy functional is defined for the specific PFC model, its extremes can be found by solving the respective ELE, 
which reads as
\begin{equation}
\frac{\delta\tilde{\mathcal{F}}}{\delta\psiSH}=\frac{\delta\tilde{\mathcal{F}}}{\delta\psiSH}\Bigg\rvert_{\psiSH_0} \,,
\label{eq19}
\end{equation}
where $\psiSH_0$ is the reduced particle number density of the unperturbed initial liquid, 
while a no-flux boundary condition is prescribed at the boundaries 
of the simulation window ($\mathbf{n}\!\cdot\!\NablaW\psiSH = 0$ and $(\mathbf{n}\!\cdot\!\NablaW)\NablaW^2\psiSH = 0$, 
where $\mathbf{n}$ is the normal vector of the boundary). For example, inserting the 1M-PFC free energy and rearranging the terms,  
one arrives at 
\begin{equation}
\big(\!-\epsilon+\big(1+\NablaW^2\big)^2\big)(\psiSH-\psiSH_0)+p(\psiSH^2-\psiSH_0^2)-(\psiSH^3-\psiSH_0^3)=0 \;.
\label{eq20}
\end{equation}
Equation \eqref{eq20} together with the boundary conditions represents a fourth-order boundary value problem (BVP).

\begin{figure}
\begin{center}
\includegraphics[width=0.85\linewidth]{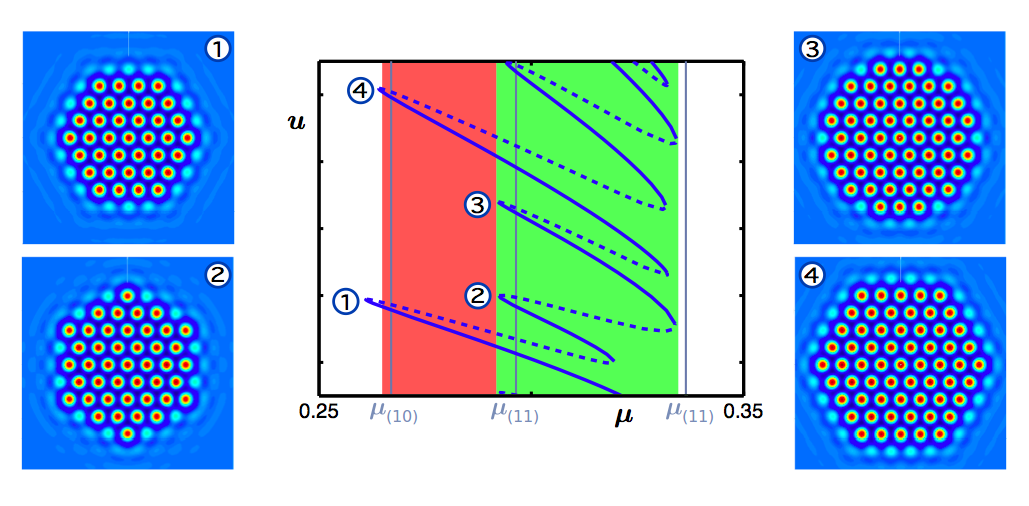}
\end{center}
\caption{\label{fig1}Multiplicity of localized hexagonal cluster solutions for the dynamic SH equation 
$\partial u/\partial t=-(1+\NablaW^{2})^{2}u-\mu u +\nu u^{2} - u^{3}$ with $\nu=1.6$. 
The central panel displays a part of the bifurcation diagram for localized hexagonal patches (for details see reference \cite{SIADS.7.1049}). 
The solid and dashed lines stand for the stable and unstable solutions, respectively. The vertical lines in grey correspond to the 
fold limits of planar (10) and (11) hexagon pulses. The red and green regions indicate where temporal self-completion does or does not occur, 
respectively. Panels 1-4 show colour plots of the hexagon patches at the inner and outer left folds. A different parametrisation is used here: 
$\mu=-\epsilon+3\psi_0^2$. 
\nocite{SIADS.7.1049}\CRP{D.\ J.\ B. Lloyd, B.\ Sandstede, D.\ Avitable, and A.\ R.\ Champneys}{Localized hexagon patterns of the planar Swift-Hohenberg equation}{SIAM J.\ Appl.\ Dyn.\ Syst.\ }{7}{2008}{1049-1100}{3}{10.1137/070707622}{The Society for Industrial and Applied Mathematics}}
\end{figure}

\textit{Multiplicity of solutions of the ELE.} 
It is worth noting that in the case of the PFC/SH-type models, a multiplicity of solutions can 
usually be found for the same BVP, defined by the boundary conditions and the ELE. This feature of the stationary solutions has been 
recently addressed in some detail in 2D for the SH \cite{SIADS.7.1049} and in 1D for the VPFC \cite{RobbinsATK2012} models. 
(Figure \ref{fig1} illustrates this phenomenon via showing the bifurcation diagram for compact hexagonal clusters in the 
SH model \cite{SIADS.7.1049}.)

\subsubsection{Binary PFC models}

\paragraph{The free energy} 

\subparagraph{Binary Swift-Hohenberg form} 
The earliest binary extension of the PFC model has been proposed in the seminal paper of Elder \etal\ \cite{ElderKHG2002} 
obtained by adding an interaction term with coefficient $A$ to the free energy of two mixed single-component SH free energies.  
Working with particle densities $\psiSH_1$ and $\psiSH_2$, this yields the free energy
\begin{equation}
\begin{split}
\tilde{\mathcal{F}} = \vintw\,\bigg( &\frac{\psiSH_1}{2}\Big(\!-\beta_1+\big(\wn_1^2+\NablaW^2\big)^2\Big)\psiSH_{1} 
+ \frac{\psiSH_1^4}{4} \\
&+ \frac{\psiSH_2}{2}\Big(\!-\beta_2+\big(\wn_2^2+\NablaW^2\big)^2\Big)\psiSH_{2} + \frac{\psiSH_2^4}{4} + A\psiSH_1\psiSH_2 \bigg) \,.
\end{split}
\label{eq21}
\end{equation}
Here, the physical properties (bulk moduli, lattice constants, \etc) of the individual species are controlled by the parameters with 
subscripts $1$ and $2$, respectively, and by the average values of the particle densities $\psiSH_1$ and $\psiSH_2$. 
It has been shown that the model can be used for studying structural phase transitions \cite{ElderKHG2002}.

\subparagraph{DFT-based binary PFC model} 
The most extensively used binary generalisation of the 1M-PFC model has been derived starting from a binary perturbative DFT, 
where the free energy is Taylor expanded relative to the liquid state denoted by $\rhoqA$ and $\rhoqB$, respectively, 
up to second order in the density differences $\Delta\rho_{\mathrm{A}}=\rho_{\mathrm{A}}-\rhoqA$ and 
$\Delta\rho_{\mathrm{B}}=\rho_{\mathrm{B}}-\rhoqB$ (up to two-particle correlations) \cite{ElderPBSG2007}:
\begin{equation}
\begin{split}
\frac{\mathcal{F}}{\kBT} = &\vint \bigg( \rho_{\mathrm{A}} \ln\!\bigg(\frac{\rho_{\mathrm{A}}}{\rhoqA}\bigg) 
-\Delta\rho_{\mathrm{A}} + \rho_{\mathrm{B}} \ln\!\bigg(\frac{\rho_{\mathrm{B}}}{\rhoqB}\bigg) - \Delta \rho_{\mathrm{B}} \bigg) \\
&\, \begin{split} -\,\frac{1}{2}\vintI{1}\!\!\!\vintI{2} 
&\Big( \Delta\rho_{\mathrm{A}}(\mathbf{r}_1)\cII_{\mathrm{AA}}(\mathbf{r}_1,\mathbf{r}_2)\Delta\rho_{\mathrm{A}}(\mathbf{r}_2)  \\ 
&\;+ \Delta\rho_{\mathrm{B}}(\mathbf{r}_1)\cII_{\mathrm{BB}}(\mathbf{r}_1,\mathbf{r}_2)\Delta\rho_{\mathrm{B}}(\mathbf{r}_2)  \\ 
&\;+ 2\,\Delta\rho_{\mathrm{A}}(\mathbf{r}_1)\cII_{\mathrm{AB}}(\mathbf{r}_1,\mathbf{r}_2)\Delta\rho_{\mathrm{B}}(\mathbf{r}_2) \Big) \,. 
\end{split} 
\end{split}
\label{eq22}
\end{equation}
It is assumed here that all two-point correlation functions are isotropic, \ie, 
$\cII_{ij}(\mathbf{r}_1,\mathbf{r}_2)=\cII_{ij}(\norm{\mathbf{r}_1-\mathbf{r}_2})$. 
Taylor expanding the direct correlation functions in Fourier space up to fourth order, one obtains 
$\cII_{ij}(\norm{\mathbf{r}_1-\mathbf{r}_2}) = \big(\cII_{ij,0} 
- \cII_{ij,2}\NablaI{2}^2 +\cII_{ij,4}\NablaI{2}^4\big)\delta(\mathbf{r}_1 - \mathbf{r}_2)$ in real space \cite{ElderPBSG2007}. 
The partial direct correlation functions $\cII_{ij}$ can be related to the measured or computed partial structure factors.

Following Elder \etal\ \cite{ElderPBSG2007}, the reduced partial particle density differences are defined as 
$\psi_{\mathrm{A}}  = (\rho_{\mathrm{A}} - \rhoqA)/\rhoL$ and $\psi_{\mathrm{B}}  = (\rho_{\mathrm{B}} - \rhoqB)/\rhoL$, 
where $\rhoL = \rhoqA + \rhoqB$. 
They have introduced then the new variables $\psi = \psi_{\mathrm{A}}  + \psi_{\mathrm{B}}$ and 
$\KDN = (\psi_{\mathrm{B}} - \psi_{\mathrm{A}}) + (\rhoqB - \rhoqA)/\rhoL$, 
obtaining fields whose amplitude expansion yields field amplitudes resembling order parameters associated with structural and composition 
changes in the conventional PF models. Expanding the free energy around $\KDN = 0$ and $\psi = 0$ one obtains
\begin{equation}
\begin{split}
\frac{\mathcal{F}}{\rhoq\,\kBT} = \vint \bigg( &\frac{\psi}{2} \big( B_{\mathrm{l}}+B_{\mathrm{s}}\big( 2R^2\Nabla^2+ R^4 \Nabla^4\big) \big)\psi  
+ s\frac{\psi^3}{3} + v\frac{\psi^4}{4} \\
&+ \gamma\KDN + w\frac{\KDN^2}{2} + u\frac{\KDN^4}{4} + \frac{L^2}{2}(\Nabla\KDN)^2 + \dotsb \bigg) \,.
\end{split}
\label{eq23}
\end{equation}
This model has been used for studying a broad range of phase transitions, including the formation of solutal dendrites, 
eutectic structures \cite{ElderPBSG2007,JPhysCondMat.22.364101}, and the Kirkendall effect \cite{PhilosMag.91.151}.

\subparagraph{Binary PFC model for surface alloying} 
A somewhat different formulation of the binary model has been proposed to model compositional patterning in monolayer aggregates of 
binary metallic systems by Muralidharan and Haataja \cite{PhysRevLett.105.126101}:
\begin{equation}
\begin{split}
\mathcal{F} = \vint \bigg( &\frac{\rho}{2} \big( \beta(c) + \big(\wn_{\mathrm{c}}^2+\Nabla^2\big)^2 \big)\rho + \frac{\rho^4}{4} + V(c)\rho  \\
&+ f_0 \bigg( \frac{w^{2}_0}{2}(\Nabla c)^2 - \theta_{\mathrm{c}} \frac{c^2}{2} 
+ \frac{\theta}{2} \big( (1+c)\ln(1+c)+(1-c)\ln(1-c) \big) \!\bigg)\!\bigg) \,. 
\end{split}
\label{eq24}
\end{equation}
In this construction, the values $c =\pm1$ of the concentration field stand for different atomic species, $V(c)$ is the atom specific 
substrate-film interaction, $\wn_{\mathrm{c}}$  incorporates the bulk lattice constant of different species, $f_0$ governs the relative strength of 
the elastic and chemical energies, while $\theta_{\mathrm{c}}$ is the critical temperature, $\theta$ the scaled absolute temperature, 
and $w_0$ tunes the chemical contribution to the interface energy between different species. 
Starting from this free-energy functional, Muralidharan and Haataja have shown, that their PFC model incorporates competing 
misfit dislocations and alloying in a quantitative way, and then employed the model for investigating the misfit- and line tension dependence 
of the domain size \cite{PhysRevLett.105.126101}.

\paragraph{The equations of motion}

\subparagraph{Dimensionless binary form in Swift-Hohenberg fashion} 
The same type of overdamped conservative dynamics has been assumed for the two types of atoms \cite{ElderKHG2002}:
\begin{equation}
\frac{\partial\psiSH_1}{\partial t} = M_1 \NablaW^2 \frac{\delta \tilde{\mathcal{F}}}{\delta \psiSH_1} 
+ \zeta_1 \quad \text{and} \quad \frac{\partial\psiSH_2}{\partial t} = M_2 \NablaW^2 \frac{\delta \tilde{\mathcal{F}}}{\delta \psiSH_2} 
+ \zeta_2 \;.
\label{eq25}
\end{equation}
Here, $M_i$ and $\zeta_i$ are the mobility and the noise term applying to species $i\in\{1,2\}$.

\subparagraph{DFT-based binary PFC model} 
In this widely used formulation of the binary 1M-PFC model, it is assumed that the same mobility $M$ applies for the two species $\mathrm{A}$ 
and $\mathrm{B}$ (corresponding to substitutional diffusion) decoupling the dynamics of the fields $\psi$ and $\KDN$. 
Assuming, furthermore, a constant effective mobility $M_{\mathrm{e}} = 2M/\rho^2$ and conserved dynamics, 
the equations of motions for the two fields have the form \cite{ElderPBSG2007} 
\begin{equation}
\frac{\partial \psi}{\partial t} = M_{\mathrm{e}} \Nabla^2 \frac{\delta \mathcal{F}}{\delta\psi} 
\quad \text{and} \quad \frac{\partial \KDN}{\partial t} = M_{\mathrm{e}} \Nabla^2 \frac{\delta \mathcal{F}}{\delta \KDN} \;.
\label{eq26}
\end{equation}
In general, the coefficients $B_{\mathrm{l}}$, $B_{\mathrm{s}}$, and $R$ in equation \eqref{eq23} depend on $\KDN$. 
A Taylor expansion of $B_{\mathrm{l}}(\KDN)$, $B_{\mathrm{s}}(\KDN)$, and $R(\KDN)$ in terms of $\KDN$ yields new coefficients 
$B_i^{\mathrm{l}}$, $B_i^{\mathrm{s}}$, and $R_i$ with $i=0,1,2,\dotsc$ that are independent of $\KDN$. 
Retaining only the coefficients $B_0^{\mathrm{l}}$, $B_2^{\mathrm{l}}$, $B_0^{\mathrm{s}}$, $R_0$, and $R_1$, and inserting the 
free energy \eqref{eq23} into equations \eqref{eq26}, one obtains 
\begin{align}
\begin{split}
\frac{\partial \psi}{\partial t} = M_{\mathrm{e}}  \Nabla^2 \Big( &\psi \big( B_0^{\mathrm{l}} + B_2^{\mathrm{l}}\KDN^2 \big)  
+ s \psi^2 + v \psi^3 \\
& + \frac{B_0^{\mathrm{s}}}{2} \Big( 2 \big(R_0+R_1 \KDN\big)^2\Nabla^2 + \big(R_0+R_1\KDN^4\big)\Nabla^4 \Big) \psi \\
& + \frac{B_0^{\mathrm{s}}}{2} \Big( 2\Nabla^2\big(\psi \big( R_0+R_1\KDN \big)^2 \big) 
+ \Nabla^4 \big( \psi \big( R_0+R_1\KDN \big)^4 \big) \!\Big)\!\Big) \,,
\label{eq27a}
\end{split}\\
\begin{split}
\frac{\partial \KDN}{\partial t} = M_{\mathrm{e}} \Nabla^2 \Big( &B_2^{\mathrm{l}}\KDN \psi^2 + \gamma + w\KDN 
+ u\KDN^3 - L^2 \Nabla^2\KDN \\
& + 2B_0^{\mathrm{s}} \psi \Big(\!\big( R_0+R_1\KDN \big)R_1\Nabla^2 + \big( R_0+R_1\KDN \big)^3 R_1\Nabla^4 \Big) \psi \Big) \,.
\label{eq27b}
\end{split}
\end{align}

\paragraph{The Euler-Lagrange equations} 
Since the derivation of the ELE is straightforward for all the binary PFC models, we illustrate it for the most frequently used version deduced 
from the binary Ramakrishnan-Yussouff-type classical perturbative DFT. The extremum of the grand potential functional 
requires that its first functional derivatives are zero, \ie,
\begin{equation}
\frac{\delta \mathcal{F}}{\delta \psi} = \frac{\delta \mathcal{F}}{\delta \psi} \bigg\rvert_{\psi_0,\KDN_0} \quad \text{and} \,
\quad \frac{\delta \mathcal{F}}{\delta \KDN} = \frac{\delta \mathcal{F}}{\delta \KDN} 
\bigg\rvert_{\psi_0,\KDN_0}  \;,
\label{eq28}
\end{equation}
where $\psi_0$ and $\KDN_0$ are the total and relative particle densities for the homogeneous initial state. 
Inserting equation \eqref{eq23} into equation \eqref{eq28}, one obtains after rearranging 
\begin{align}
\begin{split}
& \Big( B_{\mathrm{l}}\KDN + B_{\mathrm{s}} R \KDN^2 \big( 2\Nabla^2 + R \KDN^2 \Nabla^4 \big) 
+ \frac{B_{\mathrm{s}}}{2} \big( 2\Nabla^2(R^2) + \Nabla^4(R^4) \big)\!\Big) (\psi-\psi_0) \\ 
&\quad= -s\big(\psi^2-\psi_0^2\big)-v\big(\psi^3-\psi_0^3\big) \;, 
\end{split} \\
\begin{split}
& L^2\Nabla^2\KDN - \frac{\partial B_{\mathrm{l}}}{\partial \KDN} \big( \KDN\psi^2 - \KDN_0 \psi_0^2 \big) 
- 2 B_{\mathrm{s}} R\:\! \frac{\partial R}{\partial \KDN}\:\! \psi \big( \Nabla^2 + R^2\Nabla^4 \big) \psi \\
& \quad= w \big( \KDN-\KDN_0 \big) + u\big( \KDN^3-\KDN_0^3 \big) \;,
\end{split}
\end{align}
where $R=R_0+R_1\KDN$. These equations are to be solved assuming no-flux boundary conditions at the border of the simulation box 
for both fields [$\mathbf{n}\!\cdot\!\Nabla\psi = 0 $, $(\mathbf{n} \!\cdot\! \Nabla)\Nabla^2 \psi=0$, $\mathbf{n} \!\cdot\! \Nabla\KDN=0$, 
and $(\mathbf{n} \!\cdot\! \Nabla)\Nabla^2\KDN=0$].

\subsubsection{\label{sec:PFCstatisch}PFC models for liquid crystals}
Another level of complexity is to consider interactions between particles, that are not any longer spherically symmetric. 
The simplest case are \textit{oriented} particles, \ie, particles with a fixed orientation along a given direction where the interaction 
$U_{2}(\vec{r}_{1}-\vec{r}_{2})$ is not any longer radially symmetric [as assumed in equation \eqref{eq:cII}]. 
The derivation described in section \ref{sec:PFC_Isotropic} was straightforwardly extended towards this case leading to a 
microscopic justification of the APFC model \eqref{eq:APFC} proposed in section \ref{sec:tfe}. 
Clearly, the resulting crystal lattices are anisotropic, which leads also to an anisotropic crystal growth. 
Much more complicated, however, is the case of \textit{orientable} particles, which can adjust 
their orientation and form liquid crystalline mesophases.

Therefore, we now address orientable anisotropic particles, that possess orientational degrees of freedom. In order to keep the effort manageable, 
the particles are assumed to be uniaxial, \ie, they are rotationally symmetric around an internal axis. Consequently, their orientations are 
described by unit vectors $\uu_{i}$ with $i\in\{1,\dotsc,N\}$ along their internal symmetry axes. Most of the statistical quantities can suitably 
be generalized towards orientational degrees of freedom by including also the orientational configuration space on top of the translational 
configurational space.

\paragraph{Statics}
For orientable particles, the one-particle density is now defined via
\begin{equation}
\rho(\vec{r},\uu ) = \bigg\langle\sum^{N}_{i=1} 
\delta(\vec{r}-\vec{r}_{i})\delta(\uu -\uu _{i})\bigg\rangle 
\end{equation}
and contains also the probability distribution of the orientations as expressed by the dependence on the orientational unit vector $\uu $. 
Its full configurational mean is given by the mean particle number density 
\begin{equation}
\rhoM=\frac{1}{4\pi V}\!\vint\!\!\uint \rho(\vec{r},\uu )=\frac{N}{V}\;,
\end{equation}
where the orientational integral denotes integration over the two-dimensional unit sphere $S_{2}$. 
Now the external potential is $U_{1}(\vec{r},\uu )$ and couples also to the particle orientation. 
In general, also the pair-interaction potential $U_{2}(\vec{r}_{1}-\vec{r}_{2},\uu _{1},\uu _{2})$ has an orientational dependence. 
The corresponding equilibrium Helmholtz free-energy functional $\mathcal{F}[\rho(\vec{r},\uu )]$ can be split as usual in three contributions
\begin{equation}
\mathcal{F}[\rho(\vec{r},\uu )] = \mathcal{F}_{\textrm{id}}[\rho(\vec{r},\uu )] + \mathcal{F}_{\textrm{exc}}[\rho(\vec{r},\uu )] 
+ \mathcal{F}_{\textrm{ext}}[\rho(\vec{r},\uu )] 
\label{eq:FEF}
\end{equation}
namely the ideal rotator-gas free-energy contribution 
\begin{equation}
\mathcal{F}_{\textrm{id}}[\rho(\vec{r},\uu )] =\!\kBT\! \vint\!\!\uint \,\rho(\vec{r},\uu) 
\big(\ln(\Lambda^{3}\rho(\vec{r},\uu ))-1\big) \,,
\label{eq:FidAllgemein}
\end{equation}
the nontrivial excess free-energy functional $\mathcal{F}_{\textrm{exc}}[\rho(\vec{r},\uu )]$, 
and the external free-energy contribution (see section \ref{subsubsec:DFT})
\begin{equation}
\mathcal{F}_{\textrm{ext}}[\rho(\vec{r},\uu )] =\! 
\vint\!\!\uint\,\rho(\vec{r},\uu ) U_{1}(\vec{r},\uu ) \;. 
\label{eq:FextAllgemein}
\end{equation}
Analogously to equation \eqref{eq:FTEg}, a functional Taylor expansion
\begin{equation}
\mathcal{F}_{\mathrm{exc}}[\rho(\vec{r},\uu )] = \mathcal{F}^{(0)}_{\mathrm{exc}}(\rhoR)
+\kBT \sum^{\infty}_{n=1}\frac{1}{n!}\mathcal{F}^{(n)}_{\mathrm{exc}}[\rho(\vec{r},\uu )]
\label{eq:FTE}
\end{equation}
with the $n$th-order contributions  
\begin{equation}
\mathcal{F}^{(n)}_{\mathrm{exc}}[\rho(\vec{r},\uu )] = -\!\vintn{1}\dotsi\!\!\vintn{n} \! \uintn{1}\dotsi\!\!\uintn{n} 
\,c^{(n)}(\mathbf{\vec{r}^{n}},\mathbf{\uu^{n}}) \prod^{n}_{i=1}\!\Delta\rho(\vec{r}_{i},\uu _{i}) \,, 
\label{eq:TET}
\end{equation}
the $n$-particle direct correlation function $c^{(n)}(\mathbf{\vec{r}^{n}},\mathbf{\uu^{n}})$, and the abbreviations 
\begin{equation}
\mathbf{\vec{r}^{n}}=(\vec{r}_{1},\dotsc,\vec{r}_{n})\;,\quad \mathbf{\uu^{n}}=(\uu_{1},\dotsc,\uu_{n})
\end{equation}
are used. Again, the constant zeroth-order contribution of the functional Taylor expansion \eqref{eq:FTE} is irrelevant and the first-order 
contribution vanishes. In the Ramakrishnan-Yussouff approximation, the functional Taylor expansion is truncated at second order.

In order to derive PFC-type models for liquid crystals, one can follow the same strategy as for spherical systems: 
first, the full density field $\rho(\vec{r},\uu )$ is parametrized by small and slowly varying space-dependent multi-component 
order-parameter fields. An insertion into the density functional together with a perturbative and gradient expansion yields a local PFC-type 
free energy. Then several coupling terms of the order-parameter components arise, whose prefactors are given by moments of the generalized direct 
fluid correlation functions. Still, the gradient expansion is more tedious, since the orientational space has a more complicated topology. 
For stability reasons, one has to assume that the coefficients of the highest-order terms in the gradients and order-parameter fields in the 
PFC model are positive in the full free-energy functional. If this appears not to be the case for a certain system, it is necessary to take further 
terms of the respective order-parameter field up to the first stabilizing order into account.

\paragraph{\label{sec:PFCIIDStatisch}Two spatial dimensions} 
We now consider first the case of two spatial dimensions both for the translational and orientational degrees of freedom. 
Here, the topology of orientations is simpler than in three spatial dimensions. Obviously, all previous expressions can be changed towards 
two dimensions by replacing the volume $V$ by an area $A$, by changing the unit sphere $S_{2}$ into the unit circle $S_{1}$, 
and by replacing $\Lambda^{3}$ by $\Lambda^{2}$ in equation \eqref{eq:FidAllgemein}. 
The orientational vector $\uu (\varphi)=(\cos(\varphi),\sin(\varphi))$ can be parametrized by a single polar angle $\varphi\in [0,2\pi)$.

\subparagraph{Derivation of the PFC free-energy functional} 
First we chose as order-parameter fields the \textit{reduced translational density}, the \textit{polarization}, and 
the \textit{nematic tensor} field. The reduced translational density is defined via
\begin{equation}
\psi(\vec{r})=\frac{1}{2\pi\rhoR}\!\uint (\rho(\vec{r},\uu)-\rhoR)\,,
\end{equation}
while the polarization is 
\begin{equation}
\vec{P}(\vec{r})=\frac{1}{\pi\rhoR}\!\uint\rho(\vec{r},\uu )\:\!\uu  
\end{equation}
and describes the local averaged dipolar orientation of the particles. Finally, the symmetric and traceless nematic tensor with the components
\begin{equation}
Q_{ij}(\vec{r})=\frac{2}{\pi\rhoR}\!\uint\rho(\vec{r},\uu )\bigg(u_{i}u_{j}-\frac{1}{2}\delta_{ij}\bigg) 
\end{equation}
describes quadrupolar ordering of the particles. Equivalently, one could decompose the polarization $\vec{P}(\vec{r})=P(\vec{r})\pp(\vec{r})$ 
into its modulus $P(\vec{r})$ and the local normalised dipolar orientation $\pp(\vec{r})$ and use the two order parameters $P(\vec{r})$ 
and $\pp(\vec{r})$ instead of $\vec{P}(\vec{r})$. Similarly, the nematic tensor can be expressed by \cite{deGennes1971,deGennesP1995}
\begin{equation}
Q_{ij}(\vec{r})=S(\vec{r})\bigg(n_{i}(\vec{r})n_{j}(\vec{r})-\frac{1}{2}\delta_{ij}\bigg)
\label{eq:QijIID}
\end{equation}
through the \textit{nematic order parameter} $S(\vec{r})$, which measures the local degree of quadrupolar orientational order \cite{BrandK1986}, 
and the \textit{nematic director} $\nn (\vec{r})=(n_{1}(\vec{r}),n_{2}(\vec{r}))$. Notice that $\nn (\vec{r})$ and $\pp(\vec{r})$ do not 
necessarily coincide.

One may expand the total density in terms of its orientational anisotropy as 
\begin{equation}
\rho(\vec{r},\uu) = \rhoR\:\!(1 + \psi(\vec{r}) + P_{i}(\vec{r})\:\!u_{i} + u_{i}\:\!Q_{ij}(\vec{r})\:\!u_{j})\,.
\label{eq:rhoIID}
\end{equation}
Inserting the parametrisation \eqref{eq:rhoIID} into equation \eqref{eq:FidAllgemein}, performing a Taylor expansion of the integrand up to fourth order 
in the order-parameter fields, which guarantees stability of the solutions, and carrying out the angular integration yields to the approximation
\begin{equation}
\mathcal{F}_{\textrm{id}}[\psi,P_{i},Q_{ij}] = F_{\textrm{id}} + \pi\rhoR\,\kB T\!\vint f_{\mathrm{id}}(\vec{r})
\label{eq:Fida}
\end{equation}
with the local scaled ideal rotator-gas free-energy density
\begin{equation}
\begin{split}
f_{\mathrm{id}} = &\;\frac{\psi}{4}(8-2P^{2}_{i}+2P_{i}Q_{ij}P_{j}-Q^{2}_{ij}) + \frac{\psi^{2}}{4}(4+2P^{2}_{i}+Q^{2}_{ij}) 
- \frac{\psi^{3}}{3} + \frac{\psi^{4}}{6} \\
&\:\!+ \frac{P^{2}_{i}}{8}(4+Q^{2}_{kl}) - \frac{P_{i}Q_{ij}P_{j}}{4} + \frac{P^{2}_{i}P^{2}_{j}}{16} + \frac{Q^{2}_{ij}}{4} 
+ \frac{Q^{2}_{ij}Q^{2}_{kl}}{64} \;,
\end{split}
\label{eq:Fidb}
\end{equation}
where $F_{\textrm{id}} = 2\pi\rhoR\,\kB T A\:\! (\ln(\Lambda^{2}\rhoR)-1)$ is an irrelevant constant. 

If the functional Taylor expansion \eqref{eq:FTE} for the excess free energy is truncated at fourth order, 
an insertion of the parametrisation \eqref{eq:rhoIID} into 
the functional \eqref{eq:TET} and a gradient expansion yield \cite{WittkowskiLB2011}
\begin{equation}
\mathcal{F}^{(n)}_{\mathrm{exc}}[\psi,P_{i},Q_{ij}] =\! 
-\!\vint f^{(n)}_{\mathrm{exc}}(\vec{r})
\label{eq:Fexc_n}
\end{equation}
with
{\allowdisplaybreaks
\begin{align}
\begin{split}
&f^{(2)}_{\mathrm{exc}} = A_{1}\psi^{2} + A_{2}(\partial_{i}\psi)^{2} + A_{3}(\partial^{2}_{k}\psi)^{2} + B_{1}(\partial_{i}\psi)P_{i} 
+ B_{2}P_{i}(\partial_{j}Q_{ij}) + B_{3}(\partial_{i}\psi)(\partial_{j}Q_{ij}) \\
&\qquad\;\;\,+ C_{1}P^{2}_{i} + C_{2}P_{i}(\partial^{2}_{k}P_{i}) + C_{3}(\partial_{i}P_{i})^{2} + D_{1}Q^{2}_{ij} 
+ D_{2}(\partial_{j}Q_{ij})^{2} \;, \raisetag{6ex}\end{split}\label{eq:Fexc_GpEa}\\[2mm]
\begin{split}
&f^{(3)}_{\mathrm{exc}} = E_{1}\psi^{3} + E_{2}\psi P^{2}_{i} + E_{3}\psi Q^{2}_{ij} + E_{4}P_{i}Q_{ij}P_{j}  
+F_{1}\psi^{2}(\partial_{i}P_{i})+F_{2}\psi P_{i}(\partial_{j}Q_{ij}) \\
&\qquad\;\;\, +F_{3}(\partial_{i}\psi)Q_{ij}P_{j} + F_{4}P^{2}_{i}(\partial_{j}P_{j}) + F_{5}(\partial_{i}P_{i})Q^{2}_{kl} 
+F_{6}P_{i}Q_{ki}(\partial_{j}Q_{kj}) \;,
\end{split}\label{eq:Fexc_GpEb}\\[2mm]
\begin{split}
&f^{(4)}_{\mathrm{exc}} = G_{1}\psi^{4} + G_{2}\psi^{2}P^{2}_{i} + G_{3}\psi^{2}Q^{2}_{ij} + G_{4}\psi P_{i}Q_{ij}P_{j}+ G_{5}P^{2}_{i}Q^{2}_{kl} 
+ G_{6}P^{2}_{i}P^{2}_{j} \\
&\qquad\;\;\,+ G_{7}Q^{2}_{ij}Q^{2}_{kl} \;.
\end{split}\label{eq:Fexc_GpEc}
\end{align}}
Obviously, this creates a much more sophisticated series of coupling terms between gradients of the different order-parameter fields. 
They are all allowed by symmetry. There are altogether $28$ coupling coefficients $A_{i}$, $B_{i}$, $C_{i}$, $D_{i}$, $E_{i}$, $F_{i}$, $G_{i}$,  
which can in principle all be expressed as moments over the microscopic fluid direct correlation functions. These explicit expressions are summarized 
in appendix \ref{A:IID}.

The general result for the two-dimensional PFC free energy constituted by equations \eqref{eq:Fexc_GpEa}-\eqref{eq:Fexc_GpEc} 
contains several special cases, known from the literature, which we now discuss in more detail.

These special cases follow from the full free-energy functional either by choosing some of the order-parameter fields as zero or as a constant 
different from zero and by taking into account the contributions of the functional Taylor expansion \eqref{eq:FTE} only up to a certain order 
$n_{\mathrm{max}}\in\{2,3,4\}$. Table \ref{tab:SpezialfaelleIID} gives an overview about the most relevant special cases.  

\begin{table}
\newcommand{\KZelle}[1]{#1}%
\newcommand{\GZelle}[1]{#1}%
\centering
\caption{\label{tab:SpezialfaelleIID}Relevant special cases that are contained in the polar PFC model for two spatial dimensions. 
$n_{\mathrm{max}}$ is the order of the functional Taylor expansion \eqref{eq:FTE}. If $n_{\mathrm{max}}$ is not specified, 
it can be arbitrary (arb.).}%
\begin{tabular}{ccccc}%
\hline
\hline
$\boldsymbol{\psi}$ & $\boldsymbol{P_{i}}$ & $\boldsymbol{Q_{ij}}$ & $\boldsymbol{n_{\mathrm{max}}}$ & \textbf{Associated model} \\
\hline
\KZelle{$0$} & \KZelle{$0$} & \KZelle{$const.$} & \KZelle{$arb.$} & \KZelle{Landau expansion in $Q_{ij}(\vec{r})$} \\
\GZelle{$0$} & \GZelle{$0$} & \GZelle{$Q_{ij}(\vec{r})$} & \GZelle{$2$} & \parbox[c]{0.4\linewidth}{\centering Landau-de Gennes free energy for 
uniaxial nematics \cite{deGennesP1995}} \\
\KZelle{$0$} & \KZelle{$0$} & \KZelle{$Q_{ij}(\vec{r})$} & \KZelle{$arb.$} & \KZelle{Gradient expansion in $Q_{ij}(\vec{r})$} \\
\KZelle{$0$} & \KZelle{$const.$} & \KZelle{$0$} & \KZelle{$arb.$} & \KZelle{Landau expansion in $P_{i}(\vec{r})$} \\
\KZelle{$0$} & \KZelle{$P_{i}(\vec{r})$} & \KZelle{$0$} & \KZelle{$arb.$} & \KZelle{Gradient expansion in $P_{i}(\vec{r})$} \\
\KZelle{$const.$} & \KZelle{$P_{i}(\vec{r})$} & \KZelle{$Q_{ij}(\vec{r})$} & \KZelle{$4$} & \KZelle{Constant-density approximation} \\
\KZelle{$\psi(\vec{r})$} & \KZelle{$0$} & \KZelle{$0$} & \KZelle{$2$} & \KZelle{PFC model of Elder \etal\ \cite{ElderPBSG2007}} \\
\KZelle{$\psi(\vec{r})$} & \KZelle{$0$} & \KZelle{$Q_{ij}(\vec{r})$} & \KZelle{$2$} & \KZelle{PFC model of L\"owen \cite{Loewen2010}} \\
\KZelle{$\psi(\vec{r})$} & \KZelle{$P_{i}(\vec{r})$} & \KZelle{$Q_{ij}(\vec{r})$} & \KZelle{$4$} & Full free-energy functional \\
\hline
\hline
\end{tabular}
\end{table}
 
The two most simple special cases are obtained for either a constant polarization $P_{i}(\vec{r})$ or a constant nematic tensor $Q_{ij}(\vec{r})$ 
with an arbitrary choice for $n_{\mathrm{max}}$, when all remaining order-parameter fields are assumed to be zero. 
These special cases are known to be Landau expansions in $P_{i}(\vec{r})$ and $Q_{ij}(\vec{r})$, respectively. 
If the polarization $P_{i}(\vec{r})$ is not constant, but space-dependent, and the other order-parameter fields vanish, 
then the functional  can be called a gradient expansion in the polarization. Analogously, a gradient expansion in the 
nematic tensor $Q_{ij}(\vec{r})$ is obtained, if it is space-dependent and all other order-parameter fields vanish. 
If additionally $n_{\mathrm{max}}=2$ is chosen, this gradient expansion becomes the Landau-de Gennes free energy for inhomogeneous 
uniaxial nematics \cite{deGennesP1995}. When only $\psi(\vec{r})$ is constant and the other order-parameter fields are space-dependent, 
we recover the case of an incompressible system. If all anisotropies are neglected, we recover with $n_{\mathrm{max}}=2$ the original PFC model 
of Elder \etal\ \cite{ElderPBSG2007}. The liquid crystalline PFC model of L\"owen \cite{Loewen2010} is obtained, if in addition to the 
translational density also the nematic tensor is space-dependent and if it is parametrized according to equation \eqref{eq:QijIID}. 
In the polar PFC model for two spatial dimensions, one can also consider the case of a space-dependent translational density $\psi(\vec{r})$, 
a space-dependent polarization $P_{i}(\vec{r})$, and a vanishing nematic tensor, that corresponds to a ferroelectric phase without 
orientational order, but such a phase was never observed in experiments up to the present day. Therefore, this case is not included in 
Tab.\ \ref{tab:SpezialfaelleIID}.

\subparagraph{\label{subsubsec:PhaseDiagram}Equilibrium bulk phase diagram} 
While there are two independent parameters for the original PFC model (see equation \eqref{eq0} and figure \ref{fig2}), 
the number of coupling parameters explodes for the general liquid crystalline PFC model as proposed in 
equations \eqref{eq:Fexc_GpEa}-\eqref{eq:Fexc_GpEc} to $27$ coupling coefficients. One of them can be incorporated into a length scale 
such that $26$ coefficients remain. Therefore, a numerical exploration of the equilibrium phase diagram requires a much higher effort. 
Maybe not too surprising, such calculations are sparse and it was only until recently that the phase diagram was calculated for the 
apolar PFC model \cite{Loewen2010}, where $\vec{P}(\vec{r})=\vec{0}$, which contains only $5$ independent parameters \cite{AchimWL2011}. 
We summarize and outline the basic findings of reference \cite{AchimWL2011} in section \ref{app:statics}.

\paragraph{\label{sec:StatikIIID}Three spatial dimensions} 
We now discuss the three-di\-men\-sio\-nal ($d=3$) PFC model for 
liquid crystals. In spherical coordinates, the three-dimensional orientation vector is
\begin{equation}
\uu (\theta,\phi)=(\sin(\theta)\cos(\phi),\sin(\theta)\sin(\phi),\cos(\theta))
\end{equation}
with the polar angle $\theta\in[0,\pi]$ and the azimuthal angle $\phi\in[0,2\pi)$. We consider here only the apolar case, 
where the polarization vanishes: $\vec P ({\vec r})=\vec{0}$. There exists indeed a zoo of realizations of apolar particles both in the molecular 
and in the colloidal regime. For suitable interactions see 
references \cite{BolhuisF1997,Loewen1994b,FrenkelMMT1984,KirchhoffLK1996,Loewen1994d,Loewen1994c,CleaverCAN1996,FukunagaTD2004,MuccioliZ2006}.

Following similar ideas as outlined in section \ref{sec:PFCIIDStatisch} for two dimensions, we define as the corresponding order parameters 
the reduced translational density 
\begin{equation}
\psi(\vec{r})=\frac{1}{4\pi\rhoR}\uint (\rho(\vec{r},\uu)-\rhoR)
\end{equation}
and the $3\!\times\!3$-dimensional symmetric and traceless nematic tensor 
\begin{equation}
Q_{ij}(\vec{r})=\frac{15}{8\pi\rhoR}\uint\rho(\vec{r},\uu)\bigg(u_{i}u_{j}-\frac{1}{3}\delta_{ij}\bigg) \,.
\end{equation} 
Again, the nematic tensor can be expressed by the nematic order-parameter field $S(\vec{r})$ and the nematic director 
$\nn (\vec{r})=(n_{1}(\vec{r}),n_{2}(\vec{r}),n_{3}(\vec{r}))$, that is here the only unit vector that denotes a preferred direction in the 
liquid crystalline system. In the three-dimensional case, the decomposition of the nematic tensor is given by \cite{deGennes1971,deGennesP1995}
\begin{equation}
Q_{ij}(\vec{r})=S(\vec{r})\bigg(\frac{3}{2}n_{i}(\vec{r})n_{j}(\vec{r})-\frac{1}{2}\delta_{ij}\bigg) \,.
\label{eq:QijIIID}
\end{equation} 
Notice that the nematic order-parameter field $S(\vec{r})$ is the biggest eigenvalue of the nematic tensor $Q_{ij}(\vec{r})$ and the 
nematic director $\nn (\vec{r})$ is the corresponding eigenvector. Accordingly, with the order-parameter fields $\psi(\vec{r})$ 
and $Q_{ij}(\vec{r})$, the one-particle density is approximated by
\begin{equation}
\rho(\vec{r},\uu ) = \rhoR\:\!(1 + \psi(\vec{r}) + u_{i}Q_{ij}(\vec{r})u_{j}) \,.
\label{eq:rhoIII}
\end{equation} 
As before, the Helmholtz free-energy functional has to be approximated by a Taylor expansion around the homogeneous reference system and by 
a gradient expansion \cite{WittkowskiLB2010}. The ideal rotator-gas free-energy functional is approximated by
\begin{equation}
\mathcal{F}_{\textrm{id}}[\psi,Q_{ij}] = F_{\textrm{id}} + \pi\rhoR\,\kB T\!\vint f_{\mathrm{id}}(\vec{r})
\end{equation}
with the local scaled ideal rotator-gas free-energy density 
\begin{equation}%
\begin{split}%
f_{\mathrm{id}} = &\;4\:\!\psi\bigg(1-\frac{\trace(\mathbf{Q}^{2})}{15}+\frac{8\trace(\mathbf{Q}^{3})}{315}\bigg) 
+ 2\:\!\psi^{2}\bigg(1+\frac{2\trace(\mathbf{Q}^{2})}{15}\bigg) - \frac{2\:\!\psi^{3}}{3} + \frac{\psi^{4}}{3} \\
&\:\!+ \frac{4\trace(\mathbf{Q}^{2})}{15}-\frac{16\trace(\mathbf{Q}^{3})}{315}+\frac{8\trace(\mathbf{Q}^{4})}{315} \;. 
\end{split}%
\label{eq:Fidbb}%
\end{equation}%
Here, $\trace(\,\cdot\,)$ denotes the trace operator and $F_{\textrm{id}} = 4\pi\rhoR V \kB T(\ln(\Lambda^{3}\rhoR)-1)$ 
an irrelevant constant. For the excess free-energy functional, the Ramakrishnan-Yussouff approximation \eqref{eq:RYg} is used together 
with equation \eqref{eq:rhoIII} involving the direct correlation function $c^{(2)}(\vec{r}_{1}-\vec{r}_{2},\uu _{1},\uu _{2})$. 
Respecting all symmetries, one finally obtains the following approximation for the excess free-energy functional
\begin{equation}
\mathcal{F}^{(2)}_{\mathrm{exc}}[\psi,Q_{ij}] = -\!\vint f^{(2)}_{\mathrm{exc}}(\vec{r})
\end{equation}
with the local scaled excess free-energy density
\begin{equation}%
\begin{split}%
&f^{(2)}_{\mathrm{exc}} = A_{1}\psi^{2} + A_{2}(\partial_{i}\psi)^{2} + A_{3}(\partial^{2}_{k}\psi)^{2} + B_{1}Q^{2}_{ij} 
+ B_{2}(\partial_{i}\psi)(\partial_{j}Q_{ij}) \\
&\qquad\;\;\, + \widetilde{K}_{1}(\partial_{j}Q_{ij})^{2} + \widetilde{K}_{2}Q_{ij}(\partial^{2}_{k}Q_{ij}) \;.
\end{split}%
\label{eq:Fexcbb}%
\end{equation}%
The seven coupling coefficients $A_{1}$, $A_{2}$, $A_{3}$, $B_{1}$, $B_{2}$, $\widetilde{K}_{1}$, $\widetilde{K}_{2}$ can be expressed as 
generalized moments of the microscopic correlation function $c^{(2)}(\vec{r}_{1}-\vec{r}_{2},\uu _{1},\uu _{2})$. 
The full expressions are summarized in appendix \ref{A:IIID}. 
We remark that $\widetilde{K}_{1}$ and $\widetilde{K}_{2}$ correspond to the 
traditional Frank constants appearing in Frank's elastic energy \cite{Chandrasekhar1992,deGennesP1995}.

Special cases of the free-energy density are summarized in Tab.\ \ref{tab:SpezialfaelleIIID}.
\begin{table}
\newcommand{\KZelle}[1]{#1}%
\centering
\caption{\label{tab:SpezialfaelleIIID}Relevant special cases, that are contained in the apolar PFC model for three spatial dimensions.}
\begin{tabular}{ccc} 
\hline
\hline
$\boldsymbol{\psi}$  & $\boldsymbol{Q_{ij}}$    & \textbf{Associated model}        \\ 
\hline
\KZelle{$0$}         & \KZelle{$const.$}                & \KZelle{Landau expansion in $Q_{ij}(\vec{r})$}     \\ 
\KZelle{$0$}         & \KZelle{$Q_{ij}(\vec{r})$}       & \KZelle{Landau-de Gennes free energy for uniaxial nematics \cite{deGennesP1995}}  \\ 
\KZelle{$\psi(\vec{r})$}  & \KZelle{$0$}                & \KZelle{PFC model of Elder \etal\ \cite{ElderPBSG2007}}      \\ 
\KZelle{$\psi(\vec{r})$}  & \KZelle{$Q_{ij}(\vec{r})$}  & \KZelle{Full free-energy functional}                     \\ 
\hline
\hline
\end{tabular} 
\end{table}
As for the PFC model for two spatial dimensions, one obtains a Landau expansion in the nematic tensor $Q_{ij}(\vec{r})$, 
if the translational density $\psi(\vec{r})$ is zero and the nematic tensor is constant so that all gradients vanish. 
When only $\psi(\vec{r})$ is constant and $Q_{ij}(\vec{r})$ is space-dependent, the Landau-de Gennes free energy for inhomogeneous 
uniaxial nematics \cite{deGennesP1995} is recovered again. Clearly, the original PFC model of Elder \etal\ \cite{ElderPBSG2007} 
for isotropic particles in three spatial dimensions can be obtained from the full free-energy functional by choosing $Q_{ij}(\vec{r})=0$. 
Finally, we note that the  full functional was recently numerically evaluated for several situations by Yabunaka and Araki in 
reference \cite{YabunakaA2011}.

\paragraph{Dynamics}
Dynamical density functional theory (DDFT) can now be used to derive the dynamics of the PFC models for liquid crystals. 
DDFT is well justified for Brownian anisotropic particles (as, for example, colloidal rods or platelets). 
The basic derivation is similar to that performed for spherical particles in section \ref{subsubsec:DE_PFC}, 
but in practice it is much more tedious. The basic derivation is performed in three steps. At first, the order-parameter fields, 
that have been chosen for the statics, are assumed to be time-dependent and the time-dependent one-particle density $\rho(\vec{r},\uu,t)$ 
is approximated in terms of these time-dependent order-parameter fields. Secondly, the chain rule for functional differentiation is used 
to express the functional derivative $\delta\mathcal{F}/\delta\rho$ of the Helmholtz free-energy functional $\mathcal{F}$ 
in terms of the functional derivatives of the free-energy functional with respect to the chosen order-parameter fields. 
Finally, the time-dependent parametrisation for the one-particle density and the time-dependent expression for the 
functional derivative $\delta\mathcal{F}/\delta\rho$ are inserted into the DDFT equation and a set of in general coupled dynamic equations 
for the single order-parameter fields is obtained by an orthogonal projection of the DDFT equation with respect to the orientation $\uu $.
  
We first consider the case of two spatial dimensions ($d=2$). The time-dependent noise-averaged one-particle number density is now
\begin{equation}
\rho(\vec{r},\uu,t) = \bigg\langle\sum^{N}_{i=1} \delta\big(\vec{r}-\vec{r}_{i}(t)\big)\delta\big(\uu -\uu _{i}(t)\big)\bigg\rangle 
\end{equation} 
and its order-parameter parametrisation is
\begin{equation}
\rho(\vec{r},\uu,t) = \rhoR\:\!(1 + \psi(\vec{r},t) + P_{i}(\vec{r},t)\:\!u_{i} + u_{i}\:\!Q_{ij}(\vec{r},t)\:\!u_{j}) \,.
\label{eq:rhoIIDdyn}
\end{equation}
The DDFT equation for orientational degrees of freedom in two spatial dimensions (without a hydrodynamic translational-rotational coupling) 
reads \cite{WensinkL2008}
\begin{equation}
\begin{split}%
\dot{\rho}(\vec{r},\uu,t) = \Nabla\!\cdot\! \bigg(\frac{\mathrm{D_{\textrm{T}}}(\uu)}{\kBT}\:\!   
\rho(\vec{r},\uu,t) \Nabla\Fdif{\mathcal{F}(T,[\rho])}{\rho(\vec{r},\uu,t)}\bigg)& \\ 
+\frac{D_{\mathrm{R}}}{\kBT}\,\partial_{\varphi} \bigg(\rho(\vec{r},\uu,t) \,\partial_{\varphi}
\Fdif{\mathcal{F}(T,[\rho])}{\rho(\vec{r},\uu,t)}\bigg)&
\end{split}%
\label{eq:DDFT}%
\end{equation}
with the translational short-time diffusion tensor
\begin{equation}
\mathrm{D_{\textrm{T}}}(\uu )=D_{\parallel}\uu \otimes\uu +D_{\perp}(\Eins-\uu \otimes\uu )\;.
\label{eq:DTensor}%
\end{equation}
Here, $D_{\parallel}$ and $D_{\perp}$ are the translational diffusion coefficients for translation parallel and perpendicular to the 
orientation $\uu  = (\cos(\phi),\sin(\phi))$, respectively,  and the symbol $\Eins$ denotes the $2\!\times\!2$-dimensional unit matrix. 
The two terms on the right-hand-side of this DDFT equation correspond to pure translation and pure rotation, respectively.

The constant-mobility approximation (CMA) is now 
\begin{equation}
\dot{\rho}(\vec{r},\uu,t) = \rhoR\:\!\Nabla\!\cdot\! \bigg(\frac{\mathrm{D_{\textrm{T}}}(\uu)}{\kBT}\:\!  
\Nabla\Fdif{\mathcal{F}(T,[\rho])}{\rho(\vec{r},\uu,t)}\bigg) +\rhoR\:\! \frac{D_{\mathrm{R}}}{\kBT}\,\partial^{2}_{\varphi} 
\Fdif{\mathcal{F}(T,[\rho])}{\rho(\vec{r},\uu,t)} \;.
\label{eq:DDFTpKM}
\end{equation}
Within the CMA, this yields for the dynamics of the order-parameter fields 
\begin{equation}
\dot{\psi} + \partial_{i} J^{\psi}_{i} = 0 \;,\qquad \dot{P}_{i} + \Phi^{P}_{i}  = 0 \;,\qquad \dot{Q}_{ij} + \Phi^{Q}_{ij} = 0 
\end{equation}
with the currents and quasi-currents 
{\allowdisplaybreaks
\begin{align}%
\begin{split}%
J^{\psi}_{i}  = &- 2\alpha_{1}\:\!\partial_{i}\Fdif{\mathcal{F}}{\psi} -2\alpha_{3}\:\!\partial_{j}\Fdif{\mathcal{F}}{Q_{ij}} \;, 
\end{split}\label{eq:JpsiCM}\\[3pt]
\begin{split}%
\Phi^{P}_{i}  = &- 2\alpha_{2}\:\!\partial^{2}_{k}\Fdif{\mathcal{F}}{P_{i}} -4\alpha_{3}\:\!\partial_{i}\partial_{j}\Fdif{\mathcal{F}}{P_{j}}
+2\alpha_{4}\Fdif{\mathcal{F}}{P_{i}} \;,
\end{split}\label{eq:PhiPCM}\\[3pt]
\begin{split}%
\Phi^{Q}_{ij} = &-4\alpha_{1}\:\!\partial^{2}_{k}\Fdif{\mathcal{F}}{Q_{ij}} -2\alpha_{3}\bigg(\!2\:\!\partial_{i}\partial_{j}\Fdif{\mathcal{F}}{\psi} 
-\delta_{ij}\:\!\partial^{2}_{k}\Fdif{\mathcal{F}}{\psi}\bigg) +8\alpha_{4}\Fdif{\mathcal{F}}{Q_{ij}} \;.
\end{split}\label{eq:PhiQCM}%
\end{align}}%
The coefficients $\alpha_{i}$ in equations \eqref{eq:JpsiCM}-\eqref{eq:PhiQCM} are defined as 
\begin{equation}
\begin{split}
\alpha_{1}=\frac{D_{\parallel}+D_{\perp}}{8\lambda} \;, \qquad \alpha_{2}=\frac{D_{\parallel}+3D_{\perp}}{8\lambda} \;, 
\qquad \alpha_{3}=\frac{D_{\parallel}-D_{\perp}}{8\lambda} \;, \qquad \alpha_{4}=\frac{D_{\mathrm{R}}}{2\lambda} \;.
\end{split}
\end{equation}
Notice that $D_{\parallel}\geqslant D_{\perp}$ holds for all types of uniaxial particles, if the vector $\uu $ for the orientation of the 
axis of symmetry is chosen properly.

Finally, we remark that in three dimensions ($d=3$) the dynamics is much more involved but can in principle be derived along similar lines 
from DDFT \cite{RexWL2007}. Recently, the CMA dynamics was considered in reference \cite{YabunakaA2011}. A main result of this reference is shown 
in section \ref{sec:Atlc}.

\subsubsection{Numerical methods}

\paragraph{The equation of motion} 
The commonly used explicit time-stepping with a finite-difference (FD) scheme has been routinely applied for solving the EOM of the 
PFC models \cite{Hirouchi20091192}. However, owing to the high order differential operators appearing in the EOM of the PFC models 
(up to $12$th order), explicit time-stepping suffers from severe constraints. Energy stable large time-step implicit FD methods have been 
developed for the PFC \cite{SJNAAM.47.2269} and MPFC \cite{SJNAAM.7.2269} equations that lead to large sets of sparse algebraic equations. 
The resultant algebraic equations can be solved using nonlinear multigrid methods \cite{Hu20095323}. As elegant alternative, pseudo-spectral methods 
can be used that combine unconditionally stable time marching while resulting in algebraic equations of the diagonal form \cite{Cheng20086241}. 
Furthermore, pseudo-spectral methods offer exponential convergence with the spatial resolution as opposed to the polynomial convergence rate of the 
FD schemes. This means that with smaller spatial resolution one obtains results comparable to those obtained with high 
spatial resolution in the real space methods. Using operator splitting techniques the domain of the spectral methods has been extended to a wide 
range of problems, including PDEs with variable coefficients \cite{Tegze20091612}. A detailed study on the application of this method to the 
PFC models shows that the respective schemes can be parallelised easily and efficiently \cite{Tegze20091612}, yielding up to $10^5$ times faster 
computations, when compared to FD schemes. Adaptive time stepping \cite{Cheng20086241} may further accelerate the computations. 
Other numerical methods (\eg, stable semi-implicit finite element discretisation combined with adaptive time stepping \cite{PhilosMagLet.87.813}) 
have also been used for PFC models, however, further investigations are needed to assess their numerical accuracy.

\paragraph{The Euler-Lagrange equation and other saddle point finding methods} 
In recent works (see, \eg, references \cite{PhysRevLett.107.175702,JPhysCondMat.22.364101}) 
the ELE has been solved using a semi-spectral 
successive approximation scheme combined with the operator-splitting technique \cite{unpub.toth.tegze}. A different approach termed the fixed 
length simplified string method has been proposed to find the minimum energy path and the nucleation barrier in reference \cite{JPhysCondMat.22.364104}.

\subsubsection{Coarse-graining the PFC models}
Soon after the appearance of the 1M-PFC model, attempts have been made to use this atomistic approach as a basis for deriving phase-field-type 
coarse-grained models using the amplitude equation method. This extension to the 1M-PFC formalism, when combined with the adaptive grid, 
has the potential to enable simulations of mesoscopic phenomena ($\mu\mathrm{m}\to\mathrm{mm}$) that are resolved down to the atomic scale, 
still incorporating all the respective physics.

\paragraph{\label{sec:Aeborgt}Amplitude equations based on renormalisation group theory} 
Goldenfeld, Athreya, and Dantzig \cite{PhysRevE.74.011601} have developed a computationally efficient approach to polycrystalline solidification, 
based on the 1M-PFC model. The nanoscale particle density distribution is reconstructed from its slowly varying amplitude and phase, 
obtained by solving the rotationally covariant equations of motion derived from renormalisation group theory. They have shown in two dimensions 
that the microscopic density distributions from their amplitude and phase equations show a very close match to the 1M-PFC result. 
In later works Athreya and co-workers \cite{PhysRevE.76.056706,PhysRevE.72.020601} have combined this approach with adaptive mesh algorithms, 
leading to a substantial acceleration of the numerical code. (Possible ways of using the renormalisation group methods have been discussed in 
references \cite{PhysRevE.76.056706,PhysRevE.79.013601}.)

\paragraph{Phenomenological amplitude equations} 
Approximate treatments based on the amplitude expansion of the free energy of the 1M-PFC models (\ie, expressing $\mathcal{F}$ in terms of 
the Fourier amplitude of the dominant density waves) have been developed. They have been used for various purposes, such as determining the 
anisotropy of liquid-solid interfacial free energies \cite{PhysRevE.81.061601,PhysRevE.79.011607}, the Asaro-Tiller-Grinfeld (ATG) morphological instability 
of a stressed crystal surface, polycrystalline growth from the melt, grain-boundary energies over a wide range of misorientation, and grain-boundary 
motion coupled to shear deformation \cite{PhysRevB.81.214201,PhysRevB.81.165421}. Yeon \etal\ \cite{PhilosMag.90.237} have used an amplitude-equation 
approach to model the evolution of a two-phase system that has been validated by investigating the Gibbs-Thomson effect in two spatial dimensions. Elder, Huang, and 
Provatas \cite{PhysRevE.81.011602} have proposed amplitude representations for the binary PFC model in the cases of triangular lattice (2D) and 
bcc and fcc (3D) structures. The respective equations of motion have been related to those of the original PF theory of binary freezing 
and elasticity, providing explicit connection between the PF and PFC approaches. The abilities of the phenomenological amplitude models 
have been demonstrated for eutectic solidification, solute migration at grain boundaries, and for the formation of quantum dots  on nanomembranes 
\cite{PhysRevE.81.011602}.

\subsection{Phase diagrams the PFC models realize}
The different versions and extensions of the PFC model, reviewed in section \ref{gaeopfcm}, lead to different phase diagrams, 
which in turn depend on the dimensionality of the system.

\subsubsection{Phase diagram of single-component and binary systems}
\begin{figure}
\begin{center}
\includegraphics[width=0.34\linewidth]{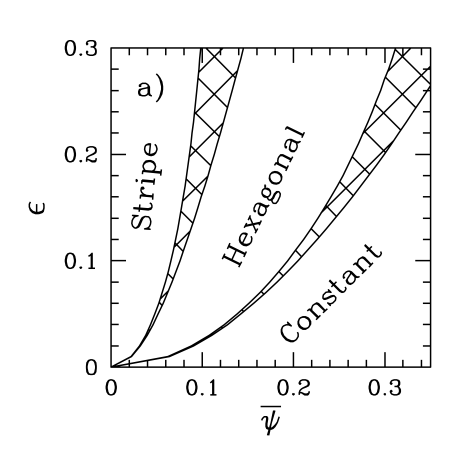}
\includegraphics[width=0.55\linewidth]{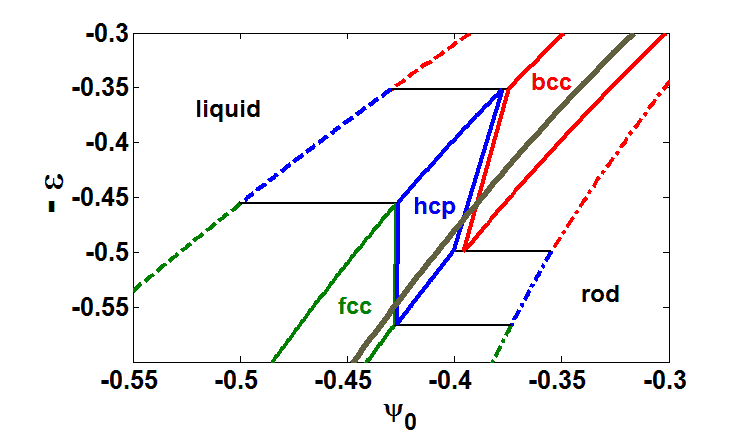}
\end{center}
\caption{\label{fig2}(a) Single-mode approximation to the phase diagram of the 1M-PFC model in two spatial dimensions with $\epsilon\equiv \beta$ and 
the average reduced particle density $\bar{\psi}$. 
\nocite{ElderKHG2002}\CRP{K.\ R.\ Elder, M.\ Katakowski, M.\ Haataja, and M.\ Grant}{Modeling elasticity in crystal growth}{Phys.\ Rev.\ Lett.\ }{88}{2002}{245701}{24}{10.1103/PhysRevLett.88.245701}{the American Physical Society}  
(b) Section of the 3D phase diagram of the 1M-PFC model evaluated by the Euler-Lagrange method described in reference \cite{JPhysCondMat.22.364101} 
with $\varepsilon\equiv \epsilon$ and the average reduced particle density $\psi_0$. 
Notice the stability domains of the bcc, hcp, and fcc phases. The homogeneous liquid is unstable right of the heavy grey line emerging from linear 
stability analysis \cite{JPhysCondMat.22.364101}. 
\nocite{JPhysCondMat.22.364101}\CRPE{G.\ I.\ T{\'o}th, G.\ Tegze, T.\ Pusztai, G.\ T{\'o}th, and L.\ Gr{\'a}n{\'a}sy}{Polymorphism, crystal nucleation and growth in the phase-field crystal model in 2D and 3D}{J.\ Phys.: Condens.\ Matter}{22}{2010}{364101}{36}{10.1088/0953-8984/22/36/364101}{Institute of Physics Publishing}}
\end{figure}

\begin{figure}
\begin{center}
\includegraphics[width=0.3\linewidth]{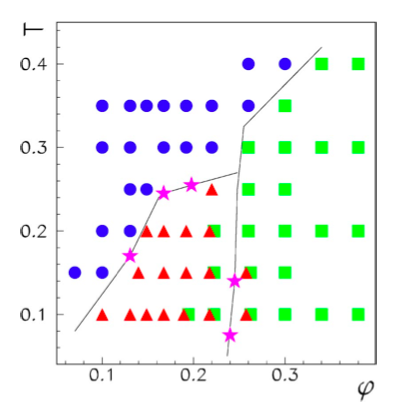}
\includegraphics[width=0.55\linewidth]{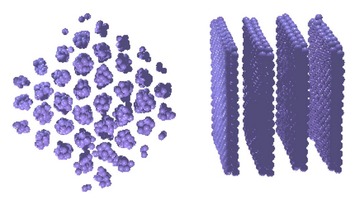}
\end{center}
{\vskip-5mm\footnotesize\hspace{0.22\linewidth}(a)\hspace{0.26\linewidth}(b)\hspace{0.27\linewidth}(c)}
\caption{\label{fig3}(a) Temperature-particle density phase diagram from MD simulations with a DLVO-type potential. 
Circles represent the disordered phase, triangles the columnar phase, and squares the lamellar phase. 
Stars stand for points, where the free energies of two phases cross. Solid lines are a guide for the eyes. 
Snapshots of the (b) triangular rod and (c) lamellar phases. 
\nocite{PhysRevE.74.010403}\CRP{A.\ de Candia, E.\ Del Gado, A.\ Fierro, N.\ Sator, M.\ Tarzia, and A.\ Coniglio}{Columnar and lamellar phases in attractive colloidal systems}{Phys.\ Rev.\ E}{74}{2006}{010403R}{1}{10.1103/PhysRevE.74.010403}{the American Physical Society}} 
\end{figure}

The phase diagrams of the 1M-PFC model \cite{ElderKHG2002} in two and three spatial dimensions are shown in figure \ref{fig2}. 
In two spatial dimensions, a single crystalline phase (the triangular phase) appears that coexists with the liquid and a striped phase \cite{ElderKHG2002}. 
Remarkably, phase diagrams of comparable structure have been predicted for weakly charged colloids with competing interactions \cite{EuroPhysLet.6.567}. 
In three spatial dimensions, as implied earlier by EOM studies \cite{PhysRevLett.103.035702}, and confirmed by full thermodynamic optimization \cite{JaatinenAN2010} 
and an equivalent method based on solving the ELE \cite{JPhysCondMat.22.364101}, stability domains exist for the homogeneous fluid, bcc, fcc, 
and hcp structures, besides the 3D version of the respective 2D structures: the rod and the lamellar structures. Interestingly, the rod and 
lamellar structures, and a phase diagram resembling to the 1M-PFC phase diagram appear in MD simulations for a Derjaguin-Landau-Verwey-Overbeek (DLVO) 
type potential \cite{PhysRevE.74.010403} (see figure \ref{fig3}), characteristic to charged colloidal systems. The 1M-PFC model prefers the formation of 
the bcc phase near the critical point.

\begin{figure}
\begin{center}
\includegraphics[width=0.41\linewidth]{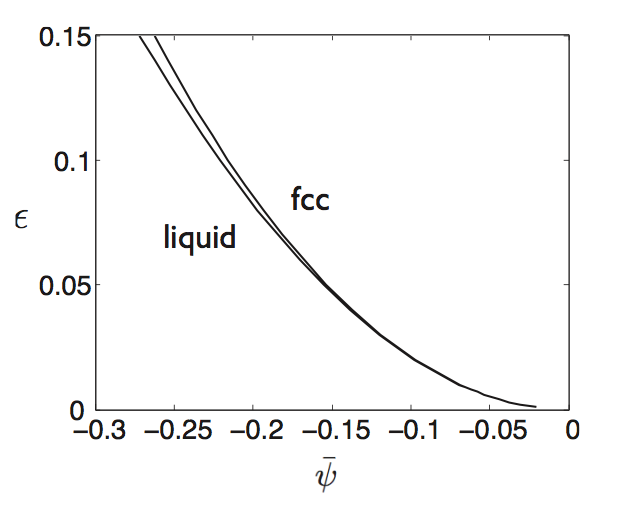}
\includegraphics[width=0.46\linewidth]{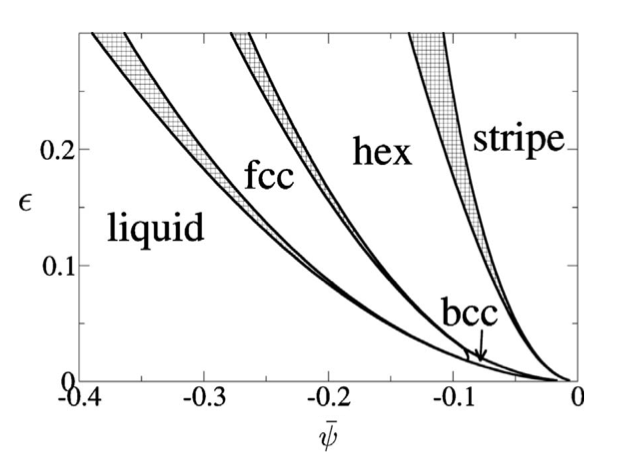}
\end{center}
{\vskip-5mm\footnotesize\hspace{0.265\linewidth}(a)\hspace{0.418\linewidth}(b)}
\caption{\label{fig4}(a) Single-mode approximations to the phase diagram of the 2M-PFC model in 3D for $R_1= 0$. (b) The same for $R_1= 0.05$.
Notice the small bcc stability domain near the critical point. 
\nocite{PhysRevE.81.061601}\CRP{K.-A.\ Wu, A.\ Adland, and A.\ Karma}{Phase-field-crystal model for fcc ordering}{Phys.\ Rev.\ E}{81}{2010}{061601}{6}{10.1103/PhysRevE.81.061601}{the American Physical Society}}
\end{figure}

In contrast, the 2M-PFC model of Wu \etal\ \cite{PhysRevE.81.061601} designed to realize fcc crystallization, suppresses the bcc phase 
[see figure \ref{fig4}(a)]. The 2M-PFC extension incorporates the 1M-PFC model as a limiting case. Interpolation between 1M-PFC and the full  
fcc limit ($R_1 = 0$), in terms of the parameter $R_1$ leads to the appearance of a bcc stability domain in the vicinity of the 
critical point [see figure \ref{fig4}(b)]. ($R_1$ is the ratio of the Fourier amplitudes for the density waves having the second and first neighbour 
RLVs as wave vector.) Whether here the appearance of the bcc phase is accompanied with the appearance of a hcp stability domain, 
as in the 1M-PFC limit, requires further investigation.

\begin{figure}
\begin{center}
\includegraphics[width=0.5\linewidth]{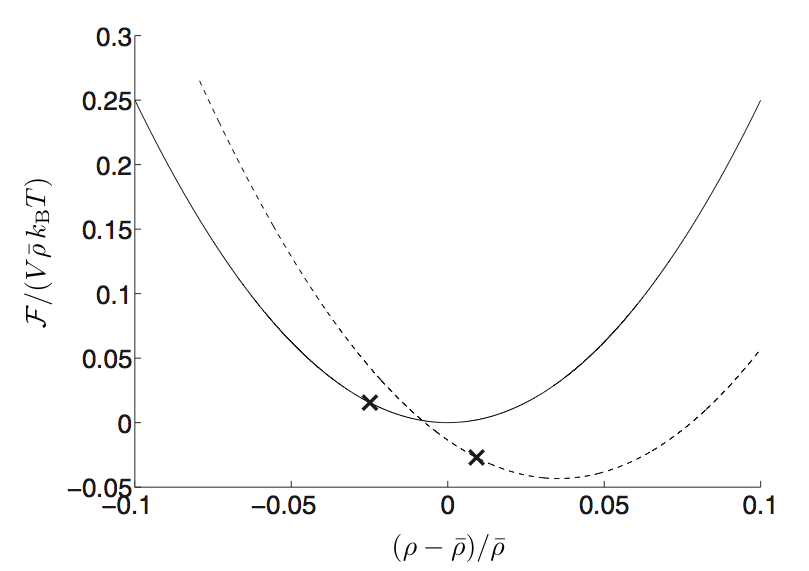}
\end{center}
\caption{\label{fig5}Free-energy density for the liquid (solid) and bcc (dashed) Fe as a function of reduced particle density in the EOF-PFC model. 
Crosses denote the equilibrium points obtained by the common tangent method. 
\nocite{JaatinenAEAN2009}\CRP{A.\ Jaatinen, C.\ V.\ Achim, K.\ R.\ Elder, and T.\ Ala-Nissila}{Thermodynamics of bcc metals in phase-field-crystal models}{Phys.\ Rev.\ E}{80}{2009}{031602}{3}{10.1103/PhysRevE.80.031602}{the American Physical Society}}
\end{figure}

The applicability of the EOF-PFC model has been demonstrated for Fe \cite{JaatinenAEAN2009}. The free energy \vs particle density curves for the 
solid and liquid phases, which were used to determine the equilibrium conditions at the melting point, are shown in figure \ref{fig5}. 
No phase diagram has been published for this model. It appears that much like the original, the 1M-PFC model, it prefers bcc freezing.

\begin{figure}
\begin{center}
\includegraphics[width=0.5\linewidth]{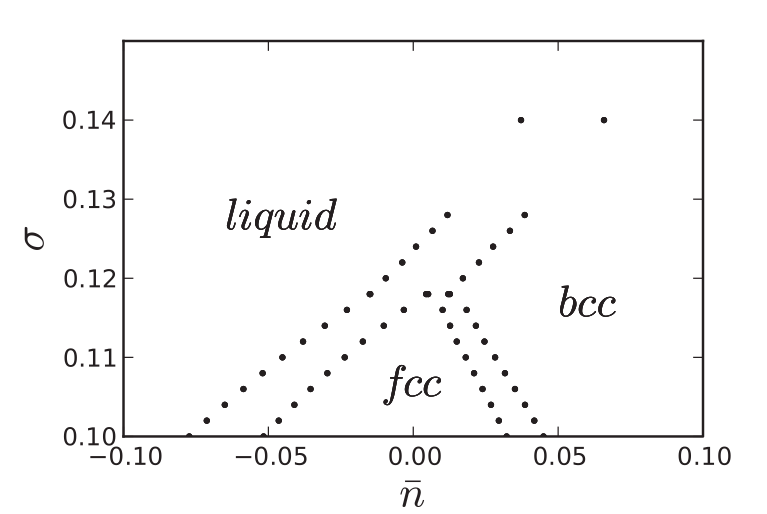}
\includegraphics[width=0.42\linewidth]{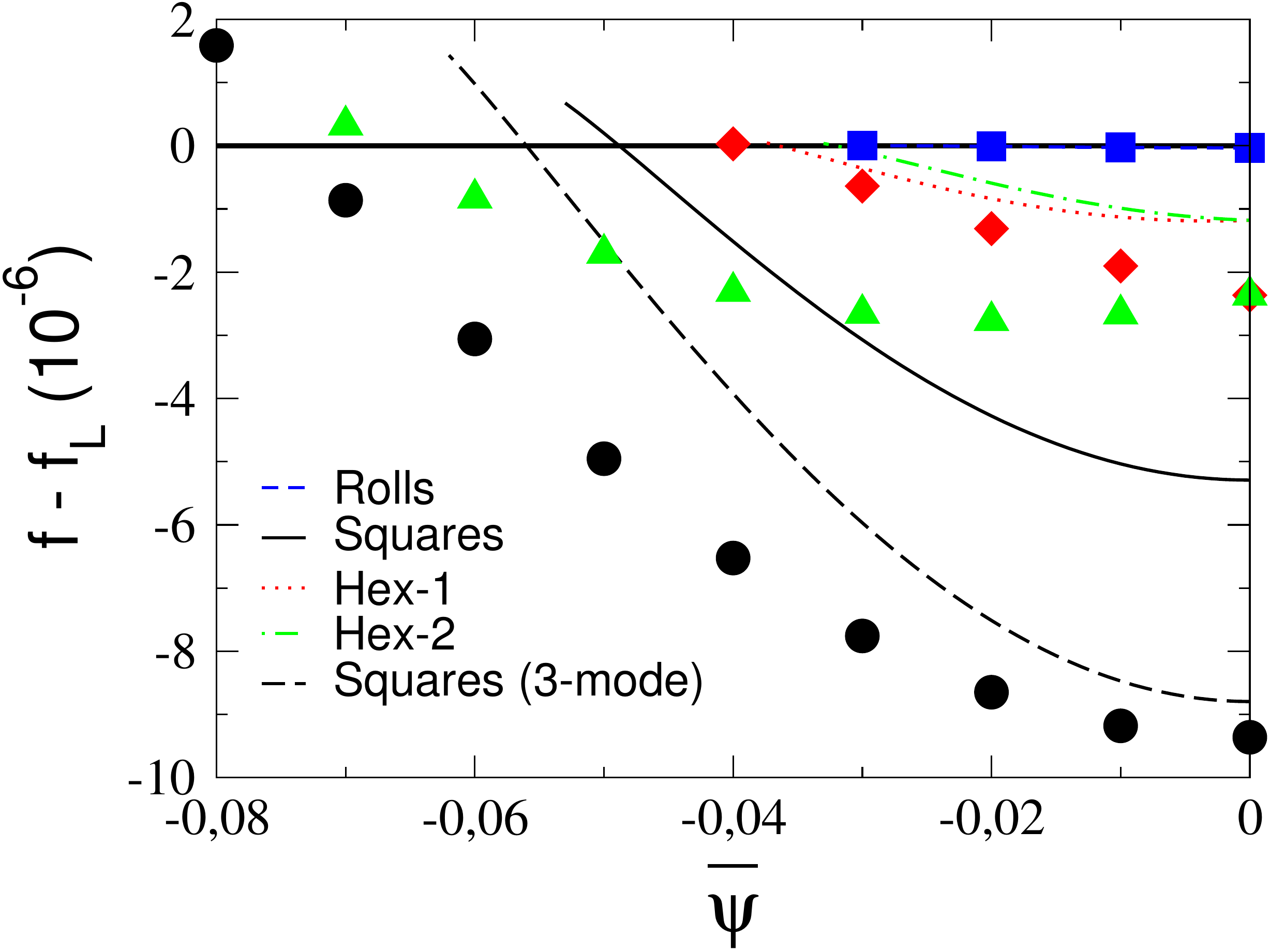}
\end{center}
{\vskip-4mm\footnotesize\hspace{0.29\linewidth}(a)\hspace{0.455\linewidth}(b)}
\caption{\label{fig6}(a) Phase diagram in the effective temperature $\sigma$-mean density $\bar{n}$-plane for a system
whose direct correlation function was manipulated so that peritectic coexistence between 
the fcc and bcc structures is realized. 
\nocite{PhysRevE.83.031601}\CRP{M.\ Greenwood, J.\ Rottler, and N.\ Provatas}{Phase-field-crystal methodology for modeling of structural transformations}{Phys.\ Rev.\ E}{83}{2011}{031601}{3}{10.1103/PhysRevE.83.031601}{the American Physical Society}  
(b) Free-energy density $f$ \vs average reduced particle density $\bar{\psi}$ when controlling the crystal structure via nonlinear resonance. 
$f_{\mathrm{L}}$ denotes the free-energy density of the liquid. The analytical solutions are plotted as lines, while the numerical simulation results 
for rolls, squares, hex-1, and hex-2 are plotted as squares, circles, diamonds, and triangles, respectively. 
\nocite{JPhysCondMat.22.364102}\CRP{K.-A.\ Wu, M.\ Plapp, and P.\ W.\ Voorhees}{Controlling crystal symmetries in phase-field crystal models}{J.\ Phys.: Condens.\ Matter}{22}{2010}{364102}{36}{10.1088/0953-8984/22/36/364102}{Institute of Physics Publishing}}
\end{figure}

An attempt to control the preferred crystal structure relies on manipulating the two-particle direct correlation function so that its peaks 
prefer the desired structural correlations \cite{PhysRevE.83.031601}. The type of phase diagram accessible for this method and the respective 
free-energy curves for coexistence between the bcc and fcc structures are displayed in figure \ref{fig6}(a). 
Starting from the observation that the nonlinearities can stabilize the square lattice \cite{ZPhyB.57.329} so that it coexists with the 
triangular phase \cite{PhysRevE.54.1560}, Wu, Plapp, and Voorhees \cite{JPhysCondMat.22.364102} have proposed a method based on nonlinear resonance 
for constructing PFC models that prefer the desired crystal structure, as they indeed demonstrated for the square lattice in two spatial dimensions: 
the relative free energies of possible competing structures are compared in figure \ref{fig6}(b).

\begin{figure}
\begin{center}
\includegraphics[width=0.57\linewidth]{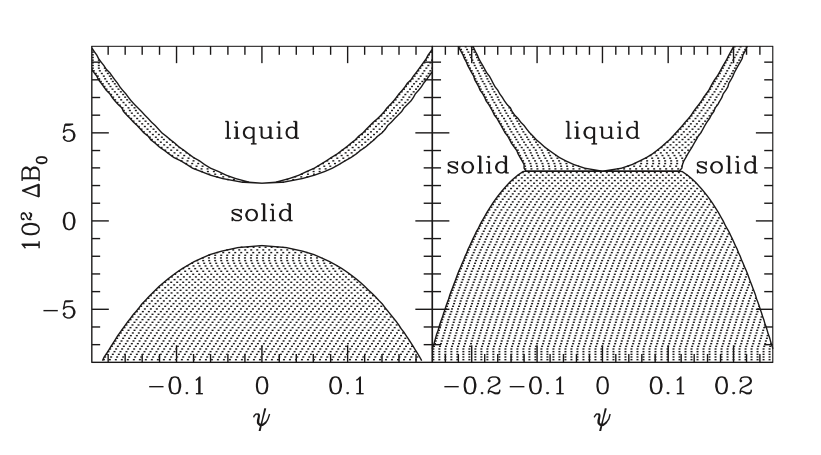}
\includegraphics[width=0.42\linewidth]{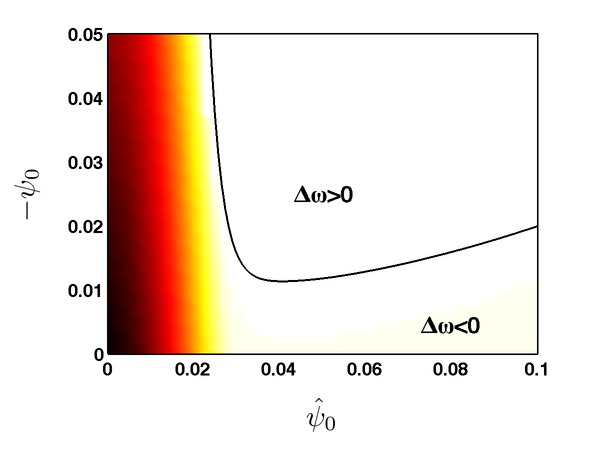}
\end{center}
{\vskip-4mm\footnotesize\hspace{0.17\linewidth}(a)\hspace{0.209\linewidth}(b)\hspace{0.355\linewidth}(c)}
\caption{\label{fig7}(a), (b) Reduced temperature $\Delta B_0$ \vs reduced particle density difference $\psi$ phase diagrams for the 
two-dimensional triangular system in the amplitude equation formalism. The filled regions correspond to regions of phase coexistence. 
\nocite{PhysRevE.81.011602}\CRP{K.\ R.\ Elder, Z.-F.\ Huang, and N.\ Provatas}{Amplitude expansion of the binary phase-field-crystal model}{Phys.\ Rev.\ E}{81}{2010}{011602}{1}{10.1103/PhysRevE.81.011602}{the American Physical Society}  
(c) Colour map for the thermodynamic driving force of eutectic solidification ($\Delta\omega$ is the grand potential density difference 
relative to the liquid) for the binary 1M-PFC model in 3D \cite{JPhysCondMat.22.364101}, as a function of the chemical 
composition $\KDN_0$ and density $\psi_0$. Notice that solidification is expected in the region, where $\Delta\omega < 0$. 
\nocite{JPhysCondMat.22.364101}\CRPE{G.\ I.\ T{\'o}th, G.\ Tegze, T.\ Pusztai, G.\ T{\'o}th, and L.\ Gr{\'a}n{\'a}sy}{Polymorphism, crystal nucleation and growth in the phase-field crystal model in 2D and 3D}{J.\ Phys.: Condens.\ Matter}{22}{2010}{364101}{36}{10.1088/0953-8984/22/36/364101}{Institute of Physics Publishing}}
\end{figure}

Elder \etal\ \cite{ElderPBSG2007} have obtained eutectic phase diagrams within the binary generalisation of the 1M-PFC model in two spatial dimensions both 
numerically and by using the single-mode approximation. Comparable phase diagrams [see figures \ref{fig7}(a) and \ref{fig7}(b)] have been reported for 
triangular phases in two spatial dimensions, and for the bcc and fcc phases in 3D, by the phenomenological amplitude equation method \cite{PhysRevE.81.011602}. 
The 3D extension of the binary 1M-PFC model has been investigated by T\'oth \etal\ \cite{JPhysCondMat.22.364101}. The map of the thermodynamic driving 
force for solidification as a function of composition and density of the initial liquid is shown in figure \ref{fig7}(c). 3D eutectic solidification 
to bcc phases of different lattice constants has indeed been observed in the domain of the largest driving force.

An essential question is how to extend the PFC framework to accommodate the physical properties of real alloys of different crystalline structures. 
A two-phase binary extension of the EOF-PFC model based on the 1M-PFC and 2M-PFC concepts might be worth exploring.

\begin{figure}
\begin{center}
\includegraphics[width=0.9\linewidth]{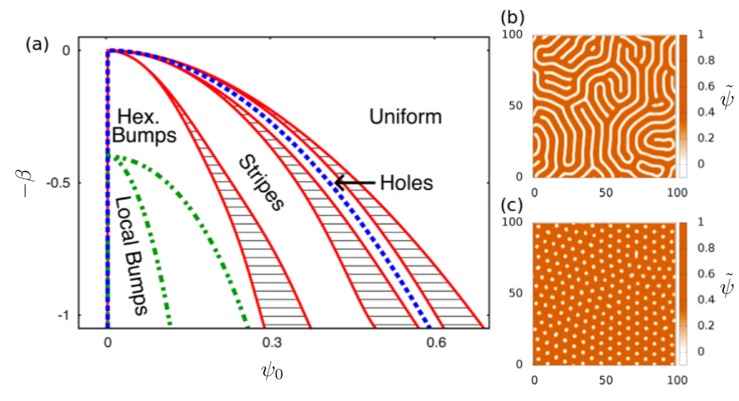}
\end{center}
\caption{\label{fig8}Phase diagram for the VPFC model in two spatial dimensions [see equation \eqref{eq13}] (a) for $\wn_0 = 1$ \cite{RobbinsATK2012}. 
The red solid lines denote coexistence curves, the green dash-dotted lines envelope the region, where localized and hexagonally ordered density peaks 
(bumps) coexist, whereas the blue dashed line indicates the linear stability limit of the spatially uniform phase. 
Snapshots of simulations for (b) stripes and (c) hexagonally ordered holes are also shown. These simulations have been performed for: 
$\wn_0 =1$, $\beta = 0.9$, $h=1500$, $M=1$ (the mobility in the EOM) and (b) $\psi_0 = 0.4$ and (c) $\psi_0 = 0.53$.
\nocite{RobbinsATK2012}\CRPE{M.\ J.\ Robbins, A.\ J.\ Archer, U.\ Thiele, and E.\ Knobloch}{Modelling fluids and crystals using a two-component modified phase field crystal model}{Phys.\ Rev.\ E}{85}{2012}{061408}{6}{10.1103/PhysRevE.85.061408}{the American Physical Society}}%
\end{figure}

The VPFC extension of the 1M-PFC model leads to a restructured phase diagram (see figure \ref{fig8}) \cite{RobbinsATK2012}, 
whose central region contains stability domains for the 2D hexagonal crystal and localized density peaks (\ZT{bumps}) that represent 
individual particles. 

\subsubsection{\label{app:statics}Phase diagram of two-dimensional liquid crystals}
For liquid crystalline systems, there are much more candidates for possible
bulk phases. Therefore the topology of the bulk phase diagram is getting more
complex. Recently, bulk phase diagrams were  computed by using the two-dimensional
apolar PFC model  of L\"owen \cite{Loewen2010}. After an appropriate scaling in energy and length,
this model reads as
\begin{equation}%
\begin{split}%
\mathcal{F}[\psi,Q_{ij}]=\!\!\vint \Big(\!&-\frac{\psi^{3}}{3}+\frac{\psi^{4}}{6} 
+(\psi-1)\frac{\psi Q^{2}_{ij}}{4}+\frac{Q^{2}_{ij}Q^{2}_{kl}}{64}\\
&+A_{1}\psi^{2} +A_{2}\psi(\partial^{2}_{k}+\partial^{2}_{k}\partial^{2}_{l})\psi\\[0.8mm]
&+B_{3}(\partial_{i}\psi)(\partial_{j}Q_{ij})+D_{1}Q^{2}_{ij}+D_{2}(\partial_{j}Q_{ij})^{2}\Big) \,.
\end{split}%
\label{eq:apolar_dimensionslos}%
\end{equation}%
Some trends of the phase diagram can directly be read of equation \eqref{eq:apolar_dimensionslos}. 
Since the parameter $D_{1}$ controls the contribution of the nematic tensor $Q_{ij}(\vec{r})$ and therefore also of the 
nematic order parameter $S(\vec{r})$, the nematic phase can be expected to be stable for large negative values of $D_{1}$. 
In the opposite case, if $D_{1}$ is large enough and positive, the term $D_{1}Q^{2}_{ij}(\vec{r})+Q^{2}_{ij}(\vec{r})Q^{2}_{kl}(\vec{r})/64$ 
dominates the free energy and only phases with $Q_{ij}(\vec{r})\propto S(\vec{r})=0$ can be stable. Crystalline phases with a non-vanishing 
nematic order can therefore only appear in a region around $D_{1}=0$. From previous work it is known that the difference $A_{1}-A_{2}/4$ has 
a big influence on the translational density field $\psi(\vec{r})$ \cite{AchimWL2011,ElderPBSG2007}. If the parameter $A_{1}$ is large and positive, 
variations of the translational density field enlarge the free energy. Similarly, gradients of the translational density field enlarge the 
free energy for large and negative values of $A_{2}$. Therefore, phases without any density modulations, \ie, the isotropic and the nematic phase, 
are preferred for positive values of the difference $A_{1}-A_{2}/4$, while all other phases with a periodic translational density field are 
preferred for negative values of this difference. Furthermore, there is a symmetry concerning the reversal of the sign of the parameter $B_{3}$ 
in the free-energy functional. From equation \eqref{eq:apolar_dimensionslos} follows directly, that the free-energy functional is invariant under 
a simultaneous change of the signs of the parameter $B_{3}$ and the nematic order-parameter field $S(\vec{r})$. Due to this symmetry, 
$B_{3}\geqslant 0$ can be assumed without loss of generality.

\begin{figure}
\centering\includegraphics[width=\linewidth]{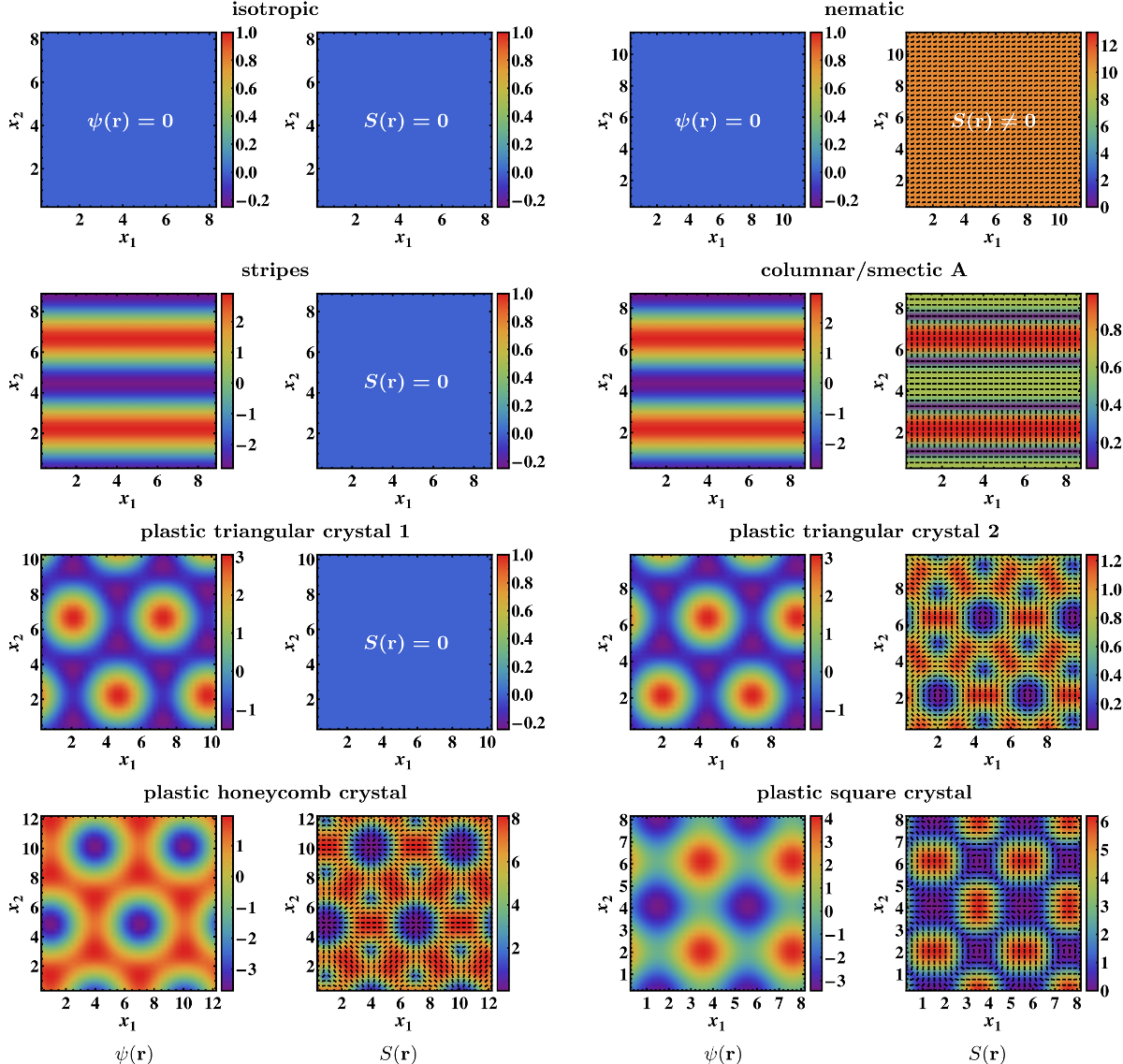}%
\caption{\label{fig:PA}Stable liquid crystalline phases. 
The contour plots show the order-parameter fields $\psi(\vec{r})$ and $S(\vec{r})$ in the $x_{1}$-$x_{2}$-plane for the isotropic and nematic phases, 
the stripe phase and columnar/smectic A phase, two plastic triangular crystals with different orientational ordering, and a plastic honeycomb crystal 
as well as a plastic square crystal. The black lines in the plots of the second and fourth column represent the director field $\nn (\vec{r})$. 
In the plots with $S(\vec{r})=0$, the director field is not shown because it is not defined. The parameters are $A_{2}=14$, $D_{2}=8$, and $B_{3}=0$ 
for the stripe phase and the plastic triangular crystal 1 and $A_{2}=14$, $D_{2}=0.8$, and $B_{3}=-4$ for all other phases. 
\nocite{AchimWL2011}\CRPE{C.\ V.\ Achim, R.\ Wittkowski, and H.\ L\"owen}{Stability of liquid crystalline phases in the phase-field-crystal model}{Phys.\ Rev.\ E}{83}{2011}{061712}{6}{10.1103/PhysRevE.83.061712}{the American Physical Society}}%
\end{figure}%

Depending on the coupling parameters, a wealth of different stable liquid crystalline phases was found (see figure \ref{fig:PA}). 
They include an \textit{isotropic} phase, which has no translational and no orientational ordering ($\psi(\vec{r})=0$ and $S(\vec{r})=0$), 
a \textit{nematic} phase with pure orientational ordering ($\psi(\vec{r})=0$ and $S(\vec{r})>0$), a \textit{columnar} phase 
(or equivalently a \textit{smectic A} phase), where the translational density and the orientation show a one-dimensional undulation, 
and various \textit{plastic crystals}. In the columnar/smectic A phase the system has positional ordering in one direction, while it is 
isotropic perpendicular to this direction. The nematic order-parameter field $S(\vec{r})$ for this phase has a similar profile to the reduced 
translational density field $\psi(\vec{r})$ with maxima of these two fields at the same positions. Near the maxima of the 
translational density $\psi(\vec{r})$, the director field $\nn (\vec{r})$ is preferentially parallel to the gradient $\partial_{i}\psi(\vec{r})$, 
while it is perpendicular to $\partial_{i}\psi(\vec{r})$ around the minima of $\psi(\vec{r})$. A similar flipping of the orientational field 
from perpendicular to parallel to the stripe direction was identified as transverse intralayer order in the three-dimensional smectic A phase of 
hard spherocylinders \cite{vanRoijBMF1995}.

\begin{figure}
\centering\includegraphics[width=\linewidth]{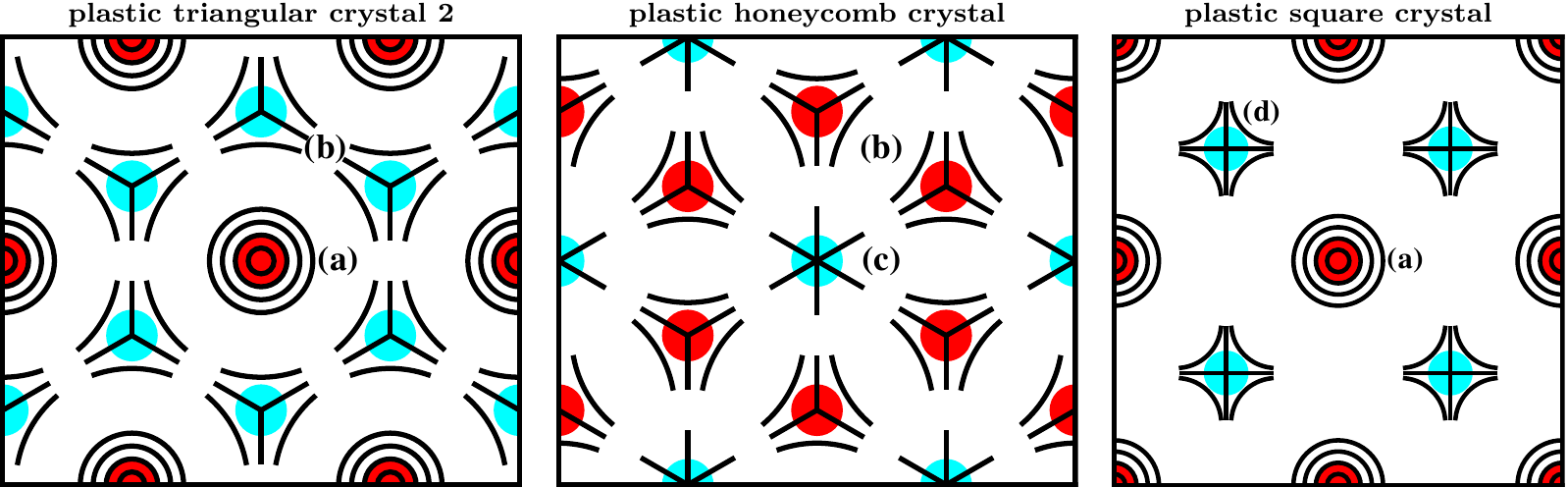}%
\caption{\label{fig:D}Topological defects in three different plastic liquid crystals in the $x_{1}$-$x_{2}$-plane (schematic). 
The defects coincide with the maxima (red discs) and minima (cyan discs) of the translational density field $\psi(\vec{r})$.  
The symbols in the plots represent the following defects: (a) vortices with the \textit{topological winding number} $m=1$, 
(b) disclinations with $m=-1/2$, (c) sources/sinks with $m=1$, and (d) hyperbolic points with $m=-1$. 
\nocite{AchimWL2011}\CRPE{C.\ V.\ Achim, R.\ Wittkowski, and H.\ L\"owen}{Stability of liquid crystalline phases in the phase-field-crystal model}{Phys.\ Rev.\ E}{83}{2011}{061712}{6}{10.1103/PhysRevE.83.061712}{the American Physical Society}}%
\end{figure}

Plastic crystals are two-dimensional modulations of the translational density and have no global (averaged) orientational direction. 
Interestingly, depending on the coupling-parameter combinations, three different types of plastic crystals are stable including two-dimensional 
triangular, square, and honeycomb lattices for the translational ordering. The orientational field of plastic crystals is schematically shown in 
figure \ref{fig:D}. It exhibits an interesting defect structure in the orientation. There are defects at the density peaks and in the interstitial 
places of the lattice. The defect structure has not yet been explored so far outside the PFC-world. It can be confirmed in microscopic computer 
simulations \cite{CremerML2012} or in real-space experiments \cite{DemiroersJvKvBI2010,GerbodeAOLEC2010,MarechalD2008}.

\begin{figure}
\centering\includegraphics[width=0.63\linewidth]{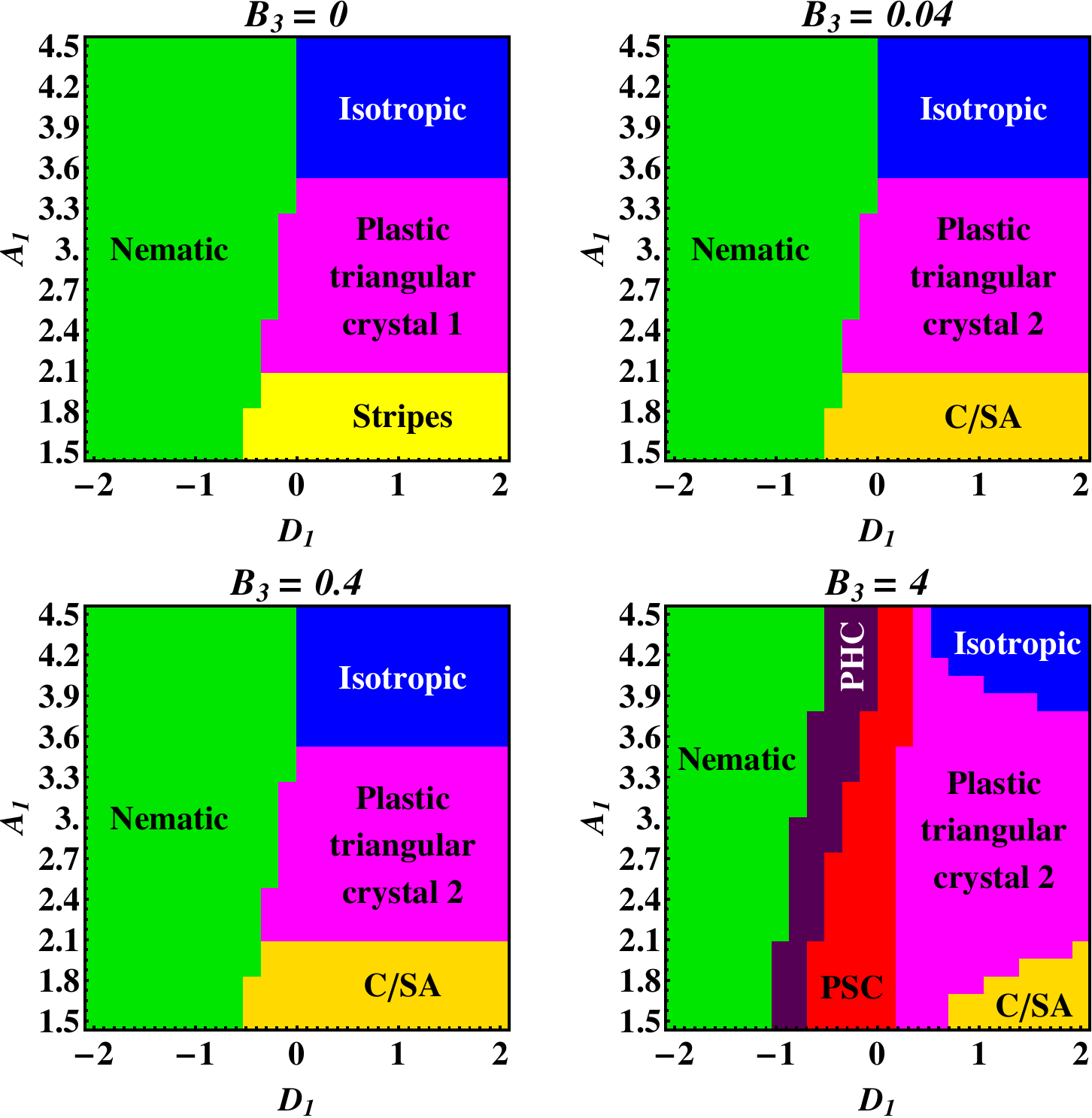}%
\caption{\label{fig:PD}Phase diagrams for the parameters $A_{2}=14$ and $D_{2}=0.8$. The relevant liquid crystalline phases are isotropic (blue), 
nematic (green), stripes (yellow), columnar/smectic A (C/SA, light orange), plastic triangular crystals (magenta), plastic honeycomb crystal 
(PHC, dark purple), and plastic square crystal (PSC, red). The cornered separation lines between different phases are due to the finite numerical 
resolution of the parameter space. 
\nocite{AchimWL2011}\CRPE{C.\ V.\ Achim, R.\ Wittkowski, and H.\ L\"owen}{Stability of liquid crystalline phases in the phase-field-crystal model}{Phys.\ Rev.\ E}{83}{2011}{061712}{6}{10.1103/PhysRevE.83.061712}{the American Physical Society}}%
\end{figure}

In a two-dimensional slice of the coupling parameter space, the phase diagram is shown in figure \ref{fig:PD}. 
Apart from the isotropic phase with $\psi(\vec{r})=0$ and $S(\vec{r})=0$, which appears for $A_{1}>A_{2}/4$ and $D_{1}>0$, 
for negative and large $D_{1}$ a nematic phase is stable. A rich topology occurs around $D_{1}=0$ including columnar phases and 
three different plastic crystals. 
However, one should bear in mind that the phase transitions shown in figure \ref{fig:PD} were assumed to be isochoric, \ie, 
without any density jump. While this is in general a good approximation for liquid crystalline phase transitions, 
it is not true in general.
Independent of the particular value of the parameter $B_{3}$, the phase transition between the isotropic and the nematic phase is  
continuous, while all other phase transitions are discontinuous.

In conclusion, rich phase diagrams with novel liquid crystalline phases were found in the apolar PFC model. 
This gives confidence that in a further step interfaces between coexisting phases and the dynamics can be explored on the basis of the PFC approach. 
It would open the way of PFC models to enter into the rich world of liquid crystals.

\subsection{Anisotropies in the PFC models}
One of the most attractive features of the PFC-type models is that the anisotropies for various physical properties follow directly form the 
crystal structure. The anisotropy of the interfacial free energy has been addressed theoretically and numerically by several authors 
\cite{PhysRevB.76.184107,PhysRevE.79.011607,PhysRevLett.103.035702,JPhysCondMat.21.464109,PhilosMag.91.123,tms2011sandiego}, 
whereas the growth anisotropy has been evaluated numerically in 3D \cite{PhysRevLett.103.035702}.

\subsubsection{Free energy of the liquid-solid interface}
\begin{figure}
\begin{center}
\includegraphics[width=\linewidth]{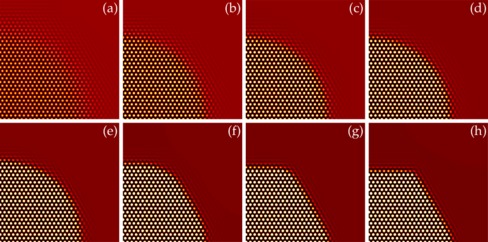}
\end{center}
\caption{\label{fig9}Equilibrium shape for the 1M-PFC model as determined by solving the EOM \vs reduced temperature $\epsilon$  
in the absence of noise \cite{PhilosMag.91.123}. (a) - (h): $\epsilon = 0.05,0.10,0.15,0.20,0.25,0.30,0.325$, and $0.35$. 
Notice that the interface thickness decreases while the anisotropy increases with an increasing distance from the critical point. 
The computations have been performed on a  $1024 \times 1024$ rectangular grid (the upper right quarter of the simulations is shown), 
whereas the crystalline fraction was $\approx 0.3$. Equilibration has been performed for a period of $106$ dimensionless time steps. 
Reduced particle density maps are shown. 
\nocite{PhilosMag.91.123}\CRPE{L.\ Gr{\'a}n{\'a}sy, G.\ Tegze, G.\ I.\ T{\'o}th, and T.\ Pusztai}{Phase-field crystal modelling of crystal nucleation, heteroepitaxy and patterning}{Philos.\ Mag.\ }{91}{2011}{123-149}{1}{10.1080/14786435.2010.487476}{Taylor \& Francis}}
\end{figure}

\begin{figure}
\begin{center}
\includegraphics[width=0.32\linewidth]{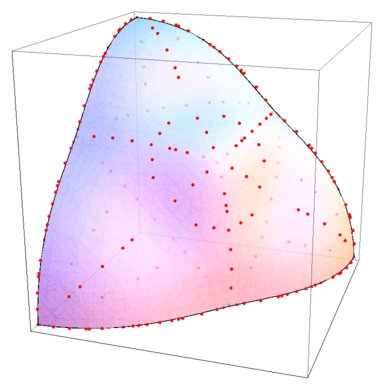}
\includegraphics[width=0.32\linewidth]{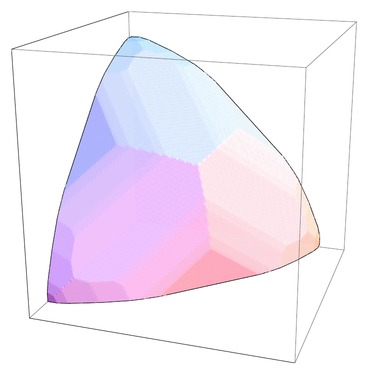}
\includegraphics[width=0.32\linewidth]{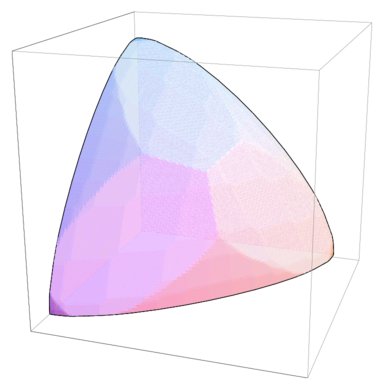}
\end{center}
{\vskip-4mm\footnotesize\hspace{0.17\linewidth}(a)\hspace{0.3\linewidth}(b)\hspace{0.3\linewidth}(c)}
\caption{\label{fig10}Anisotropy of the bcc-liquid interfacial free energy at $\epsilon = 0.3748$ evaluated numerically for the 
single-component 1M-PFC model and the respective equilibrium shapes \cite{tms2011sandiego}. 
(a) Gamma plot. Red dots represent the directions for which the interfacial free energy has been evaluated. 
The surface is a seventh-order cubic harmonics fit. Heavy dots are on the viewer's side of the surface. 
Light dots are on the other side of the surface. 
(b) Wulff shape evaluated from the red dots in the gamma plot. 
(c) The Wulff shape evaluated from a seventh-order cubic harmonics fit to the gamma plot.} 
\end{figure}

\paragraph{Numerical results} 
Backofen and Voigt \cite{JPhysCondMat.21.464109} have determined the anisotropy of the crystal-liquid interfacial free energy for small 2D clusters 
from simulations performed using the 1M-PFC model. They have observed a strong dependence of the interfacial free energy on the distance from the 
critical point. The results have been fitted with the formula of Stashevich \etal\ \cite{PhysRevB.71.245414} from low temperature expansion. 
Remarkably, the anisotropy shows strong size-dependence, when reducing the cluster size to a few particles. 
Comparable results have been obtained by Gr\'an\'asy \etal\ \cite{PhilosMag.91.123,C0SM00944J} for the equilibrium shapes in a broader 
reduced temperature range (see figure \ref{fig9}). T\'oth \etal\ \cite{JPhysCondMat.22.364101} evaluated the free energy and thickness of flat 
interfaces as a function of $\epsilon$ via solving the ELE for the equilibrium $(10\bar{1})$ triangular crystal-liquid interface, and have shown 
that mean-field critical exponents apply. A similar approach has been applied in 3D for a large number of orientations of the 
flat bcc-liquid interface forming at $\epsilon = 0.3748$. The orientation dependence of the interfacial free energy and the respective Wulff plot 
are presented in figure \ref{fig10} \cite{tms2011sandiego}. Apparently, the rhombic-dodecahedral equilibrium shape \cite{PhysRevLett.103.035702} 
obtained from simulations performed using the EOM for the same $\epsilon$ has been a growth form. It became also evident that the usual 
cubic harmonic expansion is not sufficient for reproducing reasonably the anisotropy even if terms up to seventh order are considered.

\paragraph{Analytical results} 
Wu and Karma \cite{PhysRevB.76.184107} have used multi-scale analysis close to the critical point ($0<\epsilon\ll 1$) to evaluate the anisotropy 
of the interfacial fee energy. They have approximated the EOM of the 1M-PFC model by a set of coupled equations describing the 
time evolution of the amplitudes of the dominant density waves. The analysis of the stationary solution led to an anisotropy that is independent 
of the reduced temperature $\epsilon$. This finding accords with those of Majaniemi and Provatas \cite{PhysRevE.79.011607}, who have used a 
simple coarse-graining technique, the local volume averaging method, for deriving amplitude equations for liquid-solid interfaces broad relative 
to the periodicity of the crystalline phase. In both studies, the temperature-independent anisotropy is a direct consequence of the approximations 
that lead to weakly fourth-order amplitude theories of the Ginzburg-Landau type, in which all material parameters can be scaled out from the 
free-energy functional \cite{PhysRevB.76.184107,PhysRevE.79.011607}. Accordingly, the anisotropy of the interfacial free energy depends only on 
the crystal structure. However, this independence of the anisotropy from temperature is unphysical, as it must vanish when the correlation length 
(the width of the liquid-solid interface) diverges in the critical point. In contrast to these phenomenological coarse-graining techniques, 
the renormalisation-group-based approaches by Athreya, Goldenfeld, and Dantzig \cite{PhysRevE.74.011601} lead to higher-order amplitude equations, 
from which the temperature cannot be scaled out; \ie, a temperature-dependent anisotropy is expected as found in the virtually exact numerical 
studies.

\subsubsection{Growth anisotropy}
Tegze \etal\ \cite{PhysRevLett.103.035702} have investigated the growth rate anisotropy in the 1M-PFC model at $\epsilon = 0.3748$ for freezing 
to the bcc, hcp, and fcc structures. Diffusion-controlled layerwise crystal growth has been observed, a mechanism that is consistent with 
two-dimensional nucleation. The predicted growth anisotropies were found to decrease with increasing thermodynamic driving force (or velocity) 
consistently with kinetic roughening expected on the basis of 1M-PFC simulations performed in two spatial dimensions \cite{C0SM00944J}.

\subsection{Glass formation} 
One of the most intriguing phenomena that may happen during solidification is glass formation, a process by which the undercooled liquid is 
transformed into an amorphous solid. An early 1M-PFC study for single-component system by Berry \etal\ \cite{PhysRevE.77.061506} 
relying on conservative overdamped dynamics indicated a first-order phase transition for this phase change, whereas the amorphous structure 
resembled closely to the glass structure obtained by embedded-atom-potential (EAP) molecular dynamics simulations for glassy 
Fe or Ni \cite{nanotech.20.295703,WANG2006s327}. 
These findings have been confirmed for the 1M-PFC model by T\'oth \etal\ \cite{PhysRevLett.108.025502} and for the EOF-PFC model fitted 
to Fe by T\'oth \etal\ \cite{JPhysCondMat.22.364101} and Gr\'an\'asy \etal\ \cite{PhilosMag.91.123}. In a more recent study, Berry, and Grant 
\cite{PhysRevLett.106.175702} have addressed glass formation in the framework of the monatomic and binary versions of the VPFC model with 
equation(s) of motion like equation \eqref{eq18} considering inertia and damping. They have shown that important aspects of glass formation can be 
reproduced over multiple time scales, including the agreement with mode-coupling theory (MCT) for underdamped liquids at low undercoolings and a 
rapidly growing dynamic correlation length that can be associated with a fragile behaviour. It appears that in the original PFC model 
glass formation takes place via a first-order phase transition, while the monatomic VPFC model behaves like poor glass formers, whereas the binary 
VPFC model displays features consistent with good glass formers.

\subsection{Phase-field-crystal modelling of foams} 
\begin{figure}
\begin{center}
\includegraphics[width=0.9\linewidth]{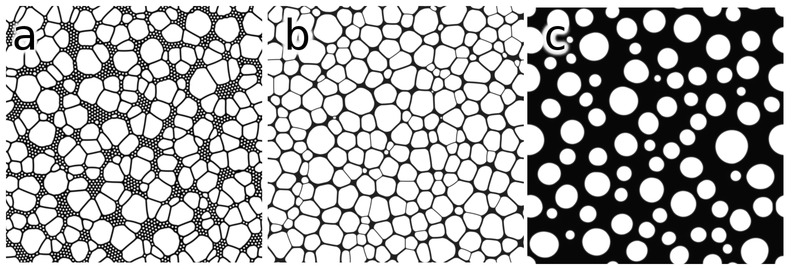}
\end{center}
\caption{\label{fig25}Foam structures predicted by the MPFC model. 
(a) Coexisting atoms and large cell foam. (b) Dry foam with no residual atoms. (c) Wet foam with circular bubbles. 
\nocite{PhysRevE.81.065301}\CRP{N.\ Guttenberg, N.\ Goldenfeld, and J.\ Dantzig}{Emergence of foams from the breakdown of the phase field crystal model}{Phys.\ Rev.\ E}{81}{2010}{065301R}{6}{10.1103/PhysRevE.81.065301}{the American Physical Society}}
\end{figure}

Working at extreme distances from the critical point, Guttenberg, Goldenfeld, and Dantzig \cite{PhysRevE.81.065301} have shown that the 1M-PFC model 
can be used to describe the formation of foams. 
Under such conditions, the free energy of the periodic phases exceeds that for a mixture of two immiscible liquids -- a situation that leads to
the appearance of a foam-like structure coarsening with time. 
Starting from this observation, the authors present a simple PFC-type continuum scalar theory of wet and dry foams (see figure \ref{fig25}).

\subsection{Coupling to hydrodynamics}
An interesting approach has been brought forward recently by Ramos \etal\ in reference \cite{PhysRevE.81.011121}.  
It couples the PFC dynamics of the form of equation \eqref{eq0} to a Stokes-type dynamics as well as to an external pinning potential. 
The general set of equations the authors arrive at is given by
\begin{equation}
\begin{split} 
\pdif{\rho}{t} &= -\Nabla\vec{g} \,, \\
\pdif{g_i}{t} &= -\rho\,\partial_i \Fdif{H_{\mathrm{int}}}{\rho} + \rho f_i -\eta g_i + \nu_i \,,
\end{split} 
\label{ramos10}
\end{equation}
where the specific configurational energy contribution $H_{\mathrm{int}}$ needs to be specified depending on the system.  
Assuming a periodic system, it is specified in reference \cite{PhysRevE.81.011121} as 
\begin{equation}
H_{\mathrm{int}} = \vint \bigg( \frac{\psi}{2} \big(-\epsilon + \big(1 + \Nabla^2\big)^2 \big) \psi 
+ \frac{\psi^4}{4} + \psi U_{\mathrm{p}} \bigg) \,.
\label{ramos12}
\end{equation}
In equations \eqref{ramos10}, $\rho(\vec{r})$ and $\vec{g}(\vec{r})$ refer to particle and momentum density, respectively.  
Moreover, $f_i$ denotes the components of an external force vector and $\nu_i(\vec{r},t)$ denotes noise components.  
Furthermore, $\eta$ is a dissipative coefficient and $U_{\mathrm{p}}$ accounts for the pinning potential.

Even if at first glance the coupled set of equations \eqref{ramos10} looks like a first step 
towards coupling of the PFC model based on equation \eqref{eq0} to a full Navier-Stokes (NS) type equation and thus 
to hydrodynamic motion, a respective extension of equations \eqref{ramos10} is not 
straightforward due to the simple relation between particle density $\rho(\vec{r})$ and momentum density $\vec{g}(\vec{r})$ assumed. 
Thus instead of providing a step towards an extension of PFC models towards 
hydrodynamics, equations \eqref{ramos10} rather provide a general framework for the 
inclusion of inertia effects in PFC models. The latter issue has also been addressed in 
references \cite{PhysRevE.74.021104,PhysRevE.79.011606,PhysRevE.79.051110},
where it was shown that the consideration of inertia terms allows to include fast degrees of freedom into the
PFC approach. Furthermore, the authors of reference \cite{PhysRevE.79.051110} have shown that inclusion of such 
fast degrees of freedom yields a two-stage relaxation process of the system. Chen \etal\ proved the 
thermodynamic consistency of such models with fast degrees of freedom \cite{moweitobe}. 
Thereby, these authors validated from the viewpoint of thermodynamic theory that such a two-stage relaxation process can 
truly be regarded as an important qualitative feature of nonequilibrium pattern formation in periodic systems. 
To do so, they developed a unified thermodynamic framework applying to both conventional PF models and PFC models
with slow and fast degrees of freedom.  Based on this framework it is now possible to validate also PFC models 
based on their thermodynamic consistency, as done previously in numerous cases for PF models (see reference \cite{emmlnpm73} for examples). 
Experimental evidence of inertia contributions to the dynamics of a non-linear 
system in the form described by the PFC model with fast degrees of freedom can indeed be found for example in
rapid solidification \cite{PhysRevE.79.051110}.

The coupling of PFC models to hydrodynamics, though, is still an open issue. It makes sense to distinguish
two cases: 

{\it Case (a):} Liquid-solid systems as, \eg, solidifying alloys, where hydrodynamic transport takes place in only the
liquid phase. This implies that such systems can be treated in a multi-scale approach, where the PFC equation for the atomic
scale dynamics in the interior of the solid is coupled to the long-range transport processes in the exterior, which -- to be numerically
efficient -- couple only to the amplitude equation of the PFC equation. The reason is that in the case of PFC models it is the 
amplitude equation, which distinguishes between solid and liquid. Such amplitude 
equation approaches have been derived and solved numerically as computationally efficient representatives of
PFC models, \eg, for diffusion limited polycrystalline grain growth \cite{PhysRevE.76.056706} based on renormalisation
group techniques (see section \ref{sec:Aeborgt} for details). Since renormalisation for PFC models of equation \eqref{eq0} type
does not directly yield NS-type equations and thus a coupling to hydrodynamics exploiting the above multi-scale
idea, this is still an open issue.

{\it Case (b)} differs from the above picture in that it applies to systems, for which interatomic respectively
intermolecular fluidic motion needs to be taken into account as, \eg, in polymers.  Then one would desire a 
PFC model, where the PFC equation itself is directly coupled to a NS-type equation. A first  
step in that direction was done by Praetorius \etal\ \cite{Praetorius2011}. They developed a PFC model with an 
advective term to simulate particles in a flowing solvent. To do so they followed the same route as Rauscher 
\etal\ \cite{RauscherDKP2007} and Penna \etal\ \cite{PennaDT2003} to derive a DDFT  
model for such systems, and approximated the latter further based on the Ramakrishnan-Yussouff 
approximation \cite{RamakrishnanY1979}. Strictly speaking the resulting PFC model applies only to potential flows. 
Furthermore, the coupling of the PFC variable $\psiPF$ into the dynamic equations for the hydrodynamic 
field is constructed simply based on numerical arguments. More rigorous generalisations of these studies might
require additional physically motivated coupling terms, as demonstrated for hydrogels in terms of the derivation 
of the corresponding PF model in reference \cite{damingtobe} (see section \ref{sec:Atlc} for details).

Just as in the PF case, one can expect efficient further advance of PFC models in the future. 
Progress will result from physical modelling concepts based upon the classical routes to patterned non-linear systems 
\cite{HohenbergH1977}, analytical mathematical expansion techniques as the renormalisation group approach 
(see section \ref{sec:Aeborgt} for details), and advanced numerical approaches \cite{Praetorius2011,Voigtn}
in close comparison to experimental techniques.

\section{\label{chap:IV}Phase-field-crystal models applied to nucleation and pattern formation in metals}
Crystal nucleation can be handled in two different ways within the framework of PFC models 
\cite{PhysRevLett.107.175702,JPhysCondMat.22.364101,PhilosMag.91.123}: 
(i) via finding the properties of the critical fluctuations (nuclei) by solving the ELE under appropriate boundary conditions 
(zero field gradients at the centre and unperturbed liquid phase in the far field), whose solution represents an 
extremum of the free-energy functional; 
(ii) by adding noise to the EOM. These approaches have their limitations. 
(i) is expected to work for small undercoolings, where the individual heterophase fluctuations do not interact. 
Furthermore, it is not immediately straightforward how one should address possible non-crystalline nucleation precursors. 
In turn, in the case of (ii) it is not clear conceptually, which fraction of the thermal fluctuations is already integrated into the free energy  
and which wavelengths should yet be added as noise to the EOM \cite{MarconiT1999,JPhysCondMat.15.V1,JPhysA.37.9325} --  
a question inherently related to the proper choice of the high frequency cutoff one needs to make to avoid an ultraviolet catastrophe 
in 3D \cite{JPhysCondMat.22.360301,PhilosMag.91.25}. Furthermore, the addition of noise to the EOM changes the free energy, the phase diagram, 
and the interfacial properties. While, in principle, correction of these is possible via parameter renormalisation 
\cite{JChemPhys.91.7265,PhysRevE.62.6116}, further study is needed in the case of PFC models. On the other hand, the original free-energy functional 
used in (i) seems to miss the effect of longer wavelength fluctuations, which could move the system out of a metastable state. 
Considering them, (i) and (ii) may be regarded as approaches that provide complementary, probably qualitative information of the 
crystallizing system, which converge when the amplitude of the noise tends to zero.

We note, however, that the results obtained via route (i) are more general than those obtained from (ii), as they follow directly from the 
free-energy functional, being therefore independent of the type of dynamics the EOM defines. Accordingly, results from (i) are valid even in cases, 
where the colloid-type diffusive dynamics is not applicable (\eg, metals). In this section, we review results obtained 
following route (i) for metals, or when pattern formation is governed by either chemical or surface diffusion. In turn, results obtained with the 
single-component PFC models with overdamped diffusive dynamics will be reviewed in section \ref{sec:pitss}.

\begin{figure}
\begin{center}
\includegraphics[width=0.5\linewidth]{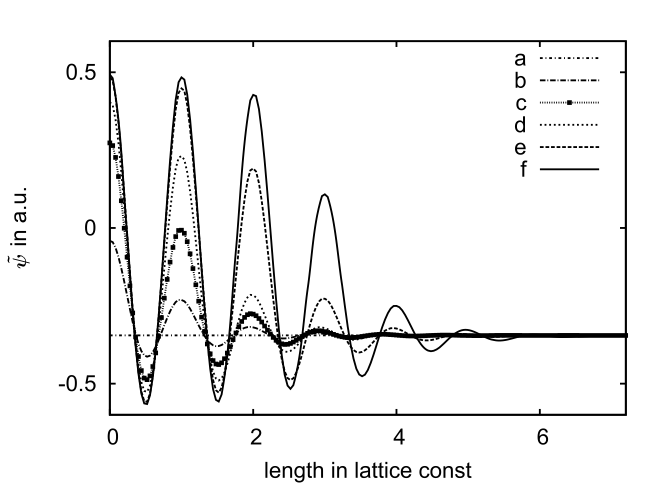}
\end{center}
\caption{\label{fig11}Particle density profiles for the critical fluctuations forming at different supersaturations 
(supersaturation decreases from a to f), as obtained by the adaptation of the string method for determining the extrema of the 
free-energy functional in the 1M-PFC model in two spatial dimensions. 
\nocite{JPhysCondMat.22.364104}\CRP{R.\ Backofen and A.\ Voigt}{A phase-field-crystal approach to critical nuclei}{J.\ Phys.: Condens.\ Matter}{22}{2010}{364104}{36}{10.1088/0953-8984/22/36/364104}{Institute of Physics Publishing}}
\end{figure}

\subsection{Properties of nuclei from extremum principles}

\subsubsection{Homogeneous nucleation} 
An adaptation of the string method to find the saddle point of the free-energy functional has been used by Backofen and Voigt 
\cite{JPhysCondMat.22.364104} for determining the properties of the critical fluctuations in the 1M-PFC model in two spatial dimensions. 
The respective density profiles are shown in figure \ref{fig11}. It is evident that at large supersaturations there are no bulk crystal properties 
at the centre of the smallest nuclei. They have also reported that small nuclei are faceted even though the large crystals are not.

\begin{figure}
\begin{center}
\includegraphics[width=0.48\linewidth]{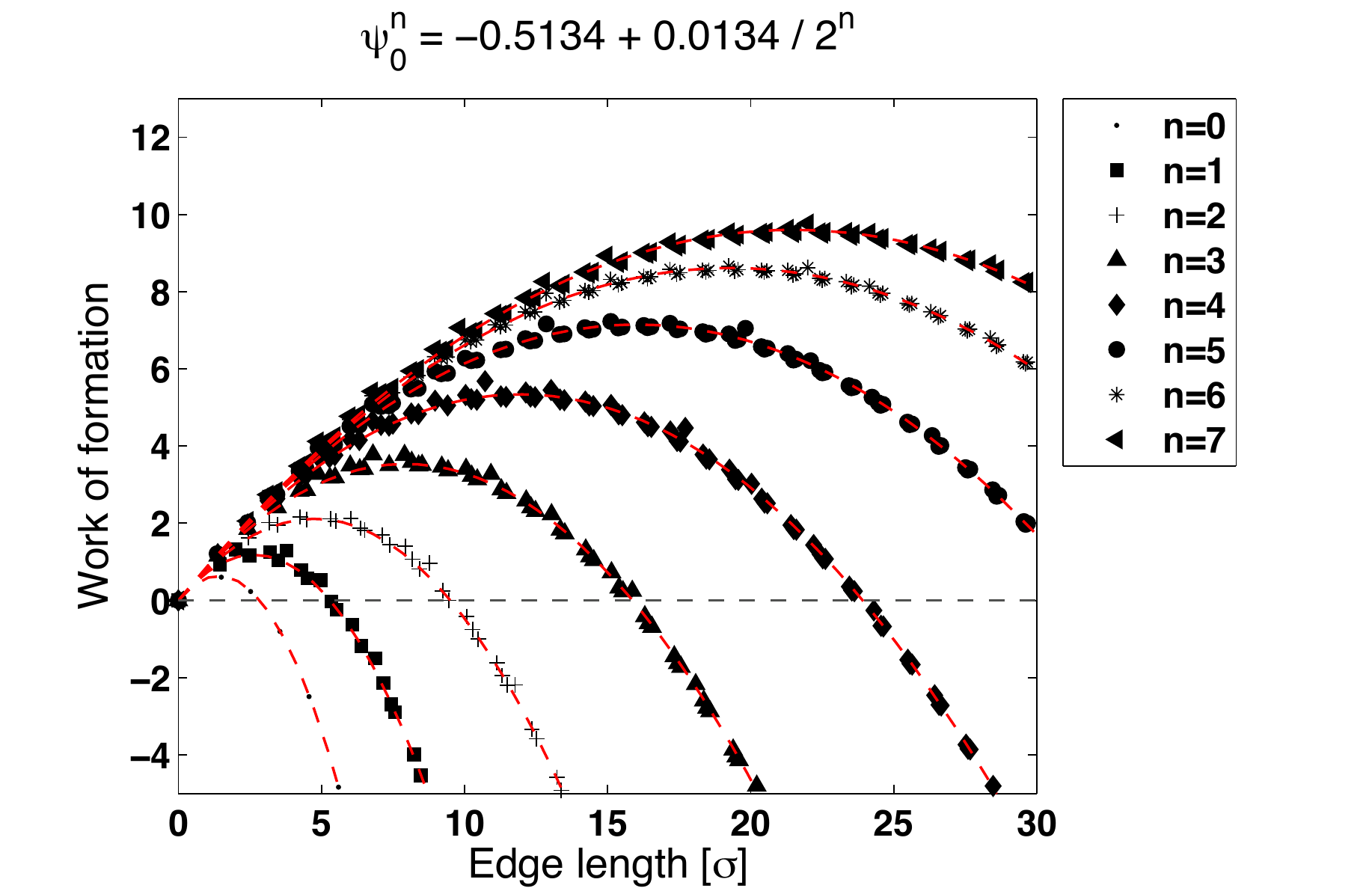}
\includegraphics[width=0.48\linewidth]{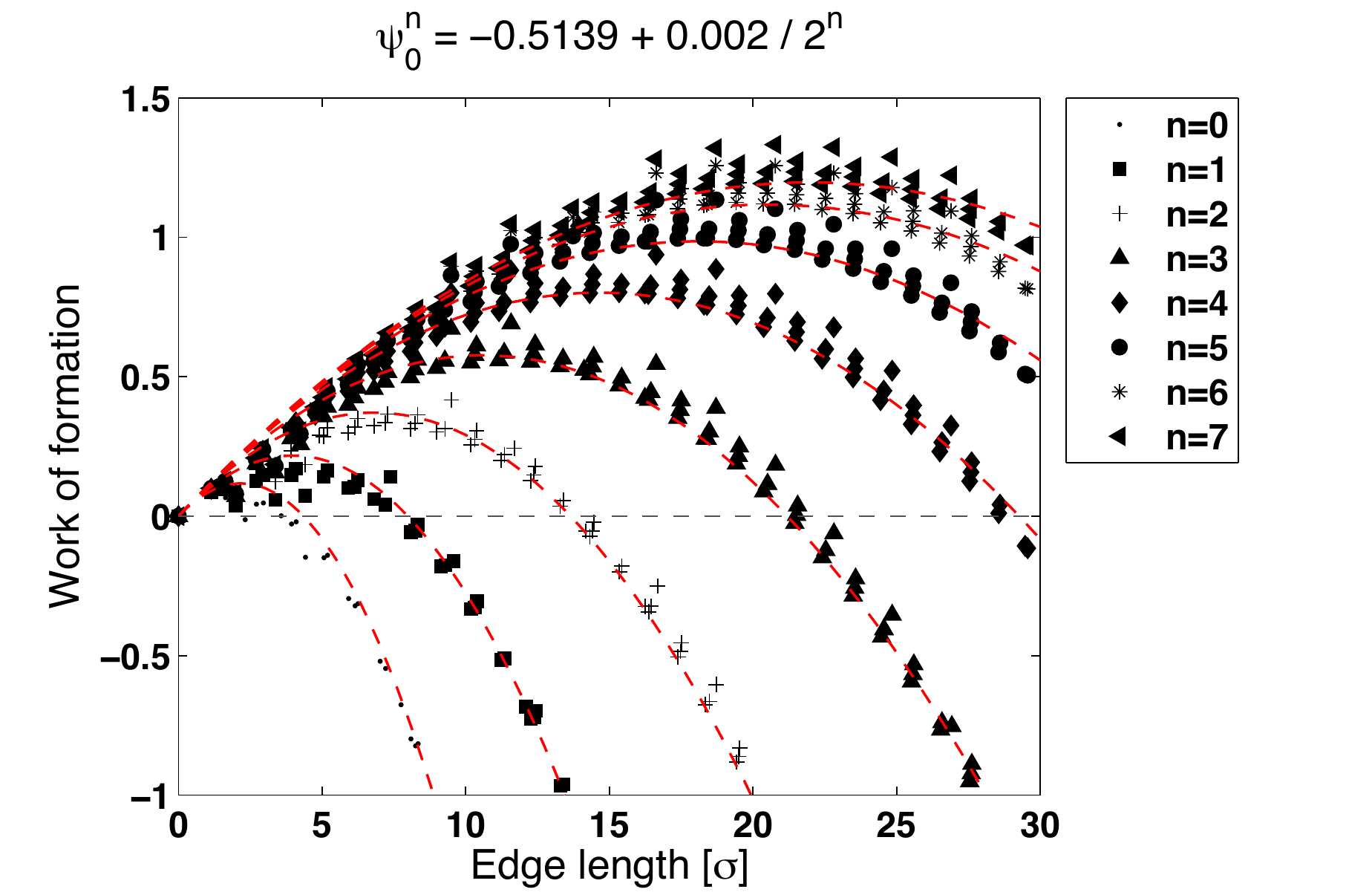}
\end{center}
{\vskip-4mm\footnotesize\hspace{0.215\linewidth}(a)\hspace{0.463\linewidth}(b)}
\caption{\label{fig12}Nucleation barrier \vs size relationship obtained by solving the ELE for faceted nuclei ($\epsilon = 0.5$) 
for the 1M-PFC model \cite{JPhysCondMat.22.364101} in two spatial dimensions: 
(a) homogeneous crystal nuclei for $\psi_0^n = -0.5134 + 0.0134/2^n$, where $n = 0,1,2,\dotsc,7$, respectively. 
(b) Heterogeneous crystal nuclei for $\psi_0^n = -0.5139 + 0.002/2^n$, where $n = 0,1,2,\dotsc,7$, respectively. 
The lattice constant of the substrate is equal to the interparticle distance in the triangular crystal. Notice the substantial reduction of the 
nucleation barrier, a monolayer adsorption layer, and the contact angle of $60^\circ$ determined by the crystal structure. 
\nocite{JPhysCondMat.22.364101}\CRPE{G.\ I.\ T{\'o}th, G.\ Tegze, T.\ Pusztai, G.\ T{\'o}th, and L.\ Gr{\'a}n{\'a}sy}{Polymorphism, crystal nucleation and growth in the phase-field crystal model in 2D and 3D}{J.\ Phys.: Condens.\ Matter}{22}{2010}{364101}{36}{10.1088/0953-8984/22/36/364101}{Institute of Physics Publishing}} 
\end{figure}

T\'oth \etal\ \cite{JPhysCondMat.22.364101} have solved in two spatial dimensions the ELE of the 1M-PFC model for the appropriate boundary conditions 
(homogeneous supersaturated liquid in the far field) to find the free-energy extrema for faceted clusters far from the critical point 
($\epsilon = 0.5$), where even the large crystals are inherently faceted. Under such conditions, the free-energy surface has many local minima 
\cite{JPhysCondMat.22.364101,SIADS.7.1049} that map out the shape of the free-energy barrier for nucleation as a function of cluster size 
[see figure \ref{fig12}(a)]. It has also been reported that the effective interfacial free energy deduced from the barrier height using the hexagonal 
classical cluster model (valid owing to the interface sharp on the atomic scale) converges towards the free energy of the planar interface as the 
supersaturation decreases \cite{JPhysCondMat.22.364101}.

T\'oth \etal\ \cite{JPhysCondMat.22.364101} have performed a similar ELE analysis in 3D to study for heterophase fluctuations forming in the 
1M-PFC model. They have evaluated the properties of crystal nuclei for bcc and fcc structures. 
It has been found that under the investigated conditions both the nucleation barrier and the driving force are fairly close for these structures, 
indicating comparable interfacial free energies and also Turnbull's coefficient\footnote{Turnbull's coefficient $C_{\mathrm{T}}$ is a reduced 
liquid-solid interfacial free energy defined via the relationship 
$\gamma_{\mathrm{ls}}=C_{\mathrm{T}}\frac{\Delta H_{\mathrm{f}}}{N_{\mathrm{A}}^{1/3}v_{\mathrm{m}}^{2/3}}$, 
where $\gamma_{\mathrm{ls}}$, $\Delta H_{\mathrm{f}}$, $N_{\mathrm{A}}$, and $v_{\mathrm{m}}$ are the total liquid-solid interfacial free energy, the molar heat of fusion, 
the Avogadro number, and the molar volume, respectively. 
$C_{\mathrm{T}}$ is expected to depend only on the crystal structure. Recent results indicate that besides structure 
the interaction potential also has influence on its magnitude.} 
for the bcc and fcc structures under conditions ($\epsilon = 0.3748$), where the thermodynamic properties of the 
crystalline phases are rather close \cite{PhysRevLett.103.035702}. This seems to contradict recent results from MD simulations performed using 
the EAP method \cite{Asta2009941}. It is worth noting, however, that Turnbull's coefficient varies with the type of 
interaction potential. For example, it is $\approx 0.55$ for EAP metals of fcc structure \cite{Asta2009941}, whereas $\approx 0.36$ has been 
deduced for the fcc Lennard-Jones (LJ) system \cite{JChemPhys.84.5759}.

\subsubsection{Heterogeneous nucleation} 
The height of the nucleation barrier is often reduced by heterogeneities (walls, floating particles, templates, \etc), 
a phenomenon termed heterogeneous nucleation \cite{tagkey2010iii}. The efficiency of the heterogeneities in instigating freezing is influenced 
by a range of microscopic properties, such as crystal structure, lattice mismatch, surface roughness, adsorption, \etc, which are often condensed 
into the contact angle used in the classical theory and coarse-grained continuum models. The atomic scale characteristics of the substrate surface, 
especially the lattice mismatch, are crucial from the viewpoint of the highly successful free-growth-limited model of particle-induced freezing 
by Greer and co-workers \cite{tagkey2010iii,Greer20002823} -- a model in which cylindrical particles, whose circular faces (of radius $R$) are 
ideally wet by the crystal, remain dormant during cooling until the radius of the homogeneous nuclei becomes smaller than $R$ and free growth 
sets in. The PFC models are especially suitable to investigate such problems as they work on the diffusive time scale \cite{ElderKHG2002} 
and can handle systems containing as many as $2.4 \times 10^7$ atoms \cite{JPhysCondMat.22.364101}.

\begin{figure}
\begin{center}
\includegraphics[width=0.3\linewidth]{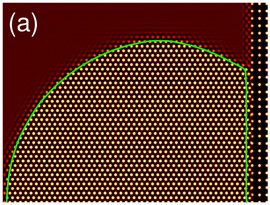}
\includegraphics[width=0.2\linewidth]{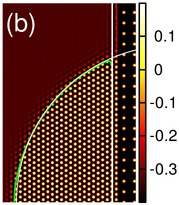}
\includegraphics[width=0.47\linewidth]{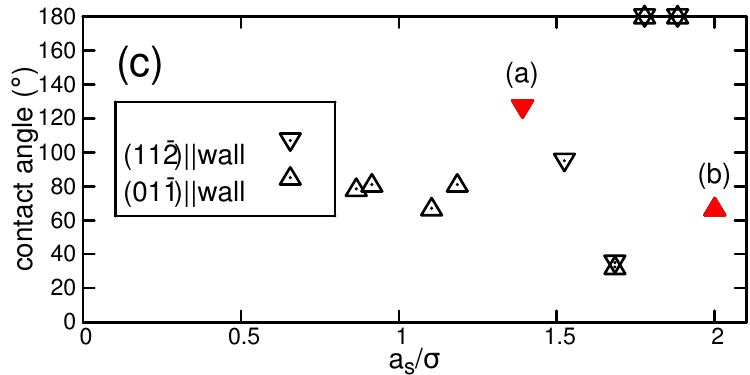}
\end{center}
\caption{\label{fig13}Heterogeneous crystal nucleation at a wall in two spatial dimensions from solving the ELE \cite{PhysRevLett.108.025502}: 
(a), (b) typical (non-faceted) nuclei obtained for $\epsilon = 0.25$, $\psi_0 = -0.341$, 
while the ratio of the lattice constant of the substrate and the particle diameter is $a_s/\sigma = 1.49$ and $2.0$, respectively, 
whereas the crystal orientations $(11\bar{2})$ and $(01\bar{1})$ are parallel with the wall. 
The intersection of the circular and linear fits (white lines) to the contour line (green line) defines the contact angle. 
(c) Contact angle \vs $a_s/\sigma$ for $\epsilon = 0.25$ and $\psi_0 = -0.341$. The red symbols indicate the states corresponding to panels 
(a) and (b). 
\nocite{PhysRevLett.108.025502}\CRPE{G.\ I.\ T{\'o}th, G.\ Tegze, T.\ Pusztai, and L.\ Gr{\'a}n{\'a}sy}{Heterogeneous crystal nucleation: the effect of lattice mismatch}{Phys.\ Rev.\ Lett.\ }{108}{2012}{025502}{2}{10.1103/PhysRevLett.108.025502}{the American Physical Society}} 
\end{figure}

\begin{figure}
\begin{center}
\includegraphics[width=0.3\linewidth]{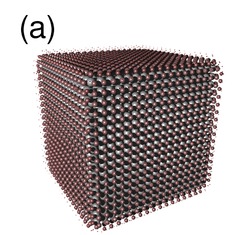}
\includegraphics[width=0.3\linewidth]{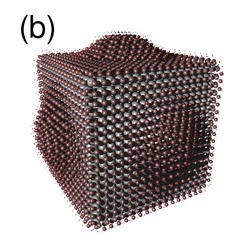}
\includegraphics[width=0.3\linewidth]{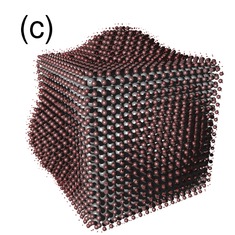}
\includegraphics[width=0.3\linewidth]{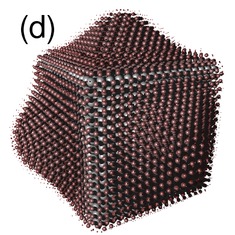}
\includegraphics[width=0.3\linewidth]{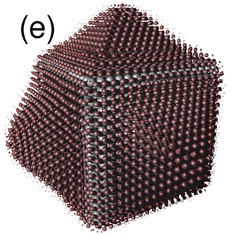}
\includegraphics[width=0.3\linewidth]{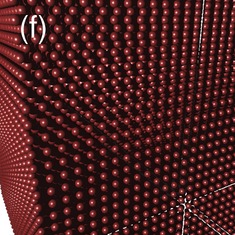}
\end{center}
\caption{\label{fig14}Free-growth limited bcc crystallization in 3D on a cube of sc structure as predicted by the ELE within the 
1M-PFC model \cite{PhysRevLett.108.025502}. Here, $\epsilon = 0.25$ and from left to right 
$\psi_0 = -0.3538,-0.3516,-0.3504,-0.3489,-0.3482$, and $-0.3480$, respectively. The linear size of the substrate is 
$L_{\mathrm{s}} = 16 a_{\mathrm{bcc}}$, while $a_{\mathrm{bcc}}$ is the lattice constant of the stable bcc structure. 
Spheres centred on density peaks are shown, whose size increases with the height of the peak. 
Colour varies with peak height, interpolating between red (minimum height) and light grey (maximum height). 
The ELE has been solved on a $256\times 256\times 256$ grid. 
\nocite{PhysRevLett.108.025502}\CRPE{G.\ I.\ T{\'o}th, G.\ Tegze, T.\ Pusztai, and L.\ Gr{\'a}n{\'a}sy}{Heterogeneous crystal nucleation: the effect of lattice mismatch}{Phys.\ Rev.\ Lett.\ }{108}{2012}{025502}{2}{10.1103/PhysRevLett.108.025502}{the American Physical Society}} 
\end{figure}

Along this line, T\'oth \etal\ \cite{JPhysCondMat.22.364101} used a \textit{periodic} external potential to incorporate a crystalline substrate 
into the ELE method for determining the properties of faceted heterogeneous nuclei. They have observed the adsorption of a monolayer of particles 
on the surface of substrate that reduced the formation energy of nuclei substantially and lead to a contact angle of $60^\circ$ determined by 
the crystal structure [see figure \ref{fig12}(b)]. In a more recent ELE study, T\'oth \etal\ \cite{PhysRevLett.108.025502} 
have shown for the 1M-PFC model 
in two spatial dimensions that the contact angle, the thickness of the crystal layer adsorbed on the substrate, and the height of the nucleation barrier vary 
non-monotonically with the lattice constant of a square-lattice substrate (see figure \ref{fig13}). They have also proven in two and three spatial dimensions that the 
free-growth-limited model of particle-induced freezing by Greer \etal\ \cite{tagkey2010iii,Greer20002823} is valid for larger nanoparticles and 
a small anisotropy of the interfacial free energy (see figure \ref{fig14}). Faceting due to either the small size of the foreign particle or a 
high anisotropy of the free energy of the liquid-solid interfacial free energy decouples free growth from the critical size of homogeneous nuclei.

\subsection{Pattern formation}
Owing to the overdamped diffusive dynamics most of the PFC models assume, diffusional instabilities that lead to fingering of the propagating 
crystal front are inherently incorporated. In the case of a single-component PFC model, diffusive dynamics means that as the growing crystal 
(of larger particle density than the liquid) consumes the particles in the adjacent liquid; the only way they can be replenished is via 
long-range diffusion from the bulk liquid. Accordingly, a depletion zone forms ahead of the growing crystal 
\cite{C0SM00944J,JPhysCondMat.22.364101,PhysRevLett.103.035702,SandomirskiALE2011}. This resembles the behaviour of colloidal suspensions, 
in which the micron-sized colloid particles move by Brownian motion in the carrier fluid. Relying on this similarity, the single-component 
PFC models can be considered as reasonable tools to address colloidal crystal aggregation. One may though get rid of this type of 
mass-diffusion-controlled dynamics when driving the system strongly enough for a diffusion-controlled to diffusionless 
transition \cite{C0SM00944J}. These phenomena are not necessarily present in the phenomenological coarse-grained PFC models, 
where the change in density upon crystallization is not always taken into account -- models that might be considered, therefore, 
as a reasonable description of metallic alloys. Diffusive dynamics is furthermore appropriate on the surface of substrates, 
where the adsorbed particles indeed move by diffusion in a periodic potential. 2D PFC models relying on the interplay of the 
inter-particle forces and a periodic external potential representing the symmetries of the substrate can be used to capture pattern formation 
on such surfaces.

\subsubsection{PFC modelling of surface patterns} 
\begin{figure}
\begin{center}
\includegraphics[width=0.85\linewidth]{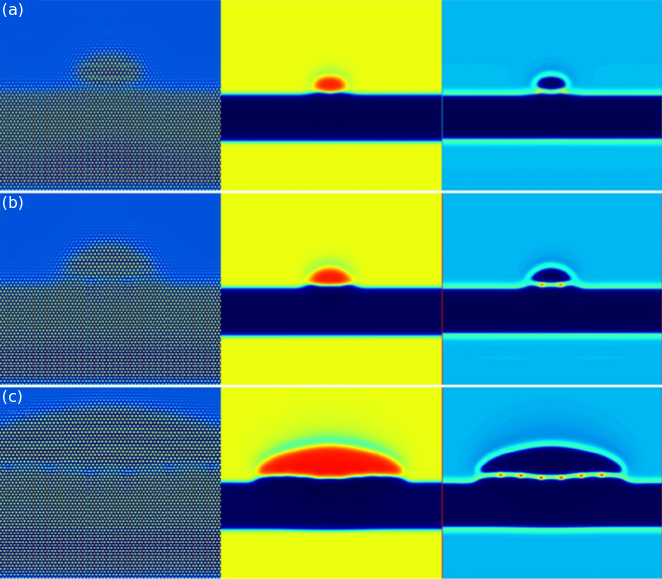}
\end{center}
\caption{\label{fig15}Growth of an island on a nanomembrane of approximately $30$ atomic layers thickness at times 
(a) $t = 60 000$, (b) $t=80 000$, (c) $t=400 000$ as predicted by the amplitude equations of reference \cite{JPhysCondMat.22.364103}. 
Similar to figure \ref{fig2}, the left, middle, and right columns correspond to the results of the total particle density, 
the difference of the particle densities for the two components, and the local free-energy density, respectively. 
Also for clarity, the particle density images have been expanded by a factor of two. 
\nocite{JPhysCondMat.22.364103}\CRP{K.\ R.\ Elder and Z.-F.\ Huang}{A phase field crystal study of epitaxial island formation on nanomembranes}{J.\ Phys.: Condens.\ Matter}{22}{2010}{364103}{36}{10.1088/0953-8984/22/36/364103}{Institute of Physics Publishing}}   
\end{figure}

A combination of computer simulations with the solution of amplitude equations for PFC models equipped with periodic potentials has been used to 
investigate a range of surface phenomena including incommensurate to commensurate transitions for triangular surface layers 
on a square-lattice substrate \cite{PhysRevE.74.021104,PhysRevE.78.031109,PhysRevE.79.011606}, 
the deposition of monolayers on quasi-crystalline substrates \cite{RottlerGZ2012}, sliding friction \cite{PhysRevE.79.011606,PhysRevE.81.011121}, 
and the formation of quantum dots/islands on nanomembranes \cite{PhysRevE.82.021605,JPhysCondMat.22.364103} (see figure \ref{fig15}).

\begin{figure}
\begin{center}
\includegraphics[width=0.3\linewidth]{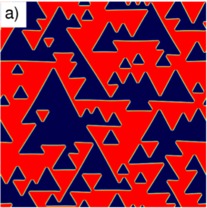}
\includegraphics[width=0.3\linewidth]{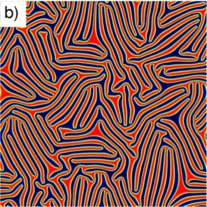}
\includegraphics[width=0.3\linewidth]{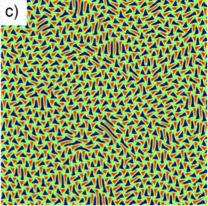}
\end{center}
\caption{\label{fig16}Surface patterns the amplitude equations predict for a decreasing dimensionless coupling $u_0$ between the layer and 
the substrate \cite{prl.unpub.elder}. From left to right: $u_0 = 12.1 \times 10^{-3}$, $3.2 \times 10^{-3}$, and $0.87 \times 10^{-3}$, 
respectively. Colouring: fcc domains are blue, hcp domains are red, and the domain walls are green. 
\nocite{prl.unpub.elder}\CRP{K.\ R.\ Elder, G.\ Rossi, P.\ Kanerva, F.\ Sanches, S.\ C.\ Ying, E.\ Granato, C.\ V.\ Achim, and T.\ Ala-Nissila}{Patterning of heteroepitaxial overlayers from nano to micron scales}{Phys.\ Rev.\ Lett.\ }{108}{2012}{226102}{22}{10.1103/PhysRevLett.108.226102}{the American Physical Society}}
\end{figure}

Muralidharan and Haataja \cite{PhysRevLett.105.126101} have extended the PFC model for describing stress-induced alloying of 
bulk-immiscible binary systems on a substrate by adding a potential energy term describing the substrate and a regular solution term. 
Fixing the model parameters to data for CoAg/Ru(0001), they demonstrated that the model captures experimentally observed morphologies. 
A similar approach has been proposed by Elder \etal\ \cite{prl.unpub.elder} using amplitude equations that allow large-scale simulations 
for stress-induced alloying in heteroepitaxial overlayers. Quantitative predictions, that are in an excellent agreement with experiments, 
have been obtained for the stripe, honeycomb, and triangular superstructures emerging in the metal/metal systems, 
Cu on Ru(0001) and Cu on Pd(111) (see figure \ref{fig16}).

\subsubsection{Pattern formation during binary solidification} 

\begin{figure}
\begin{center}
\includegraphics[width=0.24\linewidth]{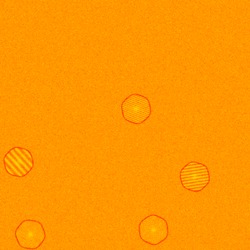}
\includegraphics[width=0.24\linewidth]{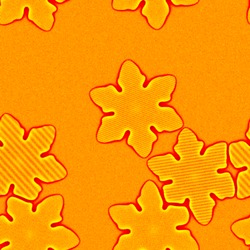}
\includegraphics[width=0.24\linewidth]{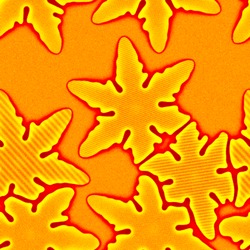}
\includegraphics[width=0.24\linewidth]{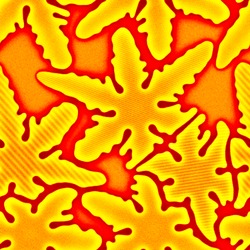}
\end{center}
{\vskip-4mm\footnotesize\hspace{0.12\linewidth}(a)\hspace{0.22\linewidth}(b)\hspace{0.22\linewidth}(c)\hspace{0.22\linewidth}(d)}
\caption{\label{fig17}Growth of five dendrites in the binary PFC model (the distribution of the field $\KDN$ is shown). 
The snapshots taken at $1000$, $5000$, $10000$ and $20000$ time steps are shown. The simulations have been performed on a 
$16384 \times 16384$ grid, using a semi-implicit spectral method \cite{Tegze20091612}.  
\nocite{JPhysCondMat.20.404205}\CRPE{T.\ Pusztai, G.\ Tegze, G.\ I.\ T{\'o}th, L.\ K{\"o}rnyei, G.\ Bansel, Z.\ Fan, and L.\ Gr{\'a}n{\'a}sy}{Phase-field approach to polycrystalline solidification including heterogeneous and homogeneous nucleation}{J.\ Phys.: Condens.\ Matter}{20}{2008}{404205}{40}{10.1088/0953-8984/20/40/404205}{Institute of Physics Publishing}}
\end{figure}

\begin{figure}
\begin{center}
\includegraphics[width=0.185\linewidth]{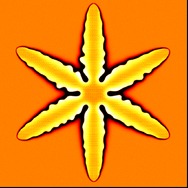}
\includegraphics[width=0.185\linewidth]{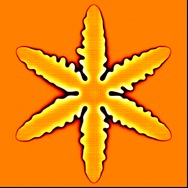}
\includegraphics[width=0.185\linewidth]{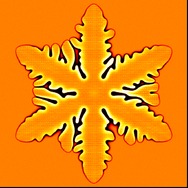}
\includegraphics[width=0.185\linewidth]{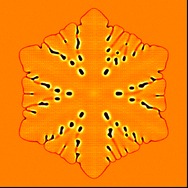}
\includegraphics[width=0.185\linewidth]{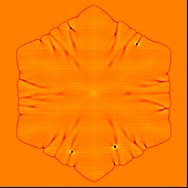}
\end{center}
{\vskip-4mm\footnotesize\hspace{0.099\linewidth}(a)\hspace{0.1668\linewidth}(b)\hspace{0.1668\linewidth}(c)\hspace{0.1668\linewidth}(d)\hspace{0.1668\linewidth}(e)}
\caption{\label{fig18}Morphological transition from dendritic needle crystals to compact hexagonal shape with increasing driving force for 
crystallization \cite{tegze.phd.thesis}. Conditions/properties are as described for dendrites in reference \cite{Tegze20091612}  
except that the initial total number densities are $\psi_0 = 0.009,0.0092,0.0094,0.0096$, and $0.0098$ (from left to right). 
The reduced number density $\KDN$ is shown. Notice the reducing contrast of the images from left to right indicating an increasing 
solute trapping. A $8192 \times 8192$ grid has been used. 
(Reproduced from \nocite{tegze.phd.thesis}G.\ Tegze, \textit{Application of atomistic phase-field methods to complex solidification problems}, 
PhD Thesis, E\"otv\"os University, Budapest, Hungary (2009).)} 
\end{figure}

\begin{figure}
\begin{center}
\includegraphics[width=0.24\linewidth]{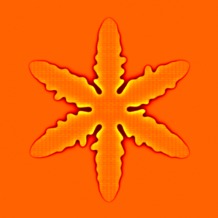}
\includegraphics[width=0.37\linewidth]{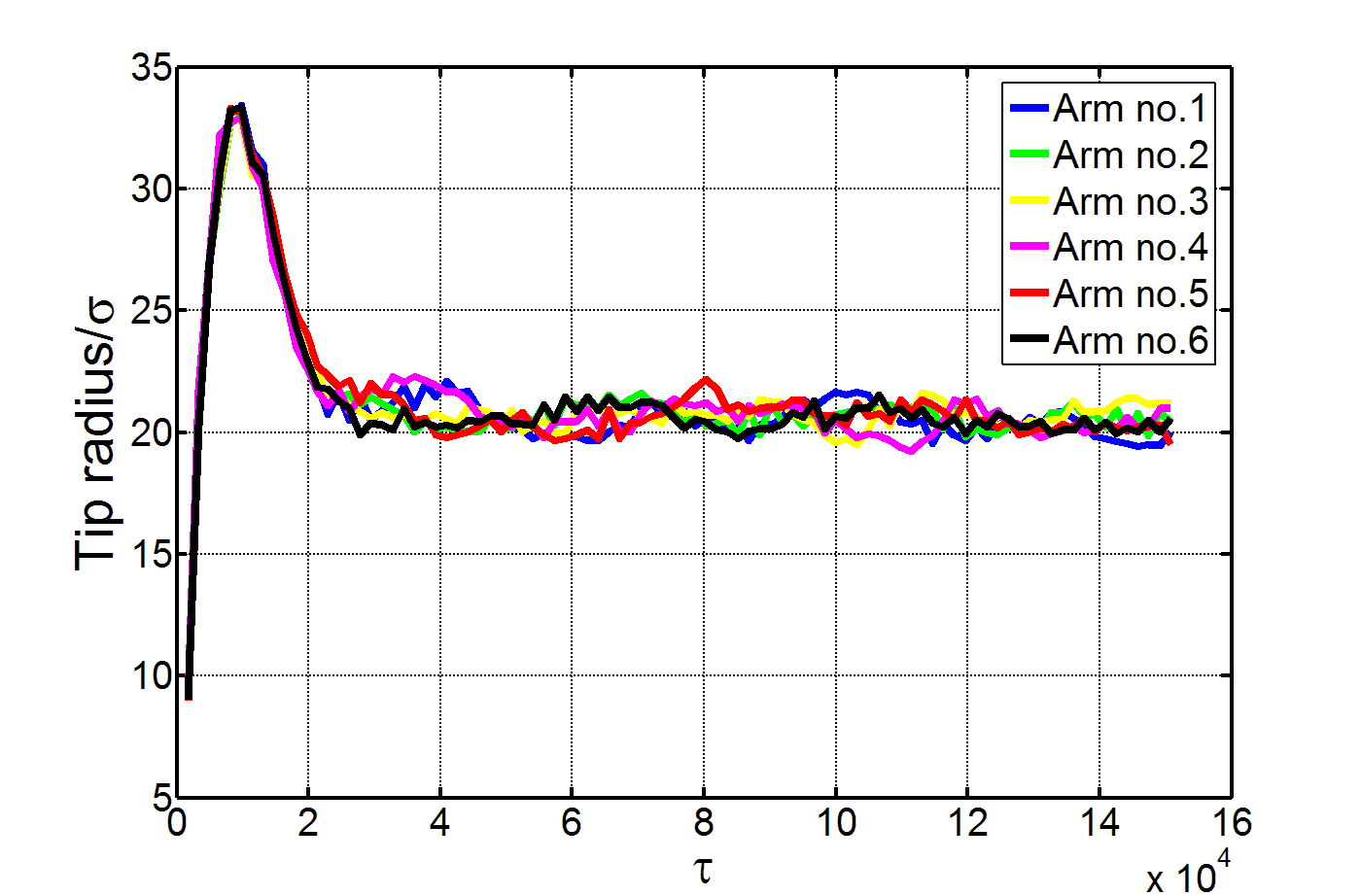}
\includegraphics[width=0.37\linewidth]{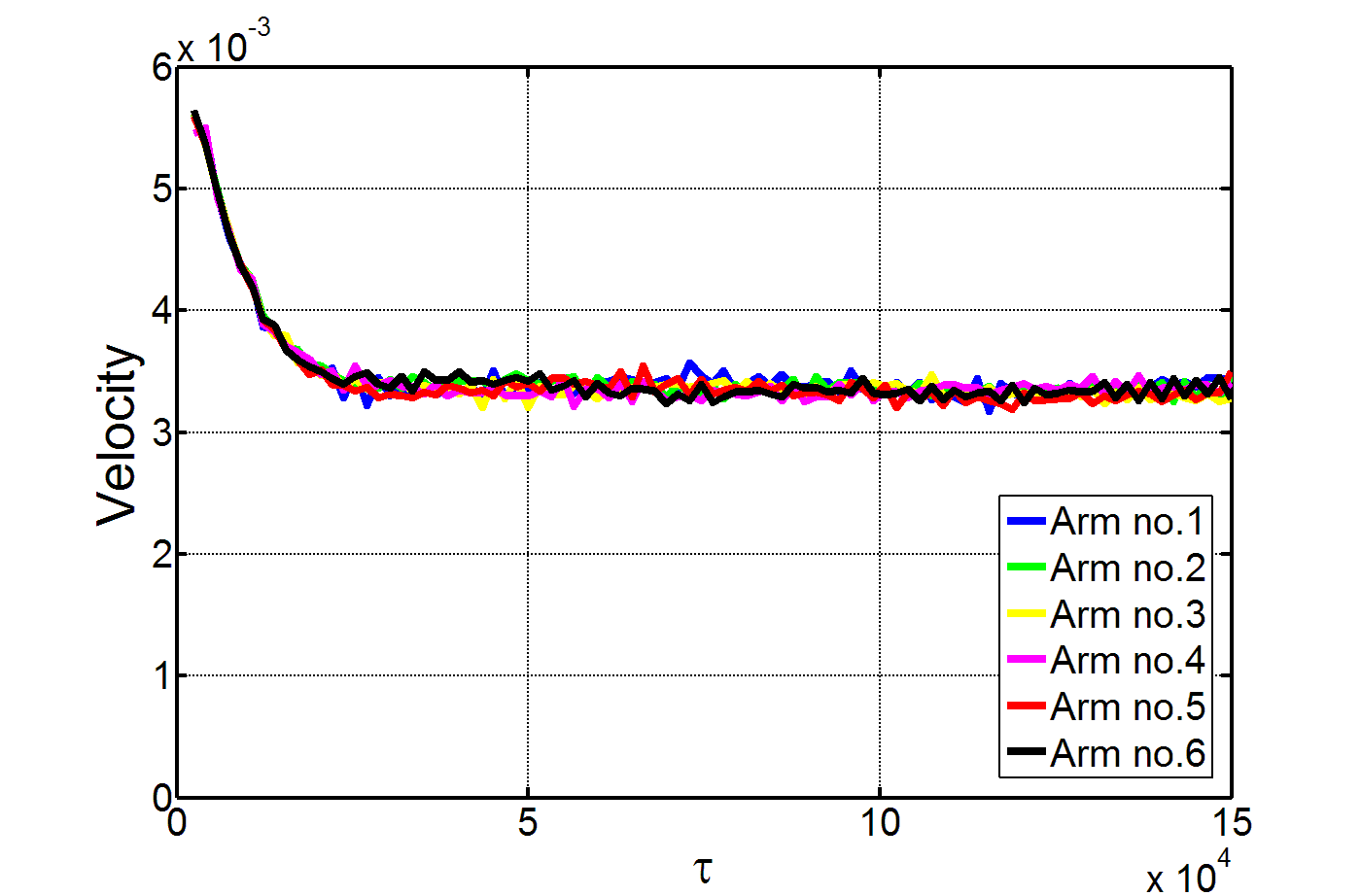}
\end{center}
{\vskip-4mm\footnotesize\hspace{0.108\linewidth}(a)\hspace{0.292\linewidth}(b)\hspace{0.351\linewidth}(c)}
\caption{\label{fig19}Analysis of a solutal dendrite grown in the binary 1M-PFC model \cite{tegze.phd.thesis}. 
The dendrite arms are numbered clockwise from the top arm. Apparently, there are no tip radius or velocity oscillations  
and steady state growth is reached after $\approx 4000$ time steps. A $8192 \times 8192$ grid has been used and $\psi_0 = 0.0092$, 
whereas other conditions as for figure \ref{fig14} are present. 
(Reproduced from \nocite{tegze.phd.thesis}G.\ Tegze, \textit{Application of atomistic phase-field methods to complex solidification problems}, 
PhD Thesis, E\"otv\"os University, Budapest, Hungary (2009).)}
\end{figure}

\paragraph{Dendritic freezing} 
The possibility of growing solutal dendrites has been first addressed within numerical 1M-PFC simulations by 
Elder \etal\ \cite{ElderPBSG2007}. Studies of the transformation kinetics for many particles including several dendrites have been 
performed using the same approach by Pusztai \etal\ \cite{JPhysCondMat.20.404205} for system sizes containing about 
$1.6 \times 10^6$ atoms (see figure \ref{fig17}). Tegze \cite{tegze.phd.thesis} has investigated the behaviour of solutal dendrites using 
binary 1M-PFC simulations of similar size. With increasing driving force obtained by increasing the total number density, 
transitions from dendritic needle crystals to compact hexagon shape crystals have been observed as in the conventional 
PF models (see figure \ref{fig18}). It has also been shown that (i) a steady state tip velocity is attained after a time, 
and (ii) tip oscillations do not occur, \ie, from the viewpoint of side branch formation the dendrite tip works like a selective amplifier 
of the fluctuations at the tip (see figure \ref{fig19}).

\begin{figure}
\begin{center}
\includegraphics[width=0.9\linewidth]{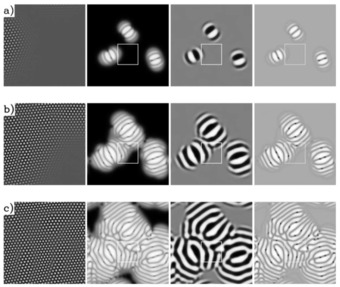}
\end{center}
\caption{\label{fig20}Time evolution of equiaxed eutectic solidification within the amplitude equation formalism proposed by Elder, 
Huang, and Provatas \cite{PhysRevE.81.011602}. Panels (a), (b), and (c) correspond to dimensionless times $30000$, $60000$, 
and $105000$, respectively. From left to right the columns display the reduced total number density in the boxed region, the 
coarse-grained number density, the reduced difference of the number densities for the two species, and the local free-energy density. 
Dislocations appear as small black dots in the local free-energy density. 
\nocite{PhysRevE.81.011602}\CRP{K.\ R.\ Elder, Z.-F.\ Huang, and N.\ Provatas}{Amplitude expansion of the binary phase-field-crystal model}{Phys.\ Rev.\ E}{81}{2010}{011602}{1}{10.1103/PhysRevE.81.011602}{the American Physical Society}}
\end{figure}

\begin{figure}
\begin{center}
\includegraphics[width=0.325\linewidth]{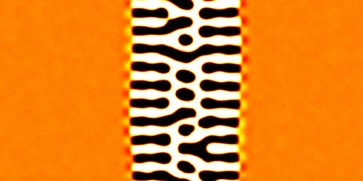}
\includegraphics[width=0.325\linewidth]{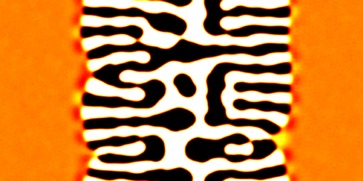}
\includegraphics[width=0.325\linewidth]{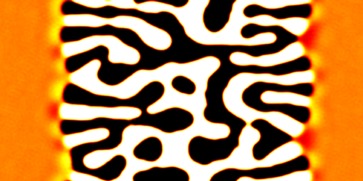}
\end{center}
{\vskip-4mm\footnotesize\hspace{0.157\linewidth}(a)\hspace{0.312\linewidth}(b)\hspace{0.312\linewidth}(c)}
\caption{\label{fig21}Snapshots of eutectic solidification on the atomistic scale in the binary 1M-PFC model in two spatial dimensions \cite{JPhysCondMat.22.364101}: 
composition $\KDN$ maps corresponding to $2 \times 10^5$, $6 \times 10^5$, and $10^6$ time steps are shown. 
White and black denote the two crystalline phases, while yellow stands for the liquid phase. The simulation has been performed on a 
$2048 \times 1024$ rectangular grid. Crystallization has been started by placing a row of supercritical crystalline clusters of 
alternating composition into the simulation window. Interestingly, the eutectic pattern evolves inside the solid region on a timescale 
comparable to the timescale of solidification. 
\nocite{JPhysCondMat.22.364101}\CRPE{G.\ I.\ T{\'o}th, G.\ Tegze, T.\ Pusztai, G.\ T{\'o}th, and L.\ Gr{\'a}n{\'a}sy}{Polymorphism, crystal nucleation and growth in the phase-field crystal model in 2D and 3D}{J.\ Phys.: Condens.\ Matter}{22}{2010}{364101}{36}{10.1088/0953-8984/22/36/364101}{Institute of Physics Publishing}}
\end{figure}

\begin{figure}
\begin{center}
\includegraphics[width=0.24\linewidth]{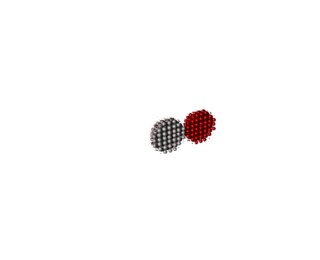}
\includegraphics[width=0.24\linewidth]{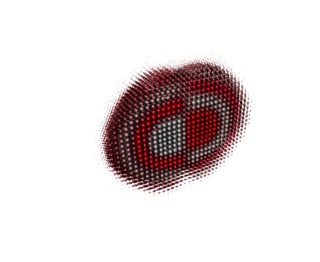}
\includegraphics[width=0.24\linewidth]{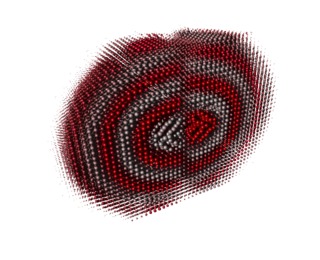}
\includegraphics[width=0.24\linewidth]{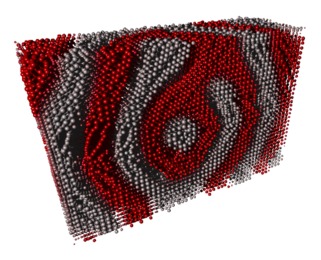}
\end{center}
{\vskip-5mm\footnotesize\hspace{0.135\linewidth}(a)\hspace{0.22\linewidth}(b)\hspace{0.22\linewidth}(c)\hspace{0.22\linewidth}(d)}
\caption{\label{fig22}Snapshots of eutectic solidification as predicted by the binary 1M-PFC model in 3D \cite{JPhysCondMat.22.364101}: 
time elapses from left to right. 
The simulation has been performed on a $450 \times 300 \times 300$ rectangular grid. Solidification has been started by placing two touching
supercritical bcc clusters of different compositions into the simulation window. Remarkably, the nanoscale solid-phase eutectic pattern roughens 
on a timescale comparable to the time of solidification. Brown and grey colours denote the terminal solutions of the two crystalline phases. 
Spheres of size reflecting the height $\psi$ of the total number density peak and coloured according to the local composition $\KDN$ are 
centred to the particle density maxima. Only half of the simulation window is shown. 
\nocite{JPhysCondMat.22.364101}\CRPE{G.\ I.\ T{\'o}th, G.\ Tegze, T.\ Pusztai, G.\ T{\'o}th, and L.\ Gr{\'a}n{\'a}sy}{Polymorphism, crystal nucleation and growth in the phase-field crystal model in 2D and 3D}{J.\ Phys.: Condens.\ Matter}{22}{2010}{364101}{36}{10.1088/0953-8984/22/36/364101}{Institute of Physics Publishing}}
\end{figure}

\paragraph{Eutectic solidification} 
Eutectic solidification in binary 1M-PFC simulations has been first observed in two spatial dimensions in the seminal paper 
by Elder \etal\ \cite{ElderPBSG2007}. The formation of lamellar eutectic grains has been explored by Elder, 
Huang, and Provatas \cite{PhysRevE.81.011602} using and approach based on amplitude equations (see figure \ref{fig20}). 
For the relatively large lattice mismatch they assumed for the two crystalline phases ($8.4\%$ in equilibrium), 
spontaneous nucleation of dislocations at the lamellar interfaces has been observed -- a phenomenon expected to modify the spacing 
selection mechanism predicted by earlier eutectic solidification theories. Larger-scale binary 1M-PFC simulations relying on a 
numerical solution of the EOMs \eqref{eq27a} and \eqref{eq27b} in two spatial dimensions imply owing to the diffusive dynamics of the total particle density in the 
binary 1M-PFC model, eutectic colonies may form even in binary systems \cite{JPhysCondMat.22.364101} (see figure \ref{fig21}). 
Here, the morphological change occurs as a result of the diffusional instability emerging from the diffusive EOM, many of the 
PFC models assume. Using the same approach in 3D, eutectic crystallization to the bcc structure has been reported by 
T\'oth \etal\ \cite{JPhysCondMat.22.364101} (see figure \ref{fig22}). These atomistic simulations indicate a remarkable time evolution 
of the eutectic pattern after solidification.

\subsection{\label{sec:pitss}Phenomena in the solid state} 
One of the most successful areas, where the PFC models make a real difference is the modelling of solid state transitions  
including grain-boundary dynamics, melting, crack formation, stress-induced morphology evolution, and the modelling of the Kirkendall effect, 
to mention a few.

\subsubsection{Dislocation dynamics and grain-boundary melting} 
\begin{figure}
\begin{center}
\includegraphics[width=0.39\linewidth]{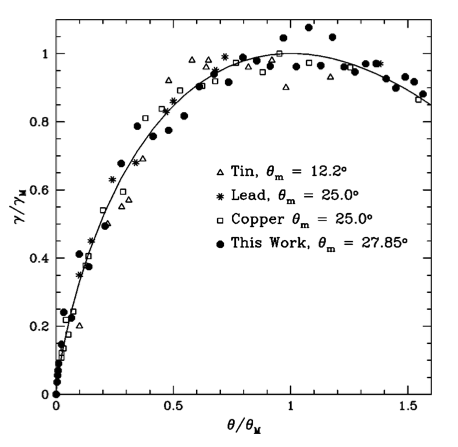}
\includegraphics[width=0.25\linewidth]{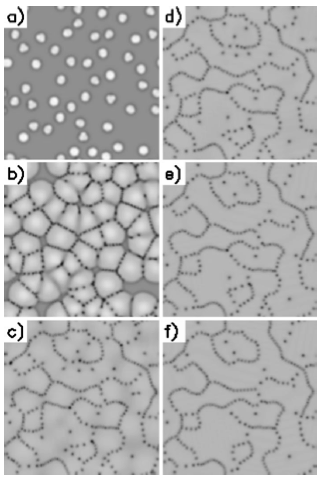}
\includegraphics[width=0.282\linewidth]{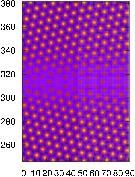}
\end{center}
{\vskip-4mm\footnotesize\hspace{0.224\linewidth}(a)\hspace{0.29\linewidth}(b)\hspace{0.28\linewidth}(c)}
\caption{\label{fig23}PFC modelling of defect and pattern formation in solids. 
From left to right: (a) grain-boundary energy \vs Read-Shockley relationship and (b) grain-boundary dynamics. 
\nocite{ElderG2004}\CRP{K.\ R.\ Elder and M.\ Grant}{Modeling elastic and plastic deformations in nonequilibrium processing using phase field crystals}{Phys.\ Rev.\ E}{70}{2004}{051605}{5}{10.1103/PhysRevB.75.064107}{the American Physical Society}   
(c) Grain-boundary melting at a large-angle grain boundary. 
\nocite{PhysRevB.78.184110}\CRP{J.\ Mellenthin, A.\ Karma, and M.\ Plapp}{Phase-field crystal study of grain-boundary premelting}{Phys.\ Rev.\ B}{78}{2008}{184110}{18}{10.1103/PhysRevB.78.184110}{the American Physical Society}}
\end{figure}

Already the first paper on the PFC method \cite{ElderKHG2002,ElderG2004} has addressed grain boundaries  
and shown that the model automatically recovers the Read-Shockley relationship between grain-boundary energy and misorientation 
(see figure \ref{fig23}). It has also been shown that PFC models are ideal for modelling grain-boundary dynamics 
\cite{ElderKHG2002,ElderG2004} (see figure \ref{fig23}) and offers the possibility to link mechanical properties with 
the grain structure \cite{ElderG2004}. Two mechanisms of dislocation glide have been observed: for high strain rates, 
continuous glide is observed, while at lower strain rate the dislocation set into a stick-slip motion \cite{StefanovicHP2006}. 
Grain-boundary melting has been addressed in several works \cite{PhysRevB.77.224114,PhysRevE.79.013601,PhysRevB.78.184110}. 
It has been reported that dislocations in low-angle to intermediate-angle grain boundaries melt similarly until an angle-dependent 
first-order wetting transition occurs, when neighbouring melted regions coalesce. In the large-angle limit, the grain-boundary energy 
becomes increasingly uniform along its length and can no longer be interpreted in terms of individual dislocations (see figure \ref{fig23}). 
The difference between high- and low-angle boundaries appears to be reflected in the dependence of the disjoining potential on the width 
of the pre-melted layer $w$: it is purely repulsive for all widths for misorientations larger than a critical angle, however, it switches 
from repulsive at small $w$ to attractive for large $w$ \cite{PhysRevB.78.184110}.

\subsubsection{Crack formation and propagation} 
\begin{figure}
\begin{center}
\includegraphics[width=0.4\linewidth]{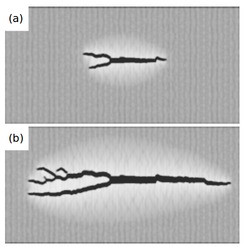}
\includegraphics[width=0.5\linewidth]{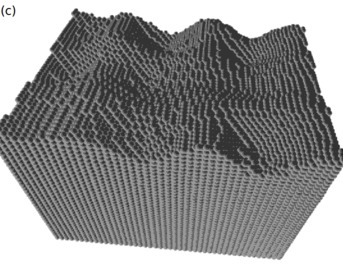}
\end{center}
\caption{\label{fig24}Crack formation (left) and strain-induced epitaxial islands (right) in the single-component 1M-PFC model. 
(a), (b) Snapshots of the energy density map taken at dimensionless times $25 000$ and $65 000$. 
\nocite{ElderG2004}\CRP{K.\ R.\ Elder and M.\ Grant}{Modeling elastic and plastic deformations in nonequilibrium processing using phase field crystals}{Phys.\ Rev.\ E}{70}{2004}{051605}{5}{10.1103/PhysRevB.75.064107}{the American Physical Society}   
(c) Gray scale image of epitaxial islands in an 1M-PFC simulation for a $4.8\%$ tensile film. 
\nocite{PhysRevLett.101.158701}\CRP{Z.-F.\ Huang and K.\ R.\ Elder}{Mesoscopic and microscopic modeling of island formation in strained film epitaxy}{Phys.\ Rev.\ Lett.\ }{101}{2008}{158701}{15}{10.1103/PhysRevLett.101.158701}{the American Physical Society}}
\end{figure}

Elder and Grant \cite{ElderG2004} have demonstrated that the 1M-PFC model can be used to model crack propagation. 
A small notch cut out of a defect-free crystal placed under $10\%$ strain in the vertical direction and filled with coexisting liquid 
has been used as a nucleation cite for crack propagation. Snapshots of crack development are shown in figure \ref{fig24}.

\subsubsection{Strain-induced morphologies} 
Huang and Elder \cite{PhysRevLett.101.158701} have studied strain-induced film instability and island formation using numerical 
1M-PFC simulations and amplitude equations (see figure \ref{fig23}). They have identified a linear regime for the island wave number scaling 
and recovered the continuum ATG instability in the weak strain limit. The ATG instability has been studied in two spatial dimensions 
by Spatschek and Karma \cite{PhysRevB.81.214201} using a different amplitude equations approach. Qualitatively similar surface roughening 
has been reported by Tegze \etal\ \cite{PhysRevLett.103.035702} for heteroepitaxial body-centred tetragonal (bct) films grown on sc crystalline 
substrates of tuned lattice constant -- a phenomenon interpreted in terms of the Mullins-Sekerka/ATG instability.

\subsubsection{Kirkendall effect} 
Elder, Thornton, and Hoyt \cite{PhilosMag.91.151} have used a simple extension of the binary 1M-PFC model incorporating unequal atomic mobilities 
to investigate different aspects of the Kirkendall effect. They have shown that the model indeed captures such phenomena as crystal 
(centre-of-mass) motion, pore formation via vacancy supersaturation, and enhanced vacancy concentration near grain boundaries.

\subsubsection{\label{sec:dst}Density/solute trapping}  
\begin{figure}
\begin{center}
\includegraphics[width=0.32\linewidth]{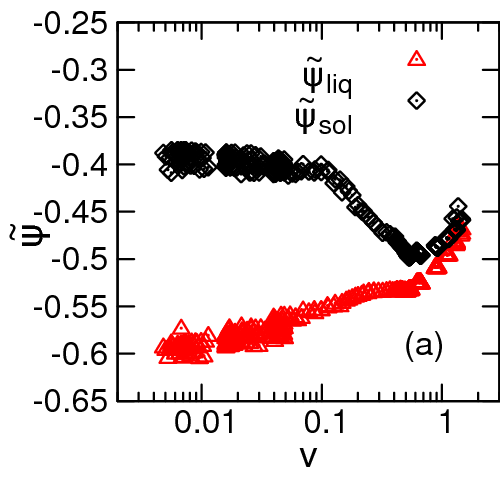}
\includegraphics[width=0.32\linewidth]{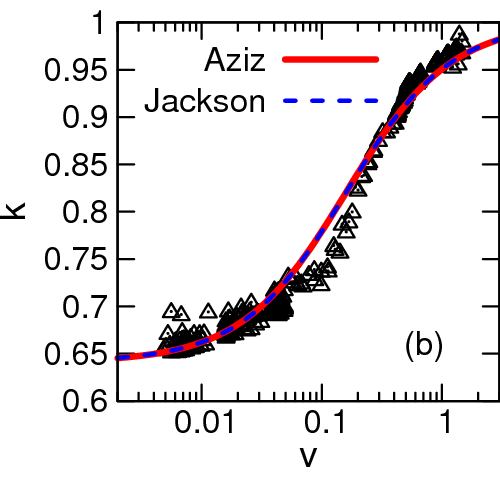}
\includegraphics[width=0.32\linewidth]{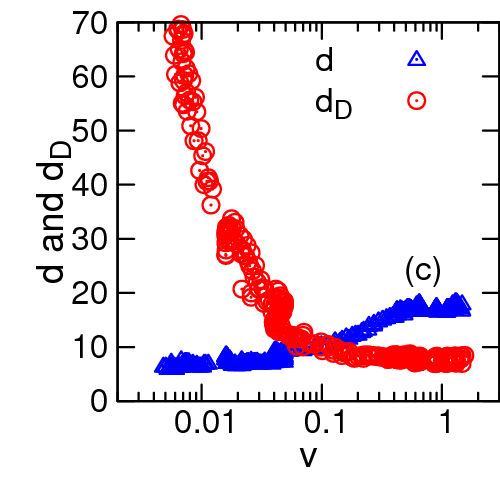}
\end{center}
\caption{\label{fig26}Density trapping as predicted by the single-component 1M-PFC model \cite{C0SM00944J,PhysRevLett.106.195502}. 
(a) Coarse-grained particle densities $\tilde{\psi}$ for the liquid and solid phases at the growth front as a function of growth velocity $v$. 
(b) Effective partition coefficient $k$ defined using the liquidus and solidus densities \vs growth velocity. 
For comparison, fits of the models by Aziz \cite{JApplyPhys.53.1158} and Jackson \etal\ \cite{Jackson2004481} are also displayed. 
(c) Comparison of the interface thickness $d$ and the diffusion length $d_{\mathrm{D}}$ as a function of growth velocity. 
\nocite{C0SM00944J}\CRPE{G.\ Tegze, L.\ Gr{\'a}n{\'a}sy, G.\ I.\ T{\'o}th, J.\ F.\ Douglas, and T.\ Pusztai}{Tuning the structure of nonequilibrium soft materials by varying the thermodynamic driving force for crystal ordering}{Soft Matter}{7}{2011}{1789-1799}{5}{10.1039/c0sm00944j}{Royal Society of Chemistry Publishing}}
\end{figure}

In recent works by Tegze \etal\ \cite{C0SM00944J,PhysRevLett.106.195502}, it has been reported for the 1M-PFC model (of diffusive dynamics) 
that at large $\epsilon$ ($= 0.5$) and high driving force a transition from diffusion-controlled to diffusionless solidification can be observed, 
during which the interface thickness increases, whereas the density difference between the crystal and the liquid decreases drastically 
(see figure \ref{fig26}). This \ZT{density trapping} phenomenon is analogous to solute trapping observed in rapid solidification of alloys 
(where due to a lack of time for partitioning, solids of nonequilibrium compositions form) and can be fitted reasonably well using the 
models of Aziz \cite{JApplyPhys.53.1158} and Jackson \etal\ \cite{Jackson2004481}. In a very recent work, Humadi, Hoyt, 
and Provatas \cite{tms2012orlando} have investigated solute trapping in the binary MPFC model. 
In agreement with the findings for density trapping, they have found that pure diffusive dynamics leads to a velocity-dependent 
partition coefficient that approaches unity for large velocities -- consistently with the model of Aziz and Kaplan \cite{Aziz19882335}. 
In contrast, the wavelike dynamics, the second-order time derivatives of the MPFC-type EOMs realize, 
leads to a solute trapping behaviour similar to the predictions of Galenko \etal\ \cite{PhysRevE.84.041143}.

\subsubsection{Vacancy/atom transport in the VPFC model} 
\begin{figure}
\begin{center}
\includegraphics[width=0.9\linewidth]{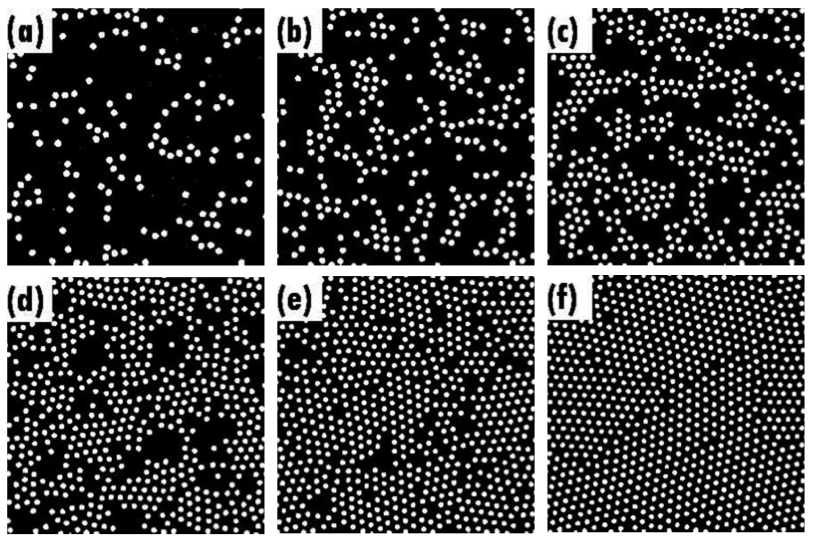}
\end{center}
\caption{\label{fig27}VPFC modelling of fluid and crystalline states of different particle densities. 
The number of atoms increases from left to right and from top to bottom. 
\nocite{PhysRevE.79.035701}\CRP{P.\ Y.\ Chan, N.\ Goldenfeld, and J.\ Dantzig}{Molecular dynamics on diffusive time scales from the phase-field-crystal equation}{Phys.\ Rev.\ E}{79}{2009}{035701R}{3}{10.1103/PhysRevE.79.035701}{the American Physical Society}}
\end{figure}

The VPFC model by Chan, Goldenfeld, and Dantzig \cite{PhysRevE.79.035701} is one of the most exciting extensions of the original PFC approach. 
The extra term added to the free energy makes particle density non-negative and allows for the formation of individual density peaks 
(\ZT{atoms} forming the fluid) and vacancies in the crystal. This, combined with the MPFC EOM \eqref{eq18}, that considers inertia 
and damping, makes it a kind of MD-like approach working on a still far longer time scale than the usual MD simulations. 
Accordingly, one can obtain configurations that look like snapshots of the fluid state (see figure \ref{fig27}) and may evaluate the structure factor 
for the fluid state, which is evidently impossible for the original PFC model. (Apparently, similar images can be obtained in the 1M-PFC model 
as a \textit{transient state} during solidification \cite{C0SM00944J}, however, with different dynamics owing to the differences in 
free energy and EOM.)

A comparison with another recent development, termed the diffusive molecular dynamics (DMD) technique, by Li \etal\ \cite{PhysRevB.84.054103} 
would be very interesting. The latter approach works on the diffusive time scale too, while maintaining atomic resolution, by 
coarse-graining over atomic vibrations and evolving a smooth site-probability representation.

\section{\label{chap:V}Phase-field-crystal modelling in soft matter physics}

\subsection{Applications to colloids}
In this section, we review results obtained using different PFC models relying on overdamped conservative dynamics --  
a reasonable approximation for colloidal crystal aggregation. We concentrate on three major areas: crystal nucleation, 
pattern formation in free growth, and pattern formation in the presence of external potentials. 

As mentioned previously, using of the EOM for simulating crystallization is not without difficulties. In the DDFT-type models, 
the system cannot leave a metastable state (\eg, the homogeneous initial fluid) unless Langevin noise representing thermal fluctuations 
is added to the EOM. This raises, however, essential questions: considering the number density an ensemble-averaged quantity, all the 
fluctuations are (in principle) incorporated into the free-energy functional. Adding noise to the EOM, a part of the fluctuations might be 
counted twice \cite{MarconiT1999,JPhysCondMat.15.V1}. If in turn the number density is viewed as being coarse-grained in time, 
there is phenomenological motivation to add a noise term to the EOM \cite{JPhysA.37.9325}. The latter approach is appealing in 
several ways: crystal nucleation is feasible from a homogeneous state and capillary waves appear at the crystal-liquid interface. 
To investigate how nucleation and growth happen on the atomistic level, a conserved noise term is usually incorporated into the 
EOM [see equations \eqref{eq14}-\eqref{eq17}]. To overcome some difficulties occurring when discretising the noise 
\cite{JPhysCondMat.22.360301,PhilosMag.91.25}, coloured noise obtained by filtering out the unphysical short wavelengths smaller than 
the inter-particle distance is often used (this removes both the ultraviolet catastrophe expected in 3D \cite{personal.karma.2009}  
and the associated dependence of the results on spatial resolution). The majority of the studies we review below follows this approach.

\subsubsection{Nucleation in colloidal crystal aggregation}

\paragraph{Homogeneous nucleation}

\subparagraph{The effect of noise}
A systematic study of the effect of the noise strength on the grain size distribution performed in two spatial dimensions by 
Hubert \etal\ \cite{JPhysCondMat.21.464108} for the original 1M-PFC model implies that grain size decreases with increasing noise amplitude, 
resulting in both a smaller average grain size and a reduced maximum grain size. They have distinguished two regimes regarding the 
cluster size distribution: for small noise amplitudes a bimodal cluster size distribution is observed, whereas for large noise amplitudes 
a monotonically decreasing distribution is reported.

\subparagraph{Phase selection in 2D and 3D} 
Mounting evidence indicates that the classical picture of crystal nucleation, which considers heterophase fluctuations of only the stable phase, 
is oversimplified. Early analysis by Alexander and McTague suggests a preference for bcc freezing in simple liquids \cite{PhysRevLett.41.702}. 
Atomistic simulations for the LJ system have verified that small heterophase fluctuations have the metastable bcc structure, 
and even larger clusters of the stable fcc structure have a bcc interface layer \cite{PhysRevLett.75.2714}, while the ratio of the two phases 
can be tuned by changing the pressure \cite{PhysRevLett.98.235502}. Composite bcc-fcc nuclei have also been predicted by 
continuum models \cite{PhysRevLett.106.045701}. Two-stage nucleation has been reported in systems that have a metastable critical point in the 
undercooled liquid (including solutions of globular proteins \cite{Galkin06062000}); the appearance of the crystalline phase is assisted by 
dense liquid droplets, whose formation precedes and helps crystal nucleation \cite{Wolde26091997}. Recent studies indicate a similar behaviour 
in simple liquids such as the LJ \cite{PhysRevLett.96.046102} or hard-sphere (HS) \cite{PhysRevLett.105.025701} fluids, where a dense liquid or 
amorphous precursor assists crystal nucleation. Analogous behaviour has been reported for colloidal systems in two spatial dimensions \cite{JACS.129.13520} and 
3D \cite{PhysRevLett.96.175701}. These findings imply that the nucleation precursors are fairly common. Since the 1M-PFC model has bcc, fcc, 
and hcp stability domains \cite{JPhysCondMat.22.364101}, the appearance of an amorphous phase and two-step nucleation has also been 
reported \cite{PhysRevE.77.061506}, and the 2M-PFC model incorporates the 1M-PFC model \cite{PhysRevE.81.061601}. This class of the 
dynamic PFC models is especially suitable for investigating phase selection during freezing of undercooled liquids.

\begin{figure}
\begin{center}
\includegraphics[width=0.19\linewidth]{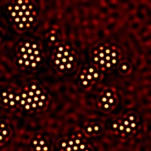}
\includegraphics[width=0.19\linewidth]{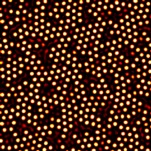}
\includegraphics[width=0.19\linewidth]{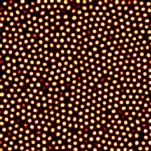}
\includegraphics[width=0.19\linewidth]{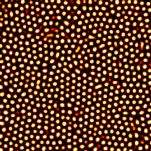}
\includegraphics[width=0.19\linewidth]{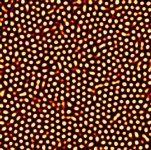}
\end{center}
{\vskip-4mm\footnotesize\hspace{0.092\linewidth}(a)\hspace{0.171\linewidth}(b)\hspace{0.171\linewidth}(c)\hspace{0.171\linewidth}(d)\hspace{0.171\linewidth}(e)}
\begin{center}
\includegraphics[width=0.19\linewidth]{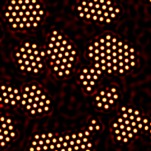}
\includegraphics[width=0.19\linewidth]{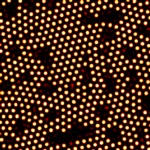}
\includegraphics[width=0.19\linewidth]{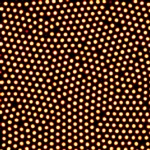}
\includegraphics[width=0.19\linewidth]{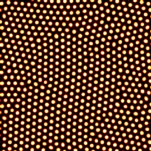}
\includegraphics[width=0.19\linewidth]{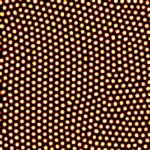}
\end{center}
{\vskip-4mm\footnotesize\hspace{0.093\linewidth}(f)\hspace{0.172\linewidth}(g)\hspace{0.172\linewidth}(h)\hspace{0.172\linewidth}(i)\hspace{0.172\linewidth}(j)}
\caption{\label{xfig1}Snapshots of early and late stages of isothermal solidification in 1M-PFC quenching simulations performed in two spatial dimensions with
initial reduced particle densities of $\psi_0 = -0.55,-0.50,-0.45,-0.40$ and $-0.35$ \cite{PhilosMag.91.123}. 
(a)-(e): Early stage: the respective reduced times are $\tau/\Delta\tau = 10000,3000,1500,1000$, and $700$. 
(f)-(j): Late stage: the same areas are shown at reduced time $\tau/\Delta\tau=60000$. 
Reduced particle density maps in $418 \times 418$ sized fractions of $2048 \times 2048$ sized simulations are shown. 
Other simulation parameters were $\epsilon = 0.75$ and $\alpha = 0.1$ (noise strength). 
\nocite{PhilosMag.91.123}\CRPE{L.\ Gr{\'a}n{\'a}sy, G.\ Tegze, G.\ I.\ T{\'o}th, and T.\ Pusztai}{Phase-field crystal modelling of crystal nucleation, heteroepitaxy and patterning}{Philos.\ Mag.\ }{91}{2011}{123-149}{1}{10.1080/14786435.2010.487476}{Taylor \& Francis}}
\end{figure}

\begin{figure}
\begin{center}
\includegraphics[width=0.49\linewidth]{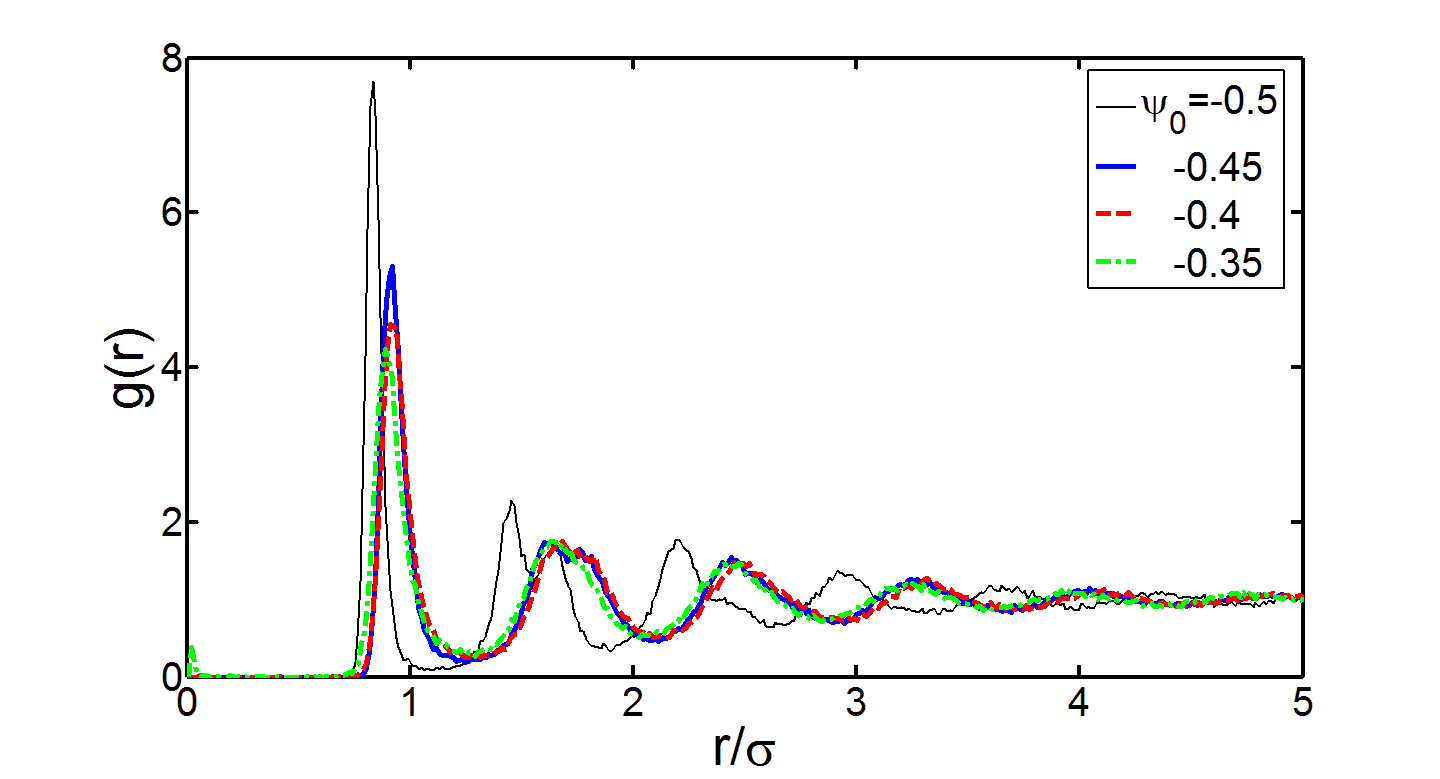}
\includegraphics[width=0.49\linewidth]{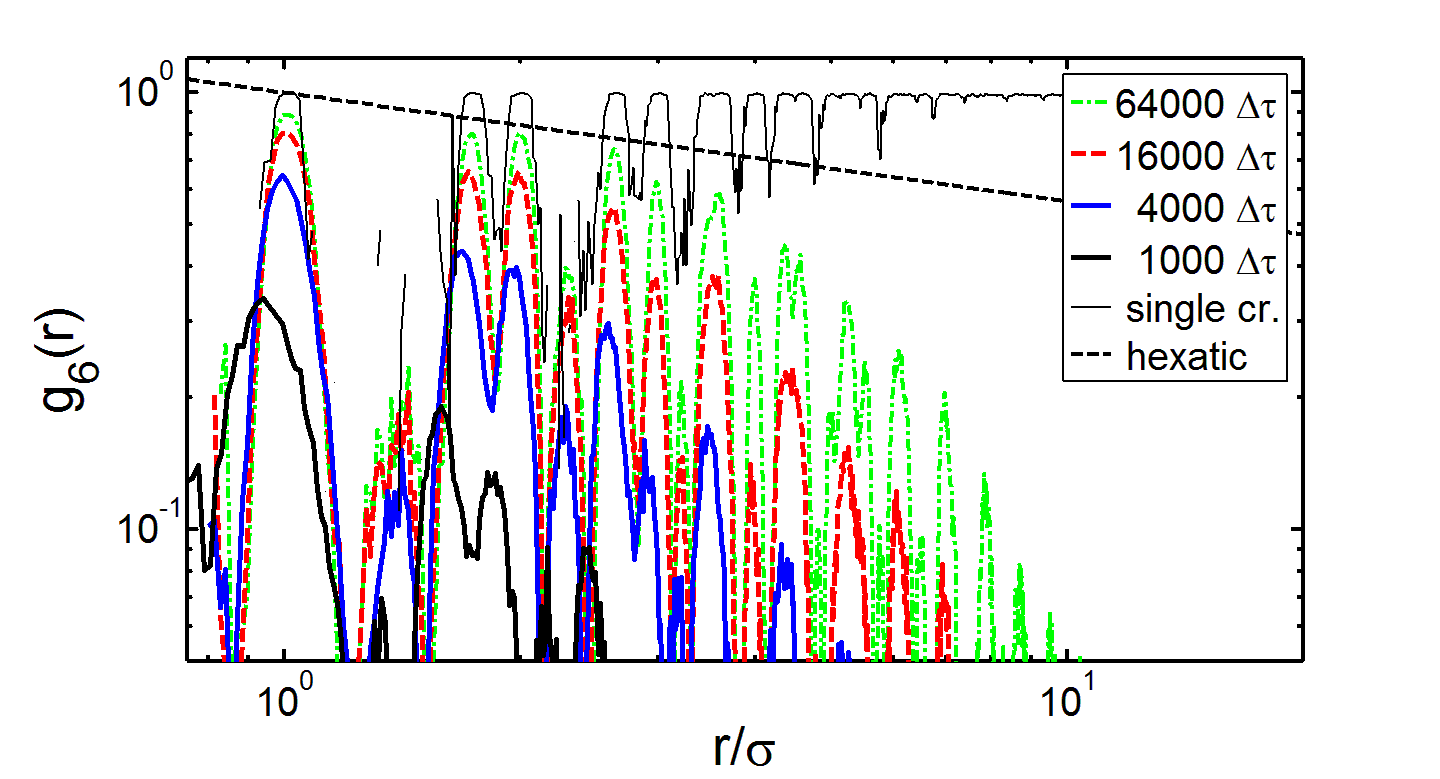}
\end{center}
\caption{\label{xfig2}Structural properties evolving after quenching in 1M-PFC simulations \cite{PhilosMag.91.123}: 
(a) pair-correlation function $g(r)$ for the early-stage solidification structures shown in figures \ref{fig10}(b)-(e). 
(b) Time evolution of the bond-order correlation function $g_{6}(r)$ for $\psi_0 = -0.4$ on log-log scale. $g_6(r)$ is shown at 
$\tau/\Delta\tau = 1000,4000,16000$, and $64000$. For comparison, the upper envelope expected for the hexatic phase and the result 
for a single crystal are also shown. These curves describe an amorphous to polycrystalline transition 
[see figures \ref{fig10}(d) and \ref{fig10}(i)]. Notice that the upper envelope of the $g_6(r)$ curves decay faster than expected 
for the hexatic phase. 
\nocite{PhilosMag.91.123}\CRPE{L.\ Gr{\'a}n{\'a}sy, G.\ Tegze, G.\ I.\ T{\'o}th, and T.\ Pusztai}{Phase-field crystal modelling of crystal nucleation, heteroepitaxy and patterning}{Philos.\ Mag.\ }{91}{2011}{123-149}{1}{10.1080/14786435.2010.487476}{Taylor \& Francis}}
\end{figure}

In two spatial dimensions, it has been shown within the framework of the 1M-PFC model that at relatively small supersaturations direct crystal nucleation takes place. 
Increasing the thermodynamic driving force, first copious crystal nucleation is observed, and at higher driving forces an amorphous precursor 
precedes crystalline nucleation \cite{PhilosMag.91.123} [see figures \ref{xfig1} and \ref{xfig2}(a)]. Similarly to quenching experiments for two spatial dimensions 
colloidal systems \cite{JPhysCondMat.20.404216}, no hexatic phase is observed in the 1M-PFC quenching simulations \cite{C0SM00944J,PhilosMag.91.123} 
[as demonstrated by the form of the radial decay of the bond-order correlation function \cite{PhilosMag.91.123}, see figure \ref{xfig2}(b)].

\begin{figure}
\begin{center}
\includegraphics[width=0.49\linewidth]{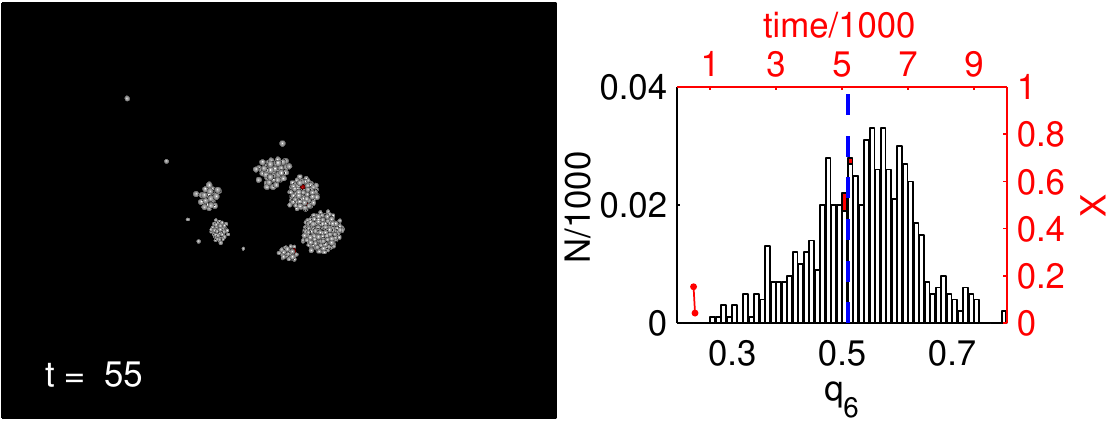}
\includegraphics[width=0.49\linewidth]{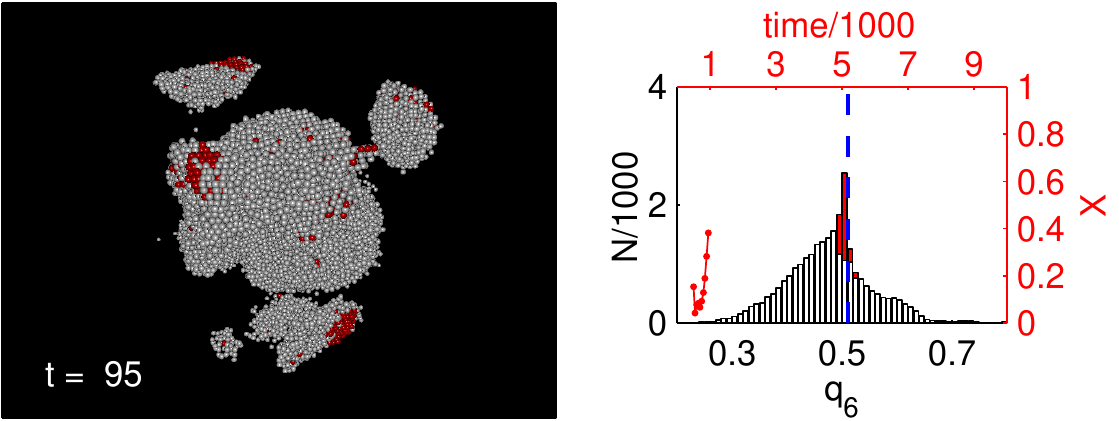}
\includegraphics[width=0.49\linewidth]{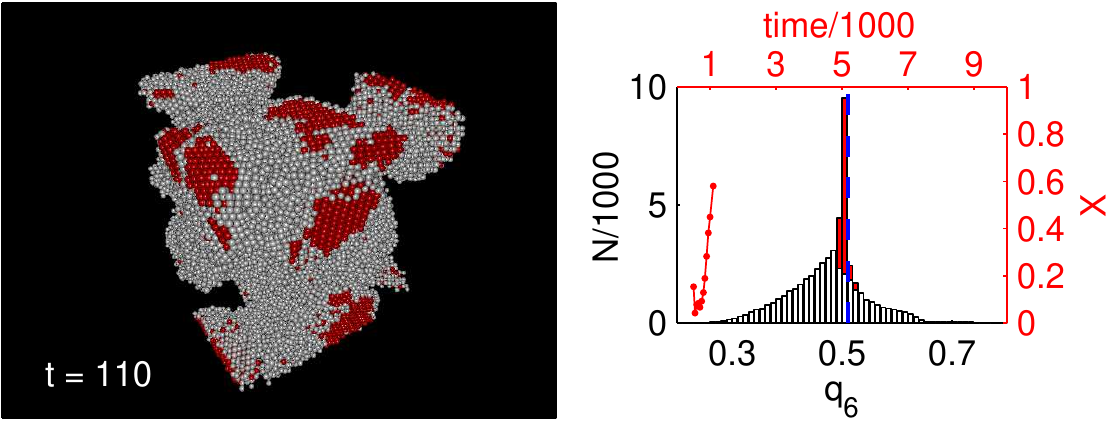}
\includegraphics[width=0.49\linewidth]{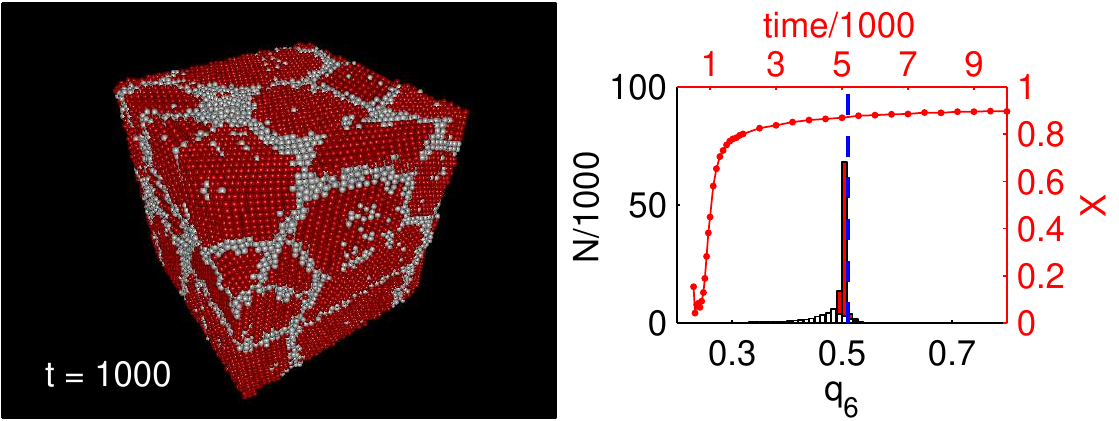}
\end{center}
\caption{\label{xfig3}Two-step nucleation in the 1M-PFC model at $\psi_0 = -0.1667$ and $\epsilon = 0.25$ \cite{PhysRevLett.107.175702}. 
Four pairs of panels are shown, where $t$ indicates the time elapsed, $N$ is the total number of particles, and $q_i$ is the bond orientational order parameter with index $i$. 
Left: snapshots of the density distribution taken at the dimensionless 
times $\tau = 57.74 t$. Spheres of the diameter of the inter-particle distance centred on density peaks higher than a threshold ($=0.15$) 
are shown. They are coloured red if $q_4 \in [0.02; 0.07]$ and $q_6 \in [0.48; 0.52]$ (bcc-like) and light grey otherwise. 
Right: population distribution of $q_6$ (histogram painted similarly) and the time dependence of the fraction $X$ of bcc-like neighbourhoods 
(solid line). 
\nocite{PhysRevLett.107.175702}\CRPE{G.\ I.\ T{\'o}th, T.\ Pusztai, G.\ Tegze, G.\ T{\'o}th, and L.\ Gr{\'a}n{\'a}sy}{Amorphous nucleation precursor in highly nonequilibrium fluids}{Phys.\ Rev.\ Lett.\ }{107}{2011}{175702}{17}{10.1103/PhysRevLett.107.175702}{the American Physical Society}}
\end{figure}

In 3D, a systematic dynamic study of the 1M/2M-PFC models by T\'oth \etal\ \cite{PhysRevLett.107.175702} shows that in these systems the first 
appearing solid is amorphous, which promotes the nucleation of bcc crystals (see figure \ref{xfig3}) but suppresses the appearance of the fcc and 
hcp phases. The amorphous phase appears to coexist with the liquid indicating a first-order phase transition between these phases in agreement 
with the observed nucleation of the amorphous state. Independent ELE studies determining the height of the nucleation barrier have confirmed that 
density and structural changes take place on different times scales \cite{PhysRevLett.107.175702}. This finding suggests that the two time scales 
are probably present independently of the type of dynamics assumed. These findings have been associated with features of the effective interaction 
potential deduced from the amorphous structure using Schommer's iterative method \cite{PhysRevA.28.3599} that shows a maximum at $r_0 \sqrt{2}$, 
where $r_0$ is the radius corresponding to the main minimum of the potential. Such a maximum in the interaction potential is expected to suppress 
crystallization to the close-packed structures fcc and hcp \cite{JChemPhys.118.2792}, whereas the multiple minima also found are expected to 
lead to coexisting disordered structures \cite{Nature.392.164}. By combining the results available for various potentials 
(LJ \cite{PhysRevLett.96.046102}, HS \cite{PhysRevLett.105.025701}, and the PFC potentials \cite{PhysRevLett.107.175702,JPhysCondMat.22.364101}), 
it appears that a repulsive core suffices for the appearance of a disordered precursor, whereas the peak at $r_0\sqrt{2}$ correlates with the 
observed suppression of fcc and hcp structures, while the coexistence of the liquid and amorphous phases seen here can be associated with multiple 
minima of the interaction potential.

3D studies, performed for bcc crystal nucleation in molten pure Fe in the framework of the EOF-PFC model 
\cite{JPhysCondMat.22.364101,PhilosMag.91.123}, lead to similar results, however, still with diffusive dynamics. 
In these simulations, the initial density of the liquid has been increased until the solidification started -- a procedure that has lead to 
an extreme compression owing to the small size and short time accessible for the simulations. While this raises some doubts regarding 
the validity of the applied approximations, the behaviour observed for the EOF-PFC Fe is fully consistent with the results obtained for the 
1M-PFC model: with increasing driving force first an amorphous precursor nucleates and the bcc phase appears inside these amorphous regions 
\cite{JPhysCondMat.22.364101,PhilosMag.91.123}. At higher driving forces the amorphous precursor appears nearly homogeneously in space  
and the bcc phase nucleates into it later. Apparently, direct nucleation of the bcc phase from the liquid phase requires a longer time than 
via the amorphous precursor, suggesting that the appearance of the bcc phase is assisted by the presence of the amorphous phase and in line with 
recent predictions by DFT \cite{PhysRevLett.96.046102} and atomistic simulations \cite{PhysRevLett.105.025701}. 
Remarkably, the interaction potential evaluated for Fe from the pair-correlation function of the amorphous structure is oscillatory and is 
qualitatively similar to the ones evaluated from experimental liquid structures \cite{shimoji.1977}.

\begin{figure}
\begin{center}
\includegraphics[width=0.9\linewidth]{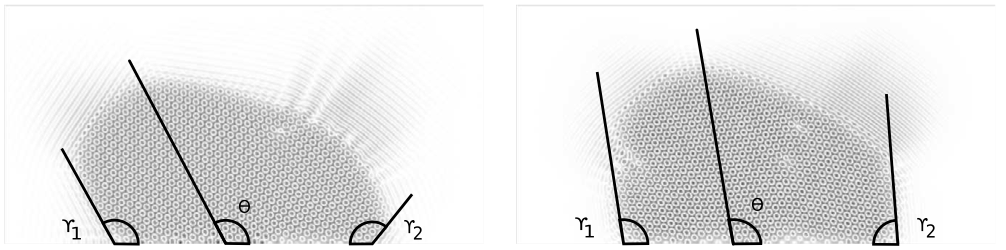}
\includegraphics[width=0.9\linewidth]{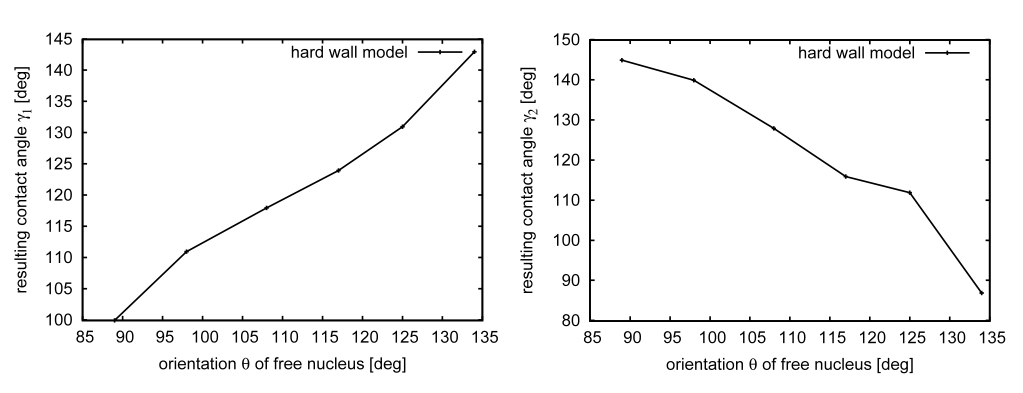}
\end{center}
\caption{\label{xfig4}Heterogeneous nuclei formed on a hard wall in the APFC model proposed in reference \cite{JPhysCondMat.21.464110} 
and the dependence of the left and right side contact angle ($\gamma_1$ and $\gamma_2$, respectively) on the crystal orientation. 
\nocite{JPhysCondMat.21.464110}\CRPE{R.\ Prieler, J.\ Hubert, D.\ Li, B.\ Verleye, R.\ Haberkern, and H.\ Emmerich}{An anisotropic phase-field crystal model for heterogeneous nucleation of ellipsoidal colloids}{J.\ Phys.: Condens.\ Matter}{21}{2009}{464110}{46}{10.1088/0953-8984/21/46/464110}{Institute of Physics Publishing}}
\end{figure}

\paragraph{Heterogeneous nucleation} 
Prieler \etal\ \cite{JPhysCondMat.21.464110} have explored crystal nucleation on an unstructured hard wall in an anisotropic version of the 
1M-PFC model, in which the particles are assumed to have an ellipsoidal shape. In particular, they have investigated how the contact angle depends 
on the orientation of the ellipsoids and the strength of the wall potential (see figure \ref{xfig4}). A complex behaviour has been observed for the 
orientational dependence, while increasing the strength of the wall potential reduced the contact angle.

\begin{figure}
\begin{center}
\includegraphics[width=0.32\linewidth]{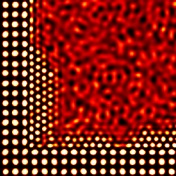}
\includegraphics[width=0.32\linewidth]{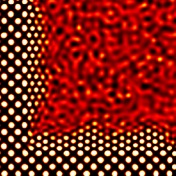}
\includegraphics[width=0.32\linewidth]{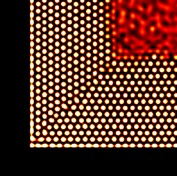}
\end{center}
{\vskip-2mm\footnotesize\hspace{0.16\linewidth}(a)\hspace{0.305\linewidth}(b)\hspace{0.305\linewidth}(c)}
\caption{\label{xfig5}Heterogeneous nucleation in rectangular inner corners of the 1M-PFC model in two spatial dimensions \cite{PhilosMag.91.123}. 
(a) Nucleation on $(01)$ surfaces of a square lattice (ratio of lattice constant of substrate to inter-particle distance $a_0/\sigma \approx 1.39$).  
(b) Nucleation on $(11)$ surfaces of a square lattice. 
(c) Nucleation on an unstructured substrate. Notice the frustration at the corner and the formation of a 
grain boundary starting from the corner at later stages. 
\nocite{PhilosMag.91.123}\CRPE{L.\ Gr{\'a}n{\'a}sy, G.\ Tegze, G.\ I.\ T{\'o}th, and T.\ Pusztai}{Phase-field crystal modelling of crystal nucleation, heteroepitaxy and patterning}{Philos.\ Mag.\ }{91}{2011}{123-149}{1}{10.1080/14786435.2010.487476}{Taylor \& Francis}}
\end{figure}

Gr\'an\'asy \etal\ \cite{JPhysCondMat.22.364102} have studied crystal nucleation in an rectangular corner of structured and unstructured 
substrates within the 1M-PFC model in two spatial dimensions. Despite expectations based on the classical theory of heterogeneous nucleation and conventional 
PF simulations \cite{PhysRevB.67.035412}, which predict that a corner should be a preferred nucleation site, in the atomistic approach 
such a corner is not a preferable site for the nucleation of the triangular crystal structure (see figure \ref{xfig5}) owing to the misfit of the 
triangular crystal structure with a rectangular corner. Crystals of different orientation nucleate on the two substrate surfaces, 
which inevitably leads to the formation of a grain boundary starting from the corner when the two orientations meet. 
The energy cost of forming the grain boundary makes the rectangular corner an unfavoured place for nucleation. In contrast, a $60^\circ$ corner 
helps the nucleation of the triangular phase. 

\subsubsection{Pattern formation in colloidal crystal aggregation}

\begin{figure}
\begin{center} 
\includegraphics[width=0.98\linewidth]{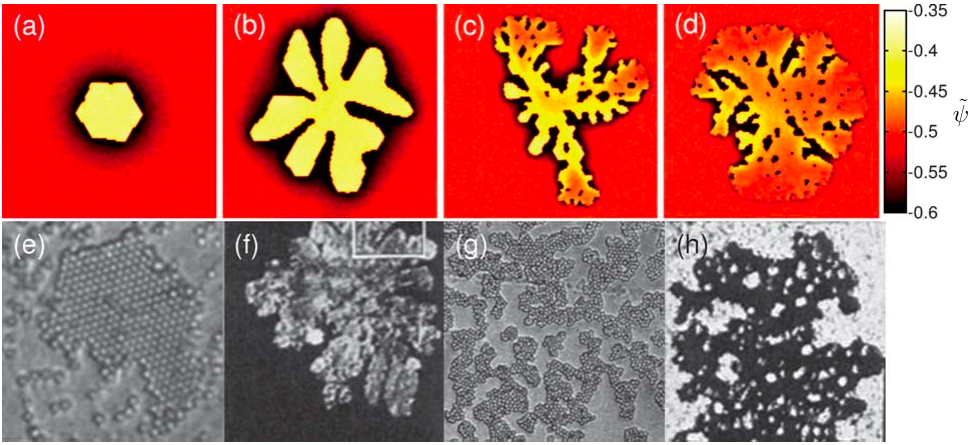}
\end{center}
\caption{\label{xfig6}Single crystal growth morphologies (a)-(d) in the 1M-PFC model \cite{C0SM00944J} (top) and experiment (bottom) (e)-(h): 
2D colloid crystals by Skjeltorp. 
\nocite{PhysRevLett.58.1444}\CRP{A.\ T.\ Skjeltorp}{Visualization and characterization of colloidal growth from ramified to faceted structures}{Phys.\ Rev.\ Lett.\ }{58}{1987}{1444-1447}{14}{10.1103/PhysRevLett.58.1444}{the American Physical Society}   
The driving force increases from left to right. In the case of the simulations, the coarse-grained particle density map is shown. 
The fractal dimensions of the single crystal aggregates evaluated from the slope of the plot $\log(N)$ \vs $\log(R_{\mathrm{g}})$ 
($N$ is the number of particles in the cluster and $R_{\mathrm{g}}$ is its radius of gyration) are: 
(a) $f_{\mathrm{d}} = 2.012 \pm 0.3 \%$, (b) $1.967 \pm 0.3\%$, (c) $1.536 \pm 0.9\%$, (d) $1.895 \pm 0.3\%$. The fast growth mode is 
recognizable via the lack of a (dark) depletion zone at the interface, whose presence is indicative to the slow mode. 
A $2048 \times 2048$ rectangular grid corresponding to $\approx 13000$ particles, or $118\,\mu\mathrm{m}\times118\,\mu\mathrm{m}$ 
(assuming $1.1\,\mu\mathrm{m}$ particles) has been used -- a size comparable to that shown by the experimental images. 
\nocite{C0SM00944J}\CRPE{G.\ Tegze, L.\ Gr{\'a}n{\'a}sy, G.\ I.\ T{\'o}th, J.\ F.\ Douglas, and T.\ Pusztai}{Tuning the structure of nonequilibrium soft materials by varying the thermodynamic driving force for crystal ordering}{Soft Matter}{7}{2011}{1789-1799}{5}{10.1039/c0sm00944j}{Royal Society of Chemistry Publishing}}
\end{figure}

\paragraph{Colloid patterns in two dimensions} 
Choosing a large value for the parameter $\epsilon$ where the liquid-solid interface is faceted, 
Tegze \etal\ \cite{C0SM00944J,PhysRevLett.106.195502} have investigated solidification morphologies as a function of the thermodynamic 
driving force. It has been found that the diffusion-controlled growth mode observed at low driving forces and characterized by faceted 
interfaces changes to a diffusionless growth mode of a diffuse liquid-solid interface that produces a crystal, whose density is comparable 
to the density of the liquid due to quenched-in vacancies (see section \ref{sec:dst}). These two modes have already been observed experimentally 
in colloidal systems \cite{Langmuir.13.3871}. It has been shown that the two modes can coexist and lead to a new branching mechanism that 
differs from the usual diffusional instability driven branching by which dendritic structures form. This new mechanism explains the 
fractal-like and porous growth morphologies \cite{PhysRevLett.58.1444} observed in 2D colloidal systems (see figure \ref{xfig6}) and may be 
relevant for the diffusion-controlled to diffusionless transition of crystallization in organic glasses.

\begin{figure}
\begin{center}
\includegraphics[width=\linewidth]{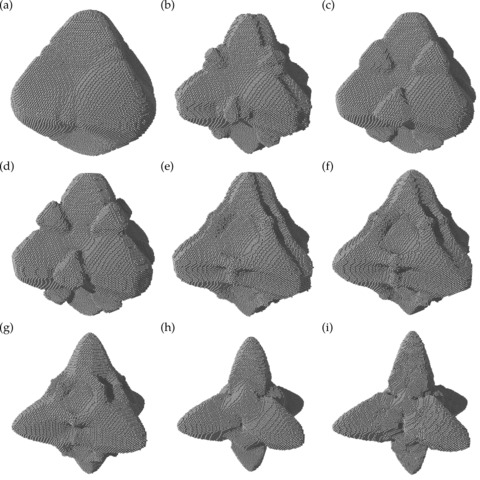}
\end{center}
\caption{\label{xfig7}3D crystal growth morphologies grown from a bcc seed in the single-component 1M-PFC model at $\epsilon = 0.3748$ in a system 
containing about $3 \times 10^6$ colloidal particles \cite{tegze.phd.thesis}. The initial fluid density decreases as 
(a) $\psi_0 = -0.015$, (b) $-0.0175$, (c) $-0.01875$, (d) $-0.02$, (e) $-0.02062$, (f) $-0. 0225$, (g) $-0.025$, (h) $-0.03$, (i) $-0.0325$. 
The simulations have been performed on a $1024 \times 1024 \times 1024$ grid. Assuming $1\,\mu\mathrm{m}$ diameter for the particles, the linear size 
of the simulation box is $\approx 0.16\,\mathrm{mm}$ -- comparable to the smaller colloidal dendrites seen in microgravity experiments 
\cite{PhysRevLett.88.015501}. 
(Reproduced from \nocite{tegze.phd.thesis}G.\ Tegze, \textit{Application of atomistic phase-field methods to complex solidification problems}, 
PhD Thesis, E\"otv\"os University, Budapest, Hungary (2009).)}
\end{figure}

\paragraph{Colloid patterns in three dimensions} 
T\'oth \etal\ \cite{JPhysCondMat.22.364101} have demonstrated first that owing to the conservative dynamics, 
the EOM of the 1M-PFC model realizes, dendritic growth forms of bcc and fcc structure evolve in the single-component theory. 
Tegze \cite{tegze.phd.thesis} and Gr\'an\'asy \etal\ \cite{tms2010seattle} have shown by simulations containing $\approx 3\times10^6$ particles 
that due to a kinetic roughening of the crystal-liquid interface that leads to interface broadening, a transition can be seen from faceted 
dendrites to compact rounded crystals (see figure \ref{xfig7}) -- a phenomenon reported earlier in experiments for dendritic growth of NH$_4$Br 
crystals \cite{EuroPhysLett.8.67}. Notice that such a kinetic effect cannot be easily incorporated into conventional PF models. 
Remarkably, as pointed out in reference \cite{JPhysCondMat.22.364101}, assuming a micrometer diameter for the \ZT{atoms}, these dendritic structures 
are comparable in size to those formed in colloid experiments in microgravity \cite{PhysRevLett.88.015501}. This is a unique situation indeed: 
an \ZT{atomistic} theory works here on the size scale of experimental dendrites.

In a recent work, Tang \etal\ \cite{Tang2011146} have performed a geometric analysis of bcc and fcc dendrites grown in the respective 
stability domains of the 1M-PFC model, and evaluated dynamic exponents characterizing dendritic growth in the $(100)$, $(110)$, and $(111)$ directions. 
They associate the relatively large values obtained for the stability constant from the geometry of the dendrite tip with the faceted morphology 
of the crystals.

\begin{figure}
\begin{center}
\includegraphics[height=0.4\linewidth]{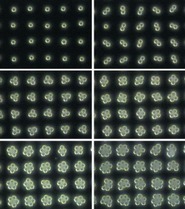}
\includegraphics[height=0.4\linewidth]{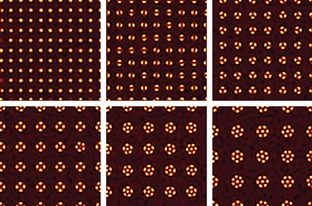}
\end{center}
{\vskip-4mm\footnotesize\hspace{0.18\linewidth}(a)\hspace{0.465\linewidth}(b)}
\caption{\label{xfig8}(a) Single and multiple occupation of a chemically patterned periodic substrate by colloidal particles as a function of increasing 
patch size in the experiments. 
\nocite{LeeZRH2002}\CRP{I.\ Lee, H.\ Zheng, M.\ F.\ Rubner, and P.\ T.\ Hammond}{Controlled cluster size in patterned particle arrays via directed adsorption on confined surfaces}{Adv.\ Mater.\ }{14}{2002}{572-577}{8}{10.1002/1521-4095(20020418)14:8$<$572::AID-ADMA572$>$3.0.CO;2-B}{Wiley}   
(b) 1M-PFC simulations \cite{PhilosMag.91.123} with increasing diameter of circular attractive potential wells. 
Reduced particle density maps are shown. The ratio of the potential well diameters relative to the single occupation case has been 
$1,1.25,1.5,2,2.13$, and $2.5$. 
\nocite{PhilosMag.91.123}\CRPE{L.\ Gr{\'a}n{\'a}sy, G.\ Tegze, G.\ I.\ T{\'o}th, and T.\ Pusztai}{Phase-field crystal modelling of crystal nucleation, heteroepitaxy and patterning}{Philos.\ Mag.\ }{91}{2011}{123-149}{1}{10.1080/14786435.2010.487476}{Taylor \& Francis}}
\end{figure}

\subsubsection{Colloid patterning} 
\begin{figure}
\begin{center}
\includegraphics[height=0.4\linewidth]{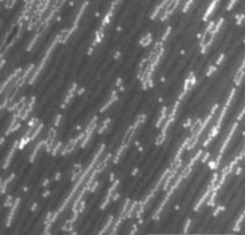}
\includegraphics[height=0.4\linewidth]{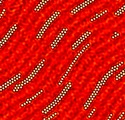}
\end{center}
{\vskip-4mm\footnotesize\hspace{0.281\linewidth}(a)\hspace{0.383\linewidth}(b)}
\caption{\label{xfig9}Patterning in experiment \vs 1M-PFC simulation: (a) single and double particle chains evolving in experiment due to 
capillary-immersion forces on the surface of a rippled substrate. 
\nocite{Langmuir.22.582}\CRP{A.\ Mathur, A.-D.\ Brown, and J.\ Erlebacher}{Self-ordering of colloidal particles in shallow nanoscale surface corrugations}{Langmuir}{22}{2006}{582-589}{2}{10.1021/la0520379}{the American Chemical Society}  
(b) The particle chains forming in the 1M-PFC simulation performed with a tilted and wavy version of the potential described in the text 
\cite{PhilosMag.91.123}. Only a fraction of the reduced particle density map is shown. 
\nocite{PhilosMag.91.123}\CRPE{L.\ Gr{\'a}n{\'a}sy, G.\ Tegze, G.\ I.\ T{\'o}th, and T.\ Pusztai}{Phase-field crystal modelling of crystal nucleation, heteroepitaxy and patterning}{Philos.\ Mag.\ }{91}{2011}{123-149}{1}{10.1080/14786435.2010.487476}{Taylor \& Francis}}
\end{figure}

Colloid patterning under the influence of periodic substrates can be realized via creating patches that are chemically attractive to the 
colloidal particles \cite{LeeZRH2002}. Depending on the size of the patches single, double, triple, \etc, occupations of the patches are 
possible (see figure \ref{xfig8}), whereas the distance of the patches may lead to the formation of various ordered patterns, as predicted by 
Langevin simulations, in which the patterned substrate is represented by appropriate periodic potentials \cite{PhysRevLett.88.248301}. 
Gr\'an\'asy \etal\ \cite{PhilosMag.91.123} has employed a 1M-PFC model supplemented with a periodic potential of circular potential wells 
arranged on a square lattice, to reproduce the patterns seen in the experiments (see figure \ref{xfig8}). Another problem, exemplifying the 
abilities of PFC simulations in modelling colloid patterning, is colloidal self-assembly under the effect of capillary-immersion forces acting 
on the colloid particles in thin liquid layers due to capillarity and a periodically varying depth of the liquid layer due to a wavy substrate surface. 
Experiments of this kind have been used to produce single and double particle chains \cite{Langmuir.22.582} and the otherwise unfavourable 
square-lattice structure \cite{ADMA:ADMA200701175}. The capillary-immersion forces can often be well represented by a potential of the form 
$U = u_1 \cos(\wn x)$, where $u_1$ is a constant, $\wn = 2\pi/\lambda$, and $\lambda$ the wavelength of the periodic potential. 
Setting $\lambda = \sigma /\sqrt{2}$, where $\sigma$ is the interparticle distance, and varying the orientation of the grooves relative to the 
crystallization (drying) front, patterns seen in the experiments \cite{ADMA:ADMA200701175} are observed to form in the 1M-PFC model: 
for grooves parallel to the front, a frustrated triangular structure of randomly alternating double and triple layers appears. 
For grooves perpendicular to the front, the particles align themselves on a square lattice with the $(1 1)$ orientation lying in the interface, 
while for a $\pi/4$ declination of the grooves the same structure forms, however, now with the $(1 0)$ face lying in the front. 
Using larger wavelengths for the potential and adding a weak transversal modulation, while starting from a homogeneous initial particle density, 
nucleation and growth of wavy single and double chains resembling closely to the experiments \cite{Langmuir.22.582} are seen \cite{PhilosMag.91.123} 
(see figure \ref{xfig9}).

Epitaxial growth on the $(100)$ surface of a sc substrate has been investigated in 3D using the 1M-PFC model by 
Tegze \etal\ \cite{PhysRevLett.103.035702}. The lattice constant $a_{\mathrm{s}}$ of the substrate has been varied in a range that incorporates 
the interatomic distance of the bulk fcc structure and the lattice constant of the bulk bcc phase, where the $(100)$ face of the sc structure 
is commensurable with the $(100)$ faces of the bulk fcc and bcc structures, respectively. A bct structure has grown, 
whose axial ratio $c/a$ varies continuously with the lattice constant of the substrate, where $c$ and $a$ are the lattice constants of the 
bct structure perpendicular and parallel to the surface of the substrate, respectively. At the matching values of $a_{\mathrm{s}}$, 
fcc and bcc structures have been observed respectively, as observed in colloid patterning experiments \cite{ADMA:ADMA200400830}. 
Analogous results have been obtained for the $(100)$ face of an fcc substrate using 1M-PFC simulations, however, 
for large lattice mismatch amorphous phase mediated bcc nucleation 
has been seen \cite{PhysRevLett.108.025502}.

\begin{figure}
\begin{center}
\includegraphics[width=0.32\linewidth]{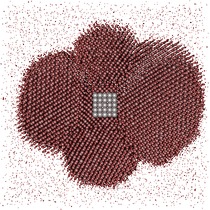}
\includegraphics[width=0.32\linewidth]{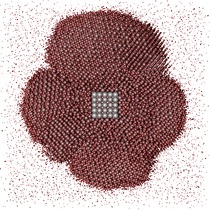}
\includegraphics[width=0.32\linewidth]{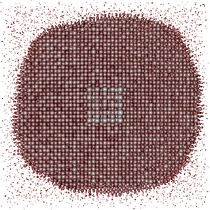}
\end{center}
{\vskip-4mm\footnotesize\hspace{0.16\linewidth}(a)\hspace{0.31\linewidth}(b)\hspace{0.3\linewidth}(c)}
\caption{\label{xfig10}In-plane snapshot of crystalline aggregates grown on $5 \times 5$ square-lattice templates of 
$a_{\mathrm{s}}/\sigma_{\mathrm{fcc}} = 1.0,1.1547$, and $1.56$ in 3D as predicted by 1M-PFC simulations \cite{C0SM01219J}. 
Here, $\sigma_{\mathrm{fcc}} = 1.056$. A $256 \times 256 \times 128$ grid has been used. The visualization is as in figure \ref{fig14}.}
\end{figure}

Optical tweezers are used widely to realize 2D periodic templates for influencing colloidal crystal aggregation in 3D \cite{PhysWorld.15.31}. 
Such templates, depending on the mismatch to the crystalline structure evolving, may instigate the formation of single-crystal or 
polycrystalline structures \cite{C0SM01219J}. Growth textures, obtained when supplementing the 1M-PFC model with a $5 \times 5$ flat 
square-lattice template (realized by a periodic potential term), show remarkable resemblance to the experiments (see figure \ref{xfig10}) 
\cite{unbup.toth.granasy}.

\subsection{\label{subsec_applic_poly}Applications to polymers}
In the field of soft matter simulations the study of polymers plays an important role. Understanding the relation between the 
statistical physics on the molecular scale and the resulting material properties is of great relevance with respect to principal questions, 
chemical and pharmaceutical applications, as well as the study of biopolymers in living cells. The investigation of polymer systems and the 
development of new polymer materials and composites with specific properties depend essentially on suitable numerical and simulation techniques. 
One of the big challenges is closing the gap between the atomistic and the microscale. Furthermore, in many polymer systems structures develop 
rather slowly and on very different length scales. Since particle-based simulations are strongly restricted, especially regarding the time scale, 
there is the demand for continuum models such as PF and PFC theories. 

Polymer systems are studied in various systems, in the form of polymer solutions and melts, in hydrogels, in glasses, and in polymer crystals. 
A principle problem of continuum models for polymers is the connectivity of the chains, which leads to long-ranging interactions between spatially 
separated atoms of the same polymer chain and, due to entanglements, also between atoms of different chains. For an implementation of these effects 
into continuum theories, appropriate simplifications are necessary. 

One major continuum method for polymer systems is the self-consistent field (SCF) theory. Starting from an equilibrium theory, it has been extended 
in order to study dynamic systems. It is therefore interesting to compare the SCF theory with the PF method, which, 
during the last years, has increasingly been applied on polymer systems, and the PFC method, which is suited for studying 
pronounced periodic structures in polymers. In this context, we will also elucidate the relation between the PF and the 
PFC method, especially for the case of polymer systems. 

In many applications of polymers, the structure and dynamics of the system are sensitive to various parameters like temperature, flow velocity, 
velocity gradients, gravity or stresses. In polyelectrolytes, also charge densities, ion concentrations, the pH value, and external fields need to be 
considered. Several of these aspects have been taken into account in PF calculations, as we will demonstrate for some examples. 
We begin with a description of the SCF theory, after which we discuss the PF and the PFC approaches to 
polymer systems.

\subsubsection{Self-consistent field theory for polymer systems}
In general, the SCF theory is a molecular-based, approximate description of many-body systems. 
For polymer blends, it has originally been developed to describe equilibrium systems \cite{Edwards1965, Helfand1971}. 
Later it was extended to dynamic self-consistent field (DSCF) theory, 
which allows investigating dynamic processes (for an review see reference \cite{Mueller2005}).

The SCF method is used to study fluctuations, pattern formation, and segregation in various kinds of polymer mixtures 
(see references \cite{Mueller2005,Schmid1998,Yang2006,Hong1981} and references therein). 
In many works, the method is applied on a mixture of different types of homopolymers 
or block copolymers built of two types of monomers A and B. In most cases, it is assumed that the polymer chains have a Gaussian 
distribution. This means that the polymer path is distributed like a random walk and corresponds to the situation 
where the persistence length is much smaller than the polymer length. 
It should be mentioned that one can also use semiflexible, wormlike polymers instead \cite{Schmid1998}. 
Assuming an incompressible polymer melt, the total polymer density $\varphi_0$ is fixed. The canonical partition 
function can be written as \cite{Schmid1998}
\begin{equation}  
\mathcal{Z}\approx \int \!\prod_{i}\!\delta\mathbf{r}_{i}\,   
P(\mathbf{r}_{i}) \exp\!\bigg(\!-\frac{\mathcal{F}_{\mathrm{FH}}}{\kBT} \bigg)\delta (\varphi_{\mathrm{A}} +\varphi_{\mathrm{B}} -\varphi_{0}) 
\label{Eq_canonical_Z_SCF_1}
\end{equation} 
Here, $\int\!\delta\mathbf{r}_{i} P(\mathbf{r}_{i})$ is a path integral of a polymer weighted 
by the Gaussian distribution $P(\mathbf{r}_i)$ and $\mathcal{F}_{\mathrm{HF}}$ is the Flory-Huggins free energy for polymers, which favours adjacent 
monomers of the same type: A-A and B-B. The delta function generates 
the incompressibility. The problematic parts of $Z$ in equation \eqref{Eq_canonical_Z_SCF_1} are the
polymer-polymer interactions (the steric and the Flory-Huggins interactions). 
They are avoided by rewriting $\mathcal{Z}$ in a form in which single polymers are 
subject to auxiliary fields that represent the effect of the other polymers. For the system with monomers A and B, two auxiliary 
fields $V$ and $W$ are introduced, where $V$ is conjugated to the total density $\varphi_{\mathrm{A}}+\varphi_{\mathrm{B}}$, 
while $W$ is conjugated to the composition $\varphi_{\mathrm{A}}-\varphi_{\mathrm{B}}$. 
Auxiliary fields are an essential concept of all SCF theories. The canonical partition function $Z$ can now be written as 
\begin{equation}  
\mathcal{Z}\approx \int \!\!\delta V\delta W \exp\!\bigg(\!-\frac{\mathcal{H}(V,W)}{\kBT} \bigg) \,,
\label{Eq_canonical_Z_SCF_2}  
\end{equation} 
where the Hamiltonian $\mathcal{H}(V,W)$ is determined by the partition functions $\mathcal{Z}_{\mathrm{sp}}$ 
of single polymers in the presence of the auxiliary fields. 
The partition functions can be calculated with the help of propagator functions. The propagator functions are the solution of diffusion 
equations \cite{Helfand1971}, which depend on $V$ and $W$. Equation \eqref{Eq_canonical_Z_SCF_2} is still 
equivalent to equation \eqref{Eq_canonical_Z_SCF_1}, but the expression is much simpler.

\paragraph{Equilibrium SCF theory} 
In principle, one can use equation \eqref{Eq_canonical_Z_SCF_2} to extract equilibrium values of the 
average concentration $\langle \varphi_A\rangle$ (and thus of $\langle \varphi_{\mathrm{B}}\rangle=\varphi_0-\langle \varphi_{\mathrm{A}}\rangle$) 
and fluctuation terms. In fact, in some cases, this has been done numerically with the help of \ZT{field MC} 
simulations \cite{Duchs2003} or complex Langevin simulations \cite{Ganesan2001}. However, for most systems the
numerical cost for treating expression \eqref{Eq_canonical_Z_SCF_1} with fully fluctuating auxiliary fields $V$ and $W$ is extremely
high. The problem can be simplified with the help of a saddle-point approximation: 
instead of integrating over the whole $V$ and $W$ space, one determines the minimum of the Hamiltonian $\mathcal{H}$ with respect to $V$ and $W$,
\begin{equation} 
\frac{\partial \mathcal{H}(V^*\!,W^*)}{\partial V} =\frac{\partial \mathcal{H}(V^*\!,W^*)}{\partial W} =0 \;,
\label{Eq_saddle_point}
\end{equation} 
which provides saddle-point fields $V^*$ and $W^*$ for given concentration 
fields $\varphi_{\mathrm{A}}$ and $\varphi_{\mathrm{B}}$. In turn, the concentration fields can be calculated from the partition function by solving the diffusion 
equation that depends on $V^*$ and $W^*$. Thus, the problem can be solved self-consistently by iterating the calculation of concentration and 
auxiliary fields \cite{Rasmussen2002}. This is the basic principle of the SCF theory. In this form, it enables calculation of equilibrium properties 
of concentration fields, including microphase separation and global phase separation for polymer blends including homopolymers 
and block copolymers. A more precise description is given by the external potential theory, which allows fluctuations of the
composition field $W$ \cite{Maurits1997}. 

\paragraph{Dynamic SCF theory} 
The dynamic self-consistent field theory (DSCF) is an extension of the SCF theory that allows to study 
nonequilibrium polymer blends. For an incompressible fluid, the saddle point values $V^*$ and $W^*$ are
determined by $\varphi_{\mathrm{A}}$ so that we can define the energy functional
\begin{equation}  
\mathcal{F}[\varphi_{\mathrm{A}}]=\mathcal{H}\big(V^*\!(\varphi_{\mathrm{A}}),W^*\!(\varphi_{\mathrm{A}})\big) \;. 
\label{Eq_F_DSCF}
\end{equation} 
The dynamics of $\varphi_{\mathrm{A}}$ is described by a Langevin equation
\begin{equation} 
\frac{\partial \varphi_{\mathrm{A}}(\mathbf{r},t)}{\partial t} 
= -\Nabla\!\cdot\!\mathbf{J}_{\varphi }(\mathbf{r},t)+\zeta(\mathbf{r},t)  
\label{Eq_Langevin_SCF} 
\end{equation} 
with a flux density $\mathbf{J}_{\varphi}(\mathbf{r},t)$ and a Gaussian random noise term $\zeta(\mathbf{r},t)$. 
The flux density is given by
\begin{equation}  
\mathbf{J}_{\varphi} = \vints \Lambda _{\varphi} (\mathbf{r},\mathbf{r}')\NablaS 
\frac{\delta\mathcal{F}}{\delta\varphi_{\mathrm{A}}(\mathbf{r}',t)} 
\label{Eq_flux}
\end{equation} 
with an Onsager coefficient $\Lambda_{\varphi}(\mathbf{r},\mathbf{r}')$, which is 
in general nonlocal, due to the forces that propagate along the polymer backbone. Nonlocal effects can be considered with 
the Rouse model \cite{Rouse1953} or the reptation model \cite{DoiE2007}. In the simplest case, the interactions 
due to the chain connectivity are neglected and a local Onsager coefficient 
$\Lambda_{\varphi}(\mathbf{r},\mathbf{r}')=k\,\varphi_{\mathrm{A}}(\mathbf{r},t)\varphi_{\mathrm{B}}(\mathbf{r},t)
\delta(\mathbf{r}-\mathbf{r}')$ with a constant $k$ is used \cite{ElderPBSG2007,RamakrishnanY1979}, leading to
\begin{equation} 
\frac{\partial \varphi_{\mathrm{A}} }{\partial t} =-k\:\!\Nabla \bigg(\varphi_{\mathrm{A}}(\varphi_{0}-\varphi_{\mathrm{A}}) 
\Nabla \frac{\delta\mathcal{F}}{\delta \varphi_{\mathrm{A}}} \bigg) + \zeta \;.  
\label{Eq_local_Onsager} 
\end{equation} 
Notice that $\mathcal{F}$ still includes the auxiliary fields and each propagation 
in time requires the iterative calculation of these fields. In the random 
phase approximation, which is valid for suitably small fluctuations, one 
can derive an approximate expression for $\mathcal{F}$, which has the form of a 
Ginzburg-Landau free energy \cite{Schmid1998}, 
\begin{equation}  
\mathcal{F}_{\mathrm{GL}} = \vint \Big(f_{\mathrm{FH}}(\varphi_{\mathrm{A}}) + b(\Nabla\psiPF)^{2} \Big) \,, 
\label{Eq_Ginzb_Landau}
\end{equation}
with the Flory-Higgins free-energy density $f_{\mathrm{FH}}(\varphi_{\mathrm{A}})$ and the parameter $b$. 

In analogy to DSCF, a dynamic version of the external potential (DEP) method can be applied. The DSCF method and the DEP theory have been used to study 
spinodal decomposition. A quantitative comparison with Monte-Carlo (MC) simulations shows a distinctly higher accuracy of the 
DEP method \cite{Reister2001}. The DSCF has been combined with the NS equation and a convection 
term has been added to the Langevin equation \cite{Honda2008,Hall2007,Hall2006}.

\subsubsection{Phase-field methods for polymer systems}

\paragraph{Conventional phase-field models} 
In recent years, there has been a growing number of articles that apply 
the PF method to polymer systems, \ie, to systems in which at least one
phase is a polymer melt or a polymer solution (see, for example, references \cite{Anders2011,Tian2011,Boffetta2010,Xu2006,Yue2005}). 
Here, the PF method is used to describe the nonequilibrium dynamics of phase boundaries. 
The method is based on a free-energy functional of a phase variable $\psiPF(\mathbf{r})$, which 
has the form \cite{Emmerich2008}
\begin{equation}  
\mathcal{F}=\vint (f(\psiPF)+f_{\mathrm{inh}}(\Nabla \psiPF)) \;, 
\label{Eq_F_PFM}
\end{equation} 
where $f$ is the free-energy density of the bulk -- usually an asymmetric double-well potential. The second part considers spatial inhomogeneities. 
A simple example of the free energy in equation \eqref{Eq_F_PFM} is the standard Ginzburg-Landau free energy
\begin{equation}  
\mathcal{F}=\vint \Big(a(\psiPF^{2}-\psiPF_{0}^{2})^{2} + b(\Nabla\psiPF)^{2} \Big) 
\label{Eq_Ginzb_Landau_PFM} 
\end{equation} 
with coefficients $a$ and $b$. If $\psiPF$ is conserved, then the dynamics is described by a Cahn-Hilliard (CH) equation \cite{Cahn1958}:
\begin{equation}  
\frac{\partial \psiPF}{\partial t} = -\Nabla \!\cdot\! \mathbf{J} \;.  
\label{Eq_Cahn_Hilliard}
\end{equation} 
Here, the flux density is given by
\begin{equation}  
\mathbf{J}=-M\:\!\Nabla \frac{\delta\mathcal{F}}{\delta\psiPF}  
\label{Eq_flux_CH}
\end{equation} 
with a mobility $M$, which in general may depend on $\psiPF$.

\paragraph{Polymer liquids} 
In a fluid polymer blend, $\psiPF$ is often 
taken as the concentration of one component and the bulk free-energy density $f(\mathbf{r})$ is the Margules activity from 
the Flory-Huggins theory \cite{Flory1942,Huggins1942}.
In many cases, a Gaussian noise term $\zeta(\mathbf{r},t)$ is added to the CH equation so that the
equation has a Langevin form 
\begin{equation}  
\frac{\partial\psiPF}{\partial t} 
=-\Nabla\!\cdot\!\mathbf{J} + \zeta = \Nabla \!\cdot\! \bigg(M\Nabla \frac{\delta\mathcal{F}}{\delta\psiPF} \bigg) + \zeta \;. 
\label{Eq_Langevin_PFM}
\end{equation}
Notice the similarity to equation \eqref{Eq_Langevin_SCF} from SCF theory. 
The expressions differ in the free-energy functional $\mathcal{F}$, which includes the auxiliary fields in the SCF theory.
Considering a flow velocity $\mathbf{v}$, equation \eqref{Eq_Langevin_PFM} can be extended to the convective CH equation with
a noise term \cite{Li2011}:
\begin{equation}  
\frac{\partial\psiPF}{\partial t} + \mathbf{v}\!\cdot\! \Nabla \psiPF =\Nabla^{2} \frac{\delta\mathcal{F}}{\delta\psiPF} + \zeta \;. 
\label{Eq_Cahn_Hilliard_flow}
\end{equation} 
Fluid dynamics is described by a continuity equation and an extended NS equation, which can generally be written as 
\begin{equation}  
\rho \bigg(\frac{\partial }{\partial t} +\mathbf{v}\!\cdot\! \Nabla \bigg)\mathbf{v}
=-\Nabla p + 2\:\!\Nabla \!\cdot\! \boldsymbol{\hat{\mu}_{\mathrm{visc}}} 
+\mathbf{k}_{\mathrm{surf}} +\mathbf{k}_{\mathrm{ext}} + \mathbf{k}_{\mathrm{elast}} \;, 
\label{Eq_Navier_Stokes}
\end{equation}
where $\rho$ is the mass density, $p$ is the pressure, $\boldsymbol{\hat{\mu}_{\mathrm{visc}}}$ is the viscous stress tensor,  
and $\mathbf{k}_{\mathrm{surf}}$ considers the surface tension from the interfaces \cite{Boffetta2010}. 
The term $\mathbf{k}_{\mathrm{ext}}$ represents external forces like gravity, while 
$\mathbf{k}_{\mathrm{elast}}\propto (r_0^{-2}\langle \mathbf{r}\,\mathbf{r}\rangle-\Eins)$ with an appropriate length scale $r_0$ 
denotes the elastic response of polymers to a flow field leading to viscoelastic behaviour \cite{Bird1987}. 
An alternative description of viscoelasticity is the Oldroid-B model \cite{Oldroyd1955},
which has been used together with the PF method to study the elongation and burst of viscoelastic droplets \cite{Beaucourt2005}
and the coalescence of polymer drops and interfaces \cite{Yue2005,Yue2006}.

The connection of the CH and the NS equation is called \ZT{model H} in the
nomenclature of Hohenberg and Halperin \cite{HohenbergH1977}. In many cases, the term $\mathbf{k}_{\mathrm{elast}}$ is neglected so that the
polymer phase is described as a Newtonian fluid. In principle, all terms on the right-hand-side (besides $p$) can be coupled to the 
PF variable $\psiPF$, since the involved material parameters may depend on the composition. 
Model H has been used to study spinodal decomposition with and without external velocity fields and Rayleigh-Taylor 
instabilities \cite{Anders2011,Boffetta2010,Badalassi2003}. 
The described method with a slightly different CH free energy has been used to study the shape 
of fluid films on a dewetting substrate \cite{MuellerKrumbhaar2001}.

\paragraph{Polymer crystals} 
Another important topic is the growth of polymer crystals in a polymer melt or solution \cite{Wang2008,Xu2006,Xu2005}. 
So far, polymer crystals have not been studied explicitly with the help of 
the PFC method. Polymer crystallization has been investigated with the 
standard PF method, in which the extension of the crystalline 
phase is characterized by a phase field $\psiPF$, while the crystalline structure 
is not considered. The dynamics of the crystal growth is described with 
the Allen-Cahn equation \cite{Allen1979}
\begin{equation}  
\frac{\partial\psiPF}{\partial t} =-\Gamma\:\!\frac{\delta\mathcal{F}}{\delta\psiPF} \;, 
\label{Eq_Allen_Cahn}
\end{equation} 
where $\mathcal{F}$ again consists of a bulk free-energy density and a gradient term. Frequently, the heat
transport is taken into account by a heat transfer equation with an extra term considering the
production of latent heat at the crystal growth front \cite{Xu2006, Xu2005}. 
In various cases, two PF variables $\psiPF_{\mathrm{mat}}$ and $\psiPF_{\mathrm{cryst}}$ are used to represent the concentration 
(or relative concentration) and the crystallization. This way the system is investigated with model C PF equations 
\cite{Tian2011,Buddhiranon2011}. 
By adding flow equations, a heat equation, and external fields, like an electrical 
field \cite{Tian2011}, a wide range of multiphase polymer systems can be studied.

\paragraph{Hydrogels}
The PF method has also been applied to hydrogels \cite{Li2011}, where the calculations consider free ions and charges fixed to the 
polymer network. The connectivity of the polymer network results in a shape elasticity as the hydrogel is swollen. This model is used to study 
the swelling of the hydrogel as the ion concentration in the surrounding is changed. Its PF variable $\psiPF(\mathbf{r},t)$ changes continuously 
(and monotonously) from $\psiPF\approxeq 0$ in the surrounding bath to $\psiPF\approxeq 1$ in the hydrogel region. 
The dynamics of $\psiPF(\mathbf{r},t)$ is described by 
\begin{equation}
\frac{\partial\psiPF}{\partial t}=-M_\psiPF \bigg(h'(\psiPF)\Big(\mu\,\varepsilon_{ij}^2
+\frac{\lambda-\tilde{\lambda}}{2}\:\!\varepsilon_{kk}^2-(p-p_{\mathrm{eq}})\Big)
+\frac{W}{V_{\mathrm{m}}}\:\!g'(\psiPF)-K_\psiPF\Nabla^{2}\psiPF\bigg) \,,
\label{Eq_hydrogel_1}
\end{equation} 
corresponding to equation \eqref{Eq_Allen_Cahn}, where $M_\psiPF$ is a mobility. 
Here, the term 
\begin{equation}
\mu\:\!\varepsilon_{ij}^2 + \frac{\lambda-\tilde{\lambda}}{2}\varepsilon_{kk}^2-(p-p_{\mathrm{eq}})
\end{equation} 
is the driving force for the phase field that depends on the 
osmotic pressure $p$ and the elasticity of the hydrogel, which is given by the shear modulus $\mu$, and the difference $\lambda-\tilde{\lambda}$ 
of the Lam\'{e} constants inside and outside the hydrogel. The deformation of the hydrogel is described by the strain 
\begin{equation}
\varepsilon_{ij}=\frac{1}{2}\bigg(\frac{\partial v_i}{\partial r_j}+\frac{\partial v_j}{\partial r_i}\bigg)  
\end{equation}
and the osmotic pressure at equilibrium is denoted by $p_{\mathrm{eq}}$. 
The function $h(\psiPF)=\psiPF^2(3-2\psiPF)$ in equation \eqref{Eq_hydrogel_1} satisfies $h(0)=0$ and $h(1)=1$. 
Its derivative $h'(\psiPF)=\dif h/\dif\psiPF$ vanishes at $\psiPF=0$ and $\psiPF=1$ so that 
the first term in equation \eqref{Eq_hydrogel_1} contributes only at the interface between hydrogel and bath. 
In the second term, the quantities $W$ and $V_{\mathrm{m}}$ denote the potential height and the molar volume of the mixture, respectively, and 
the function $g(\psiPF)=\psiPF^2(1-\psiPF^2)$ provides a double-well potential with minima at the bulk phases. 
Finally, the parameter $K_{\psi}$ controls the diffusion of the phase field. 
The sum of the second and the third term in equation \eqref{Eq_hydrogel_1} is the functional derivative of the free-energy functional 
in equation \eqref{Eq_Ginzb_Landau_PFM}, if we set $\psiPF\to\psiPF-1/2$, $\psiPF_0=1/2$, $a=W/V_{\mathrm{m}}$, and $b=K_{\psiPF}/2$.

Assuming suitably slow swelling dynamics, mechanical equilibrium provides a relation between strain $\varepsilon_{ij}$ and pressure $p$: 
\begin{equation}
\frac{\partial}{\partial x_j}\Big( h(\psiPF)(2\mu\:\!\varepsilon_{ij} 
+\lambda\:\!\varepsilon_{kk}\:\!\delta_{ij})+(1-h(\psiPF))\tilde{\lambda}\:\!\varepsilon_{kk}\:\!\delta_{ij}
-h(\psiPF)(p-p^{\mathrm{eq}})\delta_{ij}\Big)=0 \;.
\label{Eq_mechanic}
\end{equation}
The osmotic pressure is induced by the average concentration difference of free ions inside and outside the hydrogel. 
If $c_i$ is the local concentration of type $i$, one has 
\begin{equation}
\frac{p(t)}{N_{\mathrm{A}}\kBT}=\sum_{i} N 
\frac{\int\!\!\dif\vec{r}\, \psiPF(\mathbf{r},t)c_i(\mathbf{r},t)}{\int\!\!\dif\vec{r}\, \psiPF(\mathbf{r},t)} - 
\frac{\int\!\!\dif\vec{r}\, (1-\psiPF(\mathbf{r},t))c_i(\mathbf{r},t)}{\int\!\!\dif\vec{r}\, (1-\psiPF(\mathbf{r},t))} \;.
\label{Eq_osmotic}
\end{equation}
With equations \eqref{Eq_mechanic} and \eqref{Eq_osmotic} the first term in equation \eqref{Eq_hydrogel_1} can be calculated, 
if the local concentrations $c_i$ of the ions are known. 
The time dependence of the ion concentrations is given by the Nernst-Planck equation
\begin{equation}
\frac{\partial c_i}{\partial t}=\Nabla\!\cdot\!\Big(D_i\Big(\Nabla c_i
+\frac{z_i}{F_{\mathrm{r}}} N_{\mathrm{A}}\kBT \:\! c_i \Nabla U_{\mathrm{el}} \Big)\!\Big) \,, 
\end{equation}
where ions of type $i$ have a concentration $c_i$, a diffusion constant $D_i$ and a valence $z_i$. 
The quantity $F_{\mathrm{r}}$ is the Faraday constant and the electric potential $U_{\mathrm{el}}$ is given by the Poisson equation 
\begin{equation}
\Nabla\!\cdot\!(\varepsilon\Nabla U_{\mathrm{el}}) 
+F_{\mathrm{r}}\bigg(\sum_i z_i c_i + z_{\mathrm{f}}c_{\mathrm{f}}\bigg) = 0 \;,
\end{equation} 
where $\varepsilon$ is the absolute permittivity of the medium and the quantities $c_{\mathrm{f}}$ and $z_{\mathrm{f}}$ are 
the concentration and the valence of the fixed charges on the polymer network, respectively. 
Since the total number of fixed charges does not change, one has
\begin{equation}
c_{\mathrm{f}}(\mathbf{r},t)=
\frac{\psiPF(\mathbf{r},t)}{\int\!\!\dif\vec{r}'\, \psiPF(\mathbf{r}',t)}\, 
c^{\mathrm{eq}}_{\mathrm{f}} \!\vints \psiPF(\mathbf{r}',t)
\end{equation}
for the concentration of fixed charges $c_{\mathrm{f}}$, which is $c^{\mathrm{eq}}_{\mathrm{f}}$ at equilibrium. 
With the help of these equations, the system, consisting of a charged polymer-network in a solvent with ions, 
has been investigated and the swelling behaviour of the hydrogel has been analysed. This section demonstrates how the PF equation can be 
coupled with other equations that control the mechanical, chemical, and electric properties of the system. 
It shows that short-range interactions at the moving hydrogel-solution interface can be coupled with long-range fields and dynamics.

\paragraph{Phase-field-crystal method for polymer systems} 
The connectivity of polymer chains makes it difficult to apply the PFC method on the atomistic structure of polymers. 
However, PFC models are a useful tool to investigate structure formation on a larger scale. One example is the investigation of density correlations 
in polymer solvents. This has been done by Praetorius and Voigt, who studied the density profile of a polymer solvent in the vicinity of moving 
colloids \cite{Praetorius2011}. The model considers soft interactions of polymer coils that flow around spherical colloids. 
The colloid diameter and the effective polymer diameter are of the same order, leading to pronounced density oscillations in the vicinity of the 
colloid. The calculations are based on a free-energy functional of the SH-type, 
\begin{equation}
\frac{\mathcal{F}[\Psi]}{\kBT}=C\!\vint \bigg(\frac{\Psi}{2}\big(-\epsilon+\big(1+\Nabla^2\big)^2\big) 
\Psi+\frac{\Psi^4}{4}+\rho\, U_{1}\bigg) \,,
\end{equation} 
where the phase field $\Psi(\mathbf{r},t)=2(\rhoN-\rho(\mathbf{r},t))$ characterizes the deviation of the local density $\rho(\mathbf{r},t)$ 
from the average density $\rhoN$, $C=1/(48\rhoN^4)$, and $U_{1}(\mathbf{r},t)$ represents the rotationally-symmetric potential inferred 
by the colloid in units of $\kBT$.
The free-energy expression is derived analogously for equation \eqref{eq8}. The calculation of a polymer solution passing a fixed colloid shows  
pronounced density modulations around the colloid with pronounced oscillations in the side, where the flow hits the colloid. 
The example shows that the PFC model can also be used to analyse order in a fluid polymer system. 

Typically, the occurrence of structured patterns results from the interplay of competing interactions. In the case of crystals there are 
typically short-range repulsions on one hand and the external pressure plus attractive forces on the other hand. 
In various soft matter systems, other types of competing interactions may lead to periodic patterns on a distinctly longer scale. 
Seul and Andelman have described many pattern formation phenomena with the help of phase modulations with a preferred wavelength \cite{Seul1995}. 
The variety of materials ranges from fluid films of dipolar molecules over ferrofluids to fluid membranes and block copolymers. 

We consider a system with two components A and B that tend to demix. The phase variable $\Psi$ defines the deviation from the homogeneous 
distribution. A Ginzburg-Landau expansion of the free energy is given by
\begin{equation}
\mathcal{F}_{\Psi}=\mathcal{F}_{0}[\Psi]+\frac{b}{2}\vint (\Nabla\Psi(\mathbf{r}))^2 
= \mathcal{F}_{0}[\Psi]+\frac{b}{2}\kint \PsiW(-\mathbf{\wn})\wn^2 \PsiW(\mathbf{\wn}) \;,
\label{Eq_Fpsi}
\end{equation} 
where $\PsiW(\mathbf{\wn})$ is the Fourier transform of $\Psi(\mathbf{r})$. 
The bulk contribution $\mathcal{F}_{0}[\Psi]$ is typically a Landau free-energy expansion. Without extra interactions, $\mathcal{F}_{\Psi}$ 
corresponds to the standard PF model. Now we assume a long-range interaction between molecules at positions $\mathbf{r}$ and 
$\mathbf{r}'$, which is proportional to a correlation function $g(\mathbf{r},\mathbf{r}')$. This provides an extra free-energy term 
\begin{equation}
\mathcal{F}_{\mathrm{g}}=\frac{C_{\mathrm{g}}}{2}\vint\!\!\vints \Psi(\mathbf{r})g(\mathbf{r},\mathbf{r}')\Psi(\mathbf{r}') 
\end{equation}
with a constant $C_{\mathrm{g}}$. 
A simple example, given in reference \cite{Seul1995}, is a 2D film of molecules A and B, in which molecules A have a dipole moment $\mu$. 
In this case, one has $C_{\mathrm{g}}=-\mu^2$ and $g(\mathbf{r},\mathbf{r}')=\norm{\mathbf{r}-\mathbf{r}'}^{-3}$ \cite{Seul1995}.
If $g(\mathbf{r},\mathbf{r}')=g(\norm{\mathbf{r}-\mathbf{r}'})$, the free energy $\mathcal{F}_{\mathrm{g}}$ can be expressed as 
\begin{equation}
\mathcal{F}_{\mathrm{g}}=\frac{C_{\mathrm{g}}}{2}\kint \PsiW(-\mathbf{\wn})\gw(\wn)\PsiW(\mathbf{\wn}) \;, 
\label{Eq_Fg}
\end{equation}
where $\gw(\wn)$ is the 2D Fourier transform of $g(\norm{\mathbf{r}-\mathbf{r}'})$. Combining equations \eqref{Eq_Fpsi} and \eqref{Eq_Fg}, one has 
\begin{equation}
\mathcal{F} = \mathcal{F}_{\Psi}+\mathcal{F}_{\mathrm{g}}=\mathcal{F}_{0}+\kint \PsiW(-\mathbf{\wn})\gcw(\wn)\PsiW(\mathbf{\wn})
\label{Eq_Ftot}
\end{equation}
with $\gcw(\wn)=C_{\mathrm{g}} \gw(\wn) + b \wn^2$. It can be shown that $\gcw(\wn)$ has a minimum for a finite $\wn_{\mathrm{m}}$ 
and that the system forms a periodic pattern with wave number $\wn_{\mathrm{m}}$. 

A free-energy expression like in equation \eqref{Eq_Ftot} is also found in the Leibler-Ohta-Kawasaki (LOK) theory for 
AB-block copolymers \cite{Leibler1980, Ohta1986}. The theory allows the investigation of microphase separation providing lamellar, 
cylindrical, and spherical domains. The theory starts with a model Hamiltonian, which integrates interactions along all pairs of polymers. 
The ansatz corresponds to that of the SCF theory \cite{Edwards1966,Helfand1971}. 
As a next step, the LOK theory applies a random-phase approximation (RPA), which leads to an expression for the 
free energy $\mathcal{F}[\Psi]=\mathcal{F}_{\mathrm{g}}+\Fho[\Psi]$, 
where $\mathcal{F}_{\mathrm{g}}$ corresponds to equation \eqref{Eq_Fg} with 
\begin{equation}
C_{\mathrm{g}}\:\! g(\mathbf{r},\mathbf{r}')=\frac{1}{c}\:\!\Gamma(\mathbf{r},\mathbf{r}')-2\:\!\chi\:\!\delta(\mathbf{r}-\mathbf{r}') \;, 
\label{Eq_LOK1}
\end{equation}
and $\Fho[\Psi]$ contains higher-order terms of $\Psi$. 
In equation \eqref{Eq_LOK1}, $\chi$ is the Flory-Huggins parameter. 
With some approximations, a simple expression for $\tilde{\Gamma}(\mathbf{\wn},\mathbf{\wn}')$, 
which is the Fourier transform of $\Gamma(\mathbf{r},\mathbf{r}')$, 
is derived \cite{Ohta1986}. This leads to a free-energy expression of the form 
\begin{equation}
\mathcal{F}_{\mathrm{g}} = \frac{1}{2\rhoN N}\frac{1}{(2\pi)^{d}}\kint \PsiW(-\mathbf{\wn}) \hw(\wn)\PsiW(\mathbf{\wn}) \;,
\end{equation}
with a polymerization degree $N$, the average density $\rhoN$, and
\begin{equation} 
\hw(\wn) = \tilde{B}\wn^2+\tilde{A}\wn^{-2}-\bar{\chi} \;.
\label{Eq_qm2}
\end{equation}
Here, $\tilde{A}=3/(N\phi^2(1-\phi)^2)$ and $\tilde{B}=N/(4\phi(1-\phi))$ with $\phi$ being the fraction of A monomers per polymer. 
Furthermore, $\bar{\chi}=2\rhoN N \chi-\hw_{0}(\phi)$ with a function $\hw_{0}(\phi)$.  
Again, minimization of $\gcw(\wn)$ provides a preferred wave number $\wn_{\mathrm{c}}=(\tilde{A}/\tilde{B})^{1/4}$, 
while $\bar{\chi}_{\mathrm{c}}=2(\tilde{A}\tilde{B})^{1/2}$ is the critical value, 
above which $\gcw(\wn_{\mathrm{c}})$ gets negative and the spatially homogeneous 
distribution gets unstable. 

In the vicinity of the critical point, $\gcw(\wn)$ can be expanded around $\wn_{\mathrm{c}}$, which leads to 
\begin{equation} 
\mathcal{F}_{\mathrm{g}}=\frac{1}{2 \rhoN N \wn_{\mathrm{c}}^2}\frac{1}{(2\pi)^{d}}\kint 
\PsiW(-\mathbf{\wn}) \Big(\!-\wn_{\mathrm{c}}^2(\bar{\chi}-\bar{\chi}_{\mathrm{c}}) 
+\tilde{B}\big(\wn^2-\wn_{\mathrm{c}}^2\big)^2\Big) + \PsiW(\mathbf{\wn}) \;.
\label{Eq_Leibler_0}
\end{equation} 
Transformed into real space, one has
\begin{equation} 
\mathcal{F}_{\mathrm{g}}=\frac{\tilde{B}}{\rhoN N \wn_{\mathrm{c}}^2}\vint  
\frac{\Psi(\mathbf{r})}{2}\bigg(\!-\frac{\wn_{\mathrm{c}}^2(\bar{\chi}-\bar{\chi}_{\mathrm{c}})}{\tilde{B}}
+\big(\wn_{\mathrm{c}}^2+\Nabla^2\big)^2\bigg)\Psi(\mathbf{r}) \;.
\label{Eq_Leibler}
\end{equation} 
If we set $\wn_{0}=\wn_{\mathrm{c}}$ and $\beta=\wn_{\mathrm{c}}^2(\bar{\chi}-\bar{\chi}_{\mathrm{c}})/\tilde{B}$,  
the integrand in equation \eqref{Eq_Leibler} corresponds to that in the 
PFC free energy in equation \eqref{eq0}, up to a term proportional to $\Psi^4$, which is included in $\Fho[\Psi]$. 
The free energy $\mathcal{F}=\mathcal{F}_{\mathrm{g}}+\Fho[\Psi]$ with $\mathcal{F}_{\mathrm{g}}$ 
as in equation \eqref{Eq_Leibler} has already been derived by Leibler. 
It is only valid in the weak-segregation case close to the critical point. 

Tsori and Andelman have investigated a copolymer system in contact with a chemically patterned substrate \cite{Tsori2001}. 
Such systems are important for the creation of bulk structures controlled by a surface structure. Considering systems reasonably close 
to the critical point, the authors have used a bulk free-energy functional of the PFC-type. The interaction with the surface is described 
by a free-energy functional 
\begin{equation}
\mathcal{F}_{\mathrm{s}}=\vint \big(\Us(\mathbf{r})\Psi(\mathbf{r})+\tau_{\mathrm{s}} \Psi^2(\mathbf{r})\big) \;. 
\end{equation}
Here, the surface potential $\Us(\mathbf{r})$ prefers A monomers in regions with $\Us(\mathbf{r})<0$. 
The $\Psi^2$-term of $\mathcal{F}_{\mathrm{s}}$ is a surface correction to the Flory parameter $\chi$. Depending on whether the system is below 
or above the critical point $\chi_{\mathrm{c}}$, the bulk phase is either homogeneous or lamellar. 
In both cases, the surface pattern modulates the bulk phase. The Fourier transformation of the surface pattern shows that Fourier modes 
with $\wn>\wn_{\mathrm{c}}$ persist only close to the surface, while modes with $\wn<\wn_{\mathrm{c}}$ can propagate over long distances. 

In the strong segregation regime, which is present in segregating systems far from the critical point, the LOK theory provides a solution 
in which the $\wn^{-2}$-term in equation \eqref{Eq_qm2} is treated more accurately, leading to
\begin{equation}
\mathcal{F} = \frac{1}{2\rhoN N}\!\vint \bigg(\tilde{B}(\Nabla\Psi)^2 
+\bar{\chi}\Psi^2+\tilde{A}\!\vints \Psi(\mathbf{r})G(\mathbf{r},\mathbf{r}')\Psi(\mathbf{r}')\bigg) + \Fho[\Psi] \;,
\label{Eq_strong_segr}
\end{equation}
where the Green's function $G(\mathbf{r},\mathbf{r}')$ obeys 
\begin{equation}
-\Nabla^2 G(\mathbf{r},\mathbf{r}') = \delta(\mathbf{r}-\mathbf{r}') \;. 
\label{Eq_Green}
\end{equation}
In the strong segregation limit the interfacial regions are small compared to the size of the phase domains. 
The LOK theory for the strong-segregation limit has been combined with the PF method in order  
to investigate fluctuations in block copolymer systems \cite{Bosse2010}. 
The dynamics of the relative concentration difference is studied with a CH equation as given in equation \eqref{Eq_Cahn_Hilliard}. 
As the free energy in equation \eqref{Eq_strong_segr} includes long-range interactions 
explicitly, the method goes beyond the standard PF theory and also beyond the standard PFC method. 
However, inserting equation \eqref{Eq_strong_segr} into equation \eqref{Eq_Cahn_Hilliard} provides a rather simple equation, 
namely 
\begin{equation}
M^{-1}\frac{\partial \Psi}{\partial t}=\Nabla^{2}\bigg(2\:\!\bar{\chi}-\frac{\tilde{B}}{\rhoN N}\Nabla^{2}\bigg)\Psi
-\frac{\tilde{A}}{\rhoN N}\Psi \;,
\end{equation}
where we have used equation \eqref{Eq_Green} and assumed that $M$ is constant. 

The combination of LOK theory and PF method allows the calculation of statistical and dynamical properties of the diblock copolymer 
system, down to the length scale of the A-B interface width. 
In reference \cite{Bosse2010} the method has been used to investigate fluctuations of the domain boundaries in a periodic structure of 
alternating A and B layers. The method can be easily transferred to other types of micro-segregations. 

Another example of pattern formation in polymer system exists in the case of fluid membranes. 
Such membranes typically consist of amphiphiles, which aggregate in a layer such that the polar end groups screen the non-polar 
hydrocarbon chains from the surrounding water. In most cases, the membranes consist of two adjacent layers that point towards each other 
with their hydrophobic chains, but there are also various examples of fluid single-layer membranes.
Seul and Andelman introduced a simple model of a fluctuating membrane consisting of two components A and B, 
which influence the bending behaviour of the membrane \cite{Seul1995}. The local concentration is characterized by a quantity 
$\Psi$, with $\Psi=1$ for pure A regions and $\Psi=0$ for pure B regions. In the most simple case, the membrane consists of a monolayer. 
The undulations of the membrane are described by a height profile $h(\mathbf{r})$, where $\mathbf{r}$ is a 2D vector in the 
$x_{1}$-$x_{2}$-plane. 
For the elastic part of the free energy the expression
\begin{equation}
\mathcal{F}_{\mathrm{c}}=\vint \bigg(\frac{\sigma}{2}(\Nabla h)^2 
+\frac{\kappa}{2}(\Nabla^{2} h)^2 + \Lambda\Psi\Nabla^{2} h\bigg) 
\label{Eq_membrane}
\end{equation}
is used. Here, $\sigma$ is the surface tension and $\kappa$ is the bending modulus. 
The coefficient $\Lambda$ in the last term considers a coupling between the curvature and the local composition $\Psi$. 
The term $\Nabla^{2} h(\mathbf{r})$ is positive for convex bending and negative for concave bending. 
Thus, the free energy depends on the bending direction. This sign-dependent bending term does not vanish if the components A and B  
are asymmetric. The total free energy of the system is given by $\mathcal{F}=\mathcal{F}_{\Psi}+\mathcal{F}_{\mathrm{c}}$ with the 
Ginzburg-Landau expression $\mathcal{F}_{\Psi}$ from equation \eqref{Eq_Fpsi}. By minimizing $\mathcal{F}$ with respect to the 
membrane shape $h(\mathbf{r})$, the authors find the expression 
\begin{equation}
\mathcal{F}=\mathcal{F}_{0} + \vint \bigg(\frac{b'}{2}(\Nabla\Psi)^{2}+\frac{\Lambda^{2}\kappa}{2\sigma^2}(\Nabla^2\Psi)^2\bigg)
\label{Eq_membrane_2}
\end{equation}
with $b'=b-\Lambda^2/\sigma$. The important aspect is here that the coupling of the fields $h(\mathbf{r})$ and $\Psi(\mathbf{r})$ 
leads to a coefficient $b'$ in front of the term $(\Nabla\Psi)^{2}$, which may be negative. For $b'<0$, the membrane shows a curvature instability, 
and periodic patterns occur in the shape and the composition of the membrane. 

Partial integration of the free energy in equation \eqref{Eq_membrane_2} leads to 
\begin{equation}
\begin{split}
\mathcal{F}&=\mathcal{F}_0+\vint \bigg(
\frac{\Lambda^2\kappa}{2\sigma^2}\Psi\Nabla^{4}\Psi
-\frac{b'}{2}\Psi\Nabla^{2} \Psi\bigg) \\
&=\mathcal{F}_0+\vint \frac{\Lambda^2\kappa}{\sigma^2}\frac{\Psi}{2}
\bigg(\!-\bigg(\frac{\sigma^2 b'}{2\Lambda^2 \kappa}\bigg)^{2}\!+\bigg(\Nabla^2-\frac{\sigma^2 b'}{2\Lambda^2 \kappa}\bigg)^{2} \Psi\bigg) \,,
\end{split}
\label{Eq_membrane_2b}
\end{equation}
where boundary terms have been neglected. 
In the relevant case of $b'<0$ the free energy can be written as 
\begin{equation}
\mathcal{F}=\mathcal{F}_0+\frac{\Lambda^2\kappa}{\sigma^2}\vint 
\frac{\Psi}{2}\Big(\!-\wn_{\mathrm{ref}}^4 + \big(\wn_{\mathrm{ref}}^2+\Nabla^{2}\big)^{2}\Big)\Psi \;, 
\label{Eq_membrane_3}
\end{equation}
with the reference wave number
\begin{equation}
\wn_{\mathrm{ref}}=\sqrt{\frac{\sigma \abs{b}}{2\Lambda^2\kappa}} \;. 
\end{equation}
If $\mathcal{F}_{0}$ is a double-well potential that can be approximated by a fourth-order polynomial in $\Psi$, 
the free energy in equation \eqref{Eq_membrane_3} corresponds to the PFC free energy in equation \eqref{eq0}. 
As a simple example we consider that $\mathcal{F}_{0}$ is a symmetric potential 
\begin{equation}
\mathcal{F}_{0}^{\mathrm{s}} = a\:\!\Big(\tilde{\Psi}^2-\tilde{\Psi}_0^2\Big)^2 \,,
\end{equation}
where $\tilde{\Psi}=\Psi-\bar{\Psi}$ is the deviation of $\Psi$ from the overall A concentration $\bar{\Psi}$. 
Since integrals over terms linear in $\Psi$ are constant, we can write the total free energy as
\begin{equation}
\mathcal{F}^{\mathrm{s}}=4\:\!a\vint \bigg(\frac{\tilde{\Psi}}{2}\Big(\!-\big(\tilde{\Psi}_0^2+R^4 \wn_{\mathrm{ref}}^2\big)
+R^4\big(\wn_{\mathrm{ref}}^2+\Nabla^2\big)^2\Big)\tilde{\Psi}+\frac{1}{4}\tilde{\Psi}^4\bigg)+C_{\mathrm{s}}
\end{equation}
with a constant $C_{\mathrm{s}}$ and
\begin{equation}
R=\bigg(\frac{\lambda^2\kappa}{4\:\! a\:\! \sigma^2}\bigg)^{1/4}.
\end{equation}
If we introduce the rescaled spatial coordinates $\mathbf{\tilde{r}}=R \mathbf{r}$ and the quantity $\wn_0=R \wn_{\mathrm{ref}}$, 
the free energy is given by 
\begin{equation}
\mathcal{F}^{\mathrm{s}}= 
4\:\!a\:\!R^2\!\vintw \, \bigg(\frac{\tilde{\Psi}}{2}\bigg(\!-\bigg(\tilde{\Psi}_0^2+\frac{\abs{b'}}{8a}\bigg)
+\bigg(\wn_0^2+\NablaW^2\bigg)^2\bigg)\tilde{\Psi}+\frac{\tilde{\Psi}^4}{4}\bigg)\!+C_{\mathrm{s}} \;.
\label{Eq_membrane_4}
\end{equation}
The integral term is proportional to the free energy in equation \eqref{eq0} with $\beta=\tilde{\Psi}_0^2+\abs{b'}/8a$.

\subsubsection{Comparison of the methods} 
The SCF theory, the PF theory and the PFC theory are three alternative methods for studying polymer systems. 
Depending on the specific problem, each of the three methods may be most suited. 
The SCF theory has originally been an equilibrium theory, which has been extended during the last years to study dynamic systems. 
Dynamic SCF uses a Langevin equation, in which the dynamics of the system is controlled by the functional derivative of a free-energy functional. 
The same type of equation, with or without a random noise term, is used in the PF and the PFC theories. The three methods differ in the 
form of the free-energy functionals.

The SCF method is more microscopically founded than the other two. The underlying concept takes into account the behaviour of single polymer chains 
in an effective mean field provided by the surrounding polymers. This way, the connectivity and the entropy of the chains are taken into account 
more explicitly than in the PF or the PFC method. The price for this accuracy is a comparably high calculation effort, 
which limits the treatable length and time scale of the studied system. If details of molecular fluctuations and the connectivity of the chains 
are crucial, the SCF method is most accurate. The PF and PFC method depend on various parameters that have to be taken from other 
calculations and experiments. If these parameters are known, the two methods are an elegant and powerful way to study the dynamics of multiphase 
systems on a large scale.

The SCF theory has been used to study microphase separations in copolymer systems. 
In the LOK theory, the free energy used in the SCF method is simplified with the help of the RPA. 
Close to the critical point, the LOK free-energy functional is of the PFC type. This way the PFC method can be used for copolymer systems 
as an approximation of the SCF theory, valid in the weak segregation limit. By adding a surface free-energy functional, 
also surface interactions have been taken into account \cite{Tsori2001}. In the more usual case of strong segregation, a more general 
LOK free-energy functional has been inserted into the PF equation \cite{Bosse2010}. The resulting model can be seen as an extended version 
of the PFC method applied on copolymers. 

Free energies of the PFC type, which lead to pronounced periodic structures with a preferred wavelength, can be found for various polymer systems. 
As pointed out in reference \cite{Seul1995}, there are two mechanisms that can stabilize patterns with a preferred wavelength. One is the competition of 
long-range and short-range forces. The other is the interplay of two order parameters, like the shape and the composition of multi-component fluid 
membranes. In many cases, one can derive a free-energy functional of the PFC type, including second- and fourth-order spatial derivatives of the 
phase field. In the case of the fluid membrane, the $\Nabla^4\Psi$ term is proportional to the bending modulus $\kappa$. 
Thus, by reducing $\kappa$, the PFC equation can be transformed smoothly into a PF equation. 
At the same time, the reference wave number $\wn_{\mathrm{ref}}$ diverges. 
This means that the equilibrium system shows a complete phase separation rather than 
a microphase separation. While the PFC is well suited for studying the dynamics of systems that form periodic structures in equilibrium or 
in a steady state, the PF method enables the study of \ZT{intermediate} pattern formation of phase-separating systems 
(where the intermediate structure can be extremely long-living). 

As shown for the case of colloids flowing in a polymer solution \cite{Praetorius2011}, the PFC method can be used to consider the steric repulsion 
of polymer coils. Here, it reveals oscillatory structures with a wavelength in the range of the coil diameter, while similar patterns are absent 
in calculations comparable with the PF theory.
The PF theory has been applied on various polymer systems, including polymer solutions, polymer melts, 
growing polymer crystals and hydrogels. It has been demonstrated that system properties like charge and ion distributions, flow velocity, 
gravitation, and hydrodynamics can be successfully included into the model. The according methods can be transferred to PFC calculations in 
polymer systems, resulting in extra terms in the free-energy functional, additional differential equations or advective PFC equations. 

The SCF theory is useful, if details of molecular fluctuations and connectivity of the chains are crucial. The PF and PFC method depend on 
various parameters that have to be taken from other calculations and experiments. If these parameters are known, the methods provide an elegant 
and powerful way to study the dynamics of multiphase systems on a large scale. The application of PFC models on polymer systems has just started and 
opened many perspectives for studying polymer systems with pronounced pattern formation.

\subsection{\label{sec:Atlc}Application to liquid crystals}
The different forms of PFC models for orientable particles as discussed in section \ref{sec:PFCstatisch}
can be applied to compute liquid crystalline systems under various circumstances.
First of all, the two-dimensional bulk calculations presented in section \ref{app:statics} clearly need 
to be extended to three spatial dimensions. Moreover, a huge application area  is
found for confined systems. For instance, boundary conditions to the director fields
set by external walls can lead to forced topological defects of the orientational order
\cite{Stark_Lubensky_2}. 
Here, PFC models for liquid crystals provide a flexible tool 
to access those numerically. Next, liquid crystalline droplets in air or vacuum provide 
another type of confinement which induces quite peculiar orientational fields (so-called 
tactoids \cite{Verhoeff2009tactoids} or smectic droplets \cite{Verhoeff2009tactoids}) and again 
the PFC approach would provide a microscopic avenue to approach those. It would 
further be interesting to generalize the PFC model on curved surfaces in order to
explore the structure of thin liquid crystalline bubbles, see, \eg, references \cite{Dzubiella,dePablo} 
for simulation predictions. Last not least the structure of interfaces between two 
coexisting liquid crystalline phases needs future attention, in particular for phases with 
positional order.

\begin{figure}
\centering\includegraphics[width=0.5\linewidth]{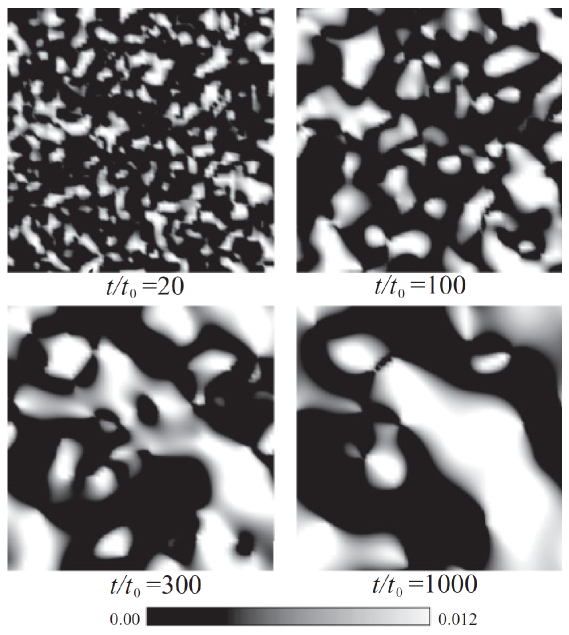}%
\caption{\label{fig:YA}Time evolution of the Schlieren pattern $\propto Q_{12}^{2}$ in the $x_{1}$-$x_{2}$-plane, 
which exhibits a dynamical coarsening. The symbol $t_{0}$ denotes a characteristic time. 
\nocite{YabunakaA2011}\CRP{S.\ Yabunaka and T.\ Araki}{Polydomain growth at isotropic-nematic transitions in liquid crystalline polymers}{Phys.\ Rev.\ E}{83}{2011}{061711}{6}{10.1103/PhysRevE.83.061711}{the American Physical Society}}
\end{figure}

A further broad range of applications have to do with dynamics. One recent example
was performed in reference \cite{YabunakaA2011}, where the CMA 
was employed to compute the dynamic coarsening of a disturbed nematic phase. The time 
evolution of such a nematic phase is shown in the Schlieren patterns in figure \ref{fig:YA}.

More applications concern the dynamics of topological defects, nucleation in liquid crystalline
systems \cite{SchillingF2004}, and the orientational dynamics induced by external switching fields
\cite{HaertelL2010}. Hence a flourishing future of many more PFC computations for liquid crystals
is still lying ahead.

\section{\label{chap:VI}Summary and outlook}
Motivated by the spectacular advances phase-field-crystal modelling has made in 
recent years, we have reviewed its present status. Besides presenting the original 
PFC model together with its derivation from dynamical density functional theory, we 
have reviewed many of its numerous extensions, including those aimed at describing 
binary solidification, vacancy transport (VPFC), anisotropic molecules (APFC),  
liquid crystals, and a quantitative description of real systems. We have reviewed, 
furthermore, a broad range of applications for metallic and soft matter systems 
(colloids, polymers, and liquid crystals), and for phenomena like the glass transition, and 
the formation of foams. We have discussed open issues such as coupling to 
hydrodynamics and the possibility of making quantitative PFC predictions for real 
materials. The main question at present is what further steps need yet to 
be made to turn the PFC-type models into even more potent modelling tools. 

To summarize the present state of affairs, it seems appropriate to recall some 
of the concluding remarks of a CECAM workshop dedicated to DDFT- and PFC-type approaches 
held in 2009 in Lausanne \cite{JPhysCondMat.22.360301}. 
It appears that despite the advances made meantime, some of the major issues 
identified there need yet further attendance. These are the following:    

\textit{(i) How to build numerically efficient, quantitative PFC models for a broad
spectrum of metallic materials?} 
The PFC models incorporate microscopic physics in a phenomenological manner. 
The respective local free-energy functional and 
the simplified dynamics lead to equations of motion that can be handled fairly 
efficiently with advanced numerical methods so that simulations containing up to 
a few times $10^7$ particles/atoms can be performed with relative ease. A major 
aim here is to develop a methodology for tuning crystal symmetry, lattice spacing, 
elastic constants, surface energy, dislocation core energy, dislocation mobility, \etc\ 
without sacrificing numerical efficiency. Along this line, methods have been proposed 
for constructing PFC free energies that allow for the tuning of the crystal structure 
\cite{PhysRevE.81.061601, PhysRevE.83.031601, JPhysCondMat.22.364102}. 
The amplitude equations represent an appealing alternative 
(see, \eg, references \cite{PhysRevE.72.020601, PhysRevE.74.011601, PhilosMag.90.237, PhysRevE.81.011602}), 
in which the density field is expressed in terms of slowly-varying amplitudes, modulated by the 
fundamental spatial periodicity of particle density. As demonstrated, this approach realizes a 
truly multi-scale approach to phase transitions in freezing liquids \cite{PhysRevE.76.056706}. 
Alternatively, one can work directly with the scaled density field of the PFC models and introduce 
additional model parameters, which can be fitted so that a required set of physical properties 
is recovered, as done in the case of pure bcc Fe \cite{JaatinenAEAN2009}.

\textit{(ii) How to construct effective, low-frequency representations from DFT/DDFT?} 
Provided that one had an accurate and predictive density functional that incorporates 
\textit{interaction potentials} between the constituent species in a multi-component system, 
it would become possible to develop an effective description that enables quantitative 
simulations for microscopically-informed continuum systems that evolve on diffusive time scales. 
However, one needs to develop first such free-energy functionals. Next, the dynamics of the 
relevant degrees of freedom should be projected out from the full DDFT description. 
It may be expected on physical grounds that the shape of a single density peak would 
relax much faster than the distance between different peak centres. Accordingly, one could \ZT{slave}  
the high-frequency modes associated with the peak shapes to the more slowly evolving 
modes with low spatial frequencies.

\textit{(iii) The role of fluctuations in DDFT and PFC modelling.} 
There is a continuing debate about the role of noise in the DDFT- and PFC-type models \cite{JPhysA.37.9325}. 
Derivations of DDFT from either the Smoluchowski level \cite{ArcherE2004} or within the projection 
operator technique \cite{EspanolL2009} lead to a deterministic equation of motion without 
any noise an approximation that becomes problematic near the critical point, or during 
nucleation, where the system has to leave a metastable free-energy minimum. In the 
former case fluctuations are needed to obtain the correct critical behaviour, whereas 
in the latter case, fluctuations are needed to establish an escape route of the system 
from a metastable phase. Other approaches treat fluctuations on a more phenomenological 
level. Often, however, the noise strength, though fundamentally correlated with the 
thermal energy, is treated as a phenomenological fitting parameter 
(see, \eg, references \cite{PhysRevE.78.031109,JPhysCondMat.21.464108}). This is a fundamental problem, 
shared by all DDFT and PFC approaches. We note that the addition of noise to the equation 
of motion in continuum models is not without conceptual difficulties (see reference \cite{PhilosMag.91.25}), 
even if noise is discretised properly during numerical solution. For example, in the presence 
of noise, the equilibrium properties of the system change. Furthermore, transformation 
kinetics generally depends on the spatial and temporal steps and in the limit of infinitely 
small steps in 3D the free energy of the PFC systems diverges, leading to an ultraviolet 
\ZT{catastrophe}. Evidently, an appropriate \ZT{ultraviolet cutoff}, \ie, filtering out the highest 
frequencies, is required to regularize the unphysical singularity. Here, a straightforward 
choice for the cut-off length is the interparticle/interatomic distance, which then removes 
the unphysical, small wavelength fluctuations 
\cite{JPhysCondMat.20.404205,PhysRevLett.103.035702,Tegze20091612,JPhysCondMat.22.364101}. 
A more elegant handling of the problem is via renormalizing the model parameters so that with noise one 
recovers the \ZT{bare} physical properties 
(as outlined for the Swift-Hohenberg model in reference \cite{PhysRevE.62.6116}). 
Further systematic investigations are yet needed to settle this issue.

\appendix
\section{Coefficients in the PFC models for liquid crystals}

\subsection{\label{A:IID}PFC model for liquid crystals in two spatial dimensions}
In the contributions \eqref{eq:Fexc_GpEa}-\eqref{eq:Fexc_GpEc} of the local scaled excess free-energy density, the coefficients 
\begin{equation}%
A_{1} = 8\,\mathrm{M}^{0}_{0}(1) \,,\qquad A_{2} = -2\,\mathrm{M}^{0}_{0}(3) \,,\qquad A_{3} = \frac{1}{8}\,\mathrm{M}^{0}_{0}(5)
\label{eq:Ca}
\end{equation}%
are associated with a gradient expansion of $\psi^{2}(\vec{r})$. These coefficients also appear in a different form in the original PFC model 
\cite{ElderPBSG2007}. The further coefficients are given by \cite{WittkowskiLB2011b,WittkowskiLB2011}
\begin{align}
B_{1} &= 4\:\!\big(\mathrm{M}^{1}_{-1}(2)-\mathrm{M}^{0}_{1}(2)\big) \,,\\ 
B_{2} &= 2\:\!\big(\mathrm{M}^{1}_{1}(2)-\mathrm{M}^{2}_{-1}(2)\big) \,,\\ 
B_{3} &= -\mathrm{M}^{2}_{-2}(3)-\mathrm{M}^{0}_{2}(3) \;,
\label{eq:Cb}%
\end{align}
\begin{equation}
C_{1} = 4\,\mathrm{M}^{1}_{0}(1) \,,\qquad C_{2} = \mathrm{M}^{1}_{0}(3) - \frac{1}{2}\,\mathrm{M}^{1}_{-2}(3) \,,\qquad 
C_{3} = -\mathrm{M}^{1}_{-2}(3) \,,
\label{eq:Cc}
\end{equation}
and 
\begin{equation}%
D_{1} = 2\,\mathrm{M}^{2}_{0}(1) \,,\qquad D_{2} = -\mathrm{M}^{2}_{0}(3) \,.
\label{eq:Cd}
\end{equation}%
So far, all these coefficients can also be obtained by using the second-order Ra\-ma\-krish\-nan-Yus\-souff functional for the excess free energy. 
The remaining coefficients, however, result from higher-order contributions in the functional Taylor expansion \cite{WittkowskiLB2011}. 
In third order, one obtains for the homogeneous terms the coefficients 
{\allowdisplaybreaks\begin{align}%
E_{1} &= 32\,\widehat{\mathrm{M}}^{00}_{00} \;, \\
E_{2} &= 16\Big(\widehat{\mathrm{M}}^{-11}_{00}+2\,\widehat{\mathrm{M}}^{01}_{00}\Big) \,, \\
E_{3} &= 8\Big(\widehat{\mathrm{M}}^{-22}_{00}+2\,\widehat{\mathrm{M}}^{02}_{00}\Big) \,, \\
E_{4} &= 8\Big(2\,\widehat{\mathrm{M}}^{-21}_{00}+\widehat{\mathrm{M}}^{11}_{00}\Big)
\label{eq:Ce}
\end{align}}%
and for the terms containing a gradient the coefficients
{\allowdisplaybreaks\begin{align}%
F_{1} &= 16\Big(\widetilde{\mathrm{M}}^{-10}_{01}-2\,\widetilde{\mathrm{M}}^{0-1}_{01}
+\widetilde{\mathrm{M}}^{00}_{01}\Big) \,, \\
F_{2} &= 16\Big(\widetilde{\mathrm{M}}^{-21}_{01}-\widetilde{\mathrm{M}}^{0-2}_{01}
+\widetilde{\mathrm{M}}^{01}_{01}-\widetilde{\mathrm{M}}^{1-2}_{01}\Big) \,, \\
F_{3} &= -16\Big(\widetilde{\mathrm{M}}^{-20}_{01}-\widetilde{\mathrm{M}}^{-21}_{01}
-\widetilde{\mathrm{M}}^{01}_{01}+\widetilde{\mathrm{M}}^{10}_{01}\Big) \,, \\
F_{4} &= -8\Big(\widetilde{\mathrm{M}}^{-1-1}_{01}-2\,\widetilde{\mathrm{M}}^{-11}_{01}
+\widetilde{\mathrm{M}}^{1-1}_{01}\Big) \,, \\
F_{5} &= -4\Big(\widetilde{\mathrm{M}}^{-2-1}_{01}-\widetilde{\mathrm{M}}^{-22}_{01}
-\widetilde{\mathrm{M}}^{-12}_{01}+\widetilde{\mathrm{M}}^{2-1}_{01}\Big) \,, \\
F_{6} &= 8\Big(\widetilde{\mathrm{M}}^{-22}_{01}-\widetilde{\mathrm{M}}^{-1-2}_{01}
+\widetilde{\mathrm{M}}^{-12}_{01}-\widetilde{\mathrm{M}}^{2-2}_{01}\Big) \,.
\label{eq:Cf}
\end{align}}%
In fourth order, only homogeneous terms are kept. The corresponding coefficients are
{\allowdisplaybreaks\begin{align}%
G_{1} &= 128\,\widehat{\mathrm{M}}^{000}_{000} \;, \\[3pt]
G_{2} &= 192\Big(\widehat{\mathrm{M}}^{-101}_{000}+\widehat{\mathrm{M}}^{001}_{000}\Big) \,, \\
G_{3} &= 96\Big(\widehat{\mathrm{M}}^{-202}_{000}+\widehat{\mathrm{M}}^{002}_{000}\Big) \,, \\
G_{4} &= 96\Big(2\,\widehat{\mathrm{M}}^{-201}_{000}+\widehat{\mathrm{M}}^{-211}_{000}+\widehat{\mathrm{M}}^{011}_{000}\Big) \,, \\
G_{5} &= 48\Big(\widehat{\mathrm{M}}^{-212}_{000}+\widehat{\mathrm{M}}^{-112}_{000}\Big) \,, \\[2pt]
G_{6} &= 48\,\widehat{\mathrm{M}}^{-111}_{000} \,, \\[5pt]
G_{7} &= 12\,\widehat{\mathrm{M}}^{-222}_{000} \,.
\label{eq:Cg}
\end{align}}%
All the coefficients from above are linear combinations of moments of the Fourier expansion coefficients of the direct correlation functions. 
These moments are defined through  
\begin{equation}
\mathrm{M}^{\mathbf{m^{n}}}_{\mathbf{l^{n}}}(\boldsymbol{\alpha}^{\mathbf{n}}) = \pi^{2n+1}\rhoR^{n+1} \Bigg(\prod^{n}_{i=1} 
\int^{\infty}_{0}\!\!\!\!\!\!\dif r_{i}\,r^{\alpha_{i}}_{i}\Bigg) \tilde{c}^{(n+1)}_{\mathbf{l^{n}}\!,\:\!\mathbf{m^{n}}}(\mathbf{r^{n}}) 
\label{eq:M}
\end{equation}
with the multi-index notation $\mathbf{X^{n}}=(X_{1},\dotsc,X_{n})$ for $X\in\{l,m,r,1,\alpha,\phi,\phi_{\mathrm{R}}\}$ and the abbreviations 
$\widehat{\mathrm{M}}^{\mathbf{m^{n}}}_{\mathbf{l^{n}}}=\mathrm{M}^{\mathbf{m^{n}}}_{\mathbf{l^{n}}}(\mathbf{1^{n}})$ and 
$\widetilde{\mathrm{M}}^{m_{1}m_{2}}_{l_{1}l_{2}}=\mathrm{M}^{m_{1}m_{2}}_{l_{1}l_{2}}(1,2)$. 
The expansion coefficients of the direct correlation functions are given by 
\begin{equation}
\tilde{c}^{(n+1)}_{\mathbf{l^{n}}\!,\:\!\mathbf{m^{n}}}\!(\mathbf{r^{n}}) 
= \frac{1}{(2\pi)^{2n}}\int^{2\pi}_{0}\!\!\!\!\!\!\dif\boldsymbol{\phi}^{\mathbf{n}}_{\mathrm{\mathbf{R}}} 
\int^{2\pi}_{0}\!\!\!\!\!\!\dif\boldsymbol{\phi}^{\mathbf{n}}\, 
c^{(n+1)}(\mathbf{r^{n}},\boldsymbol{\phi}^{\mathbf{n}}_{\mathrm{\mathbf{R}}},\boldsymbol{\phi}^{\mathbf{n}})
e^{-\ii(\mathbf{l^{n}}\cdot\boldsymbol{\phi}^{\mathbf{n}}_{\mathrm{\mathbf{R}}} + \mathbf{m^{n}}\cdot\boldsymbol{\phi}^{\mathbf{n}})} 
\label{eq:ec}
\end{equation}
with $\vec{r}_{1}-\vec{r}_{i+1}=R_{i}\uu (\varphi_{\mathrm{R}_{i}})$, $\uu _{i}=\uu (\varphi_{i})$, 
$\phi_{\mathrm{R}_{i}}=\varphi_{1}-\varphi_{\mathrm{R}_{i}}$, and $\phi_{i}=\varphi_{1}-\varphi_{i+1}$.

When the system is apolar, the modulus $P(\vec{r})$ of the polarization $\vec{P}(\vec{r})$ is zero and its orientation $\pp(\vec{r})$ 
is not defined, while the direction $\nn (\vec{r})$ associated with quadrupolar order still exists. Then, symmetry considerations lead 
to the equalities 
\begin{equation}
\tilde{c}^{(2)}_{-1,1}(R)=\tilde{c}^{(2)}_{1,0}(R)\;,\qquad \tilde{c}^{(2)}_{-1,2}(R)=\tilde{c}^{(2)}_{1,1}(R)\;,\qquad 
\tilde{c}^{(2)}_{-2,2}(R)=\tilde{c}^{(2)}_{2,0}(R)
\end{equation}
between expansion coefficients of the direct pair-correlation function and to the equations 
\begin{equation}
\mathrm{M}^{1}_{-1}(2)=\mathrm{M}^{0}_{1}(2)\;,\qquad \mathrm{M}^{2}_{-1}(2)=\mathrm{M}^{1}_{1}(2)\;,\qquad 
\mathrm{M}^{2}_{-2}(2)=\mathrm{M}^{0}_{2}(2)
\end{equation}
for the generalized moments. A consequence of these equations is that the coefficients $B_{1}$ and $B_{2}$ vanish and $B_{3}$ becomes more simple.

\subsection{\label{A:IIID}PFC model for liquid crystals in three spatial dimensions}
The coefficients in equation \eqref{eq:Fexcbb} appear in three different groups. The first group consists of the three coefficients 
\begin{equation}%
A_{1} = 8\,\Omega_{000}(0) \,,\qquad A_{2} = -\frac{4}{3}\,\Omega_{000}(2) \,,\qquad A_{3} = \frac{1}{15}\,\Omega_{000}(4) \;, 
\end{equation}%
that are already known from the original PFC model \cite{ElderPBSG2007} and belong to the gradient expansion of the translational density. 
In the next group, the two coefficients 
\begin{equation}%
B_{1} = \frac{16}{15\sqrt{5}}\,\Omega_{220}(0) \,,\qquad B_{2} = -\frac{16}{15}\,\Omega_{022}(2) \;, 
\end{equation}%
that go along with the nematic tensor and the coupling of its gradient with the gradient of the translational density, are collected. 
The last group contains the Frank constants 
\begin{equation}%
\widetilde{K}_{1} = \frac{16}{15}\sqrt{\frac{2}{35}}\,\Omega_{222}(2) \,,\qquad \widetilde{K}_{2} = \frac{8}{45\sqrt{5}}\,\Omega_{220}(2)
+\frac{1}{3}\widetilde{K}_{1} \;,
\end{equation}%
that appear in the Frank free-energy density \cite{Chandrasekhar1992,deGennesP1995}. 
All these coefficients are expressed in terms of the generalized moments 
\begin{equation}
\Omega_{l_{1}l_{2}l}(n)=\pi^{3/2}\,\rhoR^{2}\!\int^{\infty}_{0}\!\!\!\!\!\!\dif r\,r^{n+2}\omega_{l_{1}l_{2}l}(r) 
\label{eq:Omega}
\end{equation}
with the expansion coefficients
\begin{equation}
\begin{split}
\omega_{l_{1}l_{2}l}(r)
&=\sqrt{\frac{4\pi}{2\:\!l+1}}\!\!\!\!\sum^{\min\{l_{1},l_{2}\}}_{m=-\min\{l_{1},l_{2}\}}\!\!\!\!\!\!\!C(l_{1},l_{2},l,m,-m,0) \\[-1pt]
&\quad\,\times\!\uintn{1}\!\!\uintn{2} \,c^{(2)}(r\mathbf{\hat{e}_{3}},\uu _{1},\uu _{2}) 
\overline{\operatorname{Y}}_{l_{1}m}(\uu_{1})\overline{\operatorname{Y}}_{l_{2}-m}(\uu _{2}) \;,
\end{split}
\end{equation}
where the symbol $C(l_{1},l_{2},l,m_{1},m_{2},m)$ denotes a Clebsch-Gordan coefficient \cite{GrayG1984}, $\operatorname{Y}_{lm}(\uu )$ 
is a spherical harmonic, $\mathbf{\hat{e}_{3}}$ stands for the Cartesian unit vector co-directional with the $x_{3}$-axis, 
and $\overline{\,\cdot\,}$ denotes complex conjugation. 

As before, equalities between generalized moments with different index-combinations, that can be derived using symmetry considerations 
\cite{WittkowskiLB2010}, have been taken into account in order to reduce the set of generalized
moments in equation \eqref{eq:Fexcbb} to its seven independent members.

\section*{Acknowledgements}
This work has been supported by the EU FP7 Projects \ZT{ENSEMBLE} (contract no.\ NMP4-SL-2008-213669) and 
\ZT{EXOMET} (contract no.\ NMP-LA-2012-280421, co-funded by ESA), by the ESA MAP/PECS project \ZT{MAGNEPHAS III}, 
and by the German Research Foundation (DFG) in the context of the DFG Priority Program 1296.










\bibliographystyle{tADP}
\bibliography{./refs}

\begin{thebibliography}{288}
\providecommand{\natexlab}[1]{#1}

\bibitem[1]{HohenbergH1977}
P.C. Hohenberg and B.I. Halperin, {\itshape Theory of dynamic critical
  phenomena}, Rev. Mod. Phys. 49 (1977), pp. 435--479.

\bibitem[2]{gunton1983}
J.D. Gunton, M. {San Miguel}, and P.S. Sahni, 1983, {T}he dynamics of first
  order phase transitions. in {\itshape Phase transitions and critical
  phenomena} 1st  ed.,   Academic Press, London, pp. 267--466.

\bibitem[3]{CrossH1993}
M.C. Cross and P.C. Hohenberg, {\itshape Pattern formation outside of
  equilibrium}, Rev. Mod. Phys. 65 (1993), pp. 851--1112.

\bibitem[4]{Seul1995}
M. Seul and D. Andelman, {\itshape Domain shapes and patterns -- the
  phenomenology of modulated phases}, Science 267 (1995), pp. 476--483.

\bibitem[5]{ChaikinL1995}
P.M. Chaikin and T.C. Lubensky {\itshape Principles of condensed matter
  physics},  1st ed.,   Cambridge University Press, Cambridge, 1995.

\bibitem[6]{pismen2006}
L.M. Pismen {\itshape Patterns and interfaces in dissipative dynamics},  1st
  ed.,   Springer, Berlin, 2006.

\bibitem[7]{cross2010}
M.C. Cross and H. Greenside {\itshape Pattern formation and dynamics in
  nonequilibrium systems},  1st ed.,   Cambridge University Press, Cambridge,
  2010.

\bibitem[8]{Klemenz1998}
C. Klemenz, {\itshape Hollow cores and step-bunching effects on (001) {YBCO}
  surfaces grown by liquid-phase epitaxy}, J. Cryst. Growth 187 (1998), pp.
  221--227.

\bibitem[9]{VasievSW1997}
B. Vasiev, F. Siegert, and C. Weijer, {\itshape Multiarmed spirals in excitable
  media}, Phys. Rev. Lett. 78 (1997), pp. 2489--2492.

\bibitem[10]{PhysRevA.15.319}
J. Swift and P.C. Hohenberg, {\itshape Hydrodynamic fluctuations at the
  convective instability}, Phys. Rev. A 15 (1977), pp. 319--328.

\bibitem[11]{Brazowskii1975}
S.A. Brazowskii, {\itshape Phase transition of an isotropic system to a
  nonuniform state}, Sov. Phys. JETP 41 (1975), pp. 85--89.

\bibitem[12]{ElderKHG2002}
K.R. Elder, M. Katakowski, M. Haataja, and M. Grant, {\itshape Modeling
  elasticity in crystal growth}, Phys. Rev. Lett. 88 (2002), p. 245701.

\bibitem[13]{Oxtoby1991}
D. Oxtoby, {\itshape Course 3: Crystallization of liquids: a density functional
  approach}, J.P. Hansen, D.~Levesque and J.~{Zinn-Justin},  eds.,  , Vol. ~1
  of {\itshape Proceedings of the {L}es {H}ouches {S}ummer {S}chool, {C}ourse
  {LI}, 3-28 {J}uly 1989}  North {H}olland, {E}lsevier {S}cience {P}ublishers
  {B}. {V}., Amsterdam, 1991, pp. 145--192.

\bibitem[14]{Singh1991}
Y. Singh, {\itshape Density-functional theory of freezing and properties of the
  ordered phase}, Phys. Rep. 207 (1991), pp. 351--444.

\bibitem[15]{Loewen1994a}
H. L{\"o}wen, {\itshape Melting, freezing and colloidal suspensions}, Phys.
  Rep. 237 (1994), pp. 249--324.

\bibitem[16]{Rosenfeld1989}
Y. Rosenfeld, {\itshape Free-energy model for the inhomogeneous hard-sphere
  fluid mixture and density-functional theory of freezing}, Phys. Rev. Lett. 63
  (1989), pp. 980--983.

\bibitem[17]{PhysRevE.74.021603}
J.F. Lutsko, {\itshape {G}inzburg-{L}andau theory of the liquid-solid interface
  and nucleation for hard spheres}, Phys. Rev. E 74 (2006), p. 021603.

\bibitem[18]{HaertelOREHL2012}
A. H\"artel, M. Oettel, R.E. Rozas, S.U. Egelhaaf, J. Horbach, and H. L\"owen,
  {\itshape Tension and stiffness of the hard-sphere crystal-fluid interface},
  Phys. Rev. Lett. 108 (2012), p. 226101.

\bibitem[19]{PhysRevE.72.020601}
N. Goldenfeld, B.P. Athreya, and J.A. Dantzig, {\itshape Renormalization group
  approach to multiscale simulation of polycrystalline materials using the
  phase field crystal model}, Phys. Rev. E 72 (2005), p. 020601.

\bibitem[20]{PhysRevE.76.056706}
B.P. Athreya, N. Goldenfeld, J.A. Dantzig, M. Greenwood, and N. Provatas,
  {\itshape Adaptive mesh computation of polycrystalline pattern formation
  using a renormalization-group reduction of the phase-field crystal model},
  Phys. Rev. E 76 (2007), p. 056706.

\bibitem[21]{HoytAK2003}
J.J. Hoyt, M. Asta, and A. Karma, {\itshape Atomistic and continuum modeling of
  dendritic solidification}, Mater. Sci. Eng. R 41 (2003), pp. 121--163.

\bibitem[22]{GranasyPW2004}
L. Gr\'an\'asy, T. Pusztai, and J.A. Warren, {\itshape Modelling
  polycrystalline solidification using phase field theory}, J. Phys.: Condens.
  Matter 16 (2004), pp. R1205--R1235.

\bibitem[23]{Karma2005}
A. Karma, {\itshape Phase field modeling}, in S.~Yip ed.,    Springer, 2005,
  pp. 2087--2103.

\bibitem[24]{Emmerich2008}
H. Emmerich, {\itshape Advances of and by phase-field modeling in
  condensed-matter physics}, Adv. Phys. 57 (2008), pp. 1--87.

\bibitem[25]{ProvatasE2010}
N. Provatas and K. Elder {\itshape Phase-field methods in materials science and
  engineering},  1st ed.,   Wiley, Weinheim, 2010.

\bibitem[26]{ElderPBSG2007}
K.R. Elder, N. Provatas, J. Berry, P. Stefanovic, and M. Grant, {\itshape
  Phase-field crystal modeling and classical density functional theory of
  freezing}, Phys. Rev. B 75 (2007), p. 064107.

\bibitem[27]{JPhysCondMat.20.404205}
T. Pusztai, G. Tegze, G.I. T\'oth, L. K\"ornyei, G. Bansel, Z. Fan, and L.
  Gr\'an\'asy, {\itshape Phase-field approach to polycrystalline solidification
  including heterogeneous and homogeneous nucleation}, J. Phys.: Condens.
  Matter 20 (2008), p. 404205.

\bibitem[28]{JPhysCondMat.22.364101}
G.I. T\'oth, G. Tegze, T. Pusztai, G. T\'oth, and L. Gr\'an\'asy, {\itshape
  Polymorphism, crystal nucleation and growth in the phase-field crystal model
  in 2{D} and 3{D}}, J. Phys.: Condens. Matter 22 (2010), p. 364101.

\bibitem[29]{tegze.phd.thesis}
G. Tegze, {\itshape Application of atomistic phase-field methods to complex
  solidification problems}, {P}h{D} {T}hesis, {E}\"otv\"os {U}niversity,
  Budapest, {H}ungary, 2009 In {H}ungarian.

\bibitem[30]{Tang2011146}
S. Tang, R. Backofen, J. Wang, Y. Zhou, A. Voigt, and Y. Yu, {\itshape
  Three-dimensional phase-field crystal modeling of fcc and bcc dendritic
  crystal growth}, J. Cryst. Growth 334 (2011), pp. 146--152.

\bibitem[31]{PhysRevE.81.011602}
K.R. Elder, Z.F. Huang, and N. Provatas, {\itshape Amplitude expansion of the
  binary phase-field-crystal model}, Phys. Rev. E 81 (2010), p. 011602.

\bibitem[32]{JPhysCondMat.22.364104}
R. Backofen and A. Voigt, {\itshape A phase-field-crystal approach to critical
  nuclei}, J. Phys.: Condens. Matter 22 (2010), p. 364104.

\bibitem[33]{PhilosMag.91.123}
L. Gr\'an\'asy, G. Tegze, G.I. T\'oth, and T. Pusztai, {\itshape Phase-field
  crystal modelling of crystal nucleation, heteroepitaxy and patterning},
  Philos. Mag. 91 (2011), pp. 123--149.

\bibitem[34]{JPhysCondMat.21.464108}
J. Hubert, M. Cheng, and H. Emmerich, {\itshape Effect of noise-induced
  nucleation on grain size distribution studied via the phase-field crystal
  method}, J. Phys.: Condens. Matter 21 (2009), p. 464108.

\bibitem[35]{PhysRevLett.107.175702}
G.I. T\'oth, T. Pusztai, G. Tegze, G. T\'oth, and L. Gr\'an\'asy, {\itshape
  Amorphous nucleation precursor in highly nonequilibrium fluids}, Phys. Rev.
  Lett. 107 (2011), p. 175702.

\bibitem[36]{PhysRevB.77.224114}
J. Berry, K.R. Elder, and M. Grant, {\itshape Melting at dislocations and grain
  boundaries: a phase-field crystal study}, Phys. Rev. B 77 (2008), p. 224114.

\bibitem[37]{PhysRevB.78.184110}
J. Mellenthin, A. Karma, and M. Plapp, {\itshape Phase-field crystal study of
  grain-boundary premelting}, Phys. Rev. B 78 (2008), p. 184110.

\bibitem[38]{C0SM00944J}
G. Tegze, L. Gr\'an\'asy, G.I. T\'oth, J.F. Douglas, and T. Pusztai, {\itshape
  Tuning the structure of nonequilibrium soft materials by varying the
  thermodynamic driving force for crystal ordering}, Soft Matter 7 (2011), pp.
  1789--1799.

\bibitem[39]{PhysRevLett.106.195502}
G. Tegze, G.I. T\'oth, and L. Gr\'an\'asy, {\itshape Faceting and branching in
  2{D} crystal growth}, Phys. Rev. Lett. 106 (2011), p. 195502.

\bibitem[40]{PhysRevB.76.184107}
K. Wu and A. Karma, {\itshape Phase-field crystal modeling of equilibrium
  bcc-liquid interfaces}, Phys. Rev. B 76 (2007), p. 184107.

\bibitem[41]{PhysRevE.79.011607}
S. Majaniemi and N. Provatas, {\itshape Deriving surface-energy anisotropy for
  phenomenological phase-field models of solidification}, Phys. Rev. E 79
  (2009), p. 011607.

\bibitem[42]{PhysRevLett.103.035702}
G. Tegze, L. Gr\'an\'asy, G.I. T\'oth, F. Podmaniczky, A. Jaatinen, T.
  {Ala-Nissila}, and T. Pusztai, {\itshape Diffusion-controlled anisotropic
  growth of stable and metastable crystal polymorphs in the phase-field crystal
  model}, Phys. Rev. Lett. 103 (2009), p. 035702.

\bibitem[43]{JPhysCondMat.21.464109}
R. Backofen and A. Voigt, {\itshape Solid-liquid interfacial energies and
  equilibrium shapes of nanocrystals}, J. Phys.: Condens. Matter 21 (2009), p.
  464109.

\bibitem[44]{tms2011sandiego}
T. Pusztai, G.I. T\'oth, G. Tegze, and L. Gr\'an\'asy, {\itshape Dendritic
  growth: phase field crystal vs. phase field simulations}, Abstract {B}ooklet,
  {TMS} {A}nnual {M}eeting, 26 {F}ebruary - 3 {M}arch 2011, {S}an {D}iego,
  {USA}  (2011), p.~31.

\bibitem[45]{tms2012orlando}
H. Humadi, J. Hoyt, and N. Provatas, {\itshape A phase field crystal study of
  rapid solidification and solute trapping in binary alloys}, Abstract
  {B}ooklet, {TMS} {A}nnual {M}eeting, 11-15 {M}arch 2012, {O}rlando, {USA}
  (2012).

\bibitem[46]{PhysRevE.77.061506}
J. Berry, K.R. Elder, and M. Grant, {\itshape Simulation of an atomistic
  dynamic field theory for monatomic liquids: freezing and glass formation},
  Phys. Rev. E 77 (2008), p. 061506.

\bibitem[47]{PhysRevLett.106.175702}
J. Berry and M. Grant, {\itshape Modeling multiple time scales during glass
  formation with phase-field crystals}, Phys. Rev. Lett. 106 (2011), p. 175702.

\bibitem[48]{PhysRevLett.105.126101}
S. Muralidharan and M. Haataja, {\itshape Phase-field crystal modeling of
  compositional domain formation in ultrathin films}, Phys. Rev. Lett. 105
  (2010), p. 126101.

\bibitem[49]{prl.unpub.elder}
K.R. Elder, G. Rossi, P. Kanerva, F. Sanches, S.C. Ying, E. Granato, C.V.
  Achim, and T. {Ala-Nissila}, {\itshape Patterning of heteroepitaxial
  overlayers from nano to micron scales}, Phys. Rev. Lett. 108 (2012), p.
  226102.

\bibitem[50]{ElderG2004}
K.R. Elder and M. Grant, {\itshape Modeling elastic and plastic deformations in
  nonequilibrium processing using phase field crystals}, Phys. Rev. E 70
  (2004), p. 051605.

\bibitem[51]{StefanovicHP2006}
P. Stefanovic, M. Haataja, and N. Provatas, {\itshape Phase-field crystals with
  elastic interactions}, Phys. Rev. Lett. 96 (2006), p. 225504.

\bibitem[52]{PhysRevLett.101.158701}
Z.F. Huang and K.R. Elder, {\itshape Mesoscopic and microscopic modeling of
  island formation in strained film epitaxy}, Phys. Rev. Lett. 101 (2008), p.
  158701.

\bibitem[53]{ChanTDDG2010}
P.Y. Chan, G. Tsekenis, J. Dantzig, K.A. Dahmen, and N. Goldenfeld, {\itshape
  Plasticity and dislocation dynamics in a phase field crystal model}, Phys.
  Rev. Lett. 105 (2010), p. 015502.

\bibitem[54]{PhysRevB.81.214201}
R. Spatschek and A. Karma, {\itshape Amplitude equations for polycrystalline
  materials with interaction between composition and stress}, Phys. Rev. B 81
  (2010), p. 214201.

\bibitem[55]{PhilosMag.91.151}
K.R. Elder, K. Thornton, and J.J. Hoyt, {\itshape The {K}irkendall effect in
  the phase field crystal model}, Philos. Mag. 91 (2011), pp. 151--164.

\bibitem[56]{PhysRevE.79.035701}
P.Y. Chan, N. Goldenfeld, and J. Dantzig, {\itshape Molecular dynamics on
  diffusive time scales from the phase-field-crystal equation}, Phys. Rev. E 79
  (2009), p. 035701R.

\bibitem[57]{AlandLV2011}
S. Aland, J. Lowengrub, and A. Voigt, {\itshape A continuum model of
  colloid-stabilized interfaces}, Phys. Fluids 23 (2011), p. 062103.

\bibitem[58]{Loewen2010}
H. L{\"o}wen, {\itshape A phase-field-crystal model for liquid crystals}, J.
  Phys.: Condens. Matter 22 (2010), p. 364105.

\bibitem[59]{WittkowskiLB2010}
R. Wittkowski, H. L{\"o}wen, and H.R. Brand, {\itshape Derivation of a
  three-dimensional phase-field-crystal model for liquid crystals from density
  functional theory}, Phys. Rev. E 82 (2010), p. 031708.

\bibitem[60]{AchimWL2011}
C.V. Achim, R. Wittkowski, and H. L{\"o}wen, {\itshape Stability of liquid
  crystalline phases in the phase-field-crystal model}, Phys. Rev. E 83 (2011),
  p. 061712.

\bibitem[61]{PhysRevE.81.065301}
N. Guttenberg, N. Goldenfeld, and J. Dantzig, {\itshape Emergence of foams from
  the breakdown of the phase field crystal model}, Phys. Rev. E 81 (2010), p.
  065301.

\bibitem[62]{EmmerichGL2011}
H. Emmerich, L. Gr{\'a}n{\'a}sy, and H. L{\"o}wen, {\itshape Selected issues of
  phase-field crystal simulations}, Eur. Phys. J. Plus 126 (2011), p. 102.

\bibitem[63]{HansenL2000}
J.P. Hansen and H. L{\"o}wen, {\itshape Effective interactions between electric
  double layers}, Ann. Rev. Phys. Chem. 51 (2000), pp. 209--242.

\bibitem[64]{Einstein1905c}
A. Einstein, {\itshape {\"U}ber die von der molekularkinetischen {T}heorie der
  {W}{\"a}rme geforderte {B}ewegung von in ruhenden {F}l{\"u}ssigkeiten
  suspendierten {T}eilchen}, Ann. Phys. (Leipzig) 322 (1905), pp. 549--560.

\bibitem[65]{FreyK2005}
E. Frey and K. Kroy, {\itshape Brownian motion: a paradigm of soft matter and
  biological physics}, Ann. Phys. (Leipzig) 517 (2005), pp. 20--50.

\bibitem[66]{MorfillI2009}
G.E. Morfill and A.V. Ivlev, {\itshape Complex plasmas: an interdisciplinary
  research field}, Rev. Mod. Phys. 81 (2009), pp. 1353--1404.

\bibitem[67]{MorfillIBL2010}
G.E. Morfill, A.V. Ivlev, P. Brandt, and H. L{\"o}wen, {\itshape
  Interdisciplinary research with complex plasmas}, G.~Bertin, F.~de~Luca,
  G.~Lodato, R.~Pozzoli and M.~Rom{\'e},  eds.,  , Melville, 2010, pp. 67--79.

\bibitem[68]{Pusey1991}
P.N. Pusey, {\itshape Course 10: {C}olloidal suspensions}, J.P. Hansen,
  D.~Levesque and J.~{Zinn-Justin},  eds.,  , Vol. ~2 of {\itshape Proceedings
  of the {L}es {H}ouches {S}ummer {S}chool, {C}ourse {LI}, 3-28 {J}uly 1989}
  North {H}olland, {E}lsevier {S}cience {P}ublishers {B}. {V}., Amsterdam,
  1991, pp. 763--942.

\bibitem[69]{Evans1979}
R. Evans, {\itshape The nature of the liquid-vapour interface and other topics
  in the statistical mechanics of non-uniform, classical fluids}, Adv. Phys. 28
  (1979), pp. 143--200.

\bibitem[70]{vanTeeffelenBVL2009}
S. {van Teeffelen}, R. Backofen, A. Voigt, and H. L{\"o}wen, {\itshape
  Derivation of the phase-field-crystal model for colloidal solidification},
  Phys. Rev. E 79 (2009), p. 051404.

\bibitem[71]{Dhont1996}
J.K.G. Dhont {\itshape An introduction to dynamics of colloids},  1st ed.,
  Elsevier, Amsterdam, 1996.

\bibitem[72]{DoiE2007}
M. Doi and S.F. Edwards {\itshape The theory of polymer dynamics},  1st ed.,
  Oxford University Press, Oxford, 2007.

\bibitem[73]{Risken1996}
H. Risken {\itshape The {F}okker-{P}lanck equation: methods of solution and
  applications},  3rd ed.,   Springer, Berlin, 1996.

\bibitem[74]{EspanolL2009}
P. Espa{\~n}ol and H. L{\"o}wen, {\itshape Derivation of dynamical density
  functional theory using the projection operator technique}, J. Chem. Phys.
  131 (2009), p. 244101.

\bibitem[75]{Archer2006}
A.J. Archer, {\itshape Dynamical density functional theory for dense atomic
  liquids}, J. Phys.: Condens. Matter 18 (2006), pp. 5617--5628.

\bibitem[76]{Archer2009}
---{}---{}---, {\itshape Dynamical density functional theory for molecular and
  colloidal fluids: a microscopic approach to fluid mechanics}, J. Chem. Phys.
  130 (2009), p. 014509.

\bibitem[77]{MarconiM2010}
U.M.B. Marconi and S. Melchionna, {\itshape Dynamic density functional theory
  versus kinetic theory of simple fluids}, J. Phys.: Condens. Matter 22 (2010),
  p. 364110.

\bibitem[78]{MarconiTCM2008}
U.M.B. Marconi, P. Tarazona, F. Cecconi, and S. Melchionna, {\itshape Beyond
  dynamic density functional theory: the role of inertia}, J. Phys.: Condens.
  Matter 20 (2008), p. 494233.

\bibitem[79]{Roth2010}
R. Roth, {\itshape Fundamental measure theory for hard-sphere mixtures: a
  review}, J. Phys.: Condens. Matter 22 (2010), p. 063102.

\bibitem[80]{RothELK2002}
R. Roth, R. Evans, A. Lang, and G. Kahl, {\itshape Fundamental measure theory
  for hard-sphere mixtures revisited: the {W}hite {B}ear version}, J. Phys.:
  Condens. Matter 14 (2002), pp. 12063--12078.

\bibitem[81]{GrafL1999}
H. Graf and H. L{\"o}wen, {\itshape Density functional theory for hard
  spherocylinders: phase transitions in the bulk and in the presence of
  external fields}, J. Phys.: Condens. Matter 11 (1999), pp. 1435--1452.

\bibitem[82]{HansenGoosM2009}
H. {Hansen-Goos} and K. Mecke, {\itshape Fundamental measure theory for
  inhomogeneous fluids of nonspherical hard particles}, Phys. Rev. Lett. 102
  (2009), p. 018302.

\bibitem[83]{Loewen2010b}
H. L{\"o}wen, 2010, {A}pplications of density functional theory in soft
  condensed matter. in {\itshape Understanding soft condensed matter via
  modeling and computation}  World Scientific Publishing, Singapore, pp. 9--45.

\bibitem[84]{PoniewierskiH1988}
A. Poniewierski and R. Ho\l{}yst, {\itshape Density-functional theory for
  nematic and smectic-{A} ordering of hard spherocylinders}, Phys. Rev. Lett.
  61 (1988), pp. 2461--2464.

\bibitem[85]{RamakrishnanY1979}
T.V. Ramakrishnan and M. Yussouff, {\itshape First-principles order-parameter
  theory of freezing}, Phys. Rev. B 19 (1979), pp. 2775--2794.

\bibitem[86]{RosenfeldSLT1997}
Y. Rosenfeld, M. Schmidt, H. L{\"o}wen, and P. Tarazona, {\itshape
  Fundamental-measure free-energy density functional for hard spheres:
  dimensional crossover and freezing}, Phys. Rev. E 55 (1997), pp. 4245--4263.

\bibitem[87]{CurtinA1986}
W.A. Curtin and N.W. Ashcroft, {\itshape Density-functional theory and freezing
  of simple liquids}, Phys. Rev. Lett. 56 (1986), pp. 2775--2778.

\bibitem[88]{OhnesorgeLW1991}
R. Ohnesorge, H. L{\"o}wen, and H. Wagner, {\itshape Density-functional theory
  of surface melting}, Phys. Rev. A 43 (1991), pp. 2870--2878.

\bibitem[89]{OhnesorgeLW1993}
---{}---{}---, {\itshape Density distribution in a hard-sphere crystal},
  Europhys. Lett. 22 (1993), pp. 245--249.

\bibitem[90]{OhnesorgeLW1994}
---{}---{}---, {\itshape Density functional theory of crystal-fluid interfaces
  and surface melting}, Phys. Rev. E 50 (1994), pp. 4801--4809.

\bibitem[91]{Evans1992}
R. Evans, 1992, Density functionals in the theory of nonuniform fluids. in
  {\itshape Fundamentals of inhomogeneous fluids} 1st  ed.,   Marcel Dekker,
  New York, pp. 85--176.

\bibitem[92]{HansenMD2006}
J.P. Hansen and I.R. {McDonald} {\itshape Theory of simple liquids},  3rd ed.,
   Elsevier Academic Press, Amsterdam, 2006.

\bibitem[93]{Kalikmanov2010}
V.I. Kalikmanov {\itshape Statistical physics of fluids: basic concepts and
  applications},  2nd ed.,   Springer, Berlin, 2010.

\bibitem[94]{Loewen2002}
H. L{\"o}wen, {\itshape Density functional theory of inhomogeneous classical
  fluids: recent developments and new perspectives}, J. Phys.: Condens. Matter
  14 (2002), pp. 11897--11905.

\bibitem[95]{MarconiT1999}
U.M.B. Marconi and P. Tarazona, {\itshape Dynamic density functional theory of
  fluids}, J. Chem. Phys. 110 (1999), pp. 8032--8044.

\bibitem[96]{MarconiT2000}
---{}---{}---, {\itshape Dynamic density functional theory of fluids}, J.
  Phys.: Condens. Matter 12 (2000), pp. 413--418.

\bibitem[97]{ArcherE2004}
A.J. Archer and R. Evans, {\itshape Dynamical density functional theory and its
  application to spinodal decomposition}, J. Chem. Phys. 121 (2004), pp.
  4246--4254.

\bibitem[98]{Archer2005}
A.J. Archer, {\itshape Dynamical density functional theory: binary
  phase-separating colloidal fluid in a cavity}, J. Phys.: Condens. Matter 17
  (2005), pp. 1405--1427.

\bibitem[99]{WensinkL2008}
H.H. Wensink and H. L{\"o}wen, {\itshape Aggregation of self-propelled
  colloidal rods near confining walls}, Phys. Rev. E 78 (2008), p. 031409.

\bibitem[100]{RexL2008}
M. Rex and H. L{\"o}wen, {\itshape Dynamical density functional theory with
  hydrodynamic interactions and colloids in unstable traps}, Phys. Rev. Lett.
  101 (2008), p. 148302.

\bibitem[101]{RexL2009}
---{}---{}---, {\itshape Dynamical density functional theory for colloidal
  dispersions including hydrodynamic interactions},  28 (2009), pp. 139--146.

\bibitem[102]{Rauscher2010}
M. Rauscher, {\itshape {DDFT} for {B}rownian particles and hydrodynamics}, J.
  Phys.: Condens. Matter 22 (2010), p. J4109.

\bibitem[103]{RexWL2007}
M. Rex, H.H. Wensink, and H. L{\"o}wen, {\itshape Dynamical density functional
  theory for anisotropic colloidal particles}, Phys. Rev. E 76 (2007), p.
  021403.

\bibitem[104]{WittkowskiL2011}
R. Wittkowski and H. L{\"o}wen, {\itshape Dynamical density functional theory
  for colloidal particles with arbitrary shape}, Mol. Phys. 109 (2011), pp.
  2935--2943.

\bibitem[105]{vanTeeffelenLHL2006}
S. {van Teeffelen}, C.N. Likos, N. Hoffmann, and H. L{\"o}wen, {\itshape
  Density functional theory of freezing for soft interactions in two
  dimensions}, Europhys. Lett. 75 (2006), pp. 583--589.

\bibitem[106]{Barrat1987}
J.L. Barrat, {\itshape Role of triple correlations in the freezing of the
  one-component plasma}, Europhys. Lett. 3 (1987), pp. 523--526.

\bibitem[107]{BarratHP1987}
J.L. Barrat, J.P. Hansen, and G. Pastore, {\itshape Factorization of the
  triplet direct correlation function in dense fluids}, Phys. Rev. Lett. 58
  (1987), pp. 2075--2078.

\bibitem[108]{BarratHP1988}
---{}---{}---, {\itshape On the equilibrium structure of dense fluids}, Mol.
  Phys. 63 (1988), pp. 747--767.

\bibitem[109]{vanRoijBMF1995}
R. {van Roij}, P. Bolhuis, B. Mulder, and D. Frenkel, {\itshape Transverse
  interlayer order in lyotropic smectic liquid crystals}, Phys. Rev. E 52
  (1995), pp. R1277--R1280.

\bibitem[110]{LikosHLL2002}
C.N. Likos, N. Hoffmann, H. L{\"o}wen, and A.A. Louis, {\itshape Exotic fluids
  and crystals of soft polymeric colloids}, J. Phys.: Condens. Matter 14
  (2002), pp. 7681--7698.

\bibitem[111]{LikosLWL2001}
C.N. Likos, A. Lang, M. Watzlawek, and H. L{\"o}wen, {\itshape Criterion for
  determining clustering versus reentrant melting behavior for bounded
  interaction potentials}, Phys. Rev. E 63 (2001), p. 031206.

\bibitem[112]{Louis2001}
A.A. Louis, {\itshape Effective potentials for polymers and colloids: beyond
  the van der {W}aals picture of fluids?}, Phil. Trans. Roy. Soc. Lond. A 359
  (2001), pp. 939--960.

\bibitem[113]{CurtinA1985}
W.A. Curtin and N.W. Ashcroft, {\itshape Weighted-density-functional theory of
  inhomogeneous liquids and the freezing transition}, Phys. Rev. A 32 (1985),
  pp. 2909--2919.

\bibitem[114]{DentonA1989}
A.R. Denton and N.W. Ashcroft, {\itshape Modified weighted-density-functional
  theory of nonuniform classical liquids}, Phys. Rev. A 39 (1989), pp.
  4701--4708.

\bibitem[115]{HansenGoosM2010}
H. {Hansen-Goos} and K. Mecke, {\itshape Tensorial density functional theory
  for non-spherical hard-body fluids}, J. Phys.: Condens. Matter 22 (2010), p.
  364107.

\bibitem[116]{RosenfeldSLT1996}
Y. Rosenfeld, M. Schmidt, H. L{\"o}wen, and P. Tarazona, {\itshape Dimensional
  crossover and the freezing transition in density functional theory}, J.
  Phys.: Condens. Matter 8 (1996), pp. L577--L581.

\bibitem[117]{Tarazona2000}
P. Tarazona, {\itshape Density functional for hard sphere crystals: a
  fundamental measure approach}, Phys. Rev. Lett. 84 (2000), pp. 694--697.

\bibitem[118]{Rosenfeld1994}
Y. Rosenfeld, {\itshape Density functional theory of molecular fluids:
  free-energy model for the inhomogeneous hard-body fluid}, Phys. Rev. E 50
  (1994), pp. R3318--R3321.

\bibitem[119]{Smoluchowski1916}
M. {von Smoluchowski}, {\itshape {\"U}ber {B}rownsche {M}olekularbewegung unter
  {E}inwirkung {\"a}u{\ss}erer {K}r{\"a}fte und deren {Z}usammenhang mit der
  verallgemeinerten {D}iffusionsgleichung}, Ann. Phys. (Leipzig) 353 (1916),
  pp. 1103--1112.

\bibitem[120]{vanTeeffelenLL2008}
S. {van Teeffelen}, C.N. Likos, and H. L\"owen, {\itshape Colloidal crystal
  growth at externally imposed nucleation clusters}, Phys. Rev. Lett. 100
  (2008), p. 108302.

\bibitem[121]{Loewen2001}
H. L{\"o}wen, {\itshape Colloidal soft matter under external control}, J.
  Phys.: Condens. Matter 13 (2001), pp. R415--R432.

\bibitem[122]{LoewenBW1989}
H. L{\"o}wen, T. Beier, and H. Wagner, {\itshape Van der {W}aals theory of
  surface melting}, Europhys. Lett. 9 (1989), pp. 791--796.

\bibitem[123]{LoewenBW1990}
---{}---{}---, {\itshape Multiple order parameter theory of surface melting: a
  van der {W}aals approach}, Z. Phys. B 79 (1990), pp. 109--118.

\bibitem[124]{Lutsko2006}
J.F. Lutsko, {\itshape First principles derivation of {G}inzburg-{L}andau free
  energy models for crystalline systems}, Physica A 366 (2006), pp. 229--242.

\bibitem[125]{deGrootM1984}
S.R.d. Groot and P. Mazur {\itshape Non-equilibrium thermodynamics},  1st ed.,
   Dover Publications, New York, 1984.

\bibitem[126]{MartinPP1972}
P.C. Martin, O. Parodi, and P.S. Pershan, {\itshape Unified hydrodynamic theory
  for crystals, liquid crystals, and normal fluids}, Phys. Rev. A 6 (1972), pp.
  2401--2420.

\bibitem[127]{Reichl1998}
L.E. Reichl {\itshape A modern course in statistical physics},  2nd ed.,
  Wiley, New York, 1998.

\bibitem[128]{PhysRevE.79.051110}
P.K. Galenko, D.A. Danilov, and V.G. Lebedev, {\itshape Phase-field-crystal and
  {S}wift{-}{H}ohenberg equations with fast dynamics}, Phys. Rev. E 79 (2009),
  p. 051110.

\bibitem[129]{PhysRevB.75.054301}
S. Majaniemi and M. Grant, {\itshape Dissipative phenomena and acoustic phonons
  in isothermal crystals: a density-functional theory study}, Phys. Rev. B 75
  (2007), p. 054301.

\bibitem[130]{EurPhysJB.66.329}
S. Majaniemi, M. Nonomura, and M. Grant, {\itshape First-principles and
  phenomenological theories of hydrodynamics of solids}, Eur. Phys. J. B 66
  (2008), pp. 329--335.

\bibitem[131]{PhysRevE.74.011601}
B.P. Athreya, N. Goldenfeld, and J.A. Dantzig, {\itshape Renormalization-group
  theory for the phase-field crystal equation}, Phys. Rev. E 74 (2006), p.
  011601.

\bibitem[132]{PhysRevE.81.061601}
K. Wu, A. Adland, and A. Karma, {\itshape Phase-field-crystal model for fcc
  ordering}, Phys. Rev. E 81 (2010), p. 061601.

\bibitem[133]{JaatinenAEAN2009}
A. Jaatinen, C.V. Achim, K.R. Elder, and T. Ala-Nissila, {\itshape
  Thermodynamics of bcc metals in phase-field-crystal models}, Phys. Rev. E 80
  (2009), p. 031602.

\bibitem[134]{GreenwoodPR2010}
M. Greenwood, N. Provatas, and J. Rottler, {\itshape Free energy functionals
  for efficient phase field crystal modeling of structural phase
  transformations}, Phys. Rev. Lett. 105 (2010), p. 045702.

\bibitem[135]{PhysRevE.83.031601}
M. Greenwood, J. Rottler, and N. Provatas, {\itshape Phase-field-crystal
  methodology for modeling of structural transformations}, Phys. Rev. E 83
  (2011), p. 031601.

\bibitem[136]{JPhysCondMat.22.364102}
K. Wu, M. Plapp, and P.W. Voorhees, {\itshape Controlling crystal symmetries in
  phase-field crystal models}, J. Phys.: Condens. Matter 22 (2010), p. 364102.

\bibitem[137]{JPhysCondMat.21.464110}
R. Prieler, J. Hubert, D. Li, B. Verleye, R. Haberkern, and H. Emmerich,
  {\itshape An anisotropic phase-field crystal model for heterogeneous
  nucleation of ellipsoidal colloids}, J. Phys.: Condens. Matter 21 (2009), p.
  464110.

\bibitem[138]{ChoudharyLEL2011}
M.A. Choudhary, D. Li, H. Emmerich, and H. L{\"o}wen, {\itshape {DDFT}
  calibration and investigation of an anisotropic phase-field crystal model},
  J. Phys.: Condens. Matter 23 (2011), p. 265005.

\bibitem[139]{ChoudharyKE2012}
M.A. Choudhary, J. Kundin, and H. Emmerich, {\itshape Phase-field crystal
  modeling of anisotropic material systems of arbitrary {P}oisson's ratio},
  Philos. Mag. Lett.  (2012) Published electronically:
  http://www.tandfonline.com/doi/abs/10.1080/09500839.2012.686173.

\bibitem[140]{PhysRevLett.108.025502}
G.I. T\'oth, G. Tegze, T. Pusztai, and L. Gr\'an\'asy, {\itshape Heterogeneous
  crystal nucleation: the effect of lattice mismatch}, Phys. Rev. Lett. 108
  (2012), p. 025502.

\bibitem[141]{PhysRevE.60.3614}
A. Karma and W.J. Rappel, {\itshape Phase-field model of dendritic
  sidebranching with thermal noise}, Phys. Rev. E 60 (1999), pp. 3614--3625.

\bibitem[142]{SIADS.7.1049}
D.J.B. Lloyd, B. Sandstede, D. Avitabile, and A.R. Champneys, {\itshape
  Localized hexagon patterns of the planar {S}wift-{H}ohenberg equation}, SIAM
  J. Appl. Dyn. Syst. 7 (2008), pp. 1049--1100.

\bibitem[143]{RobbinsATK2012}
M.J. Robbins, A.J. Archer, U. Thiele, and E. Knobloch, {\itshape Modeling the
  structure of liquids and crystals using one- and two-component modified
  phase-field crystal models}, Phys. Rev. E 85 (2012), p. 061408.

\bibitem[144]{deGennes1971}
P.G. {de Gennes}, {\itshape Short range order effects in the isotropic phase of
  nematics and cholesterics}, Mol. Cryst. Liq. Cryst. 12 (1971), pp. 193--214.

\bibitem[145]{deGennesP1995}
P.G. {de Gennes} and J. Prost {\itshape The physics of liquid crystals},  2nd
  ed.,   Oxford University Press, Oxford, 1995.

\bibitem[146]{BrandK1986}
H.R. Brand and K. Kawasaki, {\itshape Gradient free energy of nematic liquid
  crystals with topological defects}, J. Phys. C 19 (1986), pp. 937--942.

\bibitem[147]{WittkowskiLB2011}
R. Wittkowski, H. L{\"o}wen, and H.R. Brand, {\itshape Polar liquid crystals in
  two spatial dimensions: the bridge from microscopic to macroscopic modeling},
  Phys. Rev. E 83 (2011), p. 061706.

\bibitem[148]{BolhuisF1997}
P. Bolhuis and D. Frenkel, {\itshape Tracing the phase boundaries of hard
  spherocylinders}, J. Chem. Phys. 106 (1997), pp. 666--687.

\bibitem[149]{Loewen1994b}
H. L{\"o}wen, {\itshape Brownian dynamics of hard spherocylinders}, Phys. Rev.
  E 50 (1994), pp. 1232--1242.

\bibitem[150]{FrenkelMMT1984}
D. Frenkel, B.M. Mulder, and J.P. McTague, {\itshape Phase diagram of a system
  of hard ellipsoids}, Phys. Rev. Lett. 52 (1984), pp. 287--290.

\bibitem[151]{KirchhoffLK1996}
{\relax Th}. Kirchhoff, H. L{\"o}wen, and R. Klein, {\itshape Dynamical
  correlations in suspensions of charged rodlike macromolecules}, Phys. Rev. E
  53 (1996), pp. 5011--5022.

\bibitem[152]{Loewen1994d}
H. L{\"o}wen, {\itshape Charged rodlike colloidal suspensions: an ab initio
  approach}, J. Chem. Phys. 100 (1994), pp. 6738--6749.

\bibitem[153]{Loewen1994c}
---{}---{}---, {\itshape Interaction between charged rodlike colloidal
  particles}, Phys. Rev. Lett. 72 (1994), pp. 424--427.

\bibitem[154]{CleaverCAN1996}
D.J. Cleaver, C.M. Care, M.P. Allen, and M.P. Neal, {\itshape Extension and
  generalization of the {G}ay-{B}erne potential}, Phys. Rev. E 54 (1996), pp.
  559--567.

\bibitem[155]{FukunagaTD2004}
H. Fukunaga, J.I. Takimoto, and M. Doi, {\itshape Molecular dynamics simulation
  study on the phase behavior of the {G}ay-{B}erne model with a terminal dipole
  and a flexible tail}, J. Chem. Phys. 120 (2004), pp. 7792--7800.

\bibitem[156]{MuccioliZ2006}
L. Muccioli and C. Zannoni, {\itshape A deformable {G}ay {B}erne model for the
  simulation of liquid crystals and soft materials}, Chem. Phys. Lett. 423
  (2006), pp. 1--6.

\bibitem[157]{Chandrasekhar1992}
S. Chandrasekhar {\itshape Liquid crystals},  2nd ed.,   Cambridge University
  Press, Cambridge, 1992.

\bibitem[158]{YabunakaA2011}
S. Yabunaka and T. Araki, {\itshape Polydomain growth at isotropic-nematic
  transitions in liquid crystalline polymers}, Phys. Rev. E 83 (2011), p.
  061711.

\bibitem[159]{Hirouchi20091192}
T. Hirouchi, T. Takaki, and Y. Tomita, {\itshape Development of numerical
  scheme for phase field crystal deformation simulation}, Comput. Mater. Sci.
  44 (2009), pp. 1192--1197.

\bibitem[160]{SJNAAM.47.2269}
S.M. Wise, C. Wang, and J.S. Lowengrub, {\itshape An energy-stable and
  convergent finite-difference scheme for the phase field crystal equation},
  SIAM J. Numer. Anal. 47 (2009), pp. 2269--2288.

\bibitem[161]{SJNAAM.7.2269}
C. Wang and S.M. Wise, {\itshape An energy stable and convergent
  finite-difference scheme for the modified phase field crystal equation}, SIAM
  J. Numer. Anal. 49 (2011), pp. 945--969.

\bibitem[162]{Hu20095323}
Z. Hu, S.M. Wise, C. Wang, and J.S. Lowengrub, {\itshape Stable and efficient
  finite-difference nonlinear-multigrid schemes for the phase field crystal
  equation}, J. Comp. Phys. 228 (2009), pp. 5323--5339.

\bibitem[163]{Cheng20086241}
M. Cheng and J.A. Warren, {\itshape An efficient algorithm for solving the
  phase field crystal model}, J. Comp. Phys. 227 (2008), pp. 6241--6248.

\bibitem[164]{Tegze20091612}
G. Tegze, G. Bansel, G.I. T\'oth, T. Pusztai, Z. Fan, and L. Gr\'an\'asy,
  {\itshape Advanced operator splitting-based semi-implicit spectral method to
  solve the binary phase-field crystal equations with variable coefficients},
  J. Comp. Phys. 228 (2009), pp. 1612--1623.

\bibitem[165]{PhilosMagLet.87.813}
R. Backofen, A. R\"atz, and A. Voigt, {\itshape Nucleation and growth by a
  phase field crystal ({PFC}) model}, Philos. Mag. Lett. 87 (2007), pp.
  813--820.

\bibitem[166]{unpub.toth.tegze}
G.I. T\'oth and G. Tegze (2012),  to be published.

\bibitem[167]{PhysRevE.79.013601}
Y. Shiwa, {\itshape Comment on ``{R}enormalization-group theory for the
  phase-field crystal equation''}, Phys. Rev. E 79 (2009), p. 013601.

\bibitem[168]{PhysRevB.81.165421}
Z.F. Huang and K.R. Elder, {\itshape Morphological instability, evolution, and
  scaling in strained epitaxial films: an amplitude-equation analysis of the
  phase-field-crystal model}, Phys. Rev. B 81 (2010), p. 165421.

\bibitem[169]{PhilosMag.90.237}
D. Yeon, Z. Huang, K.R. Elder, and K. Thornton, {\itshape Density-amplitude
  formulation of the phase-field crystal model for two-phase coexistence in two
  and three dimensions}, Philos. Mag. 90 (2010), pp. 237--263.

\bibitem[170]{EuroPhysLet.6.567}
G.F. Kendrick, T.J. Sluckin, and M.J. Grimson, {\itshape Macrocrystal phases in
  charged colloidal suspensions}, Europhys. Lett. 6 (1988), p. 567.

\bibitem[171]{JaatinenAN2010}
A. Jaatinen and T. {Ala-Nissila}, {\itshape Extended phase diagram of the
  three-dimensional phase field crystal model}, J. Phys.: Condens. Matter 22
  (2010), p. 205402.

\bibitem[172]{PhysRevE.74.010403}
A. {de Candia}, E. {Del Gado}, A. Fierro, N. Sator, M. Tarzia, and A. Coniglio,
  {\itshape Columnar and lamellar phases in attractive colloidal systems},
  Phys. Rev. E 74 (2006), p. 010403R.

\bibitem[173]{ZPhyB.57.329}
M. Bestehorn and H. Haken, {\itshape Transient patterns of the convection
  instability: a model-calculation}, Z. Phys. B 57 (1984), pp. 329--333.

\bibitem[174]{PhysRevE.54.1560}
C. Kubstrup, H. Herrero, and C. {P\'erez-Garc\'ia}, {\itshape Fronts between
  hexagons and squares in a generalized {S}wift-{H}ohenberg equation}, Phys.
  Rev. E 54 (1996), pp. 1560--1569.

\bibitem[175]{CremerML2012}
P. Cremer, M. Marechal, and H. L{\"o}wen (2012),  to be published.

\bibitem[176]{DemiroersJvKvBI2010}
A.F. Demir{\"o}rs, P.M. Johnson, C.M. {van Kats}, A. {van Blaaderen}, and A.
  Imhof, {\itshape Directed self-assembly of colloidal dumbbells with an
  electric field}, Langmuir 26 (2010), pp. 14466--14471.

\bibitem[177]{GerbodeAOLEC2010}
S.J. Gerbode, U. Agarwal, D.C. Ong, C.M. Liddell, F. Escobedo, and I. Cohen,
  {\itshape Glassy dislocation dynamics in 2{D} colloidal dimer crystals},
  Phys. Rev. Lett. 105 (2010), p. 078301.

\bibitem[178]{MarechalD2008}
M. Marechal and M. Dijkstra, {\itshape Stability of orientationally disordered
  crystal structures of colloidal hard dumbbells}, Phys. Rev. E 77 (2008), p.
  061405.

\bibitem[179]{PhysRevB.71.245414}
T.J. Stasevich, H. Gebremariam, T.L. Einstein, M. Giesen, C. Steimer, and H.
  Ibach, {\itshape Low-temperature orientation dependence of step stiffness on
  (111) surfaces}, Phys. Rev. B 71 (2005), p. 245414.

\bibitem[180]{nanotech.20.295703}
V.V. Hoang, {\itshape Molecular dynamics simulation of liquid and amorphous
  {Fe} nanoparticles}, Nanotechnology 20 (2009), p. 295703.

\bibitem[181]{WANG2006s327}
Y. Wang, X. Wang, and H. Wang, {\itshape Atomic simulation on evolution of
  nano-crystallization in amorphous metals}, Trans. Nonferrous Met. Soc. China
  16 (2006), pp. s327--s331.

\bibitem[182]{PhysRevE.81.011121}
J.A.P. Ramos, E. Granato, S.C. Ying, C.V. Achim, K.R. Elder, and T.
  {Ala-Nissila}, {\itshape Dynamical transitions and sliding friction of the
  phase-field-crystal model with pinning}, Phys. Rev. E 81 (2010), p. 011121.

\bibitem[183]{PhysRevE.74.021104}
C.V. Achim, M. Karttunen, K.R. Elder, E. Granato, T. Ala-Nissila, and S.C.
  Ying, {\itshape Phase diagram and commensurate-incommensurate transitions in
  the phase field crystal model with an external pinning potential}, Phys. Rev.
  E 74 (2006), p. 021104.

\bibitem[184]{PhysRevE.79.011606}
C.V. Achim, J.A.P. Ramos, M. Karttunen, K.R. Elder, E. Granato, T. Ala-Nissila,
  and S.C. Ying, {\itshape Nonlinear driven response of a phase-field crystal
  in a periodic pinning potential}, Phys. Rev. E 79 (2009), p. 011606.

\bibitem[185]{moweitobe}
M. Chen, D. Li, and H. Emmerich (2012),  to be published.

\bibitem[186]{emmlnpm73}
H. Emmerich, in {\itshape The diffuse interface approach in material science:
  thermodynamic concepts and applications of phase-field models}, Lecture Notes
  in Physics Monographs, M 73  Springer, Berlin, 2003, pp. 1--186.

\bibitem[187]{Praetorius2011}
S. Praetorius and A. Voigt, {\itshape A phase field crystal approach for
  particles in a flowing solvent}, Macromol. Theory Simul. 20 (2011), pp.
  541--547.

\bibitem[188]{RauscherDKP2007}
M. Rauscher, A. Dom{\'{\i}}nguez, M. Kr{\"u}ger, and F. Penna, {\itshape A
  dynamic density functional theory for particles in a flowing solvent}, J.
  Chem. Phys. 127 (2007), p. 244906.

\bibitem[189]{PennaDT2003}
F. Penna, J. Dzubiella, and P. Tarazona, {\itshape Dynamic density functional
  study of a driven colloidal particle in polymer solutions}, Phys. Rev. E 68
  (2003), p. 061407.

\bibitem[190]{damingtobe}
D. Li, H.L. Yang, and H. Emmerich, {\itshape Phase field model simulations of
  hydrogel dynamics under chemical stimulation}, Coll. Polym. Sci. 289 (2011),
  pp. 513--521.

\bibitem[191]{Voigtn}
R. Backofen, M. Gr\"af, D. Potts, S. Praetorius, A. Voigt, and T. Witkowski,
  {\itshape A continuous approach to discrete ordering on {S}$^2$}, Multiscale
  Model. Simul. 9 (2011), pp. 314--334.

\bibitem[192]{JPhysCondMat.15.V1}
H. L{\"o}wen, {\itshape Density functional theory: from statics to dynamics},
  J. Phys.: Condens. Matter 15 (2003), pp. V1--V3.

\bibitem[193]{JPhysA.37.9325}
A.J. Archer and M. Rauscher, {\itshape Dynamical density functional theory for
  interacting {B}rownian particles: stochastic or deterministic?}, J. Phys. A
  37 (2004), pp. 9325--9333.

\bibitem[194]{JPhysCondMat.22.360301}
M. Haataja, L. Gr\'an\'asy, and H. L\"owen, {\itshape Classical density
  functional theory methods in soft and hard matter}, J. Phys.: Condens. Matter
  22 (2010), p. 360301.

\bibitem[195]{PhilosMag.91.25}
M. Plapp, {\itshape Remarks on some open problems in phase-field modelling of
  solidification}, Philos. Mag. 91 (2011), pp. 25--44.

\bibitem[196]{JChemPhys.91.7265}
G.H. Fredrickson and K. Binder, {\itshape Kinetics of metastable states in
  block copolymer melts}, J. Chem. Phys. 91 (1989), pp. 7265--7275.

\bibitem[197]{PhysRevE.62.6116}
N.A. Gross, M. Ignatiev, and B. Chakraborty, {\itshape Kinetics of ordering in
  fluctuation-driven first-order transitions: simulation and theory}, Phys.
  Rev. E 62 (2000), pp. 6116--6125.

\bibitem[198]{Asta2009941}
M. Asta, C. Beckermann, A. Karma, W. Kurz, R. Napolitano, M. Plapp, G. Purdy,
  M. Rappaz, and R. Trivedi, {\itshape Solidification microstructures and
  solid-state parallels: recent developments, future directions}, Acta Mater.
  57 (2009), pp. 941--971.

\bibitem[199]{JChemPhys.84.5759}
J.Q. Broughton and G.H. Gilmer, {\itshape Molecular dynamics investigation of
  the crystal-fluid interface. {VI}. {E}xcess surface free energies of
  crystal-liquid systems}, J. Chem. Phys. 84 (1986), pp. 5759--5768.

\bibitem[200]{tagkey2010iii}
K.F. Kelton and A.L. Greer (eds.)  {\itshape Nucleation in condensed matter:
  applications in materials and biology},  1st ed.,   Pergamon Press, Oxford,
  2005.

\bibitem[201]{Greer20002823}
A.L. Greer, A.M. Bunn, A. Tronche, P.V. Evans, and D.J. Bristow, {\itshape
  Modelling of inoculation of metallic melts: application to grain refinement
  of aluminium by {A}l-{T}i-{B}}, Acta Mater. 48 (2000), pp. 2823--2835.

\bibitem[202]{SandomirskiALE2011}
K. Sandomirski, E. Allahyarov, H. L{\"o}wen, and S. Egelhaaf, {\itshape
  Heterogeneous crystallization of hard-sphere colloids near a wall}, Soft
  Matter 7 (2011), pp. 8050--8055.

\bibitem[203]{PhysRevE.78.031109}
J.A.P. Ramos, E. Granato, C.V. Achim, S.C. Ying, K.R. Elder, and T.
  {Ala-Nissila}, {\itshape Thermal fluctuations and phase diagrams of the
  phase-field crystal model with pinning}, Phys. Rev. E 78 (2008), p. 031109.

\bibitem[204]{RottlerGZ2012}
J. Rottler, M. Greenwood, and B. Ziebarth, {\itshape Morphology of monolayer
  films on quasicrystalline surfaces from the phase field crystal model}, J.
  Phys.: Condens. Matter 24 (2012), p. 135002.

\bibitem[205]{PhysRevE.82.021605}
Z.F. Huang, K.R. Elder, and N. Provatas, {\itshape Phase-field-crystal dynamics
  for binary systems: derivation from dynamical density functional theory,
  amplitude equation formalism, and applications to alloy heterostructures},
  Phys. Rev. E 82 (2010), p. 021605.

\bibitem[206]{JPhysCondMat.22.364103}
K.R. Elder and Z.F. Huang, {\itshape A phase field crystal study of epitaxial
  island formation on nanomembranes}, J. Phys.: Condens. Matter 22 (2010), p.
  364103.

\bibitem[207]{JApplyPhys.53.1158}
M.J. Aziz, {\itshape Model for solute redistribution during rapid
  solidification}, J. Appl. Phys. 53 (1982), pp. 1158--1168.

\bibitem[208]{Jackson2004481}
K.A. Jackson, K.M. Beatty, and K.A. Gudgel, {\itshape An analytical model for
  non-equilibrium segregation during crystallization}, J. Cryst. Growth 271
  (2004), pp. 481--494.

\bibitem[209]{Aziz19882335}
M.J. Aziz and T. Kaplan, {\itshape Continuous growth model for interface motion
  during alloy solidification}, Acta Metall. 36 (1988), pp. 2335--2347.

\bibitem[210]{PhysRevE.84.041143}
P.K. Galenko, E.V. Abramova, D. Jou, D.A. Danilov, V.G. Lebedev, and D.M.
  Herlach, {\itshape Solute trapping in rapid solidification of a binary dilute
  system: a phase-field study}, Phys. Rev. E 84 (2011), p. 041143.

\bibitem[211]{PhysRevB.84.054103}
J. Li, S. Sarkar, W.T. Cox, T.J. Lenosky, E. Bitzek, and Y. Wang, {\itshape
  Diffusive molecular dynamics and its application to nanoindentation and
  sintering}, Phys. Rev. B 84 (2011), p. 054103.

\bibitem[212]{personal.karma.2009}
A. Karma, personal communication.

\bibitem[213]{PhysRevLett.41.702}
S. Alexander and J. McTague, {\itshape Should all crystals be bcc? {L}andau
  theory of solidification and crystal nucleation}, Phys. Rev. Lett. 41 (1978),
  pp. 702--705.

\bibitem[214]{PhysRevLett.75.2714}
P.R. {ten Wolde}, M.J. {Ruiz-Montero}, and D. Frenkel, {\itshape Numerical
  evidence for bcc ordering at the surface of a critical fcc nucleus}, Phys.
  Rev. Lett. 75 (1995), pp. 2714--2717.

\bibitem[215]{PhysRevLett.98.235502}
C. Desgranges and J. Delhommelle, {\itshape Controlling polymorphism during the
  crystallization of an atomic fluid}, Phys. Rev. Lett. 98 (2007), p. 235502.

\bibitem[216]{PhysRevLett.106.045701}
G.I. T\'oth, J.R. Morris, and L. Gr\'an\'asy, {\itshape
  {G}inzburg-{L}andau-type multiphase field model for competing fcc and bcc
  nucleation}, Phys. Rev. Lett. 106 (2011), p. 045701.

\bibitem[217]{Galkin06062000}
O. Galkin and P.G. Vekilov, {\itshape Control of protein crystal nucleation
  around the metastable liquid-liquid phase boundary}, Proc. Nat. Acad. Sci. 97
  (2000), pp. 6277--6281.

\bibitem[218]{Wolde26091997}
P.R. {ten Wolde} and D. Frenkel, {\itshape Enhancement of protein crystal
  nucleation by critical density fluctuations}, Science 277 (1997), pp.
  1975--1978.

\bibitem[219]{PhysRevLett.96.046102}
J.F. Lutsko and G. Nicolis, {\itshape Theoretical evidence for a dense fluid
  precursor to crystallization}, Phys. Rev. Lett. 96 (2006), p. 046102.

\bibitem[220]{PhysRevLett.105.025701}
T. Schilling, H.J. Sch\"ope, M. Oettel, G. Opletal, and I. Snook, {\itshape
  Precursor-mediated crystallization process in suspensions of hard spheres},
  Phys. Rev. Lett. 105 (2010), p. 025701.

\bibitem[221]{JACS.129.13520}
T.H. Zhang and X.Y. Liu, {\itshape How does a transient amorphous precursor
  template crystallization}, J. Am. Chem. Soc. 129 (2007), pp. 13520--13526.

\bibitem[222]{PhysRevLett.96.175701}
H.J. Sch\"ope, G. Bryant, and W. {van Megen}, {\itshape Two-step
  crystallization kinetics in colloidal hard-sphere systems}, Phys. Rev. Lett.
  96 (2006), p. 175701.

\bibitem[223]{JPhysCondMat.20.404216}
P. Dillmann, G. Maret, and P. Keim, {\itshape Polycrystalline solidification in
  a quenched 2{D} colloidal system}, J. Phys.: Condens. Matter 20 (2008), p.
  404216.

\bibitem[224]{PhysRevA.28.3599}
W. Schommers, {\itshape Pair potentials in disordered many-particle systems: a
  study for liquid gallium}, Phys. Rev. A 28 (1983), pp. 3599--3605.

\bibitem[225]{JChemPhys.118.2792}
J.P. Doye, D.J. Wales, F.H.M. Zetterling, and M. Dzugutov, {\itshape The
  favored cluster structures of model glass formers}, J. Chem. Phys. 119
  (2003), pp. 2792--2799.

\bibitem[226]{Nature.392.164}
O. Mishima and H.E. Stanley, {\itshape Decompression-induced melting of ice
  {IV} and the liquid-liquid transition in water}, Nature 392 (1998), pp.
  164--168.

\bibitem[227]{shimoji.1977}
M. Shimoji {\itshape Liquid metals: an introduction to the physics and
  chemistry of metals in the liquid state},  1st ed.,   Academic Press, London,
  1977.

\bibitem[228]{PhysRevB.67.035412}
M. Castro, {\itshape Phase-field approach to heterogeneous nucleation}, Phys.
  Rev. B 67 (2003), p. 035412.

\bibitem[229]{Langmuir.13.3871}
W.B. Russel, P.M. Chaikin, J. Zhu, W.V. Meyer, and R. Rogers, {\itshape
  Dendritic growth of hard sphere crystals}, Langmuir 13 (1997), pp.
  3871--3881.

\bibitem[230]{PhysRevLett.58.1444}
A.T. Skjeltorp, {\itshape Visualization and characterization of colloidal
  growth from ramified to faceted structures}, Phys. Rev. Lett. 58 (1987), pp.
  1444--1447.

\bibitem[231]{tms2010seattle}
L. Gr\'an\'asy, G. Tegze, G.I. T\'oth, F. Podmaniczky, and T. Pusztai,
  Phase-field crystal modeling of nucleation, patterning, and early-stage
  growth in colloidal systems in two and three dimensions; Abstract Booklet,
  TMS Annual Meeting, 14-18 February 2010, Seattle, USA.

\bibitem[232]{EuroPhysLett.8.67}
J. Maurer, P. Bouissou, B. Perrin, and P. Tabeling, {\itshape Faceted dendrites
  in the growth of {NH}$_4${Br} crystals}, Europhys. Lett. 8 (1989), p.~67.

\bibitem[233]{PhysRevLett.88.015501}
Z. Cheng, P.M. Chaikin, J. Zhu, W.B. Russel, and W.V. Meyer, {\itshape
  Crystallization kinetics of hard spheres in microgravity in the coexistence
  regime: interactions between growing crystallites}, Phys. Rev. Lett. 88
  (2001), p. 015501.

\bibitem[234]{LeeZRH2002}
I. Lee, H. Zheng, M.F. Rubner, and P.T. Hammond, {\itshape Controlled cluster
  size in patterned particle arrays via directed adsorption on confined
  surfaces}, Adv. Mater. 14 (2002), pp. 572--577.

\bibitem[235]{PhysRevLett.88.248301}
C. Reichhardt and C.J. Olson, {\itshape Novel colloidal crystalline states on
  two-dimensional periodic substrates}, Phys. Rev. Lett. 88 (2002), p. 248301.

\bibitem[236]{Langmuir.22.582}
A. Mathur, A. Brown, and J. Erlebacher, {\itshape Self-ordering of colloidal
  particles in shallow nanoscale surface corrugations}, Langmuir 22 (2006), pp.
  582--589.

\bibitem[237]{ADMA:ADMA200701175}
J. Sun, Y.Y. Li, H. Dong, P. Zhan, C.J. Tang, M.W. Zhu, and Z.L. Wang,
  {\itshape Fabrication and light-transmission properties of monolayer square
  symmetric colloidal crystals via controlled convective self-assembly on 1{D}
  grooves}, Adv. Mater. 20 (2008), pp. 123--128.

\bibitem[238]{ADMA:ADMA200400830}
N.V. Dziomkina, M.A. Hempenius, and G.J. Vancso, {\itshape Symmetry control of
  polymer colloidal monolayers and crystals by electrophoretic deposition on
  patterned surfaces}, Adv. Mater. 17 (2005), pp. 237--240.

\bibitem[239]{PhysWorld.15.31}
K. Dholakia, G. Spalding, and M. MacDonald, {\itshape Optical tweezers: the
  next generation}, Phys. World 15 (2002), pp. 31--35.

\bibitem[240]{C0SM01219J}
M. Hermes, E.C.M. Vermolen, M.E. Leunissen, D.L.J. Vossen, P.D.J. {van
  Oostrum}, M. Dijkstra, and A. {van Blaaderen}, {\itshape Nucleation of
  colloidal crystals on configurable seed structures}, Soft Matter 7 (2011),
  pp. 4623--4628.

\bibitem[241]{unbup.toth.granasy}
G.I. T\'oth and L. Gr\'an\'asy (2012),  to be published.

\bibitem[242]{Edwards1965}
S.F. Edwards, {\itshape Statistical mechanics of polymers with excluded
  volume}, Proc. Phys. Soc. (London) 85 (1965), pp. 613--624.

\bibitem[243]{Helfand1971}
E. Helfand and Y. Tagami, {\itshape Theory of interface between immiscible
  polymers}, J. Polymer Sci. B Polymer Lett. 9 (1971), p. 741.

\bibitem[244]{Mueller2005}
M. M{\"u}ller and F. Schmid, {\itshape Incorporating fluctuations and dynamics
  in self-consistent field theories for polymer blends}, Adv. Polym. Sci. 185
  (2005), pp. 1--85.

\bibitem[245]{Schmid1998}
F. Schmid, {\itshape Self-consistent-field theories for complex fluids}, J.
  Phys.: Condens. Matter 10 (1998), pp. 8105--8138.

\bibitem[246]{Yang2006}
Y.L. Yang, F. Qiu, P. Tang, and H.D. Zhang, {\itshape Applications of
  self-consistent field theory in polymer systems}, Sci. China B Chem. 49
  (2006), pp. 21--43.

\bibitem[247]{Hong1981}
K.M. Hong and J. Noolandi, {\itshape Conformational entropy effects in a
  compressible lattice fluid theory of polymers}, Macromol. 14 (1981), pp.
  1229--1234.

\bibitem[248]{Duchs2003}
D. D{\"u}chs, V. Ganesan, G.H. Fredrickson, and F. Schmid, {\itshape
  Fluctuation effects in ternary ab+a+b polymeric emulsions}, Macromol. 36
  (2003), pp. 9237--9248.

\bibitem[249]{Ganesan2001}
V. Ganesan and G.H. Fredrickson, {\itshape Field-theoretic polymer
  simulations}, Europhys. Lett. 55 (2001), pp. 814--820.

\bibitem[250]{Rasmussen2002}
K. Rasmussen and G. Kalosakas, {\itshape Improved numerical algorithm for
  exploring block copolymer mesophases}, J. Polymer Sci. B Polymer Phys. 40
  (2002), pp. 1777--1783.

\bibitem[251]{Maurits1997}
N.M. Maurits and J.G.E.M. Fraaije, {\itshape Mesoscopic dynamics of copolymer
  melts: from density dynamics to external potential dynamics using nonlocal
  kinetic coupling}, J. Chem. Phys. 107 (1997), pp. 5879--5889.

\bibitem[252]{Rouse1953}
P.E. Rouse, {\itshape A theory of the linear viscoelastic properties of dilute
  solutions of coiling polymers}, J. Chem. Phys. 21 (1953), pp. 1272--1280.

\bibitem[253]{Reister2001}
E. Reister, M. M{\"u}ller, and K. Binder, {\itshape Spinodal decomposition in a
  binary polymer mixture: dynamic self-consistent-field theory and {M}onte
  {C}arlo simulations}, Phys. Rev. E 64 (2001), p. 041804.

\bibitem[254]{Honda2008}
T. Honda and T. Kawakatsu, {\itshape Hydrodynamic effects on the disorder to
  order transitions of diblock copolymer melts}, J. Chem. Phys. 129 (2008), p.
  114904.

\bibitem[255]{Hall2007}
D.M. Hall, T. Lookman, G.H. Fredrickson, and S. Banerjee, {\itshape Numerical
  method for hydrodynamic transport of inhomogeneous polymer melts}, J. Comp.
  Phys. 224 (2007), pp. 681--698.

\bibitem[256]{Hall2006}
---{}---{}---, {\itshape Hydrodynamic self-consistent field theory for
  inhomogeneous polymer melts}, Phys. Rev. Lett. 97 (2006), p. 114501.

\bibitem[257]{Anders2011}
D. Anders and K. Weinberg, {\itshape A variational approach to the
  decomposition of unstable viscous fluids and its consistent numerical
  approximation}, Z. Angew. Math. Mech. 91 (2011), pp. 609--629.

\bibitem[258]{Tian2011}
H. Tian, J. Shao, Y. Ding, X. Li, and X. Li, {\itshape Numerical studies of
  electrically induced pattern formation by coupling liquid dielectrophoresis
  and two-phase flow}, Electrophoresis 32 (2011), pp. 2245--2252.

\bibitem[259]{Boffetta2010}
G. Boffetta, A. Mazzino, S. Musacchio, and L. Vozella, {\itshape
  Rayleigh-{T}aylor instability in a viscoelastic binary fluid}, J. Fluid Mech.
  643 (2010), pp. 127--136.

\bibitem[260]{Xu2006}
H. Xu and C.T. Bellehumeur, {\itshape Modeling the morphology development of
  ethylene copolymers in rotational molding}, J. Appl. Polym. Sci. 102 (2006),
  pp. 5903--5917.

\bibitem[261]{Yue2005}
P.T. Yue, J.J. Feng, C. Liu, and J. Shen, {\itshape Diffuse-interface
  simulations of drop coalescence and retraction in viscoelastic fluids}, J.
  Non-Newtonian Fluid Mech. 129 (2005), pp. 163--176.

\bibitem[262]{Cahn1958}
J.W. Cahn and J.E. Hilliard, {\itshape Free energy of a nonuniform system. {I}.
  {I}nterfacial free energy}, J. Chem. Phys. 28 (1958), pp. 258--267.

\bibitem[263]{Flory1942}
P.J. Flory, {\itshape Thermodynamics of high polymer solutions}, J. Chem. Phys.
  10 (1942), pp. 51--61.

\bibitem[264]{Huggins1942}
M.L. Huggins, {\itshape Theory of solutions of high polymers}, J. Am. Chem.
  Soc. 64 (1942), pp. 1712--1719.

\bibitem[265]{Li2011}
Y.C. Li, R.P. Shi, C.P. Wang, X.J. Liu, and Y. Wang, {\itshape Predicting
  microstructures in polymer blends under two-step quench in two-dimensional
  space}, Phys. Rev. E 83 (2011), p. 041502.

\bibitem[266]{Bird1987}
R.B. Bird, C.F. Curtiss, O. Hassager, and R.C. Armstrong {\itshape Dynamics of
  polymeric liquids},  2nd ed., ,  Vol. 2: {K}inetic theory,   Wiley, New York,
  1987.

\bibitem[267]{Oldroyd1955}
J.G. Oldroyd, {\itshape The effect of interfacial stabilizing films on the
  elastic and viscous properties of emulsions}, Proc. Roy. Soc. Lond. A 232
  (1955), pp. 567--577.

\bibitem[268]{Beaucourt2005}
J. Beaucourt, T. Biben, and C. Verdier, {\itshape Elongation and burst of
  axisymmetric viscoelastic droplets: a numerical study}, Phys. Rev. E 71
  (2005), p. 066309.

\bibitem[269]{Yue2006}
P. Yue, C. Zhou, and J.J. Feng, {\itshape A computational study of the
  coalescence between a drop and an interface in {N}ewtonian and viscoelastic
  fluids}, Phys. Fluids 18 (2006), p. 102102.

\bibitem[270]{Badalassi2003}
V.E. Badalassi, H.D. Ceniceros, and S. Banerjee, {\itshape Computation of
  multiphase systems with phase field models}, J. Comp. Phys. 190 (2003), pp.
  371--397.

\bibitem[271]{MuellerKrumbhaar2001}
H. {M{\"u}ller-Krumbhaar}, H. Emmerich, E. Brener, and M. Hartmann, {\itshape
  Dewetting hydrodynamics in 1+1 dimensions}, Phys. Rev. E 63 (2001), p.
  026304.

\bibitem[272]{Wang2008}
D. Wang, T. Shi, J. Chen, L. An, and Y. Jia, {\itshape Simulated morphological
  landscape of polymer single crystals by phase field model}, J. Chem. Phys.
  129 (2008), p. 194903.

\bibitem[273]{Xu2005}
H.J. Xu, R. Matkar, and T. Kyu, {\itshape Phase-field modeling on morphological
  landscape of isotactic polystyrene single crystals}, Phys. Rev. E 72 (2005),
  p. 011804.

\bibitem[274]{Allen1979}
S.M. Allen and J.W. Cahn, {\itshape Microscopic theory for antiphase boundary
  motion and its application to antiphase domain coarsening}, Acta Metall. 27
  (1979), pp. 1085--1095.

\bibitem[275]{Buddhiranon2011}
S. Buddhiranon, N. Kim, and T. Kyu, {\itshape Morphology development in
  relation to the ternary phase diagram of biodegradable pdlla/pcl/peo blends},
  Macromol. Chem. Phys. 212 (2011), pp. 1379--1391.

\bibitem[276]{Leibler1980}
L. Leibler, {\itshape Theory of microphase separation in block co-polymers},
  Macromol. 13 (1980), pp. 1602--1617.

\bibitem[277]{Ohta1986}
T. Ohta and K. Kawasaki, {\itshape Equilibrium morphology of block copolymer
  melts}, Macromol. 19 (1986), pp. 2621--2632.

\bibitem[278]{Edwards1966}
S.F. Edwards, {\itshape The theory of polymer solutions at intermediate
  concentrations}, Proc. Phys. Soc. (London) 88 (1966), pp. 265--280.

\bibitem[279]{Tsori2001}
Y. Tsori and D. Andelman, {\itshape Diblock copolymer ordering induced by
  patterned surfaces}, Europhys. Lett. 53 (2001), pp. 722--728.

\bibitem[280]{Bosse2010}
A.W. Bosse, {\itshape Phase-field simulation of long-wavelength line edge
  roughness in diblock copolymer resists}, Macromol. Theory Simul. 19 (2010),
  pp. 399--406.

\bibitem[281]{Stark_Lubensky_2}
T.C. Lubensky, D. Pettey, N. Currier, and H. Stark, {\itshape Topological
  defects and interactions in nematic emulsions}, Phys. Rev. E 57 (1998), pp.
  610--625.

\bibitem[282]{Verhoeff2009tactoids}
A.A. Verhoeff, R.H.J. Otten, P. {van der Schoot}, and H.N.W. Lekkerkerker,
  {\itshape Shape and director field deformation of tactoids of plate-like
  colloids in a magnetic field}, J. Phys. Chem. B 113 (2009), pp. 3704--3708.

\bibitem[283]{Dzubiella}
J. Dzubiella, M. Schmidt, and H. L\"owen, {\itshape Topological defects in
  nematic droplets of hard spherocylinders}, Phys. Rev. E 62 (2000), pp.
  5081--5091.

\bibitem[284]{dePablo}
J.A. {Moreno-Razo}, E.J. Sambriski, N.L. Abbott, J.P. {Hernandez-Ortiz}, and
  J.J. {de Pablo}, {\itshape Liquid-crystal-mediated self-assembly at
  nanodroplet interfaces}, Nature 485 (2012), pp. 86--89.

\bibitem[285]{SchillingF2004}
T. Schilling and D. Frenkel, {\itshape Self-poisoning of crystal nuclei in
  hard-rod liquids}, J. Phys.: Condens. Matter 16 (2004), pp. S2029--S2036.

\bibitem[286]{HaertelL2010}
A. H{\"a}rtel and H. L{\"o}wen, {\itshape Fundamental measure density
  functional theory for hard spherocylinders in static and time-dependent
  aligning fields}, J. Phys.: Condens. Matter 22 (2010), p. 104112.

\bibitem[287]{WittkowskiLB2011b}
R. Wittkowski, H. L{\"o}wen, and H.R. Brand, {\itshape Microscopic and
  macroscopic theories for the dynamics of polar liquid crystals}, Phys. Rev. E
  84 (2011), p. 041708.

\bibitem[288]{GrayG1984}
C.G. Gray and K.E. Gubbins {\itshape Theory of molecular fluids: fundamentals},
   1st ed.,   Oxford University Press, Oxford, 1984.

\end{thebibliography}


\section*{List of abbreviations}
\begin{longtable}{@{}p{0.25\linewidth}p{0.75\linewidth}@{}}
2D & two spatial dimensions \\
3D & three spatial dimensions \\ 
1M-PFC model & single-mode PFC model \\
2M-PFC model & two-mode PFC model \\
APFC model & anisotropic PFC model \\
ATG instability & Asaro-Tiller-Grinfeld instability \\
bcc crystal structure & body-centred cubic crystal structure \\
bct crystal structure & body-centred tetragonal crystal structure\\
BVP & boundary value problem \\
CH equation & Cahn-Hilliard equation \\
CMA & constant-mobility approximation \\ 
DDFT & dynamical DFT \\
DEP method & dynamic external potential method \\
DFT & density functional theory \\
DLVO potential & Derjaguin-Landau-Verwey-Overbeek potential \\
DMD simulation & diffusive MD simulation \\
DSCF theory & dynamic SCF theory \\
EAP-MD simulation & embedded-atom-potential MD simulation \\
ELE & Euler-Lagrange equation \\
EOF-PFC model & eighth-order fitting PFC model \\
EOM & equation of motion \\
fcc crystal structure & face-centred cubic crystal structure \\
FD scheme & finite-difference scheme \\
FMT & fundamental-measure theory \\ 
GRP-PFC model & Greenwood-Rottler-Provatas PFC model \\
hcp crystal structure & hexagonal close packed crystal structure \\
HS potential & hard-sphere potential \\
LJ potential & Lennard-Jones potential \\
LOK theory & Leibler-Ohta-Kawasaki theory \\
MC simulation & Monte-Carlo simulation \\
MCT & mode-coupling theory \\
MD simulation & molecular-dynamics simulation \\
Model A & relaxational dynamical equation for a non-conserved order\newline parameter \\
Model B & relaxational dynamical equation for a conserved order\newline parameter \\
Model C & coupled relaxational dynamical equations for a non-conserved and a conserved order parameter \\
Model H & dynamical equation for a conserved order parameter coupled to flow \\
MPFC model & modified PFC model \\ 
NS equation & Navier-Stokes equation \\
Oldroid-B model & constitutive model for the flow of viscoelastic fluids \\
PFC model & phase-field-crystal model \\ 
PFC1 model & dynamical equation for the original PFC model without CMA \\ 
PFC2 model & dynamical equation for the original PFC model with CMA \\ 
PF model & phase-field model \\
RLV & reciprocal lattice vector \\
RPA & random-phase approximation \\
sc crystal structure & simple cubic crystal structure\\
SCF theory & self-consistent field theory \\
SH model & Swift-Hohenberg model \\
VPFC model & vacancy PFC model 
\end{longtable}

\label{lastpage}
\end{document}